# Introduction to High-Temperature Superconductivity for Solid State Chemists

Zenji Hiroi

*Institute for Solid State Physics, University of Tokyo, Kashiwa 5-1-5, Chiba 277-8581, Japan*

ABSTRACT
Superconductivity is one of the most amazing properties that metallic conductors exhibit. Electrical resistance is completely eliminated below the critical temperature ($T_c$), which is the most important parameter in superconductivity. Since the discovery of copper oxide superconductors 40 years ago, many solid state chemists have made significant contributions to the field by discovering new compounds and producing high-quality samples for physical measurements. However, superconductivity research remains challenging for most solid state chemists because it requires knowledge of complicated solid state physics. This manuscript aims to provide a simple, intuitive introduction to superconductivity using only fundamental physics concepts that solid state chemists are familiar with. The author investigates a wide range of materials and classifies them according to the superconductivity mechanisms that may drive them. Specifically focusing on a series of copper oxide superconductors with the highest $T_c$ at ambient conditions, the remarkable material dependence of $T_c$ and the underlying, unconventional superconductivity mechanism that leads to the high $T_c$ are thoroughly examined. Although our understanding of cuprate superconductivity is still fragmented, the author believes that once the branches and leaves are removed, the story will be fairly simple, similar to the phonon-based superconductivity mechanism revealed by the BCS theory. Furthermore, potential strategies for raising the $T_c$ of cuprates and other superconductors are discussed. The author hopes that this article will pique interest in superconductors in young solid state chemists and encourage them to pursue the discovery of still unknown and unexplored room-temperature superconductors in the future.

## 1. Introduction

Materials science entails observing the phenomena displayed by various materials, comprehending the underlying organizing principle, and developing new materials. Many atoms form chemical bonds by exchanging electrons, resulting in molecules or solids such as crystals and glasses at low temperatures. Solid state chemistry studies the chemical properties of solids. The kind of atoms involved, the type of chemical bonding, and the arrangement of atoms within the solid all influence its properties. Furthermore, interesting physical properties emerge when the highest-energy electrons are trapped in an atom or spread out and move around in a solid. As a result, a diverse set of phenomena appear, some with practical applications. Superconductivity is the most remarkable property of metallically conductive materials.

### 1.1. Superconductivity and superconducting materials

Superconductivity is an intriguing phenomenon. A material's electrical resistance changes dramatically at a critical temperature ($T_c$) and disappears completely below it. Since its discovery by Kamerlingh Onnes in 1911, it has piqued scientists' interest for more than a century, resulting in numerous material advances and fundamental investigations into physical properties [1, 2]. Superconductivity is a phase transition in solids caused by an abundance of conducting electrons. In general, many-particle systems frequently change state due to interparticle interactions, such as intermolecular attraction forces that convert a collection of $H_2O$ molecules in liquid water into solid ice, with molecules aligned on a regular basis at temperatures lower than freezing. Similarly, in many-electron systems, electron interactions cause a variety of electronic instabilities that must be removed, yielding specific ground states. Superconductivity is a common option for many metallic conductors.

Superconductivity has been used in a variety of applications because it enables low-loss power transmission and the generation of strong magnetic fields using coils capable of carrying large currents. This is because Joule heat cannot be produced in the zero-resistance state of superconductivity below $T_c$. However, because the $T_c$ values of all materials are currently lower than room temperature, refrigeration using liquid helium or nitrogen, or mechanical cooling equipment, is required. One of the primary goals of modern science is to develop a superconductor with a higher $T_c$, preferably above room temperature.

$T_c$ has gradually increased since the discovery of superconductivity in Hg at 4.2 K in 1911, as shown in Fig. 1, but remained sluggish for a long time after the discovery of $Nb_3Ge$ at 23.2 K in 1973. As a result, the so-called BCS wall for $T_c$ was believed to exist around 30 to 40 K [3]. Bednorz and Müller's discovery of copper oxide superconductors in 1986 opened the door to a new era, with $T_c$ rapidly increasing in a matter of years [4]. The author, who received his doctorate during this "superconductivity fever", was naturally drawn to the field of study and has enjoyed conducting research at the intersection between solid state chemistry and physics. Since then, numerous superconductors have been discovered, including iron-based compounds in 2008 and, more recently, nickel oxides in 2024. Furthermore, research in solid state physics has focused on a variety of distinct superconductors with low $T_c$ but novel superconducting mechanisms.

1.2. Solid state chemistry and physics

Solid state physics investigates superconductivity, while solid state chemistry creates new superconductors. The discovery of new compounds has always helped advance superconductivity research [5]. Because copper oxide superconductors exhibit distinct structure–property relationships based on complex crystal structures and non-stoichiometric chemical compositions, they have attracted the interest of many solid state chemists, resulting in the discovery of numerous superconductors. While physics studies a wide range of substances in order to deduce the underlying principles that govern the phenomena, chemistry is obsessed with each substance as an individual; physicists excel at understanding phenomena, whereas chemists excel at comprehending complex substances. When two are combined, materials science makes significant progress. The history of superconductivity research over the last 40 years is a prime example.

Recent advances in physical property measurement devices, on the other hand, have made it relatively easy for chemists to assess basic superconducting properties using commercially available tools. Quantum Design's Physical Property Measurement System (PPMS) and Magnetic Property Measurement System (MPMS) are widely used as measurement standards. As a result, chemists can confidently characterize superconductivity on their own and feel connected to the phenomenon. Furthermore, based on the physical characterization feedback, they conduct new material searches and produce high-quality samples.

Understanding superconductivity requires advanced physics concepts, but this is not always the case when searching for and studying superconducting materials rather than superconducting phenomena. Nevertheless, being able to visualize basic physics concepts improves one's understanding of superconductivity. Electrons are "protean" quantum mechanics residents who behave as both particles (primarily quasiparticles in solids) and waves: the former is described by position coordinates in real space, while the latter by a wavenumber vector defined by a wavelength and a propagation direction in momentum space. Chemists prefer simple real-space imaging, whereas physicists rely heavily on momentum space; they could be distinct species living in real and reciprocal spaces and speaking different languages.

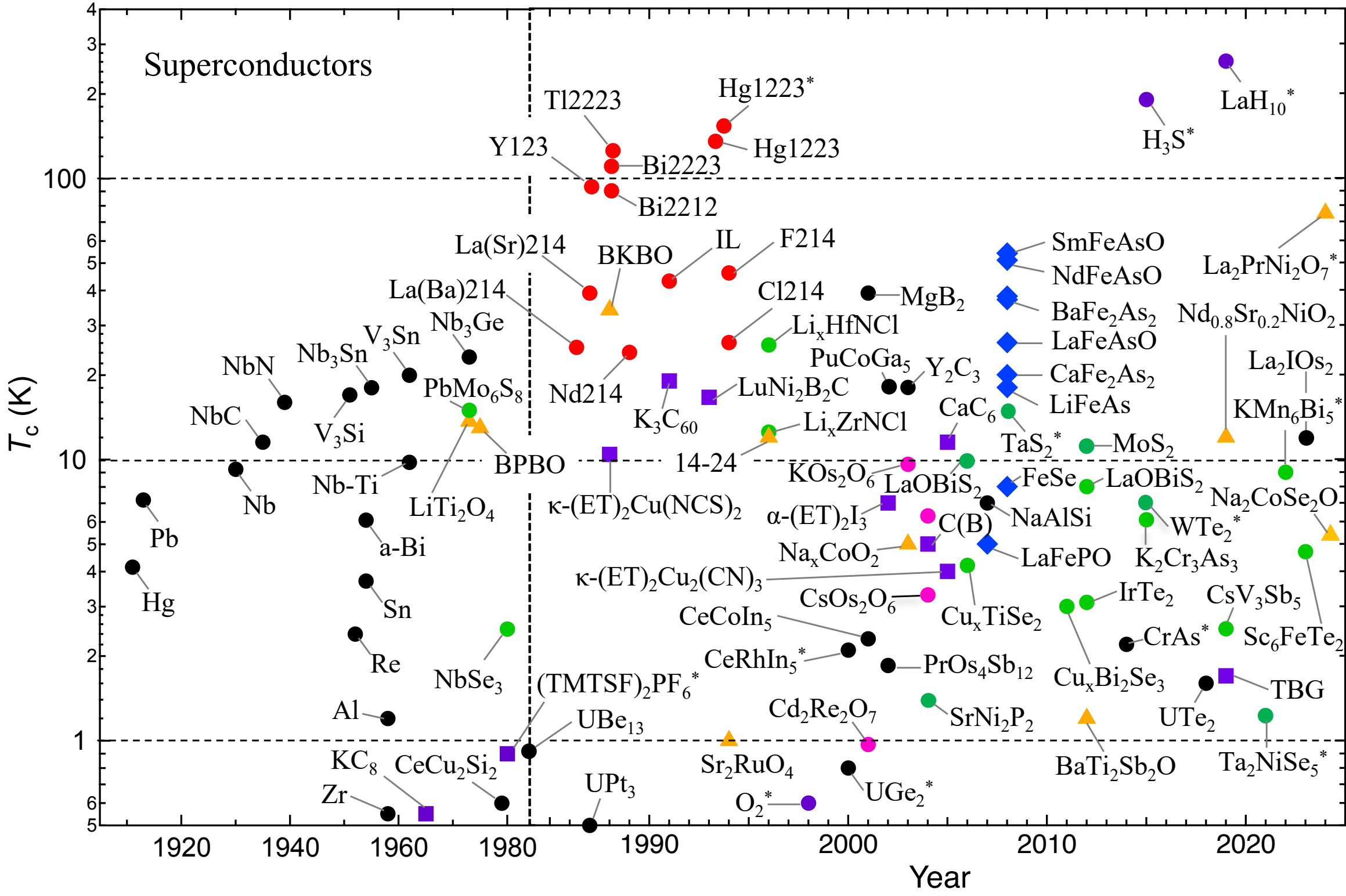

**Fig. 1**. Evolution of the superconducting transition temperature ($T_c$), from Hg at 4.2 K in 1911 to $La_2PrNi_2O_7$ at 75 K in 2024 [17]. The time scale shifts in 1985, before and after the discovery of copper oxide superconductors in 1986 [4]. The figure depicts typical superconductors, and Tables 1 and 2 provide additional information about them. 'a-Bi' denotes amorphous Bi, while 'C(B)' and 'TBG' refer to boron-doped diamond and twisted bilayer graphene, respectively. Compounds marked with an asterisk exhibit superconductivity under high pressure. Several important materials with $T_c$ lower than 0.5 K are missing: the perovskites $SrTiO_{3-\delta}$ ($T_c$ = 0.25 K) in 1964 [18], $(SN)_x$ (0.26 K) in 1975 [8, 19], β-$YbAlB_4$ (0.08 K) in 2008 [5, 20], 12CaO·7$Al_2O_3$ (0.2 K) in 2007 [21], and the semimetal Bi (0.53 mK) in 2017 [22].

Our intuitive understanding of molecular structures and their electronic states is aided by the traditional concept of chemical bonding, which includes two-center–two-electron, multi-center–multi-electron, and $sp^n$ hybridized orbital bonds [6]. However, such simple chemical bonds cannot be strictly realized in actual molecules or solids: the former requires the molecular orbital method to describe the wavefunction spread across the entire molecule, while the latter requires consideration of the wavefunction spread across the entire crystal. Superconductivity is a common property of the latter type of electron, making it difficult to create an appropriate particle image. Nevertheless, it is still possible to convey an image that emphasizes the electron's particle nature, which is frequently helpful for interpreting the phenomenon. When dealing with concepts such as chemical bonding, however, keep in mind that a chemical image that appeals to intuition sacrifices precision. In this paper, we describe how the properties of electrons responsible for superconductivity change from waves to particles as $T_c$ rises. Fortunately, chemists can understand high-temperature superconductivity better than conventional low-temperature superconductivity.

1.3. Manuscript's objective

The Bardeen–Cooper–Schrieffer (BCS) theory, discussed in Section 2.3 [1], provided a clear picture of superconductivity through bold approximations. Superconductivity is based on the formation of electron pairs known as Cooper pairs [7]. To overcome Coulomb repulsion, two negatively charged electrons must be drawn together by attraction, also known as "glue". In the BCS theory, phonons, which are vibrations of nearby atoms or ions in a crystal, serve as a glue to hold the electron pair together. It is thought that glues, rather than phonons, play a major role in unconventional superconductivity, such as copper oxide superconductivity. However, their exotic mechanisms remain unresolved, and no simple representation comparable to the phonon model has been established. The current, seemingly chaotic situation prevents solid state chemists from joining the field. In general, $T_c$ increases with the strength of the glue's attraction; therefore, identifying the type of glue and understanding the mechanism of attraction generation is critical for achieving a high $T_c$. The author believes that even exotic superconductivity should have a straightforward explanation, similar to the phonon mechanism.

The goal of this manuscript is to assist solid state chemists work with superconductors without requiring a thorough understanding of solid state physics. Furthermore, it would be great if this manuscript could help young physics students get started with their superconductivity research. The author will make every effort to provide a straightforward description of superconductivity using as few equations and simplified concepts as possible. There are many textbooks on the physics of superconductivity [1, 8-10]. There are also general reviews of superconducting materials [5, 11-15]. The former, of course, focuses on the physics aspects of superconductivity, while the latter describes the chemical properties of typical superconductors, such as cuprates with chemical decorations and crystalline defects that cause carrier doping. This manuscript will distinguish itself by focusing on key, well-known superconductors and their superconductivity mechanisms, thereby demonstrating the essence and allure of the superconductivity phenomenon. The author will respond to the solid state chemistry question of why copper oxides have such high $T_c$ and why $T_c$ varies so widely among compounds. He will discuss the conditions needed to achieve a high $T_c$ and the path to room-temperature superconductors [16], even if this may be a pipe dream. The author hopes that after reading this paper, solid state chemists who thought that they had a firm grasp on the subject will challenge the search for new superconductors. If they continue their efforts, they will almost certainly discover a higher $T_c$ material or superconductivity with an unknown mechanism. It would be fantastic if the younger generation pursued it further.

The first two chapters of this manuscript provide readers with a basic introduction to superconductivity, while the third chapter describes the general properties of superconducting materials. The fourth chapter summarizes experimental data on copper oxide superconductors and how to derive meaningful conclusions from them, the fifth on other superconductors, and the sixth on prospects for the discovery of novel superconductors. Readers familiar with superconductivity can skip the first half of the article. However, because the author worked hard to write the article in a manner that required little prior knowledge, it is worth reading as an educational resource. Readers who worked on copper oxide superconductors in the middle of the article may feel uneasy or even enraged by the discussion. Nevertheless, the author believes he has contributed to a greater understanding of the subject. Despite the lack of specificity in the conclusion and the absence of notable novel proposals, the author hopes that the format of this manuscript will be useful in future research development. Those with limited time (everyone is now too busy) may be able to grasp what the author is attempting to say in this paper by simply looking at the figures and captions, which are written in detail despite their repetition in the main texts. Furthermore, the author has organized the content into as many detailed chapters and sections as possible, allowing for some skimming. He hopes that this article will contribute to future superconductivity research.

## 2. Superconductivity fundamentals

2.1. "Nonfree" electrons in a crystalline solid

2.1.1. Free electrons

Since electrons are fermions with half-integer spin 1/2 and obey Fermi statistics, they can hold two positions in a single quantum state, including spin degree of freedom. According to the Pauli exclusion principle, electrons in a single atom occupy atomic orbital levels in ascending energy sequence, with the highest "frontier" orbital containing zero, one, or two electrons (Fig. 2). When $N$ atoms join to form a crystal, their frontier orbitals can overlap because they are the most distant from the atomic nucleus. As a result, the atomic orbital level widens, creating an electron cloud (band) with energy spread $W$ that spans the crystal. When the original atomic orbital is a relatively expanded s and p orbitals, the large overlap causes a wide band, while concentrated d and f orbitals produce a narrow band. When the frontier orbital is occupied by zero or two, the band is empty or completely filled, preventing electron movement (band insulator); when it is one, half of the band is occupied, resulting in a half-filled band in which electrons can conduct in a metallic state. As mentioned in the following sections, strongly correlated electron systems with the feature of original orbitals appear only near this half-filled state. In the band picture, partial electron occupancy is also an option, and a single band can accommodate any number of electrons less than 2 $N$. In addition, by adding a small number of electrons to an empty band or removing some from a filled band, electron and hole carriers are generated, respectively, contributing to electrical conduction. This is carrier injection via semiconductor doping, which enables precise control of electrical conductivity in many electronic devices.

Propagating electrons in solids are therefore viewed as spatially extended waves, with their states defined in momentum space by a wavenumber vector $\boldsymbol{k}$ with an inverse wavelength dimension, rather than the real-space position coordinate [9]. A dispersion relation curve in momentum space is formed by aligning points representing electrons of varying momentum and increasing kinetic energy (Fig. 2b); the energy spread represents the bandwidth $W$. The wavelength $\lambda$ of focus can range from infinity to the atomic spacing $a$, but it is difficult to describe a wave with a long wavelength in real space. In momentum space, however, an infinite wavelength can easily be set to the origin at $k = 0$, and the range between $k = -\pi/a$ and $\pi/a$ encompasses all states of interest based on 2 $N$ electrons. Momentum space is preferred due to its convenience.

Many electrons in a solid move through successive levels along the dispersion curve, with kinetic energies ranging from zero to the Fermi energy $E_F$ at Fermi wavenumber $k_F$ at absolute zero temperature. One electron in real space is equivalent to one wave in momentum space. The actual electron number is calculated by integrating electron density in real space; in momentum space, it is obtained by integrating the density of state (DOS) up to $E_F$, which represents the number of waves per energy. The outermost electrons at $E_F$ form a three-dimensional Fermi surface that extends in all directions in momentum space. Fig. 3 depicts a simple Fermi surface made up of s-electrons, yielding a sphere in momentum space. The anisotropic shape of the orbitals, directional chemical bonding, and number of occupied electrons all contribute to the Fermi surface's overall complexity. "Frontier electrons" exist near the Fermi level and govern the electrical conductivity of materials.

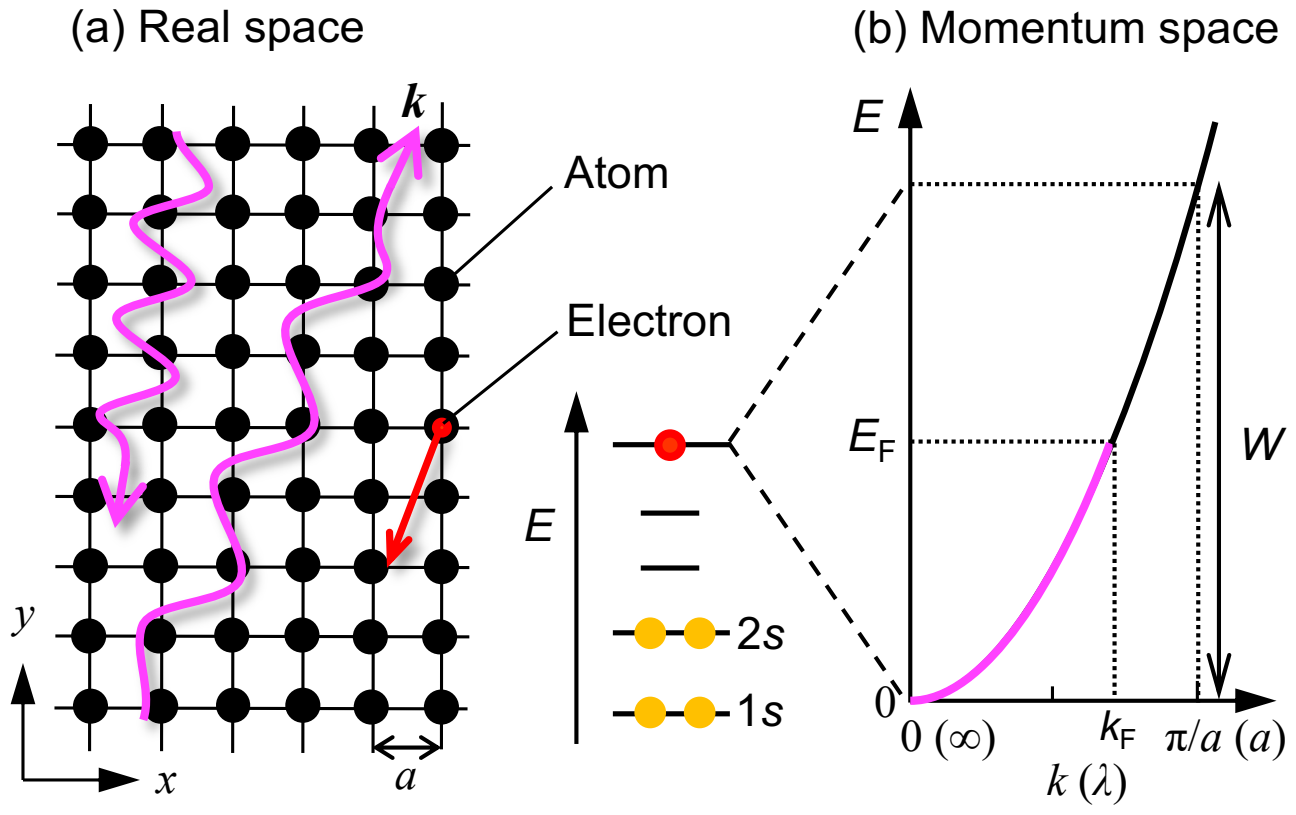

**Fig. 2.** Electrons propagating through a crystal. In real space (a), an electron (red ball) in the highest atomic orbital moves across the crystal, forming a wave (magenta wavy arrow) with wavelength $\lambda$ and wavevector $\boldsymbol{k}$. The kinetic energy increases with the magnitude $k$, resulting in the dispersion curve with an energy spread of $W$, as depicted in momentum space (b). Electrons in a crystal can have electronic states ranging from zero energy at $k = 0$ ($\lambda = \infty$) to Fermi energy $E_F$ at Fermi wavenumber $k_F$ at absolute zero. Electrons at $E_F$ propagate in all directions, forming a three-dimensional Fermi surface, as typically illustrated in Fig. 3 for the isotropic case, which governs crystal transport properties.

2.1.2. Fermi liquid instability

Fermi surfaces of 'Fermi gases' or populations of independent electrons are stable down to absolute zero, but they become unstable when interactions are introduced, resulting in a 'Fermi liquid' composed of nonfree electrons (Fig. 3) [9]. For example, a negatively charged electron and a positively charged cation have a Coulomb attraction. As mentioned in Section 2.3, the electron–phonon (e–ph) interaction can produce an effective attraction between electrons. As a result, at a critical temperature, a phase transition to superconductivity occurs, destroying the original Fermi surface (more detail on this later). Superconductivity is common because it exists unless when the e–ph interactions are extremely small or very large. Electrons become immobile in low-dimensional systems with large e–ph interactions after coupling with a wavelength-appropriate lattice deformation. This results in insulators rather than superconductors or normal metals, and it is known as a charge density wave (CDW) insulator because of the spatial modulation of the charge distribution.

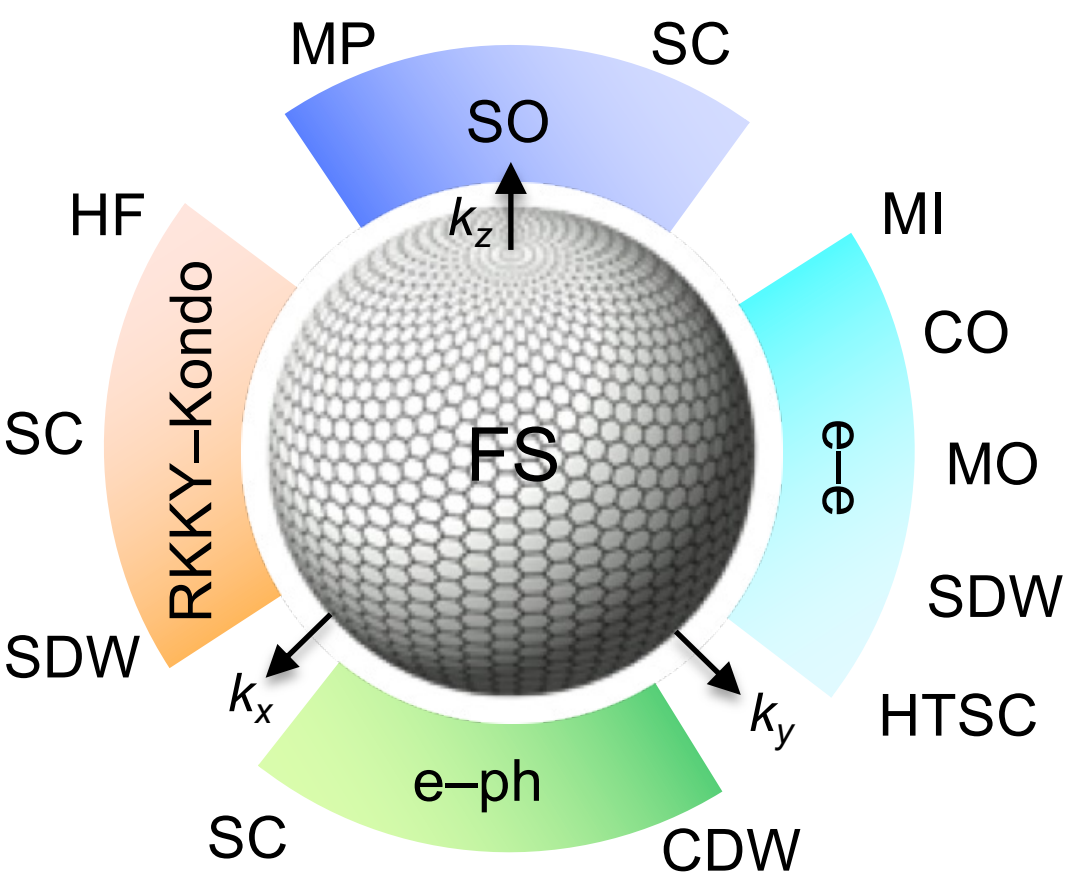

**Fig. 3.** Fermi liquid instability leading to a variety of phases with long-range orders of specific degrees of freedom. The central sphere depicts the Fermi surface (FS) of isotopic free electrons in the momentum space $k_x$–$k_y$–$k_z$. Focus interactions include electron–phonon (e–ph), electron–electron (e–e), Ruderman–Kittel–Kasuya–Yosida versus Kondo (RKKY–Kondo), and spin–orbit (SO). These interactions destabilize the Fermi surface, resulting in a variety of symmetry-breaking phases appearing: superconductor (SC) to charge-density-wave (CDW) insulator with increasing e–ph interactions; high-temperature superconductor (HTSC), spin-density-wave (SDW), magnetic order (MO), charge order (CO), and Mott insulator (MI) with increasing e–e interactions; heavy fermion

(HF), SC, and SDW with increasing RKKY and decreasing Kondo interactions; itinerant or localized multipole (MP) order and SC for SO interaction. Superconductivity is a common order resulting from various types of Fermi liquid instability. It should be noted that while e–ph interactions occur at all times, other interactions can take precedence in defining the system. The resulting types of superconductivity carry the flavors of the original perturbation.

Coulomb repulsion between electrons (e–e), also known as electron correlation, is another major cause of Fermi liquid instability. Strong electron correlations in chemistry frequently indicate strong attractive bonding between electrons, whereas in physics, they prevent electrons from approaching one another. If the electron correlation is too strong, the electrons repel each other and are unable to migrate, resulting in a Mott insulator (MI), as mentioned in the following section. The MI is magnetically active because halted electrons behave as spins, specifically magnetic moments, which exhibit a variety of long-range orders (LRO) as they cool. The other resulting ground states include charge order (CO), which is equivalent to CDW in a weak electron correlation, magnetic order (MO) of localized spins, and spin-density wave (SDW) order with extended spins. When electrons are moderately correlated, they exhibit exotic and occasionally high-$T_c$ superconductivity (HTSC).

Two more Fermi surface destabilizing effects are the Ruderman–Kittel–Kasuya–Yosida (RKKY) interactions and the Kondo effect, which act against each other between localized f-electron spins and conduction electrons [23], as well as the spin–orbit (SO) interaction, which combines an electron's spin and orbital degrees of freedom to transform into multipole degrees of freedom [24, 25]. The former produces unusual electronic states known as 'heavy fermions (HF)' and associated exotic superconductivity, while the latter produces 'spin–orbit-coupled metals' with multipole (MP) order and associated superconductivity. As a result, any electronic instability can cause superconductivity; in other words, superconductivity is a universal ground state in metallic conductors, regardless of the type of perturbation. It is important to note that e–ph interactions occur in all systems, but they can be overridden by others in exotic superconductors.

Solid state physics studies the various LROs associated with each of the four electronic instabilities. This will lead to a better understanding of the materials' inherent interactions and properties. Electronic phase transitions caused by these orders with decreasing temperature are interpreted as spontaneous symmetry breaking: equivalency (symmetry) in terms of a specific degree of freedom exists at high temperatures but vanishes when the temperature falls below a critical point [26]. Fluctuations in prior order, or the spatial and temporal creation and annihilation of ordered domains, can serve as distinct glues for Cooper pairing in superconductivity while retaining the original perturbation characteristics, as mentioned in Chapter 5.

### 2.1.3. Strongly correlated electron systems

Band theory describes how numerous electrons behave in a solid. It is based on a simplification known as the mean-field approximation or one-electron approximation; because the many-body problem cannot be solved exactly, an approximation is always necessary. A single electron of interest is assumed to move in a virtual field that averages the effects of interactions with neighboring atoms and other electrons. Moving electrons in this mean field behave almost like free electrons, but gain a greater effective mass than in the absence of interaction. Weak electron–phonon interactions and electron correlations are treated as such, resulting in nearly free electrons with higher effective masses, also known as mass renormalization.

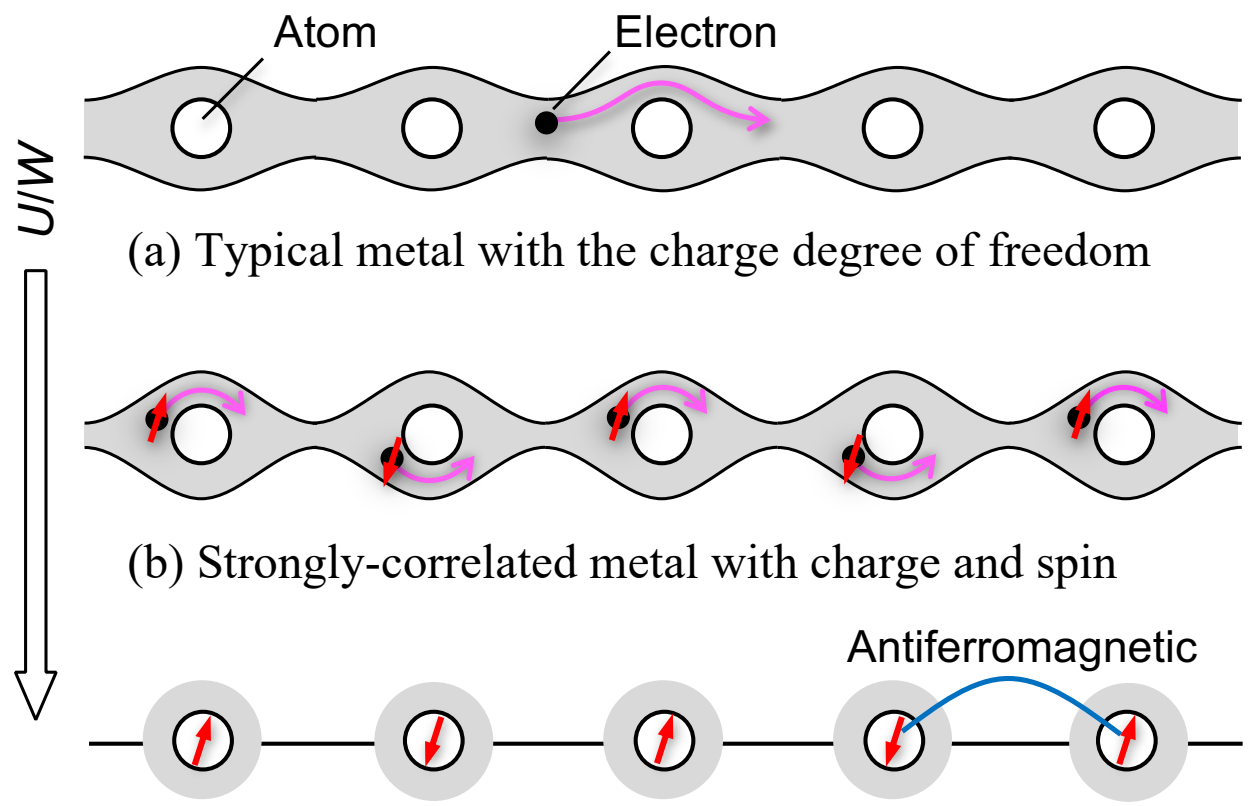

**Fig. 4.** Cartoon illustrating how electron correlations affect the crystal electronic states. As the parameter $U/W$ increases downward, where $U$ and $W$ are the magnitudes of electron correlation and bandwidth, respectively, a typical metal (a) with nearly free electrons carrying the charge degree of freedom in a broad band made from expanded orbitals, such as the s-orbital, transforms into a Mott insulator (c), with each electron localized and acting as a spin in the narrow half-filled band with one electron per atom. A strongly correlated metal (b) lies between them, allowing an electron with both charge and spin degrees of freedom to move against Coulomb repulsion in a narrow band of unexpanded orbitals, such as the d-orbital. Localized spins in Mott insulators and nearly localized spins in strongly correlated metals typically interact antiferromagnetically with neighboring spins, as observed in the $CuO_2$ plane of copper oxide superconductors.

When there are fewer than two electrons per atom, a conventional metal can form in a crystal, as depicted in the cartoon of Fig. 4a. When s-orbital atoms align, they form an s-band with significant orbital overlap and a wide bandwidth. In the mean field, an electron can move nearly freely without being influenced by other specific electrons.

Electron correlation, unlike electron–phonon interaction, is a many-body effect that can cause the band theory's mean-field approximation to fail [27, 28]. When electron correlation ($U$) increases or bandwidth ($W$) decreases, electrons are unable to move freely. When there is one electron per atom in a half-filled band, increasing $U/W$ creates a Mott insulator (Fig. 4c), which confines all electrons to the atom. The electron cannot hop to another site because two electrons on the same atom cause a significant increase in on-site energy by $U$. The electron transitions from a wave with wave number vector $\boldsymbol{k}$ in momentum space to a particle with real space position $\boldsymbol{r}$. Localized electrons in the Mott insulator turn spins, replacing the charge degree of freedom found in conventional metals [28].

Strong antiferromagnetic interactions between localized spins frequently yield antiferromagnetic Mott insulators.

The term 'strongly correlated electron systems' (SCES) refers to the intermediate regime (see Fig. 4b). A line of d-orbitals, for example, forms a narrow band because they overlap less than s-orbitals. Unlike the mean-field approximation, Coulomb repulsion from the surrounding electrons cannot be averaged in a background potential, and each electron feels it explicitly. As a result, electrons in the strongly correlated metal push and shove each other, causing them to move barely. There, electron's wave and particle characters compete, resulting in protean electrons. In addition to the charge degree of freedom, the spin degree of freedom has been partially restored. Thus, strongly correlated electrons have both degrees of freedom, and their entanglement can produce physical properties that are not found in conventional metals or semiconductors. Copper oxide superconductivity is the most prominent of these phenomena.

When $U/W$ increases, strongly correlated electron systems emerge; however, when $W$ decreases, the same situation occurs even in weakly correlated electron systems with low $U$. The obtained insulator may not require a Mott insulator. Electrons interact more strongly with the lattice just before localization, causing lattice distortion and aiding in insulator stabilization [29]. As a result, the insulator is formed through electron–phonon interaction rather than electron–electron interaction. When the two effects are competing, it is critical to interpret the observed phenomena with caution.

2.2. Superconducting properties

2.2.1. Zero electrical resistance

The absence of electrical resistance is the fundamental and most important feature of superconductivity; however, many textbooks emphasize the magnetic properties as the principle of superconductivity phenomena rather than why resistance is eliminated. In general, electrical resistance occurs when electrons accelerated by an applied electric field collide with a specific irregularity. Actual crystals are imperfect, with varying degrees of impurities and crystalline defects. When they act as a scattering source, they disrupt electron mobility, resulting in electrical resistance. As depicted in Fig. 5a, a scattered electron (red ball) travels a shorter distance in the direction of the voltage $V$ than an unscattered electron (orange ball). According to Ohm's law ($V = RI$), the resistance $R$ increases as the current $I$ decreases. Impurity scattering is the primary cause of electrical resistance at $T = 0$, also known as residual resistivity $\rho_0$ (Fig. 6). If a crystal were perfect without defects, its electrical resistance would approach zero as $T$ decreases; however, achieving a true zero-resistance state at any temperature requires a superconducting transition. It is generally challenging to demonstrate that a physical quantity is zero. A superconductor's electrical resistivity is regarded as nearly zero since it has been found to be less than $10^{-24}$ Ω cm, which is $10^{-18}$ times lower than the value of copper at room temperature [30]. At high temperatures, atoms' thermal vibrations cause irregularities in their periodic arrangement. This also causes electron scattering and increases electrical resistance, a process known as electron–phonon scattering. Higher temperatures generate more phonons, which raises resistance via electron–phonon interactions.

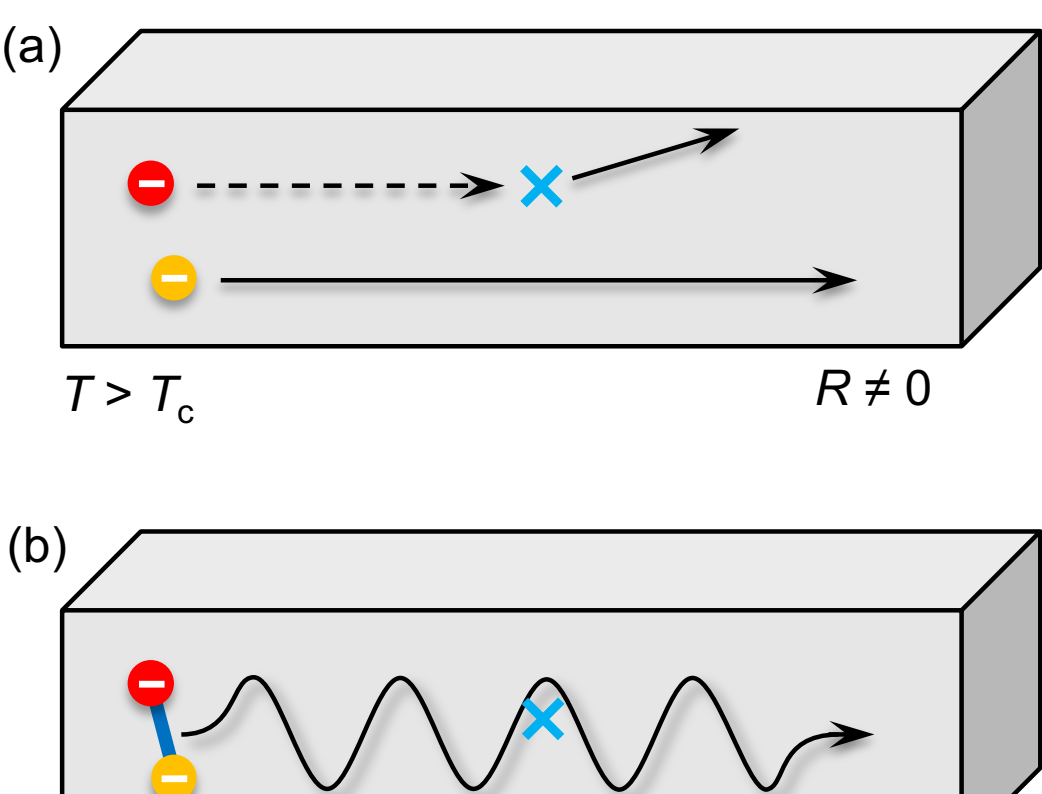

**Fig. 5.** Cartoons depicting how the electrical resistance $R$ occurs in a solid (a) and how zero resistance is attained (b). In the normal conducting state at high temperatures above $T_c$, a single electron is easily scattered by a crystal defect (blue cross), resulting in a finite resistance. At low temperatures below $T_c$, a pair of electrons (Cooper pair) is produced in the superconducting state and is not scattered by a defect unless both electrons are scattered simultaneously or the pair is broken by enhanced scattering by defects, both cases of which are unusual, resulting in zero resistance.

Figure 6 depicts a schematic comparison of the electrical resistivity between a superconductor (such as Pb) and a normal conductor (Au). At higher temperatures, excited phonons scatter more electrons, leading to higher electrical resistivity than $\rho_0$. Pb has a higher electrical resistivity above $T_c$ than Au, and it increases faster with temperature. This demonstrates a stronger interaction between electrons and phonons. The strong electron–phonon interaction in Pb causes superconductivity with a $T_c$ of 7.2 K, whereas Au with fewer interactions remains in a normal conducting state until the experimental limit temperature. As a result, the low-temperature properties of Pb and Au differ significantly, with resistivity clearly distinguishing the two compounds; the balance distinguishes not between gold and lead, but resistivity does!

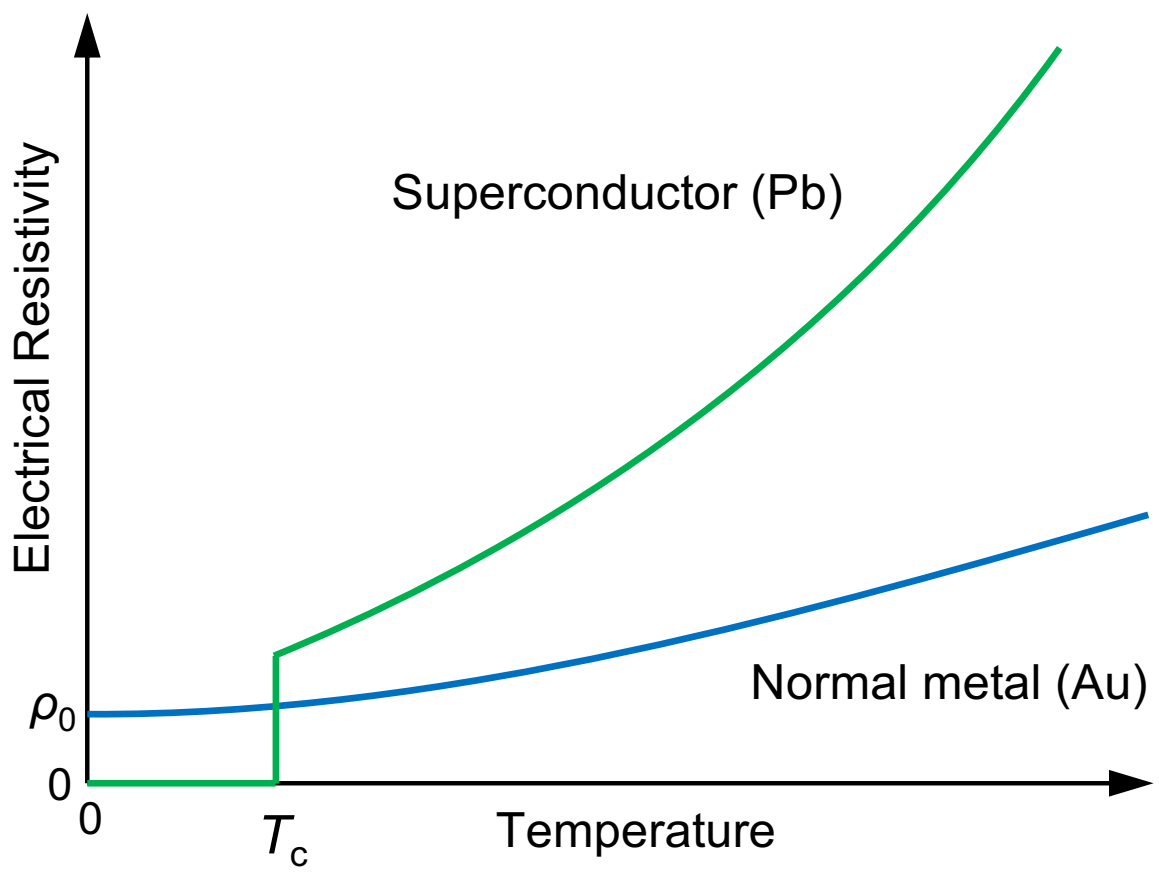

**Fig. 6.** Electrical resistivity of a superconductor, such as Pb ($T_c$

= 7.2 K), versus a normal metal, such as Au. The former has a higher resistivity that rises faster with heating than the latter, indicating larger electron–phonon interactions that result in a superconducting state with zero resistivity below $T_c$. In contrast, the latter's resistivity approaches the residual resistivity $\rho_0$ at $T = 0$, with no drop due to superconducting transition.

What happens to electron motion as Cooper pairs form, causing superconductivity? Electrons move in pairs, as depicted in Fig. 5b, so they are less likely to be scattered by defects or phonons. If one of them is about to be scattered, the other will intervene to prevent the couple from divorcing. Pair breaking necessitates both electrons being scattered at the same time, which is extremely unlikely and nearly impossible. Another way for breaking superconductivity is to have a scatterer powerful enough to break the pair; however, scattering by ordinary defects, with the exception of magnetic impurities, is insufficiently strong; thus, superconductivity can be realized stably in many real-world materials. The simple depiction in Fig. 5b describes that zero resistivity occurs when the pair's wave nature dominates the particle nature of the single electron, thereby preventing scattering.

2.2.2. Additional experimental results and points to note

A superconducting transition can also be investigated using a variety of other physical quantities [1, 8, 30]. Because most superconductors are magnetically hostile and completely exclude low magnetic fields, the Meissner effect causes exceptionally large diamagnetism. In general, superconductivity is lost when the magnetic field exceeds a critical value in type-I superconductors with low $T_c$. In type-II superconductors with relatively high $T_c$, once the magnetic field penetrates as quantized magnetic flux, its number grows with increasing magnetic field until superconductivity fails at the upper critical field $B_{c2}$. Superconductivity is a second-order phase transition in electron systems that causes a $\Delta C$ jump at $T_c$ in heat capacity (the second-order derivative of free energy). The jump magnitude scales with pairing strength: $\Delta C/\gamma k_B T_c = 1.43$ for weak-coupling superconductivity, and higher for strong-coupling superconductivity (see Fig. 9a), where $k_B$ represents the Boltzmann constant and $\gamma$ is the Sommerfeld constant, which scales with DOS at the Fermi level. A large diamagnetic signal from the Meissner effect directly supports superconductivity, while a heat capacity anomaly only indicates the presence of a phase transition.

We will address the $T_c$ criteria below, as some previously reported values appear to be based on unreasonable assumptions. A high-quality superconductor sample with a homogeneous composition, uniform structure, and few defects should exhibit a sharp superconducting transition with temperature. In this case, the resistive offset temperature at which zero resistance is achieved, the magnetic onset temperature at which diamagnetism develops, and the midpoint of the heat capacity jump are all nearly identical, indicating that $T_c$ is distinct. However, because the transition broadens in an inhomogeneous sample, the resistance transition's midpoint temperature, rather than the offset, is empirically equivalent to the other two temperatures, which should be defined as average $T_c$ in typical superconductors with some heterogeneity.

The onset temperature at which electrical resistivity begins to deviate from the high-temperature curve should not be referred to as $T_c$. This is due to a gradual decrease in electrical resistance at temperatures far above $T_c$ in low-dimensional superconductivity, which is accompanied by significant superconducting fluctuations, such as two-dimensional (2D) superconductivity of cuprate superconductors [1]. These onset temperatures represent the temperature at which superconducting "seeds" start to grow locally in a crystal due to strong fluctuations; the actual phase transition takes place at lower temperatures. Furthermore, such onset temperatures are difficult to experimentally determine and are always arbitrary. $HgBa_2Ca_2Cu_3O_{10+\delta}$ (Hg1223), the cuprate superconductor with the highest $T_c$, was reported to exhibit a notable increase in $T_c$ under high pressure, reaching 164 K at 30 GPa [31]. Although this finding has received widespread attention for setting a record, the $T_c$ was clearly overestimated because it was determined by the onset of an electrical resistance transition. Subsequent research on Hg1223 revealed zero resistance at 153 K and 15 GPa [32, 33]. This is the highest $T_c$ value recorded for copper oxide superconductors.

Electrical resistivity measurements using the four-terminal method are usually the most common and simplest way to detect superconductivity; however, they should be interpreted with caution if the sample's homogeneity is in doubt. When a filamentary superconducting path is established between the voltage terminals, resistance is eliminated. In the case of a mostly insulating material, current flows selectively through the low-resistance region, making resistance appear to fall significantly below $T_c$. Even in a homogeneous single crystal, a specific experimental setup of current terminals can cause uneven current distribution, reducing the current flowing between the voltage terminals and resulting in a lower voltage drop and apparent resistance. For this reason, superconductivity was erroneously suggested in early studies on a molecular conductor with high electrical anisotropy [34]. This is primarily because electrical resistivity is not a bulk physical quantity that varies according to sample volume. Any claim of bulk superconductivity must be supported by volume-dependent measurements, such as magnetic susceptibility and heat capacity [35].

Magnetic susceptibility measurements can be used to estimate two types of superconducting volume fractions: the shielding fraction, which is obtained by heating in a weak magnetic field (for example, 10 Oe) after cooling to the lowest temperature in a zero field, and the Meissner fraction, which is then measured upon cooling in a small magnetic field. The shielding fraction will be 100% (or higher due to demagnetization effects) if the entire sample becomes superconducting; however, this is also true when a superconducting current flows only around the sample's periphery in the absence of bulk superconductivity. The Meissner fraction is smaller than the shielding fraction because the pinning effect suppresses the magnetic flux exclusion. Empirically, bulk superconductivity requires shielding and Meissner fractions of greater than 10% and a few percent, respectively. Otherwise, the main phase identified by powder X-ray diffraction measurements is unlikely to be a superconductor. However, careful judgment is required because complex chemical aspects such as compositional deviations and crystal defects frequently cause inhomogeneity in the main phase, with only a small portion exhibiting superconducting properties. Furthermore, demonstrating superconductivity based on a specific electronic state at the surface or interface is challenging, but it may be possible with advanced microscopy and spectroscopy.

Koichi Kitazawa referred to the suspicious high-$T_c$ superconductors as "Unidentified Superconducting Object (USO)" [12, 35], which translates to 'lie' in Japanese. Because the majority of them relied solely on electrical resistivity measurements, the significance of including bulk measurements was highlighted. Unfortunately, there have been additional reports of USOs since then. Even when a sample with a high $T_c$ signal is found, the superconducting volume fraction is frequently low, making it difficult to determine what is superconducting. Many researchers, including the author, have encountered this challenge and struggled with synthetic experiments. Every effort has been made to determine the chemical composition and crystal structure of an unknown superconductor in order to convert it into a "Identified Superconducting Object".

### 2.3. Basic concept of the BCS theory

#### 2.3.1. Cooper pair formation via phonons

Cooper pairs are formed via phonons according to the BCS mechanism, as schematically illustrated in Fig. 7. Consider two electrons embedded in a positively charged ion crystal (which does not exist, but we assume a hypothetical positive charge remains after a neutral atom releases a conduction electron): one with momentum $\boldsymbol{k}$ and up spin ($\boldsymbol{k}\uparrow$) and another with momentum $-\boldsymbol{k}$ in the opposite direction and down spin ($-\boldsymbol{k}\downarrow$). The first $\boldsymbol{k}\uparrow$ electron travels through the lattice, attracting nearby ions via Coulomb interaction (a phonon is created) and being scattered to a different direction ($\boldsymbol{k}'\uparrow$), while maintaining its spin state (Fig. 7b). Because atoms are $10^4$ times heavier than electrons, the electron flees, leaving a positive charge region that persists for a while (the retardation effect). The second electron ($-\boldsymbol{k}\downarrow$) is drawn to the excess positive charge region and scatters into the $-\boldsymbol{k}'\downarrow$ state, following the momentum conservation law (Fig. 7c). The lattice returns to its original state at the end of the process, with the previously created phonon absorbed and vanished. As a result, the two electrons transition from $\boldsymbol{k}\uparrow$ and $-\boldsymbol{k}\downarrow$ to $\boldsymbol{k}'\uparrow$ and $-\boldsymbol{k}'\downarrow$ states. This virtual creation and annihilation of phonons is thought to generate an effective attractive force that allows two electrons to couple. It should be noted that this process does not actually occur; it is simply a convenient way of explaining why a pair is generated.

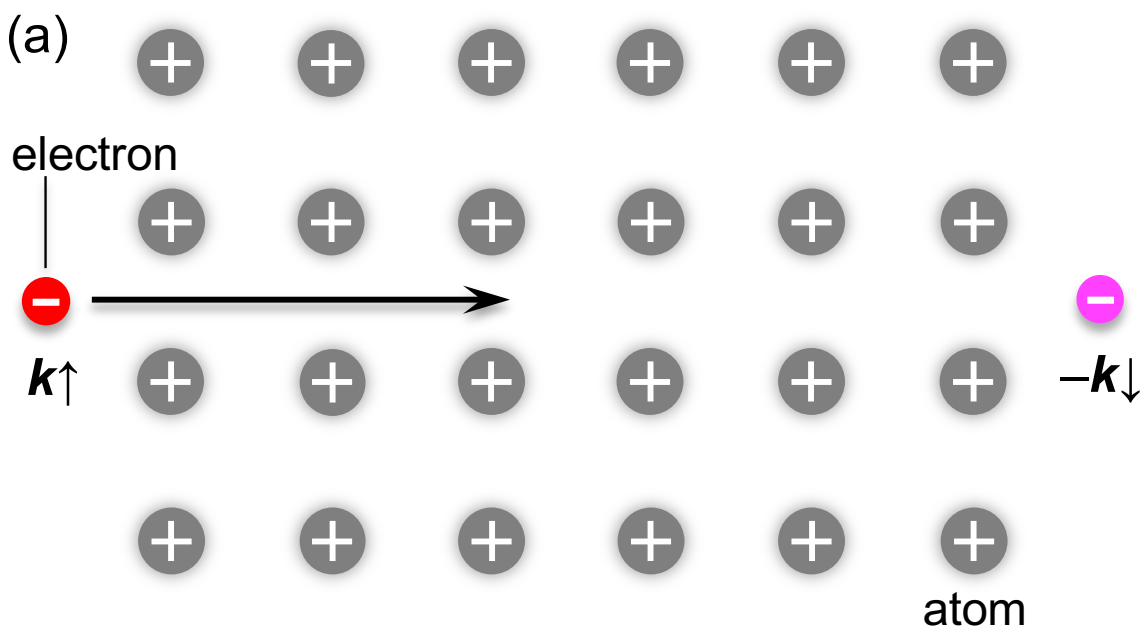

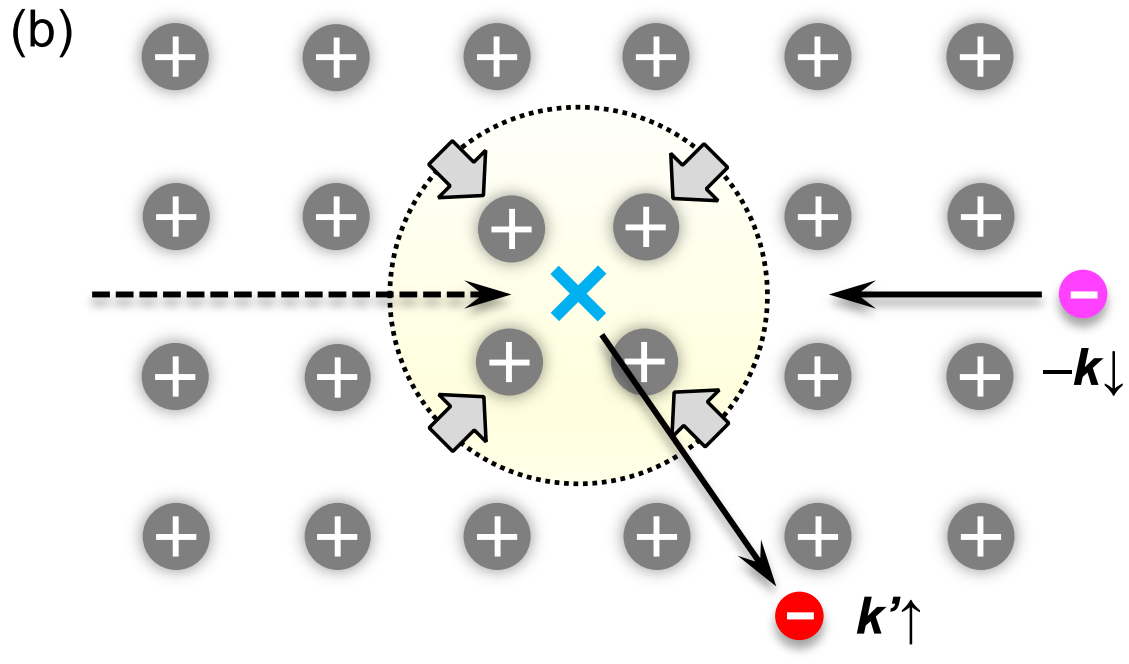

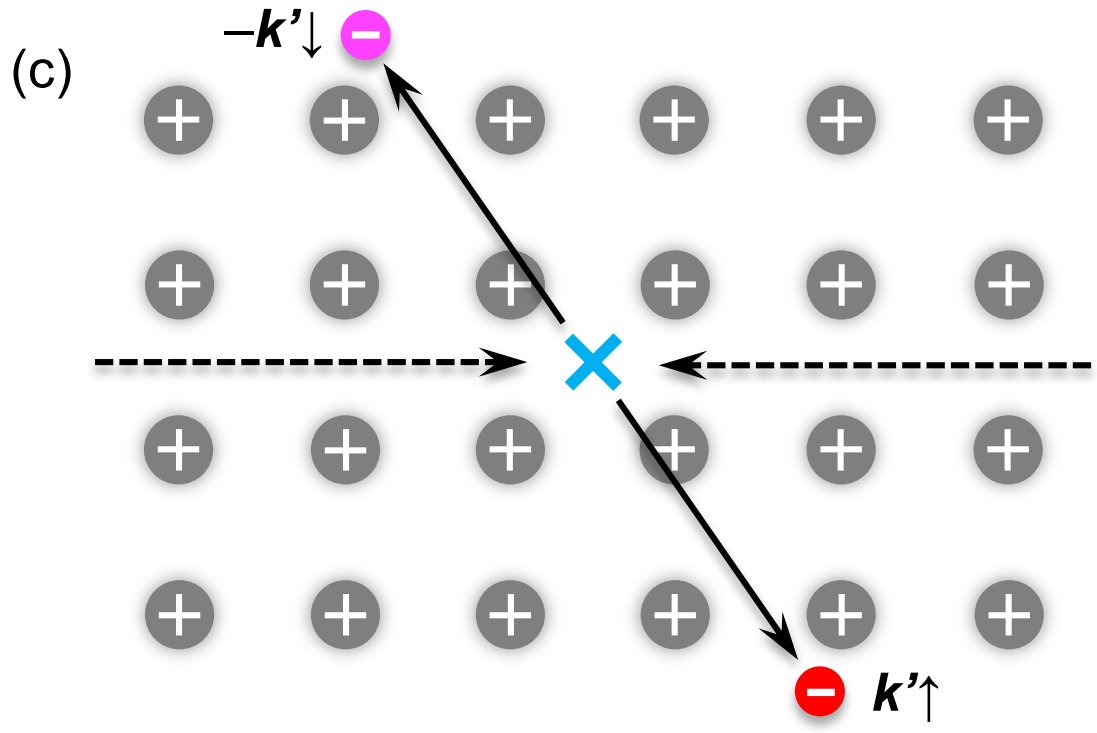

**Fig. 7.** Schematic representation of Cooper pairing via electron–phonon interactions in BCS superconductivity. (a) Consider two electrons, $\boldsymbol{k}\uparrow$ and $-\boldsymbol{k}\downarrow$, with opposite momenta and spins in the initial state. They conduct in a crystal made up of atoms that are presumed to be positively charged after electron donation. (b) When the first electron $\boldsymbol{k}\uparrow$ (red ball) passes through the crystal, it attracts the surrounding atoms via Coulomb interaction and scatters to $\boldsymbol{k}'\uparrow$. As a result, a positively charged region forms (a phonon is created) and persists for some time due to the retardation effect (the atom is much heavier than the electron). The second electron $-\boldsymbol{k}\downarrow$ (magenta ball) is then drawn towards the positively charged region. (c) The second electron scatters to $-\boldsymbol{k}'\downarrow$, restoring the lattice to its initial state (phonon absorbed). This virtual process of phonon creation and annihilation induces effective coupling between two electrons, resulting in a Cooper pair in superconducting state.

#### 2.3.2. BCS image simplified

Figure 8 captures the essence of the BCS mechanism in momentum space; a 2D crystal with an energy-independent DOS is used for simplicity. Consider a spherical Fermi surface with free electrons packed to the Fermi level at absolute zero. As mentioned in Section 2.1.1, this Fermi surface remains stable in the absence of interactions between electrons. Cooper investigated the effect of adding two electrons with $\boldsymbol{k}\uparrow$ and $-\boldsymbol{k}\downarrow$, both slightly more energetic than the $E_F$ (Fig. 8a) [7]. It is assumed that electron–phonon interaction occurs only between two electrons and not with any other electrons. Previous studies have revealed that electron–phonon interactions are most effective in these types of pairs [36]. This model's Schrödinger equation can be solved exactly because it is a simple two-body problem. Equation 1 calculates the pair energy as follows:

$$E = 2E_{\mathrm{F}} - 2\hbar\omega_0 \exp\left[\frac{-2}{N(E_{\mathrm{F}})V}\right], \qquad \text{Eq. 1}$$

where $\hbar\omega_0$ is the phonon energy, $N(E_F)$ is the DOS at the Fermi level, and $V$ is a positive constant that indicates the strength of the pairing attraction.

The second term in Eq. 1 is always negative. Thus, two coupled electrons require less energy than two independent electrons. In addition, the newly added electron pair has a lower energy than $2E_F$, implying that both equivalent electrons are less energetic than the $E_F$. However, this is an odd conclusion because the electron, a Fermi particle, is unable to enter the already clogged Fermi sphere. This means that the two extra electrons defy Fermi statistics and behave like a single Bose particle, resulting in the transition from Fermi to Bose statistics. In superconductivity, the complex Bose particle is referred to as a Cooper pair. This transformation is the foundation of superconductivity.

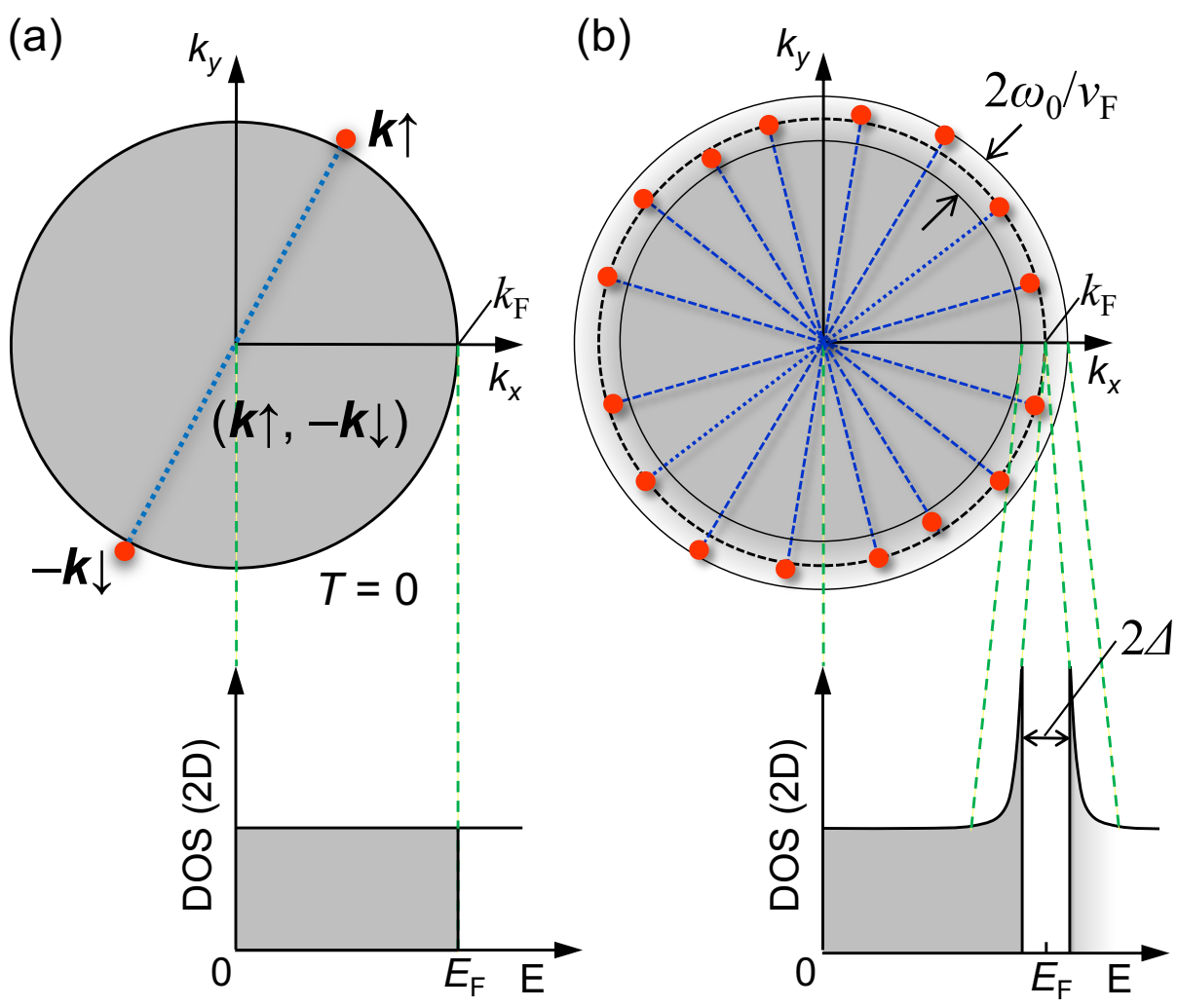

**Fig. 8.** Basic concept of the BCS theory. For the sake of simplicity, we consider a two-dimensional electron system with a circular Fermi surface rather than a sphere and an energy-independent DOS profile (Fig. 12). (a) Electrons with $\boldsymbol{k}$↑ and −$\boldsymbol{k}$↓ that couple via electron–phonon interactions are added just above the Fermi circle of free electrons (Fermi gas) [7]. The pair transforms into a boson with a lower energy than $2E_F$, allowing the two electrons to enter the Fermi circle. (b) In a Fermi liquid, electron–phonon interactions have the potential to destroy the Fermi surface. In the thin surface area between $E_F - \omega_0/v_F$ and $E_F + \omega_0/v_F$, electrons find counterparts and form pairs to lower their energies, similar to (a). The electron pairs are thought to be complex bosons that Bose–Einstein condense into Cooper pairs at temperatures below $T_c$, resulting in superconductivity. The bottom figure depicts the energy spectrum of divorced Cooper pairs, which corresponds to the spectrum obtained via tunneling electron spectroscopy measurements. The DOS profile shows a superconducting gap of $2\Delta$ around $E_F$, which represents the energy required for pair breaking.

The BCS theory predicts that many electrons within the $E_F \pm \hbar\omega_0/v_F$ ($v_F$ is the Fermi velocity: $v_F = \hbar k_F/m$) window pair up (Fig. 8b), similar to the two electrons studied by Cooper (Fig. 8a) [37]. Because the Schrödinger equation for this many-body system cannot be rigorously solved, the theory makes the four bold assumptions below: (1) Electrons attract one another only when they have energy in the $E_F \pm \hbar\omega_0/v_F$ window, (2) the attraction is isotropic without $k$-dependence, (3) the attractive interaction $V$ is small (weak-coupling approximation), and (4) the DOS near $E_F$ can be fixed to the value $N(E_F)$ at $E_F$, ignoring energy dependence. These simplifications reveal that electrons around the Fermi circle transform into Cooper pairs. The bottom picture in Fig. 8b depicts a DOS profile for electrons (quasiparticles) generated following pair breaking, which corresponds to the DOS observed in tunneling electron spectroscopy. Compared to the original DOS in Fig. 8a, the states within $E_F \pm \hbar\omega_0/v_F$ disappear, and sharp peaks that do not obey the Fermi statistics arise above and below the gap. This results in a superconducting gap of $2\Delta$, as described by the following equation:

$$2\Delta = 4\hbar\omega_0 \exp\left[\frac{-1}{N(E_{\mathrm{F}})V}\right]. \qquad \text{Eq. 2}$$

The wavefunctions of charge-carrying pairs (bosons) overlap to become in phase at $T_c$ (Bose–Einstein condensation (BEC), see Section 2.4.2), and then transform into Cooper pairs in the superconducting state, contributing to zero resistance without scattering, as depicted in Fig. 5b. To induce resistivity, break the pair by injecting energy beyond the superconducting gap ($2\Delta$). This is not possible with normal impurity scattering, so the superconducting state remains stable. As a result, when the weak electron–phonon interaction perturbs the Fermi surface, it is destroyed by opening a superconducting gap, as described in Section 2.1.2 of Fermi liquid instability.

BCS theory states that the superconducting gap is proportional to $T_c$ and equals $3.5k_BT_c$. High $T_c$ requires a large gap, which is accomplished by combining a high phonon frequency, a large DOS, and a strong electron–phonon interaction (Eq. 2). It should be noted that the superconducting gap is caused by many-body interactions, as opposed to the single-particle state gap found in conventional semiconductors or insulators. The superconducting gap refers to the amount of energy needed to convert a Cooper pair carrying supercurrent into two independent electrons carrying normal current. Furthermore, it declines with heating and vanishes at $T_c$. In contrast, semiconductors have a constant gap, and their electrical conductivity is determined by thermally excited carriers across the gap.

The BCS theory's success stems from its ability to reduce the complex many-body problem to a simple two-body problem by making bold assumptions while retaining the essence of superconductivity. As a result, we can easily imagine Cooper pairs as the foundation of superconductivity, as well as the reason for the zero resistance depicted in Fig. 5. As mentioned in Section 2.2.1, phonons interfere with electron motion, causing increased electrical resistance at higher temperatures. In the superconducting state, however, they act as a pairing glue, binding electrons together and producing zero resistance; the stronger the interaction, the higher the $T_c$. The key lies in the creation and annihilation of virtual phonons, as depicted in Fig. 7; in normal scattering, a single actual phonon exchanges momentum or energy with one electron, generating electrical resistance.

2.3.3. $T_c$ representation

Substituting $2\Delta = 3.5k_BT_c$ into Eq. 2 yields

$$T_c = 1.13\hbar\omega_0 \exp\left[\frac{-1}{N(E_F)V}\right] \propto \omega_0 \exp[-1/\lambda], \qquad \text{Eq. 3}$$

where $\omega_0$ represents the phonon's characteristic energy $\omega_{ph}$ and $V$ represents the electron–phonon interaction strength. As $\lambda$ [$N(E_F)V$] decreases, the exponential term becomes extremely small, but it approaches unity as $\lambda$ increases. At $\lambda$ infinity, $T_c$ is equal to $1.13\hbar\omega_0$ and never exceeds it. In other words, the exponential term serves as a reduction factor. Typical solids have an $\omega_{ph}$ of 300–400 K. Research on A15 compounds reveals an exponential term of no more than 0.1 [3]. This means that the maximum $T_c$ is limited to 30–40 K, also known as the BCS wall.

Equation 3 is for the weak-coupling case with a small $V$, while McMillan–Allen–Dynes' Equation 4 is for strong-coupling cases [3, 38]. The equation goes as follows:

$$T_c = \frac{\omega_{ln}}{1.2} \exp\left[\frac{-1.04(1+\lambda)}{\lambda-\mu^*(1+0.62\lambda)}\right], \qquad \text{Eq. 4}$$

where $\omega_{ln}$ is the logarithmic mean of phonon energy and $\mu^*$ is the Coulomb interaction constant. The former represents the average phonon energy, which is introduced to account for the fact that the phonons that support the force of attraction have varying energies. The latter is difficult to estimate through experimentation or theory, but it is usually assumed to be around 0.1. Using $\mu^*$ to estimate $T_c$ can produce significant errors. Recent advancements allow for more reliable evaluation of $\mu^*$ [39]. Equation 4 accurately describes the $T_c$ of many conventional superconductors [40].

As mentioned in Section 2.1, Fermi liquid instabilities are diverse, and not only electron–phonon interactions, but also Coulomb interactions, which appear to be repulsive between two bodies, can provide the source of attraction via multibody effects, resulting in superconductivity. The attraction in cuprate superconductors, for example, is caused by an antiferromagnetic spin background produced by electron correlations. In general, $T_c$ can be calculated by generalizing Eq. 3, where $\omega_0$ is the energy scale of the interaction or elementary excitation driving the attraction and $\lambda$ is the strength of electron interaction via excitation. A new glue demonstrates unconventional superconductivity that extends beyond the phonon mechanism and could have a higher $T_c$.

2.4. Key concepts in superconductivity

Before delving into superconductivity phenomena, it is important to grasp a few physical concepts. To provide solid state chemists with an intuitive explanation, we will only cover Cooper pair size and shape in this article.

2.4.1. Cooper pair size: Superconducting coherence length

The Ginsburg–Landau coherence length ($\xi$) is the characteristic length for type-II superconductivity. In a moderate magnetic field, superconductivity partially brakes and returns to its normal state in the $\xi$ region near the vortex core. At the upper critical field ($B_{c2}$), superconductivity completely vanishes when the magnetic field is strong enough to allow vortices to overlap over a distance approaching $\xi$. Consequently, $\xi$ is calculated using $B_{c2}$, with a higher value indicating a smaller $\xi$.

Alternatively, the symbol $\xi$ represents the Cooper pair's size. A high $\xi$ value indicates a large Cooper pair with weak attraction, while a low value indicates a small Cooper pair with strong attraction. The former corresponds to weak-coupling superconductivity in BCS theory, whereas the latter is compatible with strong-coupling superconductivity. Weak-coupling phonon-based superconductors typically have spatially isotropic $\xi$ values of 10 to 100 nm. Strong-coupling cuprate superconductivity has anisotropic $\xi$, ranging from 2–3 nm in the $CuO_2$ plane to less than 0.1 nm perpendicular to it. This results in 2D superconductivity, with the Cooper pair confined to the plane [1].

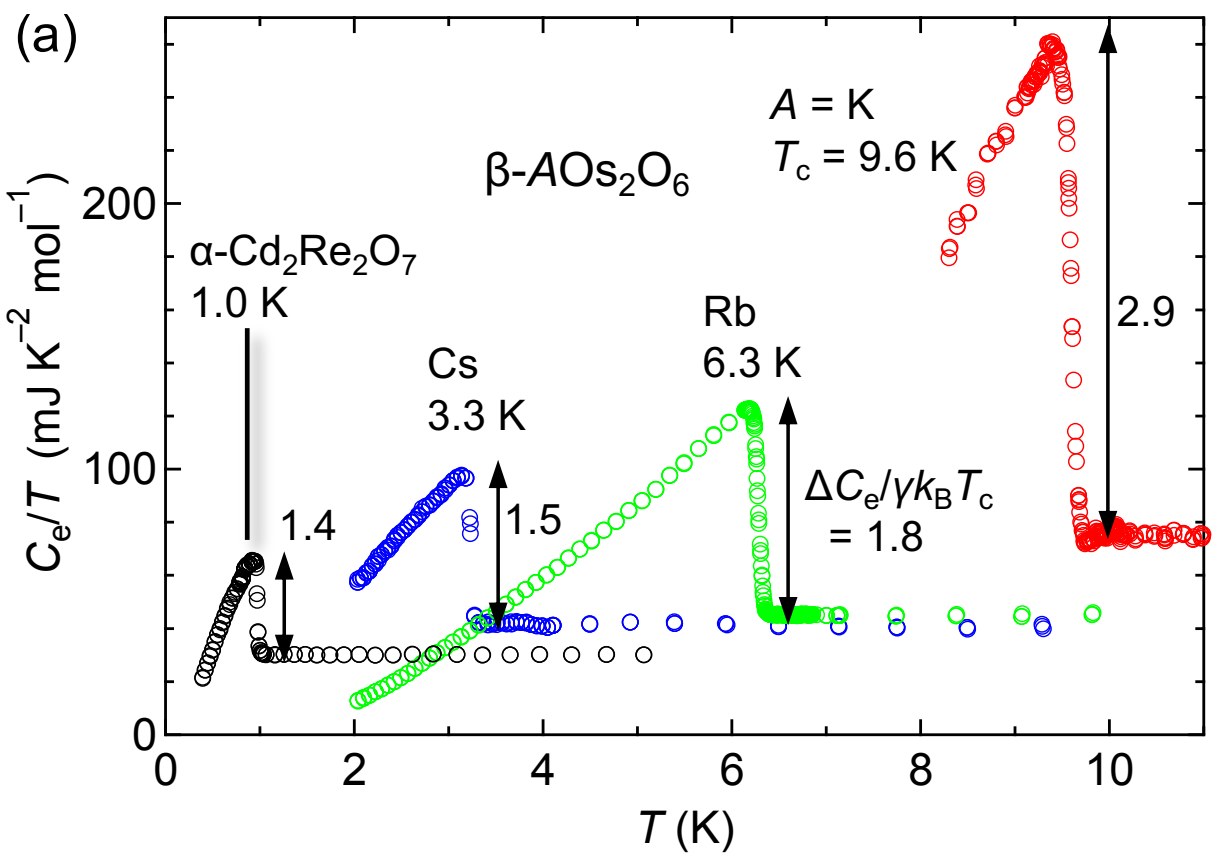

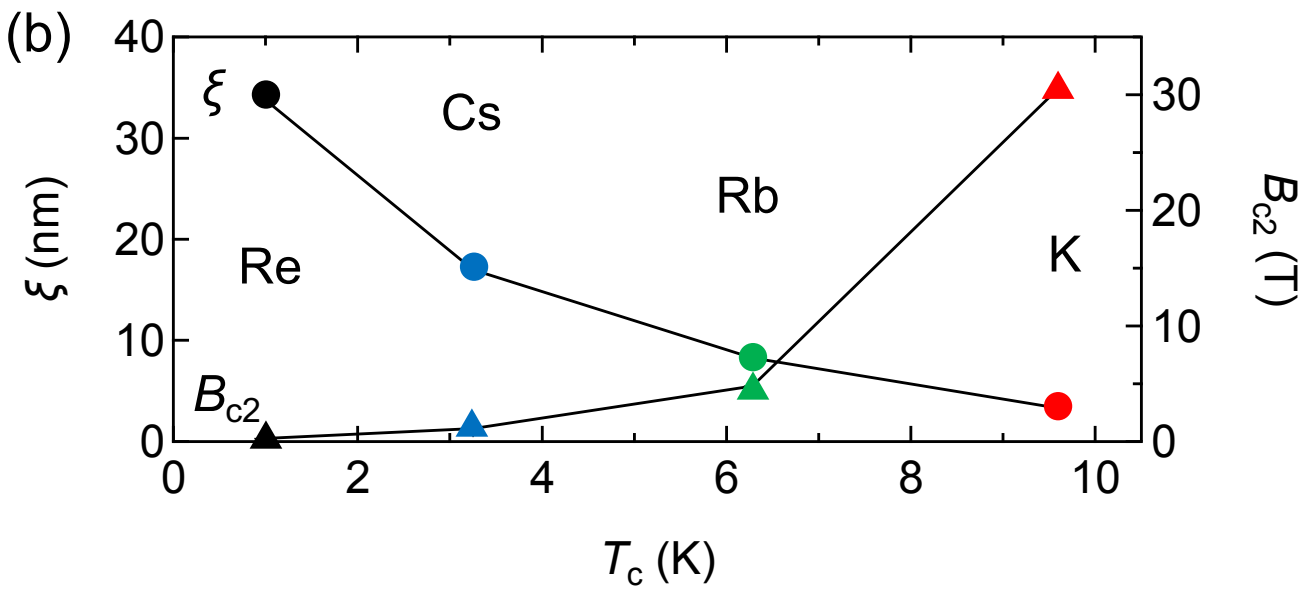

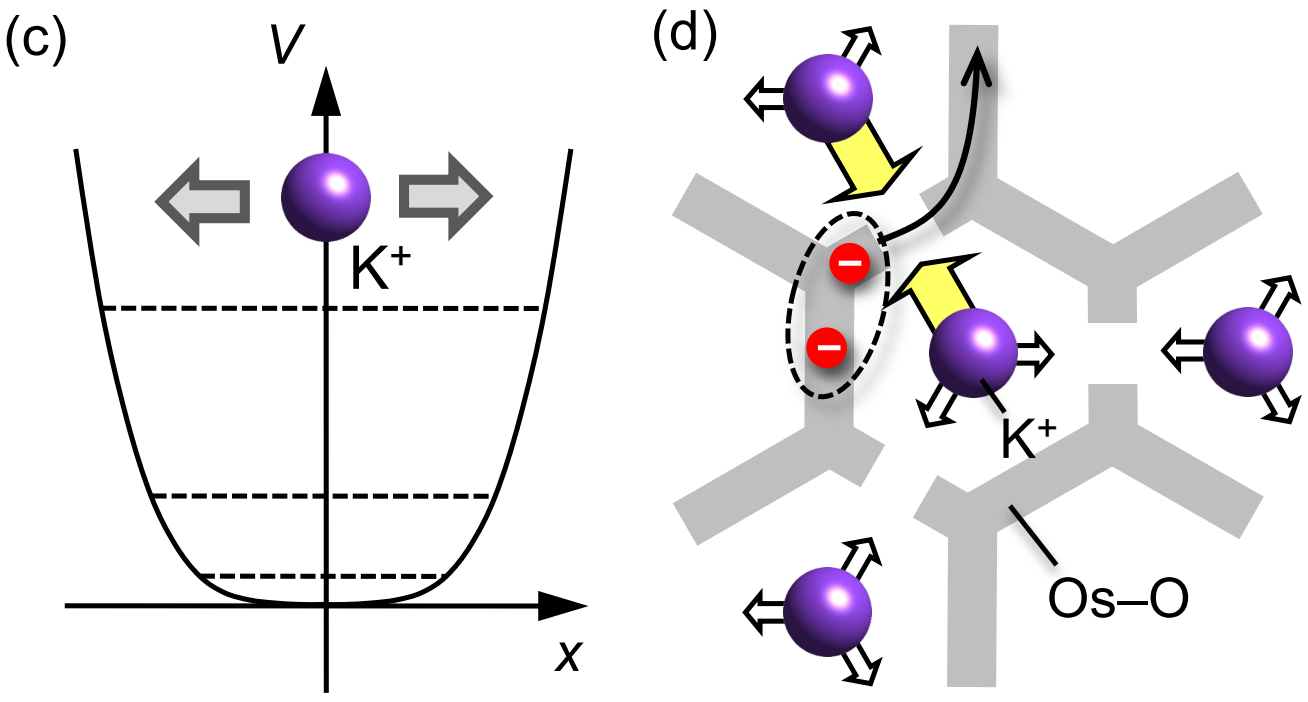

**Fig. 9.** β-pyrochlore osmium oxide superconductor $AOs_2O_6$ [42]. (a) The electronic heat capacity $C_e$ divided by $T$ reveals superconducting transitions at $T_c$ = 3.3, 6.3, and 9.6 K for A = Cs, Rb, and K, respectively. In comparison, $Cd_2Re_2O_7$, an α-pyrochlore oxide superconductor, has a $T_c$ of 1.0 K [43]. The two-directional arrow represents the magnitude of the jump at $T_c$ ($\Delta C/\gamma k_BT_c$), which indicates the evolution from weak-coupling for $Cd_2Re_2O_7$ and $CsOs_2O_6$ to strong-coupling

superconductivity for $KOs_2O_6$. The data for $KOs_2O_6$ below 8.2 K have been deleted to conceal a sharp, intense peak caused by the structural transition at 7.65 K, which appears to be linked to an unknown change in K-ion rattling. (b) The superconducting coherence length ($\xi$) and upper critical field ($B_{c2}$) are plotted against $T_c$. (c) The small K ion in the Os–O cage has a distinct anharmonic potential that differs from the nearly harmonic potential of the large Cs and Rb ions, as well as the majority of other atoms in crystals, including $Cd_2Re_2O_7$. (d) An illustration of strong-coupling superconductivity in the K compound, in which a Cooper pair is generated by a strong electron–phonon interaction caused by large excursions of the rattling K ions in real space. In the strong-coupling limit, a real-space pairing image may be appropriate.

β-pyrochlore osmium oxide (β-$AOs_2O_6$) exhibits distinct superconducting properties that vary with A-ion size [41, 42]. Figure 9a displays the heat capacity data. The $T_c$ reaches 3.3 K (A = Cs), 6.3 K (Rb), and 9.6 K (K). The jump magnitude at $T_c$ ($\Delta C/\gamma k_B T_c$) for Cs is 1.49, comparable to the BCS theoretical value of 1.43, 1.83 for Rb, and 2.87 for K, which is nearly twice as large. As a result, superconducting properties change from weak to strong coupling in this order. The $B_{c2}$ data indicate $\xi$ values of 17 nm (Cs), 8.3 nm (Rb), and 3.3 nm (K) (Fig. 9b). The attraction thus grows in this order, resulting in smaller Cooper pairs and higher $T_c$ values.

The high attraction and $T_c$ in $KOs_2O_6$ are attributed to anharmonic local vibrations of K ions, known as "rattling" with large excursions in the cage-like structure formed by Os–O bonds [42]. While the large Cs ions fit perfectly into the cage and may contribute to superconductivity as regular phonons, the small K ions vibrate like a baby's "rattler" with unusually large amplitudes in the cage's flat bottom potential (Fig. 9c). Harmonic oscillators can approximate normal spring-linked atoms such as Cs ions, but K ions are essentially anharmonic oscillators. As depicted in Fig. 9d, the first conduction electron strongly attracts the surrounding K ions, while the second electron is drawn to the positive charge of the gathered K ions. Consequently, real-space Cooper pairings may offer a better approximation than $k$-space pairings. $T_c$ rises as the superconducting gap widens due to the strong attraction of electrons and rattling K ions. In Eq. 3, $\omega_0$ decreases from Cs to K, while $\lambda$ increases more effectively, raising $T_c$. A more elaborate mechanism involves conventional nonlocal electron–phonon coupling, which is thought to be enhanced by K rattling [42].

2.4.2. BCS–BEC crossover

Weak-coupling superconductivity in $CsOs_2O_6$ is represented by BCS-type Cooper pairing. In other words, two electrons weakly couple, forming spatially extended Cooper pairs. In contrast, $KOs_2O_6$ exhibits strong-coupling superconductivity. At the extremes of strong coupling, electron pairs are thought to form in real space, making chemists easier to understand. Remember that pair size and attraction strength are inextricably linked. Nevertheless, selecting the appropriate initial image will provide a shortcut to a conclusion.

Superconductivity in the strong coupling limit is commonly regarded as BEC superconductivity. The BCS–BEC crossover, which originated in the study of 'cold atom gas' [44, 45], is the concept that bridges the gap between weak-coupling and strong-coupling superconductivity. The universe contains two types of particles: fermions (e.g., electrons and $^3$He atoms) and bosons (e.g., $^4$He atoms or phonons), which follow Fermi and Bose statistics, respectively. A fermion has an odd number of spin angular momentum, while a boson has an even number; one quantum state can be occupied by either a single fermion or multiple bosons.

The thermal de Broglie wavelength, $\lambda_{th} = h/(2\pi m k_B T)^{1/2}$, determines a boson's size, which increases with decreasing temperature. When bosons grow at low temperatures and overlap at $T_B$, their wavefunctions align in phase, creating a spread quantum state described by a single wavefunction, known as the BEC. The neutral atoms of $^{87}$Rb and $^{23}$Na are Bose particles, exhibiting BEC at extremely low temperatures below a few μK [46]. In contrast, $^{40}$K is a Fermi particle that exhibits BEC at low temperatures following pairing at high temperatures [47]. Bosonic $^4$He atoms have a BEC below 2.17 K and become superfluid, while fermionic $^3$He atoms pair up to form complex bosons and eventually exhibit superfluidity at around 1 mK; however, neither becomes superconducting due to a lack of charge. Cooper electron pairs are clearly linked to the $^{40}$K and $^3$He cases.

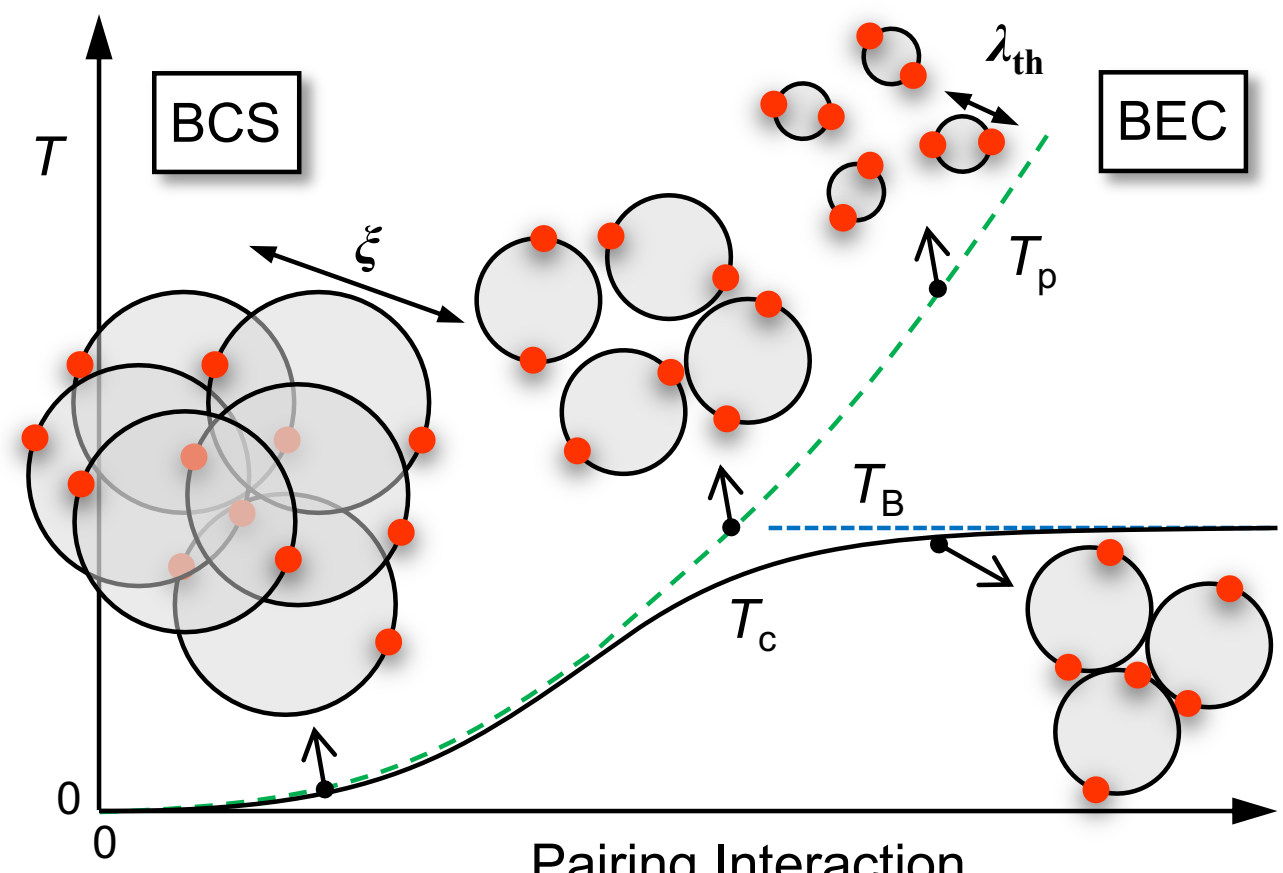

**Fig. 10.** Evolution of electron pairings from the BCS to the BEC regime, with increasing pairing interaction, based on research on the cold fermionic atom gas system [44, 45, 48]. Two red balls on a shaded circle represent an electron pair, with their orientations mimicking wavefunction phase. Increased pairing interaction reduces circle diameter ($\xi$), leading to a smaller pair. In the BCS regime with weak interaction on the left, large bosonic pairs form at $T_p$ and almost immediately transform into Cooper pairs with phase coherence when they overlap at $T_c \sim T_p$ in the superconducting state. In the BEC regime with strong interaction on the right, small bosonic pairs form at high temperatures below $T_p$ and grow upon cooling as the thermal de Broglie wavelength ($\lambda_{th}$) increases. Superconductivity occurs when wavefunctions overlap and share a phase at $T_c \sim T_B$. This evolution, known as the BCS–BEC crossover, is applicable to any system, regardless of pairing interactions, and serves as a general guide to high-temperature superconductivity.

Figure 10 depicts a general phase diagram based on electron pair attraction strength, which can be experimentally controlled in a cold atom gas system [44, 45, 48]. $T_p$, the temperature at which a pair forms, rises monotonically as the attraction force increases. When the attraction is strong enough on the right side of Fig. 10, small pairs form at high temperatures, but their wavefunctions are too small to overlap. As the temperature falls, the bosonic pairs' wavefunctions broaden due to the $\lambda_{th}$ extension and overlap at $T_B$, resulting in BEC. BEC in 2D happens when a boson's area ($2\pi\lambda_{th}^2$) equals the inverse of the pair density ($2/n_s$), where $n_s$ represents the electron density per unit plane. Consequently, $k_B T_B = (h^2/m)(n_s/2)$, and $T_c$ is directly proportional to the boson density. It is important to emphasize that $T_B$ is solely determined by $n_s$, rather than attraction strength or pair size. In contrast, when the attraction is weak on the left side of Fig. 10, large pairs form at low temperatures at $T_p$, and their already extended wavefunctions overlap immediately, resulting in BEC. At $T_c \sim T_p$, pair formation and superconductivity (BEC) are almost simultaneous, supporting the BCS superconductivity image. In contrast to the BEC regime, $T_c$ rises with the strength of the attraction in the BCS regime, as shown by Eq. 3.

The BCS–BEC crossover model describes the relationship between BCS and BEC superconductivity in terms of pairing force strength [45, 48]. Their continuous connection demonstrates that the BCS pairing in momentum space and the BEC pairing in real space are essentially identical. The question then becomes which view of the actual Cooper pair is more appropriate; actual superconducting states fall somewhere in the middle. As we will see in Chapter 4, the BCS–BEC crossover in cuprate superconductivity occurs with hole doping and is a key concept for understanding the mechanism. Furthermore, Figure 10 illustrates that high $T_c$ is expected in the BEC superconducting regime and that increasing the pairing attraction is critical for high $T_c$ (Section 6.3). Therefore, the BCS–BEC crossover concept is also important when developing a strategy to raise $T_c$.

### 2.4.3. Cooper pair shape: Superconducting gap symmetry

In addition to their size, Cooper pairs' shape is critical for understanding the underlying superconducting mechanism. It is determined by the symmetry of the superconducting gap. In phonon superconductivity, the superconducting gap is evenly distributed across the Fermi surface in momentum space (Fig. 11a). Because the gap size is uniform in all directions, the attractive force for Cooper pairing must be isotropic. As depicted at the bottom of Fig. 11a, place one electron of a Cooper pair at the origin of real space and consider how the other electron is distributed. The isotropic gap implies that the bond length remains constant in all directions, yielding a probability distribution identical to the hydrogen 1s atomic orbital. Consequently, this type is known as s-wave superconductivity. The interaction between electrons must always be attractive because the pair wavefunction grows in amplitude as two electrons approach each other within this s-wave symmetry. When an s-wave superconducting gap is observed, the superconductivity mechanism relies on an attractive interaction, regardless of the conditions, such as electron–phonon interactions.

When the Coulomb interaction between negatively charged electrons is strong, s-waves are obviously undesirable. Cooper pairs with p-wave or d-wave symmetry are accordingly adapted. Because the pair wavefunctions have an origin node, the counter electron cannot approach the central electron, resulting in reduced Coulomb energy loss. Even a repulsive two-body interaction, as discussed in Chapter 4 for copper oxide superconductivity, can cause an attractive force that produces Cooper pairs via many-body effects. Cuprates have a zero $dx^2–y^2$ superconducting gap along the <110> direction, including the origin node (Fig. 11b). This indicates that Cooper pairing cannot occur along the <110> direction. As a result, the Cooper pair looks like a $dx^2–y^2$ type clover [49]. $CeCoIn_5$ exhibits a similar d-wave superconductivity, while $UPt_3$ has p-wave Cooper pairs [50]. An unusual f-wave pairing has been predicted but not yet confirmed for $Na_{0.35}CoO_2 \cdot 1.3H_2O$ [51].

The shape of the superconducting gap and Cooper pairs reveals the origin of the pairing forces that drives the superconductivity mechanism [52]. Recent advances in angle-resolved photoemission spectroscopy (ARPES) [53-55] and scanning tunneling spectroscopy (STS) experiments [56] have enabled the experimental determination of superconducting gap symmetry.

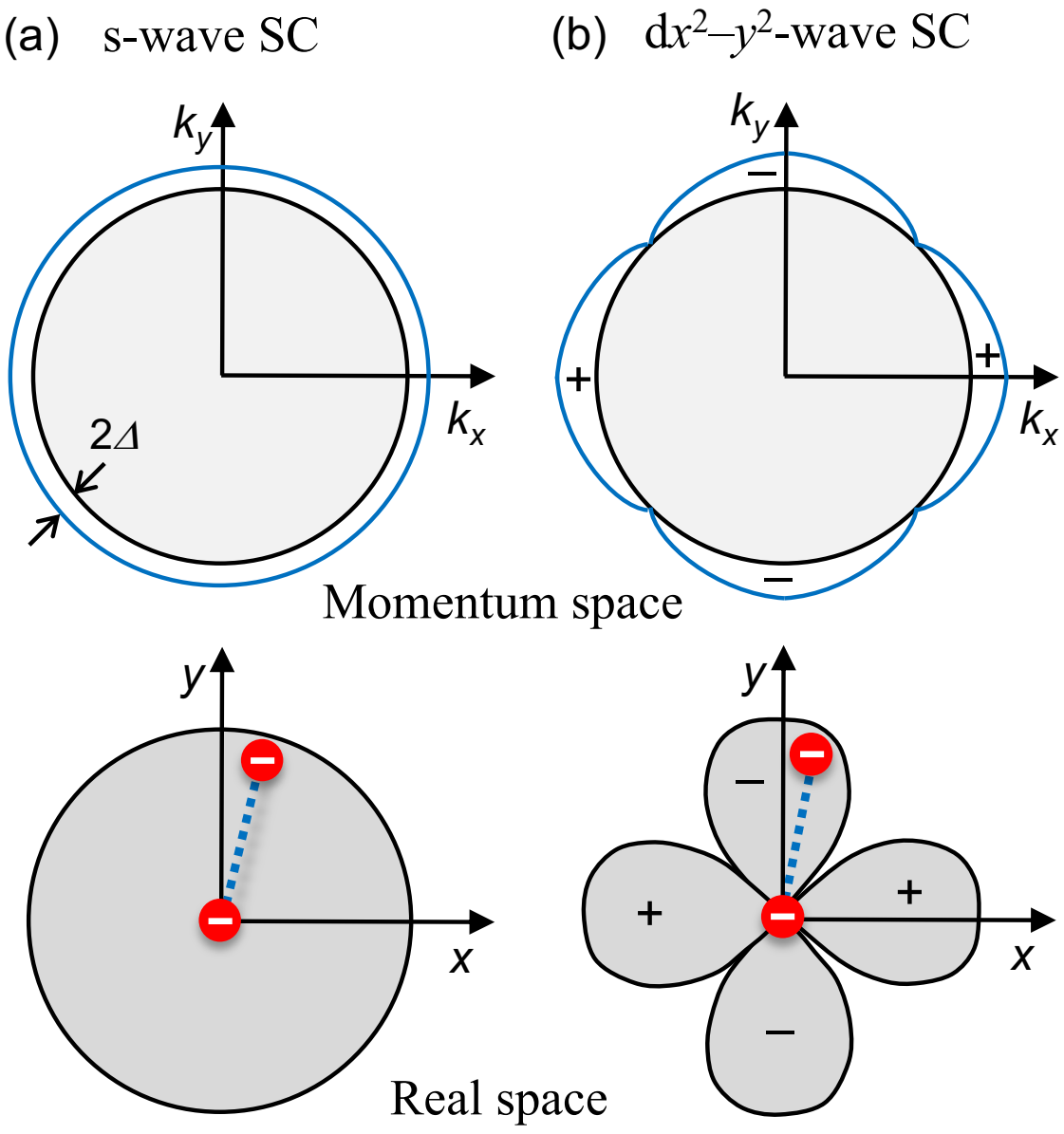

**Fig. 11.** Superconducting gaps in momentum space (above) and Cooper pair wavefunctions in real space (below) for (a) s-wave and (b) $dx^2–y^2$-wave superconductivity, respectively. The superconducting gap opens isotropically in the s-wave and reverses sign across the node at <110> in the $dx^2–y^2$-wave. A Cooper pair in real space is represented by two red balls (electrons) connected by a dashed line (attraction interaction). The distribution shows the probability of finding one electron while leaving the other at the origin.

More complex gap symmetries have also been investigated. When a crystal structure has spatial inversion symmetry, the gap symmetry is simply classified as s, p, or d waves, as previously stated. In noncentrosymmetric crystals, however, they can combine to form complex Cooper pairs, as seen in a variety of noncentrosymmetric superconductors [57-59]. However, while this type of symmetry argument guarantees the possibility, the degree of mixing is unknown, and in actual superconductors, one type is often dominant, so the mixture has little effect on superconducting properties. On the other hand, complex compounds frequently have multiple Fermi surfaces. $MgB_2$ with two types of Fermi surfaces exhibits two distinct s-wave gaps [60, 61]. Iron pnictide superconductors may have s-wave gaps with different signs [62, 63]. Anisotropic gaps are formed by some compounds due to their complex Fermi surface geometry.

Cooper pairs, in addition to the aforementioned orbital shapes, have two spin orientation-dependent magnetic states: singlet when electron spins are antiparallel and triplet when they are parallel. A singlet is used for s- and d-wave Cooper pairs with even-function orbitals, while a triplet is used for p-wave pairs with odd parity. Crystal surfaces and interfaces are expected to support a wide range of combinations [64]. Magnetic fields can disrupt singlet superconductivity while preserving triplet superconductivity [65]. Antiferromagnetic or spin-independent interactions, such as electron–phonon interactions, are thought to drive singlet superconductivity, while ferromagnetic interactions may cause triplet superconductivity. Therefore, determining the spin component of the Cooper pair is also critical for understanding the underlying superconductivity mechanism.

FFLO superconductivity is considered a unique superconducting state [66]. In BCS superconductivity, the Cooper pair has ($\boldsymbol{k}\uparrow$, $-\boldsymbol{k}\downarrow$) with zero center-of-mass momentum. However, FFLO superconductivity has finite momentum $\boldsymbol{q}$ in ($\boldsymbol{k}\uparrow + \boldsymbol{q}/2$, $-\boldsymbol{k}\downarrow + \boldsymbol{q}/2$). The latter is more unstable than the former due to its higher kinetic energy. Nevertheless, it appears when the energy of up (down) spin electrons decreases (increases) in a strong magnetic field and the spin polarization energy increases sufficiently. The FFLO superconducting state in $CeCoIn_5$ and other materials under strong magnetic fields has been extensively studied [67].

## 3. General characteristics of superconducting materials

Numerous superconductors have been discovered thus far, as illustrated in Fig. 1; see Cava's 2000 materials history review [12]. Table 1 describes the characteristics of copper oxide superconductors, and Table 2 lists other types of superconductors. Because all superconductors cannot be listed, the selection is based on the author's preference. He apologizes to those who had studied the unlisted ones. This chapter focuses on general aspects of superconductors rather than specific materials. When phonons or other attraction bases are present, combined with a typical size DOS at the Fermi level, the Fermi surface becomes unstable, resulting in superconductivity as the ground state, the most stable state at absolute zero.

Although this paper focuses on crystalline solids, there are superconductors with non-periodic atomic arrangements. For example, Bi shows superconductivity at 0.53 mK in a crystal [22] and 6.1 K in its amorphous form [68], while $Au_{64}Ge_{22}Yb_{14}$, a quasicrystal approximant, is a superconductor with $T_c$ = 0.68 K [69]. As depicted in Fig. 7, the phonons employed in the BCS mechanism are virtual and do not necessarily propagate as defined by a single point in momentum space. Cooper pairing can also achieve BEC-like superconductivity by creating and annihilating localized phonons, as depicted in Fig. 9 for rattling superconductivity. In any case, the phonons needed for pairing are absorbed and disappear as soon as they are created, so whether they propagate or remain localized is irrelevant. Therefore, superconductivity can occur in non-periodic systems without the requirement for a regular lattice. However, the type of phonons used to mediate pairing influences the strength of the electron–phonon interaction, and thus the height of $T_c$. Because many superconductors, particularly those with high $T_c$, have distinct crystal structures that are intimately linked to the superconductivity mechanism, we will concentrate on crystalline superconductors.

None-superconducting materials have a low DOS (or low electron density in BEC superconductivity), weak electron attraction, or a competing ground state (Fig. 3). The extremely low $T_c$ in crystalline Bi is due to the exceptionally small DOS or carrier density in its semimetallic band structure [22], whereas the high $T_c$ in amorphous Bi may be caused by the crystal structure disorder broadening the DOS energy profile, which occasionally results in a larger DOS at the Fermi level. Competing ground states can replace superconductivity when the Fermi liquid instability combines with the lattice instability to open an insulator gap in CDW, or when strong electron correlations result in magnetic long-range order. Superconductivity that carries the order's remnants is commonly observed after suppressing the competing order with carrier doping or pressure. The examples will be summed up in Chapter 5.

### 3.1. Elements

Many elements exhibit low $T_c$ superconductivity. $T_c$ values for metal elements range from 0.4 mK for Li to 9.2 K for Nb [70]. Under high pressure, many non-superconductors transform into superconductors, including Ca, which has the highest $T_c$ = 29 K of any single element at $P$ = 125 GPa [71]. At such ultrahigh pressures, even solidified oxygen shows 0.6 K superconductivity [72]. Hydrogen may crystallize at higher pressures exceeding a few hundreds of GPa and is predicted to be a superconductor with $T_c$ higher than room temperature [73], but this has yet to be confirmed experimentally.

High-pressure superconductivity could be caused by increased DOS or enhanced electron–phonon interactions as a result of structural deformation or electronic structure modifications. Alternatively, superconductivity under pressure replaces a competing order, such as magnetic order. Carrier doping in an insulator is also an effective route for achieving superconductivity. Diamond, an insulator with a large band gap of 5.5 eV, can be converted into a superconductor with $T_c$ = 4–7 K by doping it with boron and adding hole carriers [74, 75]. It is not surprising that superconductivity occurs following the formation of a nonmagnetic metallic state via pressure or carrier doping.

### 3.2. Complex compounds

Compound superconductors tend to have more complex crystal structures than elemental superconductors or alloys. Their crystal structures highlight the importance of two components: the conducting path and the block that fills the remaining space. Chemical bonds composed of p or d orbitals form various networks due to their anisotropy. When these orbitals dominate the electronic state near the Fermi level, electrical conduction occurs primarily via the network. The dimensions of the conduction path vary according to orbital connectivity. Superconductivity occurs in three dimensions (3D) for many intermetallic compounds, perovskite oxides, pyrochlore oxides, etc., in two dimensions (2D) for intercalated graphite, transition metal dichalcogenides, copper oxides, iron pnictides, molecular compounds, etc., and in one dimension (1D) for polymeric sulfur nitride $(SN)_x$ [19], $NbSe_3$ [76], etc.; however, all crystals are three dimensional in nature and only pseudo-low-dimensional due to anisotropy in the crystal and electronic structures.

Conduction paths and void-filling blocks combine to create a wide range of composite structures. Conduction pathways are typically comprised of covalent or metallic chemical bonds, whereas blocks are made of ionic bonds. Although the conduction path is obviously responsible for superconductivity, block characteristics are critical to understanding the superconductivity mechanism and controlling physical properties. A trick in the block causes Cooper pairing, and chemically modifying the block alters the number of carriers in the conduction path while also putting chemical pressure to it. In β-pyrochlore oxides, 3D conduction occurs in a cage-like structure made of covalent Os–O bonds, with A cations doping electrons and acting as an attraction source (rattling). The 2D conduction path in copper oxides occurs in covalent $CuO_2$ planes, with carriers generated by a chemical modification of the ionic-bonded block layer sandwiched between them. In $CeCu_2Si_2$, larger Ge can replace Si, resulting in a lattice expansion and negative pressure on the conduction path [77].

The dimensionality of conduction paths plays an important role in electronic system stability. Figure 3 depicts a spherical Fermi surface for a 3D isotropic path, while the 2D and 1D paths yield cylindrical (Fig. 12) and planar Fermi surfaces, respectively. Because such low-dimensional Fermi surfaces are unstable and easily destroyed by electron–phonon or other perturbations, novel electronic ordered states, including superconductivity, are expected to emerge. In low-dimensional systems, the Peierls instability distorts the lattice for an additional period determined by band filling, trapping electrons in the lattice modulation and preventing their migration in a CDW insulator [9].

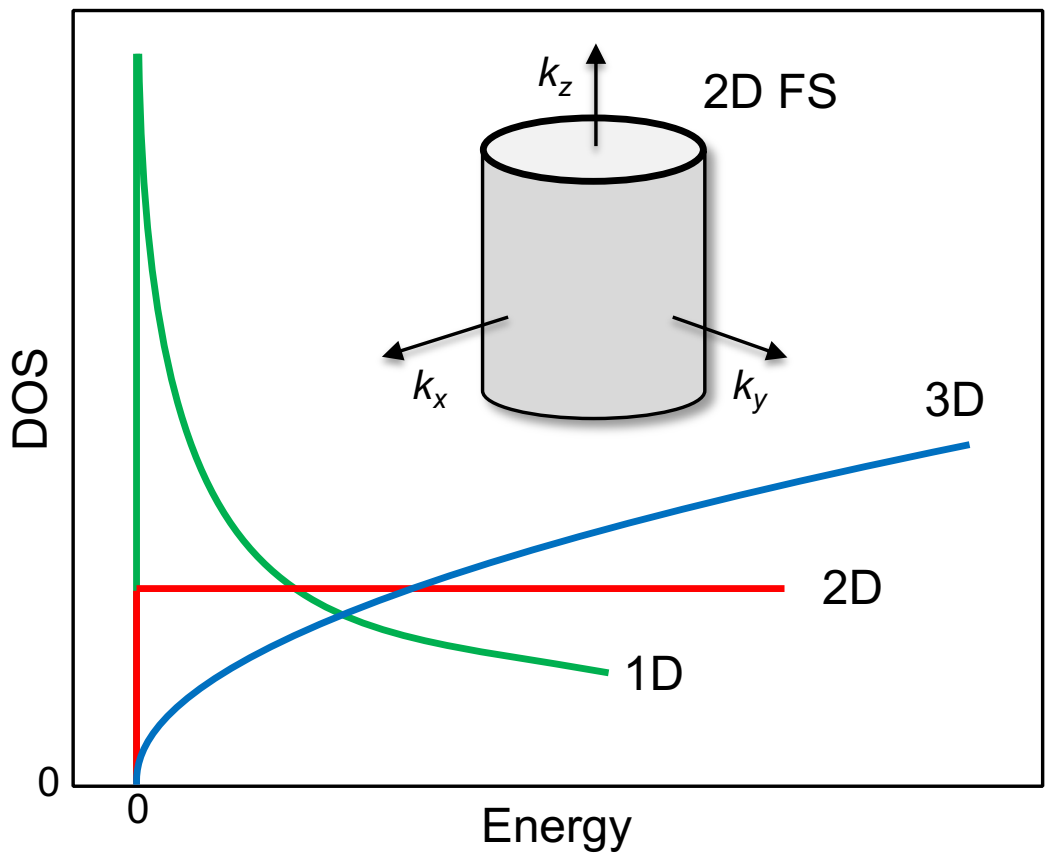

**Fig. 12.** DOS profiles for Fermi gas in 1D, 2D, and 3D. Zero energy is placed at the bottom of the band. The inset depicts a cylindrical Fermi surface (FS) for 2D electrons. The energy dependence of DOS is proportional to $E^{1/2}$ and $E^{-1/2}$ for 3D and 1D, respectively, while the 2D DOS is flat.

The dimensionality of the conduction path influences $T_c$ height. Figure 12 shows that the DOS distribution in the 3D band gradually rises from the band edge with an energy dependence of $E^{1/2}$; in 2D, it is flat with no energy dependence; and in 1D, it diverges at the band edge with an energy dependence of $E^{-1/2}$ before shrinking in the center. Fermi energy is always found near the band edge in compound semiconductors and superconductors doped with carriers. Therefore, the DOS at $E_F$ increases in the order 3D, 2D, and 1D, potentially leading to a higher $T_c$ (Eq. 3). The majority of known high-$T_c$ superconductors, however, have 2D structures. Although 1D is advantageous concerning DOS size, it is unlikely to have a high $T_c$ due to large fluctuations, making any LRO unstable [78]. Furthermore, as explained in the following section, randomness has a greater influence on 1D than on 2D and 3D.

The arrangement of atoms that comprise the conduction path is especially important in SCES. When electrons are about to localize on atoms, as shown in Fig. 4, the original atomic arrangement has a significant impact on the electronic properties. For example, when strongly correlated electrons from atoms forming a 2D kagome lattice (an arrangement of equilateral triangles connected at their vertices) are responsible for conduction, a perfectly flat band is expected to form, resulting in ferromagnetism (flat-band ferromagnetism) and nearby superconductivity [79]. Furthermore, in 2D triangular lattices (an arrangement of equilateral triangles connected by edges) and 3D pyrochlore lattices (an arrangement of tetrahedra connected by vertices), magnetic order is suppressed due to geometrical frustration effects between localized spins [80], and exotic electronic states including superconductivity emerge instead (Section 5.3.3). Spin liquid states and superconductivity are predicted to appear in a 1D ladder lattice (a network of multiple chains) [81]. Even in weakly correlated electron systems, the lattice plays an important role. The honeycomb lattice of carbon-based graphene contains zero-mass Dirac electrons with linear band dispersion, allowing carriers to be extremely mobile [82]. The emphasis on low-dimensional lattices in actual 3D crystal structures is one guiding principle for the development of materials with novel physical properties.

### 3.3 Chemical decorations and randomness introduced

Many unconventional superconductors have an insulating "parent" phase nearby. Carrier doping via chemical modifications frequently produces superconductivity. It is critical to remember that chemical modifications always disrupt the system; pressure, in contrast, is a clean method of phase control. Although general superconductivity textbooks and reviews rarely address this issue, the author believes it is important to consider when discussing the chemical trend of $T_c$ in actual doping-induced superconductors. The effects of disorder are most visible in low-dimensional systems, where electrons are more difficult to reroute to avoid a defect due to fewer pathways. When 2D superconductivity is achieved in thin films of Bi and Pb (with a small film thickness relative to the superconducting coherence length), $T_c$ decreases as the thickness decreases, followed by a clear transition from superconductor to insulator [83]. This is thought to be the result of disorder, which worsens as it approaches two dimensions.

In general, elemental substitutions are used to modify conventional semiconductor conductivity [84]. In a silicon wafer crystal, neighboring Al and P substitutions for Si generate holes and electrons, respectively. The extra charge from randomly placed impurities creates a local potential, scattering doped carriers. However, the scattering sources in these band insulators are effectively masked by the highly mobile light carriers that surround them. As a result, the base material's electrical conductivity can be easily controlled using a small amount of impurity doping (less than 1 ppm), and the resulting low-density carriers govern the electronic properties, allowing the fabrication of a variety of semiconductor devices. Furthermore, modulated doping in multilayers can reduce impurity scattering by spatially separating the conducting and impurity-containing layers, producing semiconductor devices with extremely high mobility (high-electron-mobility transistor: HEMT) [84].

Controlling the properties of SCES, unlike conventional semiconductors with weak electron correlations, requires relatively large number of carriers (at least 1%) [28]. In the typical cuprate superconductor $La_{2-x}Sr_xCuO_4$ (La214), hole doping above 2% results in a metallic state (Fig. 13). The limited bandwidth and strong electron correlation effects obstruct electron movement, yielding low mobility and poor screening (Fig. 4). To move freely, carriers must be relatively dense. A high impurity density and low screening effects, particularly near the insulator–metal phase boundary, have a significant impact on the electronic state due to randomness (Section 4.5.2). As a result, randomness effects are critical to understanding the SCES properties.

Impurities in heavily doped superconductors cause strong scattering, breaks up Cooper pairs and lowers $T_c$ [85]. To avoid $T_c$ reduction caused by impurity scattering, complex compounds require smart material design. The goal is to effectively separate the block and conduction paths, minimize chemical substitution in the block, and keep the conduction path as clean as possible, similar to modulated doping in HEMT. For example, in $BaPb_{1-x}Bi_xO_3$ (BPBO), substituting Pb in the conduction path yields a Bi–O 3D network with a $T_c$ of 13 K [14], whereas in $Ba_{1-x}K_xBiO_3$ (BKBO), substituting K in the block Ba raises the $T_c$ to 30 K in the cleaner Bi–O path [86].

For carrier doping in copper oxides, elemental substitution or the introduction of excess oxygen into the block layer is employed. Even with spatial isolation, impurity atoms provide a sizable random potential for carriers to move along the $CuO_2$ plane in the conduction path, disrupting the superconducting state and lowering $T_c$ [87-89]. The effect is highly dependent on crystal structure and varies greatly across materials. Understanding the materials science of cuprate superconductors and deducing the common mechanism of superconductivity, as covered in the following chapter, requires careful consideration of randomness effects. In other words, once randomness effects are dealt with properly, the true superconductivity mechanism will emerge!

## 4. Copper oxide superconductors

The discovery of copper oxide superconductors inspired solid state chemists to pursue superconductivity research, thereby bridging the gap between solid state chemistry and physics. Since the first report 'Possible high $T_c$ superconductivity in the Ba-La-Cu-O system' by Bednorz and Müller in 1986 [4], many superconductors were synthesized in just seven years until 1993, reaching $T_c$ of 135 K [90]: a remarkable speed compared to the 62 years it took from the first 4.2 K for mercury to 22 K for $Nb_3Ge$ (Fig. 1); see the article on the exciting era of "superconductivity fever" [91], as well as the article on the period 20 years later [92].

Copper oxide superconductors have been covered in numerous articles, reviews, and books, including solid state chemistry reviews [12, 93-98] and physics perspectives [28, 53, 99-103]. Figure 13 depicts a typical phase diagram found in many publications. Superconductivity occurs when the $Cu^{2+}$-containing antiferromagnetic insulator $La_2CuO_4$ ($Nd_2CuO_4$) is doped with holes (electrons) via chemical substitution in $La_{2-x}Sr_xCuO_4$ ($Nd_{2-x}Ce_xCuO_4$). There are many compounds for hole doping with $T_c$ values as high as 135 K, while compounds for electron doping are far less common, with $T_c$ as low as 40 K. As a result, much research, including this manuscript, has focused on the former, with only a passing mention of the latter.

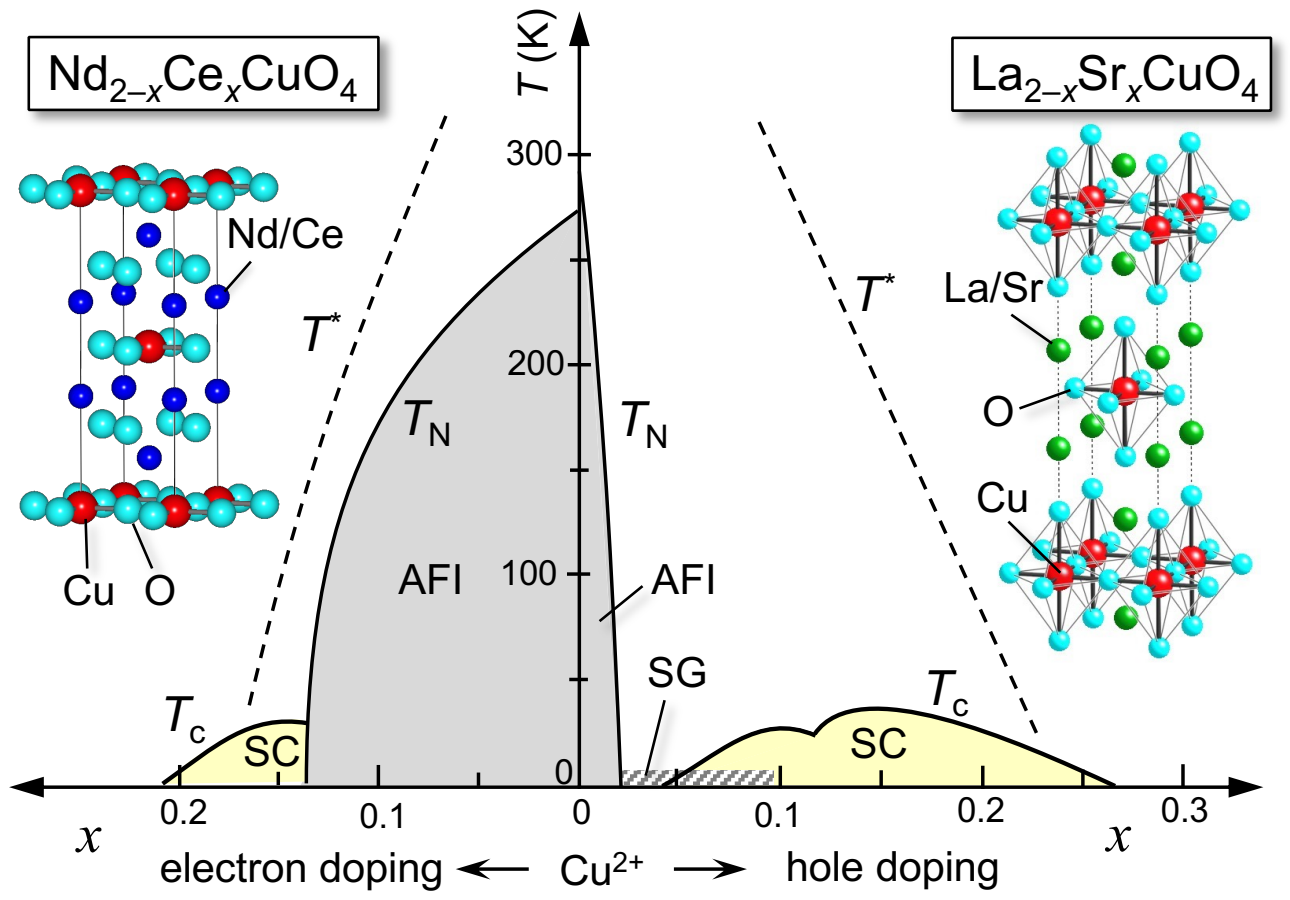

**Fig. 13.** Typical phase diagram for copper oxide superconductivity, with $La_{2-x}Sr_xCuO_4$ on the right and $Nd_{2-x}Ce_xCuO_4$ on the left. The Sr and Ce substitutions introduce holes and electrons into the parent insulating phases with $Cu^{2+}$, respectively, resulting in superconductivity at specific doping levels. The terms used are superconductivity (SC), antiferromagnetic insulator (AFI), spin glass (SG), superconducting critical temperature ($T_c$), antiferromagnetic ordering temperature ($T_N$), and pseudogap temperature ($T^*$). At the end of this chapter, the phase diagram will be compared to the ideal in Fig. 40.

Despite numerous discussions about superconductivity mechanisms, there is still no consensus on a single superconductivity mechanism 40 years later. An article published in 2006 summarizes the arguments advanced by each of the twelve distinguished theorists [104]. It was suggested there that cuprate superconductivity, unlike conventional superconductivity, may lack a comprehensive theory that explains all experimental facts without contradictions (of course, each theorist believes he or she is correct). Indeed, the vast majority of observed experimental results are perplexing and frequently contradictory, making it appear almost impossible to construct a plausible story that explains everything. However, many of these experimental findings are likely to have been influenced by external factors. Even the well-known phase diagram in Fig. 13 requires caution because it contains some extrinsic modifications from the ideal case, which will be discussed later (Fig. 40). Among the theorists' claims in the 2006 article, Vojta says, 'Progress will only be made by discerning primary from secondary effects.' [104]. The author totally agrees.

As a solid state chemist, the author prefers to select experimental results that he believes are reliable and significant, and then uses them to construct an intuitive and straightforward understanding of the superconductivity mechanism, with a particular emphasis on explaining $T_c$'s material dependence. Physicists may find the current arguments overly speculative, or even suspicious. While physicists are prone to getting bogged down in individual arguments, the author believes that chemists, due to their inability to understand the details, are better at drawing broad inferences and can occasionally get closer to the more important path.

There are numerous copper oxide superconductors, each with a unique solid state chemistry (Table 1). $YBa_2Cu_3O_{7-\delta}$ (Y123), for example, exhibits an incredibly rich structure–property relationship (Fig. 14b) [105]. Soon after its discovery, the author started studying it and was astounded to find that the $T_c$ of prepared samples varied dramatically depending on the firing temperature and reaction gas used in an electric furnace. Knowing why, Y123 is still a fascinating chemical compound. While physicists may find these compounds' diverse properties bothersome, chemists will find them fascinating. However, the goal of this manuscript is not to provide an exhaustive list of each compound's properties. Rather, the emphasis is on extracting common features through physical reasoning, comprehending the general mechanism of superconductivity, and obtaining hints for higher $T_c$. Consequently, individual material considerations are minimized here. Readers who are interested can consult additional references [12, 96].

Because the author is most interested in Chapter 4, which is lengthy enough to warrant a detailed explanation, he will begin by summarizing the story's progression. Section 4.1 describes crystal chemistry, which includes similarities in chemical composition and crystal structure, as well as differences between block layers. Section 4.2 goes over the electronic state of the $CuO_2$ plane, what happens during hole doping, and what causes the attraction that generates Cooper pairs in cuprates. Section 4.3 explains a simple superconductivity mechanism. Section 4.4 discusses the experimental data necessary to understand the material dependence of $T_c$ from the perspective of solid state chemistry, with a focus on the relationship between $T_c$ and the number of holes $p$ doped on the $CuO_2$ plane. After assessing the validity of the previously assumed $T_c$–$p$ relationship, we will seek an alternative. The optimal hole concentration $p_o$ at which $T_c$ achieves the highest $T_{co}$ is also examined for several compounds, revealing a strong correlation between $T_{co}$ and $p_o$. It should be noted that many previous studies have failed to distinguish between the $p$-dependent $T_c$ and the unique, material-specific parameter $T_{co}$ when discussing chemical trends, resulting in confusion. Section 4.5 emphasizes the importance of two factors: apical oxygen and randomness effects, which are both required to consistently combine the majority of experimental data. Section 4.6 examines the material dependence of $T_{co}$, highlighting key NMR and ARPES findings in multilayer systems with three or more $CuO_2$ planes in the conduction layers. Section 4.7 illustrates the phase diagram of ideal $CuO_2$ planes. Finally, Section 4.8 covers electron-doped systems and other orders that compete with superconductivity, while Section 4.9 concludes.

Sections 4.1–4.3 explain the essence of cuprate superconductivity and its significance in general superconductivity research, which is sufficient to readers who want to grasp the feature. Readers should go through the entire section to get a better understanding of the structure–property relationship and strategies for dealing with high $T_c$. The author welcomes feedback from physicists in this field, including questions, comments, and criticism.

### 4.1. The solid state chemistry of copper oxide superconductors

Solid state chemistry entails experimental investigations into materials, including synthesis, chemical composition analysis, microstructure observation, crystal structure determination, and physical property measurements [6]. Copper oxide superconductors are synthesized using a variety of methods, including traditional solid-state synthesis, high-pressure synthesis, and thin-film fabrication; however, due to material dependence, we will not cover them here. In this section, we will organize the crystal chemistry of the block layers as well as the chemical modifications required for carrier doping. This will lay the groundwork for the interpretation of $T_c$'s material dependence, which will be discussed at the end of the chapter and serves as the foundation for understanding the superconducting mechanism.

#### 4.1.1. Fundamental chemical compositions and crystal structures

Copper oxide superconductors are named empirically, unlike most inorganic compounds based on IUPAC rules to arrange elemental symbols in order of electronegativity [106]. The characteristic metal element's symbol appears first, followed by a list of other metal elements with Cu at the end, and finally anions (Table 1), such as $YBa_2Cu_3O_{7-\delta}$, $Bi_2Sr_2CaCu_2O_{8+\delta}$, and $HgBa_2Ca_2Cu_3O_{10+\delta}$. In addition, for convenience, they are abbreviated as Y123, Bi2212, and Hg1223, respectively, with the first metal symbol followed by the metal element ratio. The simplified notation does not include oxygen content, which is automatically determined by each metal element's formal charge and composition ratio when the $CuO_2$ plane contains $Cu^{2+}$ ions ($\delta$ = 0.5 for Y123, 0 for others). Nonstoichiometry is known to slightly alter these ideal compositions. Exceptions to the naming rules include La214 [La(Sr)214 or La(Ba)214 when identifying the substitution elements] and Nd214, which stand for $La_{2-x}Sr_xCuO_4$ ($La_{2-x}Ba_xCuO_4$) and $Nd_{2-x}Ce_xCuO_4$, respectively. Furthermore, F214 and Cl214 refer to isostructural compounds with F and Cl atoms as key elements, respectively.

Figure 14 depicts the crystal structures of typical copper oxide superconductors. They have a conduction layer composed of one or $n$ $CuO_2$ planes separated by atoms such as Ca or Y, which alternately stacks with various block layers [107] or charge reservoir layers [12] along the $c$ axis (Fig. 15). Compounds with conduction layers of $n$ = 1, 2, 3, and so on are denoted by the symbols C1, C2, C3, and so on, with "C" standing for the $CuO_2$ plane or conduction layer. The conduction layer actually consists of $(2n - 1)$ atomic sheets, including $n$ $CuO_2$ planes and intervening $(n - 1)$ atomic sheets. Because the conduction layer is common, each compound is identified by its block layer, as explained in the following section. The parent phase, which contains divalent Cu ions on the $CuO_2$ plane, is a Mott (charge-transfer) insulator, as described in Section 4.2. Chemical modifications, such as elemental substitutions or the addition of excess oxygen, can alter the charge of the block layer. The excess charge, which is generated to meet the charge neutrality condition specified in the compositional formula, is supplied to the $CuO_2$ planes, where it serves as the hole or electron carrier responsible for metallic conductivity and superconductivity.

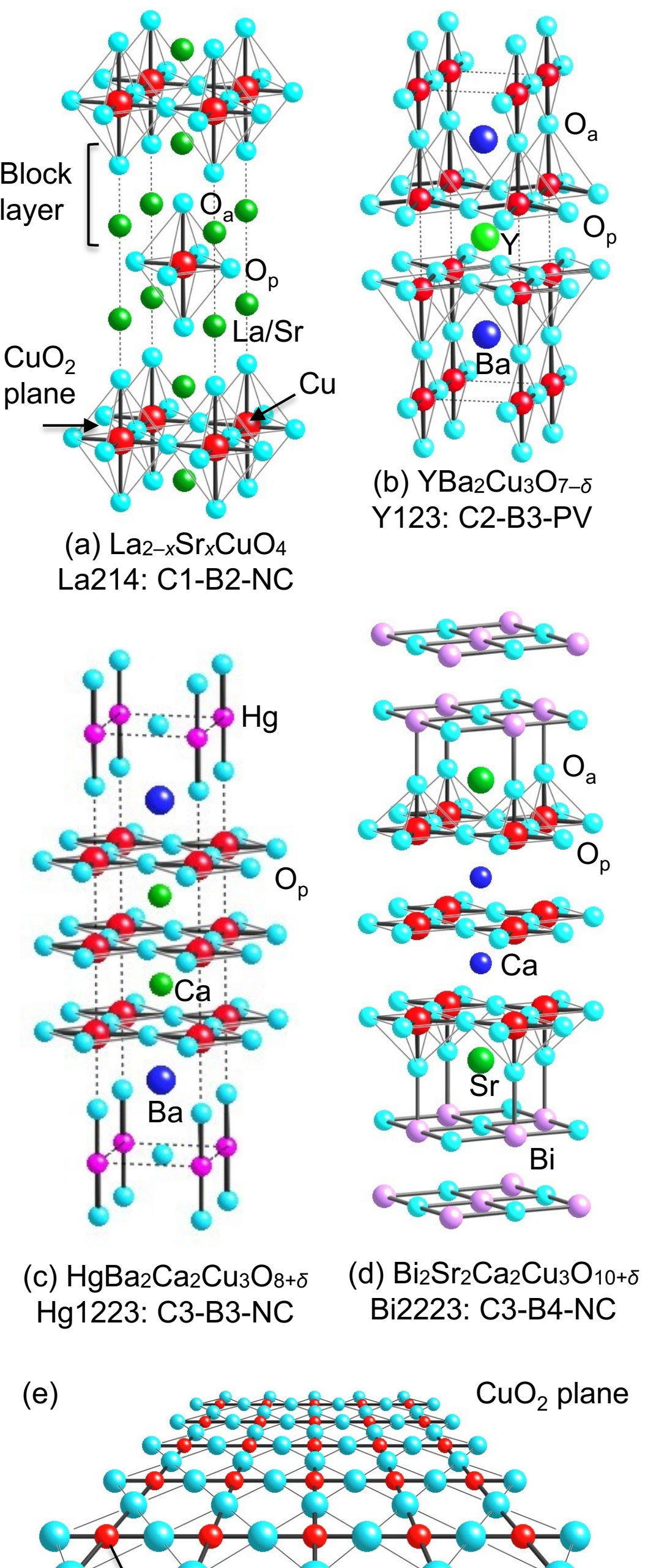

**Fig. 14.** Typical crystal structures of copper oxide superconductors with vertical direction along the $c$ axis: (a) $La_{2-x}Sr_xCuO_4$ (La214), (b) $YBa_2Cu_3O_{7-\delta}$ (Y123), (c) $HgBa_2Ca_2Cu_3O_{8+\delta}$ (Hg1223), and (d) $Bi_2Sr_2Ca_2Cu_3O_{10+\delta}$ (Bi2223). The common $CuO_2$ plane, where superconductivity occurs, is depicted in (e). $O_a$ and $O_p$ are the apical and in-plane oxide atoms of the $CuO_6$ octahedron, $CuO_5$ pyramid, or $CuO_4$ square (only $O_p$ exists). The $O_a$ in (c) Hg1223 is rather bonded to Hg to form the $HgO_2$ dumbbell, resulting in three $CuO_2$ planes stacked and the highest $T_c$ among copper oxide superconductors.

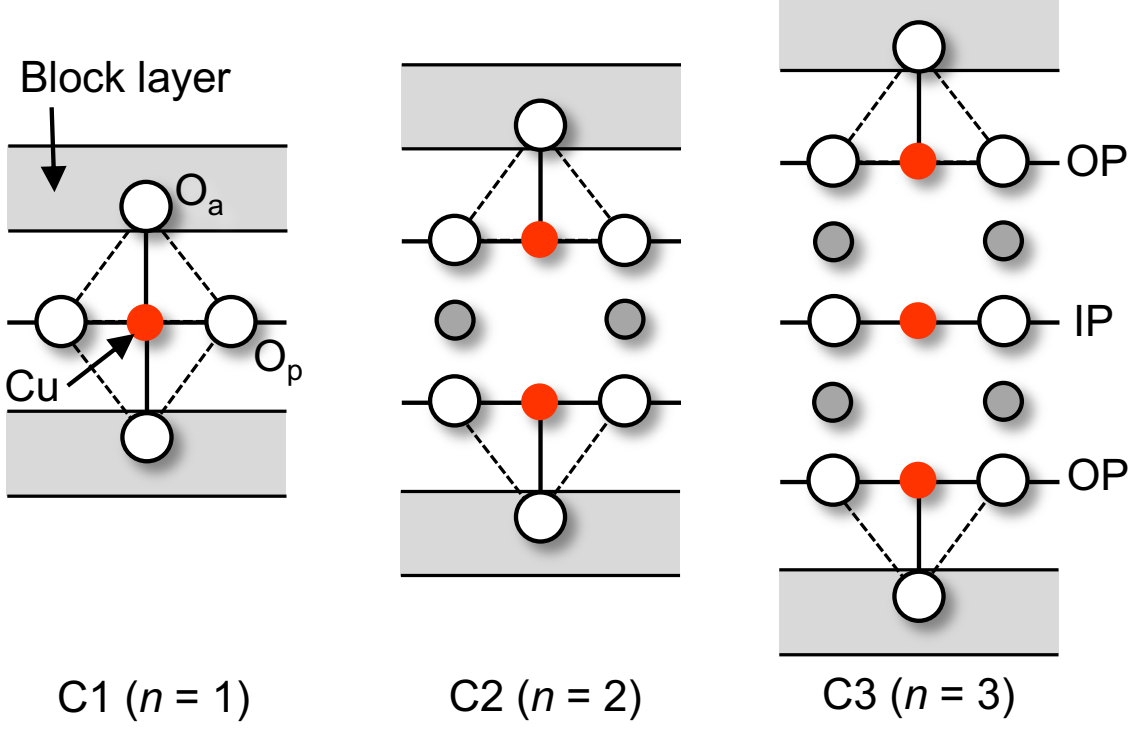

**Fig. 15.** Fundamental structure of the conduction layer in copper oxide superconductors. In the $n = 1$ compound (C1), the copper atom is octahedrally coordinated by six oxide atoms: four $O_p$ atoms in the $CuO_2$ plane and two apical $O_a$ atoms in the block layer. In the $n = 2$ compound (C2), a pair of pyramidally coordinated Cu atoms are separated by small cations such as Ca or Y atoms, while in the $n = 3$ compound (C3), an extra $CuO_2$ plane (inner plane: IP) without $O_a$ is inserted between the outer planes (OP).

| $n$ \ $m$ | 1 | 2 | 3 | 4 | 5 |
|---|---|---|---|---|---|
| 1 | IL | La214<br>Nd214<br>F214<br>Cl214 | Hg1201<br>Tl1201<br>Sr0201-$CO_3$ | Bi2201<br>Tl2201 | |
| 2 | | La2126<br>Ba0212<br>Sr0212 | Y123<br>Hg1212<br>Tl1212<br>Cu1212<br>Pb1212 | Y124<br>Bi2212<br>Tl2212 | Pb2213 |
| 3 | | Ba0223<br>Sr0223 | Hg1223<br>Tl1223<br>Cu1223 | Bi2223<br>Tl2223 | |
| 4 | | Ba0234 | Hg1234<br>Tl1234<br>Cu1234 | Tl2234 | |
| 5 | | Ba0245 | Hg1245<br>Cu1245 | | |

**Fig. 16.** Structure types composed of a C$n$ conduction layer containing $n$ $CuO_2$ planes and a B$m$ block layer containing $m$ cation sheets. All copper oxide superconductors are classified as C$n$-B$m$. Multilayer Hg and Ba series compounds with $n$ larger than 5 will be listed below the table.

Several classification schemes have been proposed for interpreting the structural chemistry of copper oxide superconductors. Tokura and coworkers introduced the block layer concept, gave an overview of all crystal structures, and explored the structure–property relationship, with a focus on carrier doping mechanisms [107]. In the present manuscript, we will look at the number $m$ of metal atom sheets contained in a single block layer, similar to how the conduction layer is classified by the number $n$ of $CuO_2$ planes. We refer to the block layer composed of $m$ metal sheets as B$m$; 'B' stands for the block layer. After combining with $n$, the structure type is now called C$n$-B$m$. Figure 16 depicts the general structure type, which consists of C$n$ and B$m$, as well as the classification of all copper oxide superconductors listed in Table 1.

4.1.2. Structural chemistry of block layers

When we view the crystal structure as a stack of metal atom sheets, we notice a feature shared by cuprate superconductors' crystal structures (Fig. 17). Metal atoms always form a body-centered-tetragonal (BCT) lattice (or a slightly deformed orthorhombic lattice) across the conduction and block layers. Specifically, two types of sheets stack alternately along the $c$ axis, with metal atoms located in the corners or centers of the sheet's unit square, which has an edge length of around 0.39 nm. Coulomb energy determines the basic crystal structure of most metal oxides, which requires lowering the electrostatic repulsion between cations in an oxygen's packing structure [108-110]. There is only one option for crystal structures made up of square planar sheets of atoms linked by fourfold chemical bonding: the BCT structure. Metal types primarily define crystal forms since their in-plane positions are fixed, and stacking positions are nearly fixed with similar interplane spacing. This is why the C$n$-B$m$ notation is useful for identifying the structure type. In addition, as explained below, the amount and position of oxide atoms in the resulting space created between metal atoms in the block layer determine the overall structure type.

Given that metal M is located at (0 0 0), the sheet has only two oxygen positions available: the square center (1/2 1/2 0) and the edge centers (1/2 0 0) and (0 1/2 0) (Figs. 17d and 17e). When fully occupied, MO and $MO_2$ sheets form for large and small M, respectively. In the block layer, the NaCl (NC) structure is created by stacking MO–M'O–MO sheets with M at (0 0 0) and M' at (1/2 1/2 1/2). $MO_2$ sheets cannot stack, however, due to oxygen atom overlap. Instead, alternating stacking, such as $MO_2$–M'O–$MO_2$, can produce a perovskite (PV) structure. When M' and M are of comparable sizes, such as Hg and Ba, NC is selected; when M is significantly smaller, such as Cu versus Ba, PV is preferred [109]. We'll now look at structure types like C$n$-B$m$-NC/PV, taking into account the block layer architecture. It should be noted that a conduction layer made up of multiple $CuO_2$ planes, such as $CuO_2$–Sr–$CuO_2$ (Fig. 17a), has an oxygen-deficient perovskite structure, with no oxygen in the middle M'O sheet. The structure of the conduction layer is apparent, so specification is unnecessary. The C$n$-B$m$-NC/PV notation is useful not only for organizing crystal structures, but also for later discussions of $T_c$'s material dependence. The number of stacked $CuO_2$ planes ($n$) is the most important factor in determining $T_{co}$, while the block layer thickness ($m$) and structural type influence the shape of the $T$–$p$ phase diagram.

Based on the preceding considerations, there are only six distinct types of block layers that can classify almost any compound, as illustrated in Fig. 17 and Table 1. The "infinite layer (IL) structure" is the most fundamental, as depicted in Fig. 17a [111, 112]. In $SrCuO_2$, for example, $CuO_2$ planes alternate with Sr sheets along the $c$-axis. To complete the $CuO_2$–Sr–$CuO_2$ stacking sequence, all oxygens from the SrO sheet of the original perovskite's $CuO_2$–SrO–$CuO_2$ stack are removed. The structure is identified as C1-B1, consisting of a single conducting $CuO_2$ plane and a single block sheet. According to the block layer concept, $n$ is one, not infinite, as is commonly believed.

B2-NC represents the block layer composed of two MO sheets stacked in the NaCl structure depicted in Fig. 17b. A typical example can be found in $La_2O_2$ in La214 (Fig. 14a), which is C1-B2-NC, also known as the T structure; however, we avoid using such a specific name in order to preserve the overall scheme. The mixed anion compounds F214 and Cl214, which contain fluorine and chlorine as key elements, have isomorphic structures. B2 includes another structure with the same metal arrangement but a different oxygen position: the $CaF_2$ (CF) type, which has an M–$O_2$–M stack. Figure 17c illustrates an $O_2$ sheet formed by the oxide atoms at (1/2 0 $z$) and (0 1/2 $z$) between the M sheets. A common example is the $Nd_2O_2$ layer in Nd214, C1-B2-CF (the T' structure). The T* structure [C1-B2-(NC–CF)] consists of alternating NC- and CF-type block layers separated by a single $CuO_2$ plane, as seen in (Nd, Ce, $Sr)_2CuO_{4-\delta}$. In addition to the C1 compounds listed above, B2 compounds with multiple $CuO_2$ planes in the conduction layer can be found in Ba- and Sr-based compounds containing $Ba_2(O_{1-y}F_y)_2$ and $Sr_2(O_{1-y}F_y)_2$ block layers.

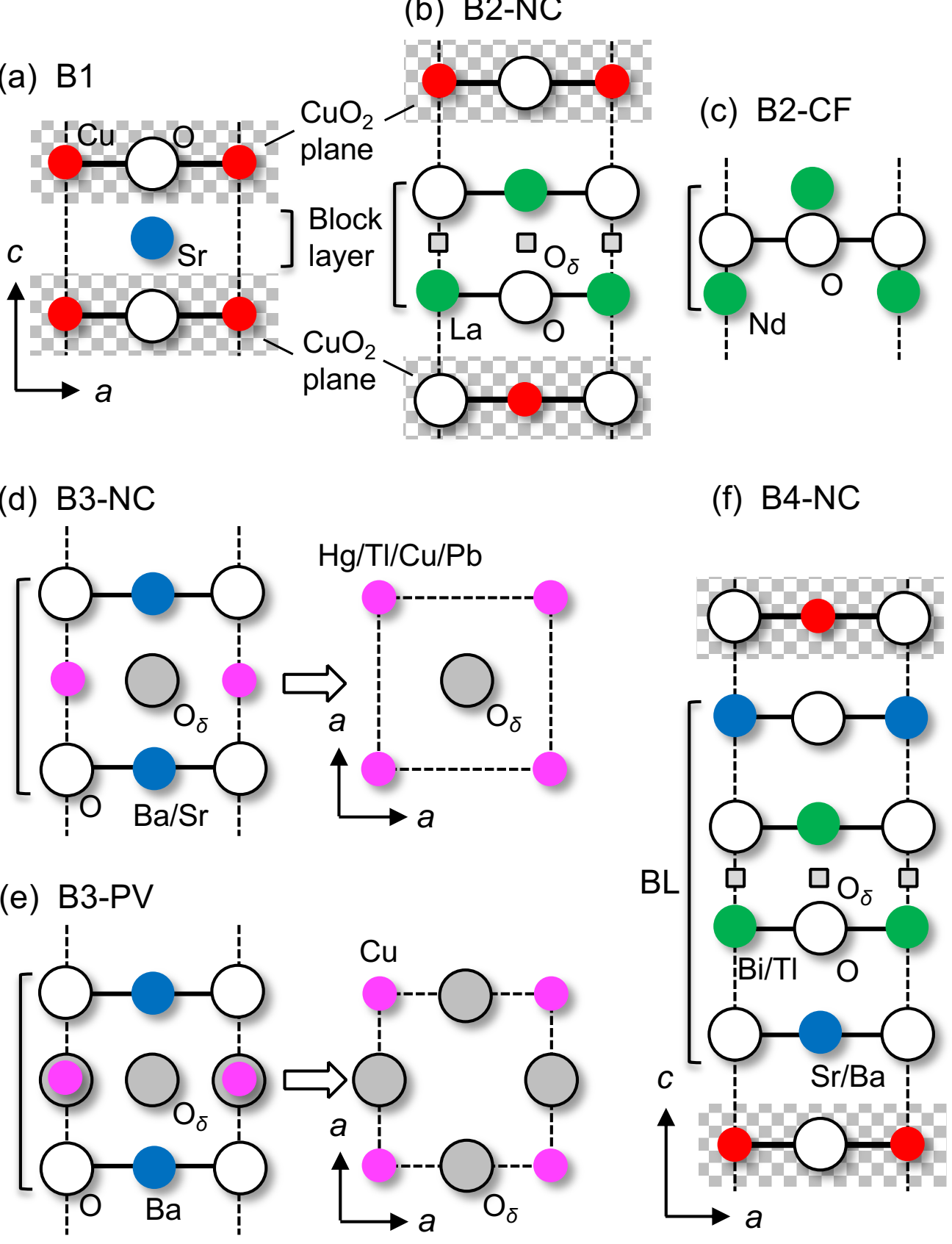

**Fig. 17.** Six distinct types of block layers. (a) The minimum block layer is made up of a single Sr sheet ($m$ = 1) sandwiched

between $CuO_2$ planes. This B1 block layer is found in $SrCuO_2$ (C1-B1), which has a “infinite-layer” (IL) structure. (b) A double-sheet rock-salt block layer (B2-NC) sandwiched between $CuO_2$ planes, like the $La_2O_2$ layer in La214 (C1-B2-NC). The $La_2O_{2+\delta}$ layer in oxygenated $La_2CuO_{4+\delta}$ contains excess oxygen $O_\delta$ at the interstitial position marked by small squares between the sheets, which corresponds to the normal oxygen position in B2-CF in (c). (c) The $CaF_2$ structure type double-sheet layer (B2-CF) is found in Nd214's $Nd_2O_2$ layer. (d) B3-NC, a triple-sheet layer of rock-salt stacking found in the Hg, Tl1 (single sheet), Cu, and Pb series of compounds. The middle sheet contains varying amounts of excess oxide atoms ($O_\delta$) at (1/2 1/2 1/2). (e) B3-PV is a triple-sheet perovskite layer composed of BaO–$CuO_\delta$–BaO, with $O_\delta$ at (1/2 0 1/2) and (0 1/2 1/2) in $YBa_2Cu_3O_{6+\delta}$. (f) The rock-salt layer, Sr(Ba)O–$[Bi(Tl)]_2O_2$–Sr(Ba)O, also known as B4-NC, is a four-sheet block layer that occurs in the Bi and Tl2 (double sheets) series. Excess oxide atoms can be incorporated into the interstitial space between the two Bi(Tl)O sheets, which corresponds to the oxygen position in the $La_2O_{2+\delta}$ layer in (b). The majority of block layers in copper oxide superconductors are classified into these six types, with the exception of Pb2213, which has a five-sheet B5-NC block layer (Fig. 16).

B3 consists of a MO–M'$O_\delta$–MO sequence with varying oxygen atoms in the middle M' sheet. Structures are classified into two types based on their in-plane positions: B3-NC has $O_\delta$ at (1/2 1/2 1/2) (Fig. 17d) and is found in Hg, Tl1, Cu, and Pb systems, such as Hg1223 (C3-B3-NC) (Fig. 14c). B3-PV has $O_\delta$ at (1/2 0 1/2) and (0 1/2 1/2) (Fig. 17e) and is found only in Y123 ($YBa_2Cu_3O_{6+\delta}$) (C2-B3-PV) (Fig. 14b). Related compounds to Y123 include Y124 (C2-B4-PV), which has two Cu-O sheets in the block layer, and Y123.5 (C2-(B3–B4)), which stacks Y123 and Y124 types alternately. The final option is B4-NC, which comprises four sheets with a rock-salt structure (Fig. 17f). The structure consists of $Bi_2O_{2+\delta}$ or $Tl_2O_{2+\delta}$ layers, which are isostructural to $La_2O_{2+\delta}$, with two additional Sr(Ba)O sheets above and below, as seen in Tl2212 (C2-B4-NC) or Bi2223 (C3-B4-NC) (Fig. 14d). There are two types of block layers in the Tl (Pb) system: the Tl1 system of B3-NC (containing a single Tl atomic sheet), such as Tl1201, and the Tl2 system of B4-NC, such as Tl2201 (containing two Tl atomic sheets).

Thicker block layers are more likely to form using rock salt structures, which are so stable in many inorganic compounds. In fact, $Pb_2Sr_2YCu_3O_{8+\delta}$ (Pb2213: C2-B5-NC) has an $m = 5$ block layer with SrO–PbO–$CuO_\delta$–PbO–SrO stacking [113]. Other B5 and larger-$m$ block layers will be accessible. To make even thicker layers, such as B6-NC, place two SrO sheets above and below the B4-NC block layer in Fig. 17f. Because the addition of a charge-neutral SrO sheet has no effect on the block layer's total plus charge, so it can be combined with any conduction layers with a comparable negative charge. In extreme cases, hole-generating NC $(Bi_2O_{2+\delta})^{(2-2\delta)+}$ block layers and $(Sr_{n-1}Cu_nO_{2n})^{(2-2\delta)-}$ conduction layers are embedded in a matrix SrO bulk crystal with a rock salt structure and stacked alternately with multiple SrO spacer sheets. On the other hand, new block layers may have a perovskite structure. In addition to the known PV-type block layer in Y123, the abundance of perovskite compounds suggests that there will be a variety of single and multiple PV block layers, as well as a PV-NC hybrid. For example, the central $CuO_\delta$ sheet in the Pb2213 block layer can have a PV type oxygen arrangement. However, for PV-type block layers, which typically contain other transition metal atoms, it is challenging to avoid Cu contamination while keeping the $CuO_2$ plane clean.

Any copper oxide superconductor will be classified according to the current naming convention, C$n$-B$m$-(structure type). This notation is applicable to most layered compounds containing other transition metals, including the superconductors $Sr_2RuO_4$ [114] in C1-B2-NC and $La_2PrNi_2O_7$ [17] in C2-B3-NC. Most compounds, with the exception of Nd214 and Y123, use NC, so the C$n$-B$m$ notation is sufficient in many cases.

4.1.3. Chemical modifications for carrier doping

Carrier injection into the $CuO_2$ plane is accomplished by chemically modifying the block layer, such as through elemental substitutions or oxygen addition. In $La_{2-x}Sr_xCuO_4$, substituting $Sr^{2+}$ atoms for some of the $La^{3+}$ atoms produces excess positive charge, which is injected into the $CuO_2$ plane as hole carriers under charge neutral conditions. In crystals, La and Sr are considered fully ionized to trivalent and divalent, respectively, because they lack electronic states near the Fermi level, whereas Cu–O covalent states dominate the conduction states at the Fermi level with variable charge. As a result, the number of holes ($p$) per Cu equals $x$, allowing for straightforward hole-doping calculations. In the same C1-B2-NC structure type, $Ca_{2-x}Na_xCuO_2Cl_2$ is hole-doped by replacing $Ca^{2+}$ with $Na^+$ in the block layer [115, 116]. In the Sr and Ba series of compounds, hole carriers are generated by partially replacing $O^{2-}$ with $F^-$ in the B2-NC block layers of $Sr_2(O, F)_2$ or $Ba_2(O, F)_2$ [117, 118].

In most other hole-doped compounds, excess oxide ions ($O_\delta$) enter the block layer and generate hole carriers in the $CuO_2$ plane. Ideally, hole carriers with $p = 2\delta$ are produced. Excess oxide ions occupy specific crystallographic sites, limiting block layers' ability to donate holes. The parent phase of La214 can absorb a small amount of oxygen to become $La_2CuO_{4+\delta}$, which exhibits superconductivity below 38 K [119]. It has a B2-NC $La_2O_{2+\delta}$ block layer, with excess oxide ions occupying interstitial sites (marked by small squares in Fig. 17b, corresponding to the B2-CF's oxygen position). Upon cooling below 250 K, a $La_2CuO_{4.03}$ sample exhibited a phase separation into $La_2CuO_{4.01}$ and $La_2CuO_{4.06}$, with the latter showing 33 K superconductivity [120], indicating a miscibility gap between them [121]. This finding demonstrates the high mobility of interstitial oxygens even at low temperatures. High oxygen pressure annealing yielded a maximum $\delta$ of 0.13 [119]. Excess oxygens are found in B4-NC block layers (Fig. 17f), which contain $Bi_2O_{2+\delta}$ and $Tl_2O_{2+\delta}$, similar to the $La_2O_{2+\delta}$ block layer. Although the maximum $\delta$ values are unknown, they appear to be higher than in $La_2O_{2+\delta}$; 0.29 and 0.20 for $Bi_2O_{2+\delta}$ (Bi2212 (ARPES)) and $Tl_2O_{2+\delta}$ (Tl2201), respectively, if the maximum hole concentration in the $T_c$–$p$ phase diagram in Fig. 25 is only originated from excess oxygen. Thus, the hole supply is limited to approximately 0.3 and 0.4–0.6 for the B2-NC and B4-NC block layers, respectively.

B3-NC (Fig. 17d) and B3-PV (Fig. 17e) accept a large number of excess oxide ions at normal sites in the middle Hg,

Tl, Pb, or Cu sheets, allowing for a wide range of hole supply to the $CuO_2$ planes. The former Hg-based B3-NC $Ba_2HgO_{2+\delta}$ block layer, with a BaO–$HgO_\delta$–BaO stacking, can have a $\delta$ value of 0.4 [122-124]. The maximum $\delta$ appears to be much larger than in the above B2 and B4 block layers. Nevertheless, Coulomb repulsion between oxide ions likely prevents $\delta$ from exceeding 0.5. Consequently, B3-NC can provide the conduction layer with a maximum of one hole, meaning that each $CuO_2$ plane in C3 has an average of 0.33 holes. In the Y123 B3-PV $Ba_2CuO_{2+\delta}$ block layer, $\delta$ ranges from 0 to 1, with the latter corresponding to half occupancy. All Cu in the conduction and blocking layers is formally divalent at $\delta = 0.5$. As $\delta$ approaches 1, excess oxygen atoms are regularly arranged to form CuO chains (Fig. 14b). The $T_c$–$\delta$ relationship is complex because the hole partitioning to the two types of Cu depends on the amount of excess oxygen and chain formations, which change the chain Cu valence. Alternate chain formation at $\delta = 0.5$ increases complexity. This results in an intriguing structure–property relationship [95, 125]. In contrast, many systems, such as $Bi_2Sr_{2-x}La_xCuO_{6+\delta}$ in C1-B4-NC Bi2201 [126] and (Pb, Cu)$Sr_2$(Y, Ca)$Cu_2O_{7-\delta}$ in C2-B3-NC Pb1212 [127], have used both excess oxygen introduction and element substitution to adjust hole concentration over a wide range, resulting in increased complexity.

In contrast to the above hole-doping systems, the B2-CF block layer in Fig. 17c only supports electron doping. For example, electrons are generated by replacing $Nd^{3+}$ ions in the $Nd_2O_2$ layer with $Ce^{4+}$ ions, which flow into the $CuO_2$ plane and cause electron-doped superconductivity (Section 4.8.1). The B2-NC and B2-CF block layers have vastly different in-plane dimensions: the former is small and compresses the adjacent $CuO_2$ planes, whereas the latter is large and pulls them. Only hole (electron) doping is possible because injected holes (electrons) increase (decrease) Cu's formal valence, shortening (stretching) the Cu–O distance in the B2-NC (B2-CF) block layer. The B1 block layer of a single $Sr^{2+}$ sheet in Fig. 17a does not have a size matching constraint and can generate electron carriers using $Nd^{3+}$ or $La^{3+}$ ions.

#### 4.1.4. Complex chemical compositions and estimating hole concentration

Many material systems fail to produce ideal stoichiometric metal compositions. During the synthesis process, defects are formed, two metal atoms may partially swap positions, or unexpected elements may be introduced. Avoiding them in multi-element compounds is challenging. At high synthesis temperatures, the thermodynamically stable composition should always deviate from the stoichiometric ratio. Gibbs free energy determines a material's stability, while the entropy term in it increases randomness by introducing more defects or foreign elements as temperatures rise. Although low-temperature synthesis can partially suppress them, the presence of chemical reaction barriers and slow diffusion cause inhomogeneous sample formation, which poses a dilemma in material synthesis. Fortunately for superconductivity, Cu in the $CuO_2$ plane is difficult to contaminate unless other 3d transition metals are present, because it is much smaller than the other atoms and prefers the unique site environment of square planner oxygen coordination [109].

In addition to excess oxygen, the middle metal sheet in B3 block layers can contain defects and other elemental substitutions. Cu can replace Tl in Tl1 compounds [128], and carbon or $CO_3$ can replace Hg in Hg compounds [129], which are both difficult to avoid. $CO_3$ occupies all metal sites of C1-B3-NC $Sr_2CuO_2CO_3$ [130]. Moreover, Au or Fe can completely replace the middle metal site [131, 132]. On the other hand, in a multi-component phase diagram, preparing a single-phase sample may require partial substitution of other components. For example, Pb substitution enables the creation of a high-quality Bi2223 sample [133], which contains five elements (Bi, Sr, Ca, Cu, and Pb) in addition to O; however, the composition is nearly impossible to control and determine experimentally.

The relationship between $T_c$ and $p$ is critical for understanding the superconductivity mechanism, which will be covered in the following sections. The complex chemical composition makes determining $p$ from the charge neutral condition difficult. Even when absolute $p$ values are difficult to calculate, relative $p$ values and the resulting $T_c$ variation can be reliably calculated (see Section 4.3.3 and Fig. 24). Solid state chemists' goals are to use chemical techniques to prepare high-quality samples, which are then characterized to obtain a consistent $T_c$–$p$ relationship. On the other hand, physical quantity measurements, with the exception of Cu NMR and ARPES data, such as the Hall coefficient, are frequently unreliable and provide limited information (Section 4.4.5). We will carefully examine the data gathered thus far and use it to develop a comprehensive understanding of copper oxide superconductivity.

### 4.2. Electronic state of the $CuO_2$ plane

#### 4.2.1. Mott (charge-transfer) insulator

The $Cu^{2+}$ ion in the parent phase has a $3d^9$ electron configuration. When six oxide ions coordinate, the $CuO_6$ octahedron stretches uniaxially due to the strong Jahn–Teller effect. As a result, one unpaired electron resides in the uppermost, nondegenerate $dx^2$–$y^2$ orbital state (Fig. 18). The $dx^2$–$y^2$ orbital's lobe extends toward the four in-plane oxide ions $O_p$ and forms a strong σ covalent bond with their 2p orbitals. In contrast, the bond with $O_a$ at the apex is weak and most likely ionic. As discussed in Section 4.5.1, the presence of oxide ions with two distinct chemical bonding characteristics has significant implications for the superconductivity mechanism.

The $CuO_2$ plane is a square lattice of Cu($O_p$)$_4$ quadrangles that are connected by vertex sharing (Figs. 14e and 19). Cu $3dx^2$–$y^2$ orbitals strongly hybridize with $O_p$ 2p orbitals, and the wavefunction spreads across the plane, forming an extended d band. Because electrons occupy half of the d band, the mean-field band diagram predicts metallicity. However, due to the strong electron correlation $U$ of the spatially confined 3d electrons at half-filling, which is the constraint that a second electron entering the same site increases the energy by $U$, the electrons repel each other and cannot move, remaining at each site and forming insulators (Mott insulator; Fig. 4c). $U$ of several eV divides the d band into two narrow bands, the lower Hubbard band ($d_{LHB}$) and the upper Hubbard band ($d_{UHB}$), with the former completely occupied by electrons and the latter empty, creating a band gap. The $CuO_2$ plane is actually a charge-transfer insulator, rather than a Mott insulator, with an energy gap between $d_{LHB}$ and $d_{UHB}$ [28]. The broad 2p band of oxygen is located just between the above Mott-split d bands, leaving an energy gap between O 2p and $d_{UHB}$. Electron excitation occurs when a charge is transferred from oxygen to copper.

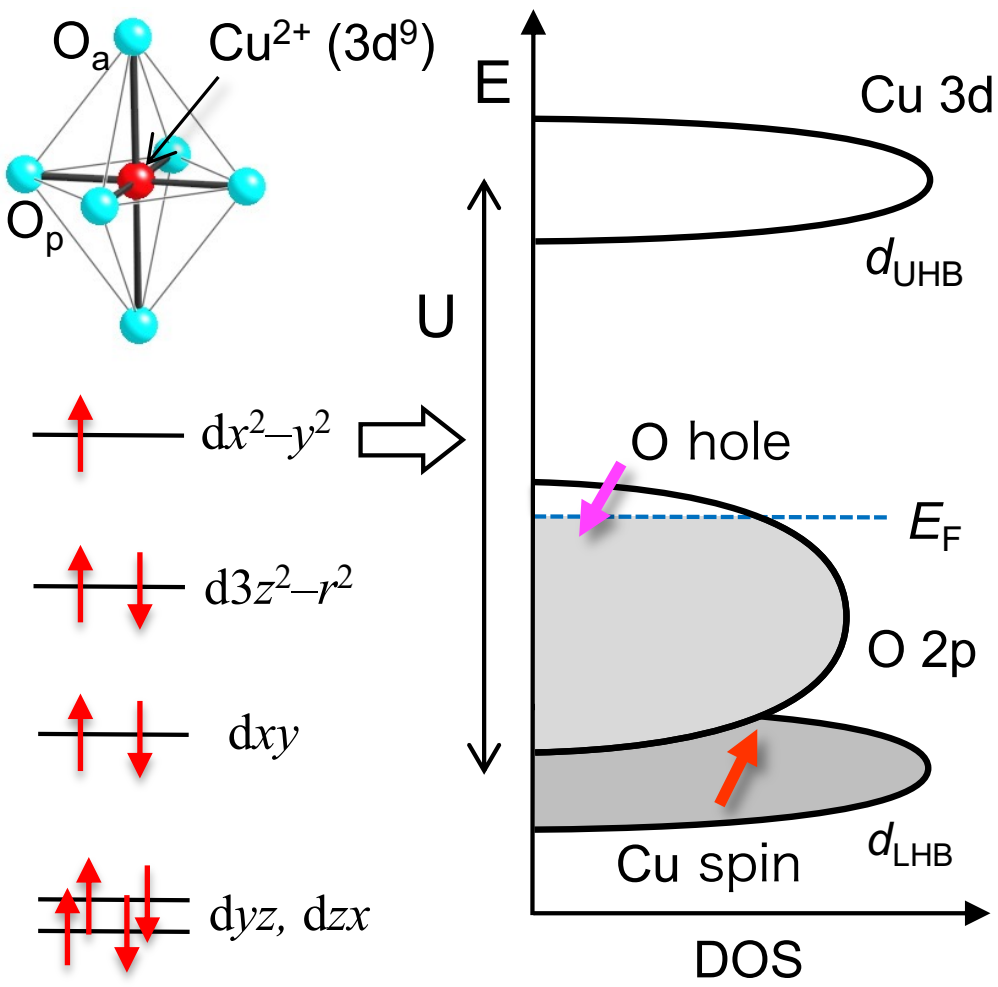

**Fig. 18.** Basic energy diagram of the $CuO_2$ plane in copper oxide superconductors. When a $Cu^{2+}$ ion with a $3d^9$ electron configuration is placed in an elongated oxygen octahedron composed of four in-plane $O_p$ atoms and two distant apical $O_a$ atoms, the unpaired electron occupies the highest d$x^2$–$y^2$ level, as shown to the left. When the d$x^2$–$y^2$ orbitals are connected to $O_p$'s 2p orbitals to form a square net in the $CuO_2$ plane, they form the extended band shown on the right. The electron correlation $U$ divides the resulting d$x^2$–$y^2$ band into two narrow bands: the lower Hubbard d band ($d_{LHB}$) and the upper Hubbard d band ($d_{UHB}$), with the former fully occupied by electrons carrying localized spins (red arrow) and the latter empty. A charge-transfer insulating gap forms between the broad, occupied $O_p$ 2p band and the empty $d_{UHB}$. A doped hole at the top of the $O_p$ band has an antiparallel spin (magenta arrow) that is tightly coupled with the Cu spin in $d_{LHB}$, resulting in a Zhang–Rice singlet (Fig. 19), which is responsible for superconductivity.

The $CuO_2$ plane acts as a 2D magnet on the square lattice because each electron in the fully occupied $d_{LHB}$ has one localized spin 1/2 (Fig. 19a). The oxygen-mediated covalent coupling of neighboring Cu spins generates a strong superexchange antiferromagnetic interaction $J$, which is estimated to be as large as 1500 K [134, 135]. Adjacent spins are forced to align up and down, but two-dimensional strong antiferromagnetic fluctuations prevent long-range order. Actual compounds with stacked $CuO_2$ planes, on the other hand, eventually transition to antiferromagnetic long-range order when in-plane correlations increase sufficiently, aided by minimal interplane coupling. The transition temperature ($T_N$) is approximately 300 K [136], which is reduced to only 20% $J$. This is obviously due to the fluctuations that are characteristic of low-dimensional systems.

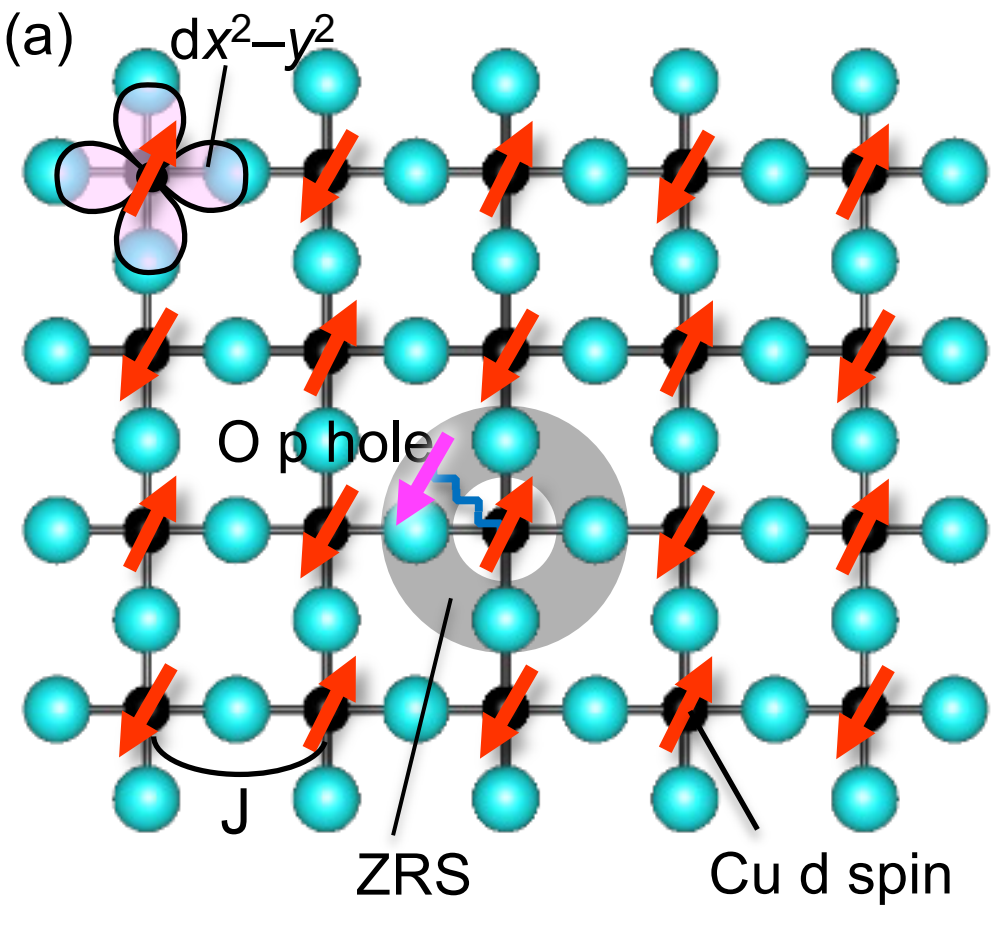

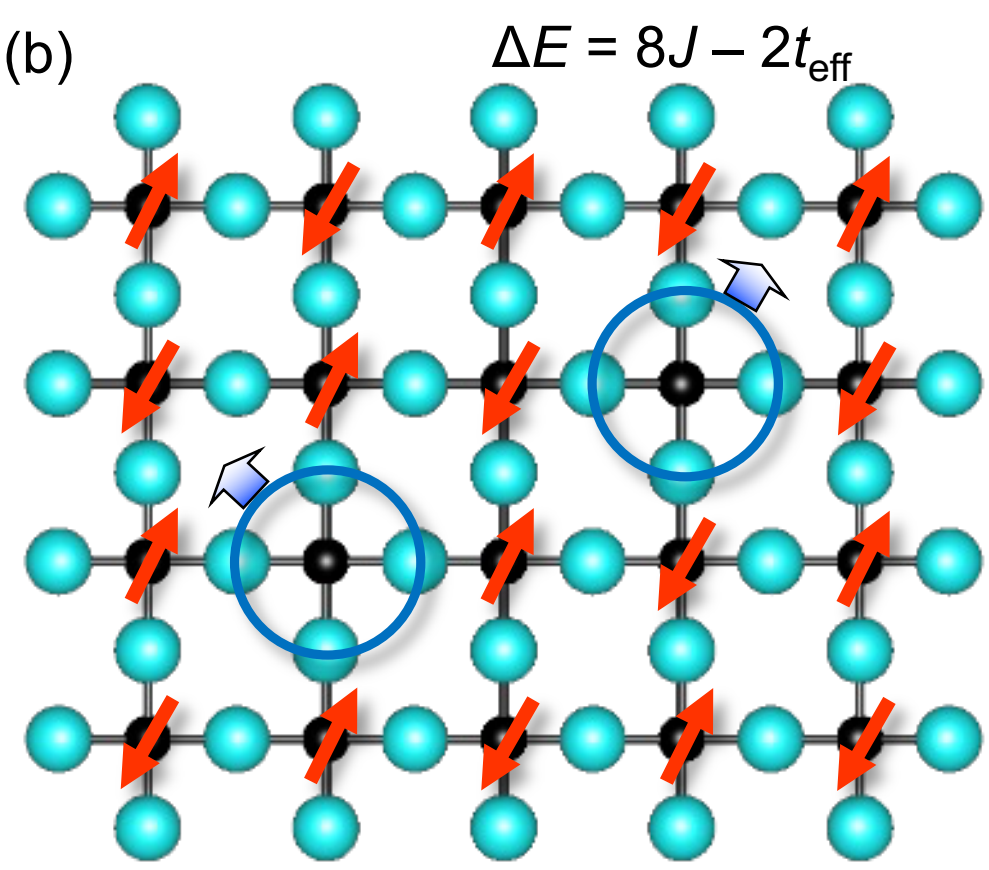

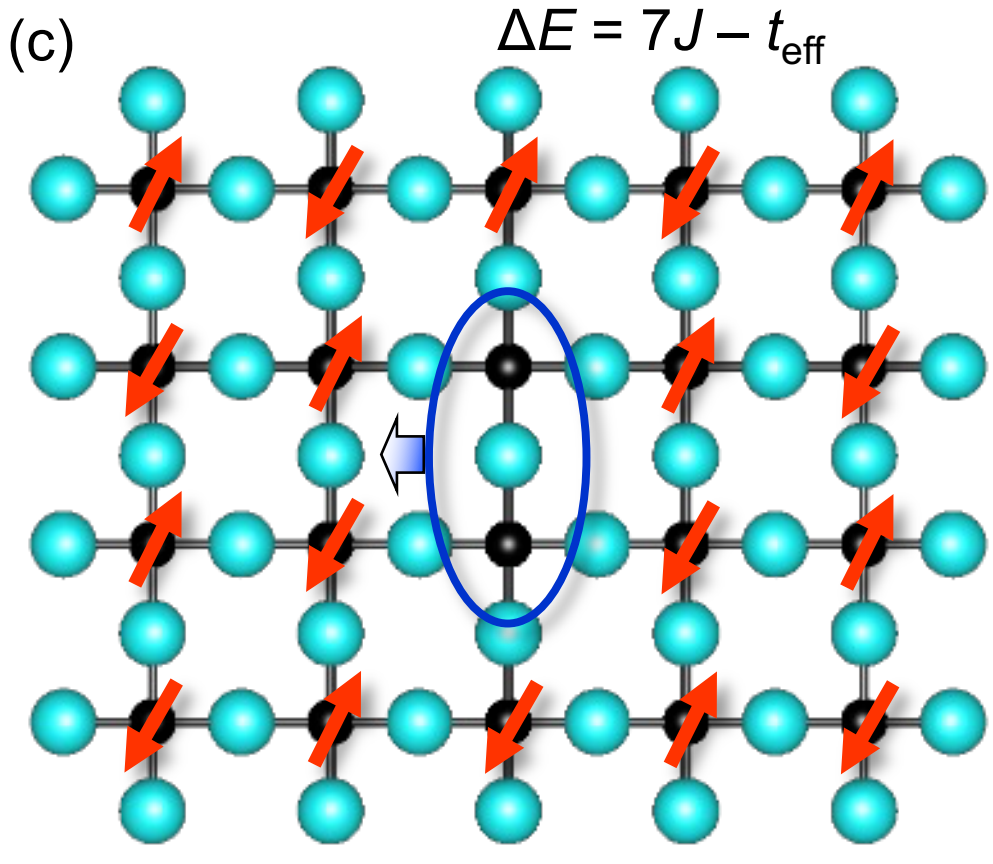

**Fig. 19.** Schematic representations of the $CuO_2$ plane with Cu spins in the d$x^2$–$y^2$ orbital that are coupled together by the antiferromagnetic interaction $J$ and arranged in antiferromagnetic order, as well as what happens when holes are introduced. (a) A doped hole on the O 2p orbital encircles a Cu spin. The O hole spin (magenta arrow) has a stronger antiferromagnetic interaction with the Cu 3d spin (red arrow) to form a Zhang–Rice singlet (ZRS). (b) Two ZRSs (blue circles) move independently in a metallic state, each losing 4$J$ bonds and gaining kinetic energy $t_{eff}$ in comparison to the insulating magnetic state. The total energy change ($\Delta E$) is $8J - 2t_{eff}$. (c) When two ZRSs are paired at nearby sites (blue oval), the magnetic energy loss is reduced to 7$J$, and the kinetic

energy gain is reduced to $t_{eff}$. Thus, when $J$ is greater than $t_{eff}$, the energy gain of $J - t_{eff}$ causes an effective attractive interaction between ZRSs, resulting in the formation of a ZRS pair, which then transforms into a Cooper pair in the superconducting state.

4.2.2. Hole doping in the $CuO_2$ plane

When a hole enters the $CuO_2$ plane, it creates a vacancy, allowing previously halted electrons to flow. In momentum space, the doped hole is located at the upper edge of the O 2p band, not the Cu 3d band (Fig. 18) [137]. As illustrated schematically in real space in Fig. 19a, an oxygen hole with a spin of 1/2 should surround a single Cu spin. Because the Cu 3d and O 2p orbitals overlap so much, the Cu spin–O spin interaction must be extremely strong antiferromagnetic direct exchange, much stronger than superexchange interactions between two nearby Cu spins. At interest temperatures, the two should be inseparable. The spins are coupled in opposite directions, producing a quantum mechanical singlet state known as the Zhang–Rice singlet (ZRS), named after two leading theorists, F. C. Zhang and T. M. Rice [138, 139]. Copper oxides' electrical conductivity is determined not by simple electron or hole carriers, but by ZRS, which combines charge and spin degrees of freedom and is unique to strongly correlated electron systems (Fig. 4).

Because of the formation of ZRSs, the Cu spin is completely obscured by the surrounding O hole spin and cannot be seen from the outside. As illustrated in Fig. 19b, "holes" are created in the spin 1/2 square net, allowing electrons or holes to hop. Thus, doped holes in the $CuO_2$ plane form ZRSs, resulting in metallic conduction. In addition, as described in Section 4.3.1, they combine to form pairs, resulting in BEC superconductivity (Fig. 19c). While ZRS is considered a simple "hole" in the first approximation, the superconducting mechanism requires it to be a singlet state composed of an oxygen p-hole spin and a copper d-electron spin.

4.2.3. Appearance of superconductivity with hole doping

Figure 20 depicts the electronic phase diagram of $La_{2-x}Sr_xCuO_4$ [140], a typical cuprate superconductor, with the characteristic temperatures for emerging phases or states plotted against hole concentration $p$, which is assumed to be equal to Sr content $x$. The parent phase, $La_2CuO_4$, is an antiferromagnetic insulator (AFI) with a $T_N$ of approximately 300 K. When $Sr^{2+}$ replaces $La^{3+}$, the $T_N$ drops rapidly as substitution and thus $p$ increases, and the AFI phase disappears around 0.02 [140-143]. Following suppression, at low temperatures, a spin glass (SG) phase forms, in which spins freeze in random orientations [144]. After additional doping, a superconducting phase (SC) appears at 0.05, and $T_c$ rises but peaks at $T_{co}$ = 39 K and the optimum hole concentration of $p_o$ = 0.16, before disappearing at $p_e$ = 0.26. The doping regimes to the left and right of $p_o$ are known as underdoped (UD) and overdoped (OD), respectively. The $T_c$ dips during the UD regime due to a competing stripy charge order, which stabilizes at $p$ = 1/8 (Section 4.8.2). Above $p_e$, the normal metallic state (even if superconducting, $T_c$ is below the experimental temperature range) persists in a broad $p$ range until 0.40 [145].

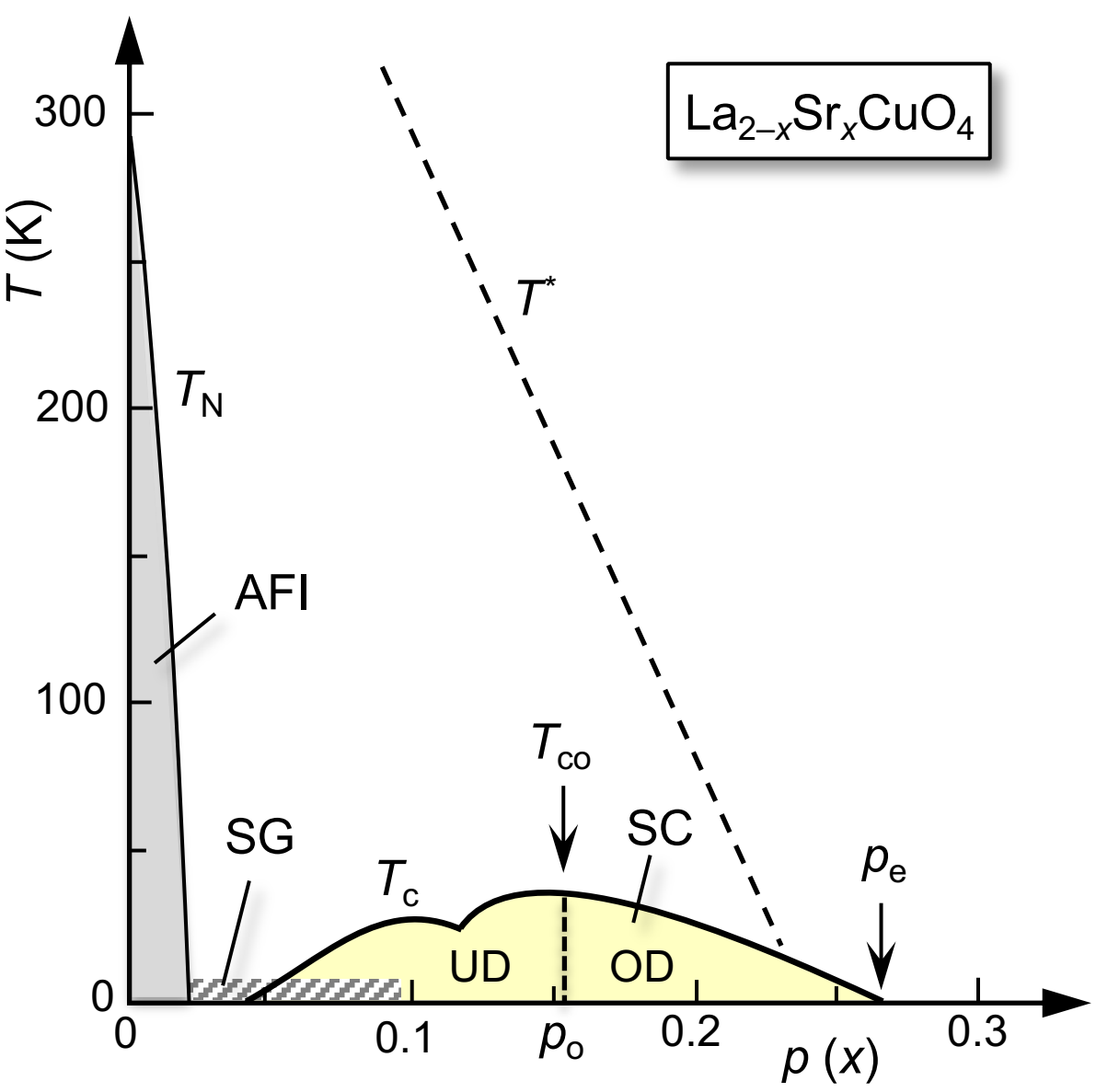

**Fig. 20.** $T$–$p$ phase diagram for La214, assuming $p = x$ [140]. At $p$ = 0.02–0.05, an antiferromagnetic insulator (AFI) phase is converted into a superconducting (SC) phase via a spin glass (SG) phase. $T_c$ reaches its highest point at $T_{co}$ = 39 K and $p_o$ = 0.16 before vanishing at $p_e$ = 0.26. The $T_c$ dome's left and right sides are referred to as the underdoped (UD) and overdoped (OD) regimes, respectively. The $T_c$ dome dip is caused by an electronic instability at $p$ = 1/8 (see Section 4.8.2). $T^*$ represents the temperature at which various measurements detect anomalies, also known as the pseudogap phenomenon.

When accounting for the differences in $T_{co}$ between compounds, it is thought that all cuprate superconductors form bell-shaped $T_c$ domes [143]; this will be discussed in Section 4.4.3. Understanding the superconductivity mechanism necessitates a proper interpretation of the $T_c$–$p$ relationship. Moreover, understanding material dependence is crucial for solid state chemistry. Except for La214, the vast majority of compounds have only a portion of the $T_c$ dome that can be experimentally accessed. This is because the block layer's chemical capacity restricts the amount of controllable hole doping (Section 4.1.3). The B4-NC block layer accepts more excess oxygens than the B2-NC type, while the B3 block layer accepts even more, increasing hole donation capacity ($p_B$) and expanding the $p$ range. Notably, the conduction layer in C2 and C3 compounds with high $T_c$s contains multiple $CuO_2$ planes, requiring a block layer with two and three times the $p_B$, respectively, to keep the $p$ constant across each plane. As $n$ increases, the controllable hole range shrinks and shifts to the low-doping side, even within the same block layer. Fortunately, we already have a large number of compounds and experimental data on them, allowing us to confidently construct a true $T_c$–$p$ relationship with careful consideration.

4.3. Superconductivity mechanism

Several superconductivity mechanisms have been proposed and debated thus far, but no clear consensus has emerged [104]. The typical phase diagram in Fig. 20 illustrates two approaches to the $T_c$ dome: from the left [101, 139, 146] and from the right [147-149]. The former emphasizes the $CuO_2$ plane's Mott (charge transfer) insulator properties while looking into superconductivity caused by hole doping. The latter accounts for superconductivity induced by spin fluctuations, which begin in a heavily doped normal metal and grow with decreasing $p$ as the material approaches a leftward antiferromagnetic order. As is always the case in physics, the question is where to place the approximation's starting point; however, reality is always in the middle, and once climbed, you will arrive at the same summit. In this manuscript, we will climb the $T_c$ dome from the left, starting with a doped Mott insulator, to create an appealing real-space pairing image for chemists.

4.3.1 Cooper pairing's driving force

The Cu spins on the undoped $CuO_2$ plane are antiferromagnetically arranged over a long distance, as depicted in Fig. 19a. Adjacent spins have antiparallel directions on average, and each spin is either thermally at a finite temperature below $T_N$ or quantum mechanically fluctuating at absolute zero, all while preserving long-range order. When an oxygen-site hole surrounds a Cu spin, it forms a ZRS (Fig. 19a). The four antiferromagnetic bonds that had previously existed around the Cu spin are then broken, resulting in a loss of $4J$ magnetic energy. A ZRS, on the other hand, can travel to a nearby site and acquire kinetic energy. Adding a second ZRS causes an $8J$ loss of magnetic energy (Fig. 19b). When two ZRSs are placed next to one another and moved together, they lose only $7J$ of magnetic energy (Fig. 19c). In other words, real-space pairing of ZRSs yields a net energy gain of $J$. However, the kinetic energy of paired holes is not doubled; instead, it remains at the level of a single hole, resulting in a deficit. As a result, real-space pairing is feasible if the magnetic energy gain outweighs the kinetic energy loss.

The kinetic energy of a ZRS is unknown, while $J$ equals 1500 K. The basic $t$–$J$ model employs the electron transfer integral $t$, which is estimated to be 4500 K [139]. The mean-field theory predicts a kinetic energy (bandwidth) of $4t$ for four bonds, which is much higher than $J$. However, for SCES like cuprates, this estimate is overly simplistic. Moreover, the kinetic energy must vary with the number of holes or vacancies in the square net. As a result, one can anticipate significantly lower effective kinetic energy $t_{eff}$, especially in the low-doped regime. On the other hand, if only $J$ were to be included in the above diagram, all doped holes would aggregate and form two distinct regions, one with holes and one without, resulting in electronic phase separation. Kinetic energy can prevent phase separation and keep ZRSs in pairs [150, 151]; trimer or larger clusters do not form because quantum mechanics prefers more stable pairs. The $t$–$J$ model predicts no phase separation at $J/t$ = 1500/4500 = 1/3 [139], but it may happen at lower $t_{eff}$. In fact, phase separation or its tendency has been observed in some copper oxide superconductors (refer to Section 4.5.2). Superconductivity occurs when $t_{eff}$ is moderately smaller than $J$, and phase separation may occur at even lower $t_{eff}$. It should be noted, however, that the observed phase separation is always accompanied by randomness effects, so it may not exist in the clean limit (Section 4.8.2).

The pairing picture in Fig. 19 assumes a rigid antiferromagnetic order, but hole doping degrades the actual LRO. At higher temperatures and hole doping levels above and to the right of the $T_N$ line, each spin's orientation is expected to vary over time and place, resulting in a zero order parameter [152]. Regardless, the discussion above may be valid. The background is made up of antiferromagnetically fluctuating copper spins. In the "sea" of fluctuating Cu spins, tightly bound oxygen hole (ZRS) pairs travel as bosons below $T_p$ and transform into Cooper pairs below $T_c$ [153]. This simple pairing diagram for copper oxide superconductivity has been referenced in several textbooks and reviews [154], but the author is unsure who created it. It is closely related to Dagotto's spin ladder superconductivity mechanism, with a comparable pairing image proposed (Section 5.3.3; Fig. 44) [155, 156]. The author wonders if Dagotto and his colleagues have applied their hypothesis to the cuprate superconductivity mechanism. Even though it is unclear how such an intuitive picture can represent the true mechanism, it gives the impression that we have understood, which is vital for chemists.

Equation 3, which calculates $T_c$ for weak-coupling superconductivity, does not necessarily apply to extremely strong-coupling superconductivity. However, we consider it after translating it with $\omega_o = J$ into simplified Eq. 5:

$$T_c = J\exp(-1/\lambda). \qquad \text{Eq. 5}$$

With $J$ set to 1500 K, the current maximum $T_c$ of 135 K corresponds to an exponential term of 0.09. In addition, the $T_c$ of 153 K at high pressures [32, 33] yields 0.10. These findings are strikingly similar to the maximum exponential term of 0.1 evaluated for the A15 compound using strong-coupling BCS theory [3]. Although this is a wild guess, it is plausible to believe that the empirically determined maximum value of $T_c$ is approximately 10% of the glue's energy. Provided that the maximum $T_c$ expected is 150 K, the most important question to ask when studying the material dependence of $T_c$ is not why $T_c$ is high in cuprate superconductors, but rather why $T_c$ is low in most of them when compared to the ideal value, and what material factors contribute to this reduction. This is the starting point for discussing $T_c$ in copper oxides, which will be discussed in detail in Section 4.5.

4.3.2 Doping dependence

The preceding discussion focused on the UD regime, which has a crossover associated with the formation of ZRS pairs at significantly higher temperatures ($T_p$ in Fig. 21) than $T_c$. Previous research has found corresponding anomalies, such as $T^*$ in Fig. 20, which indicate pseudogap formation [157] (Section 4.8.2.2). As the temperature falls in the UD regime of Fig. 21, tiny preformed pairs form at $T_p$, and their wavefunctions expand, causing them to overlap and be in phase at the BEC temperature $T_B$. As a result, BEC superconductivity emerges at $T_c \sim T_B$ [45]. Because 2D $T_B$ is proportional to the number of particles, or in this case the number of ZRS pairs ($p$/2) (Section 2.4.2), $T_c$ should rise proportionally to $p$. In practice, however, superconductivity does not develop until $p$ exceeds a specific threshold, and the $T_c$–$p$ relation is far from linearity, as shown for La214 in Fig. 20. This deviation is attributed to the randomness of real materials [87, 89], which will be discussed in Section 4.5.2.

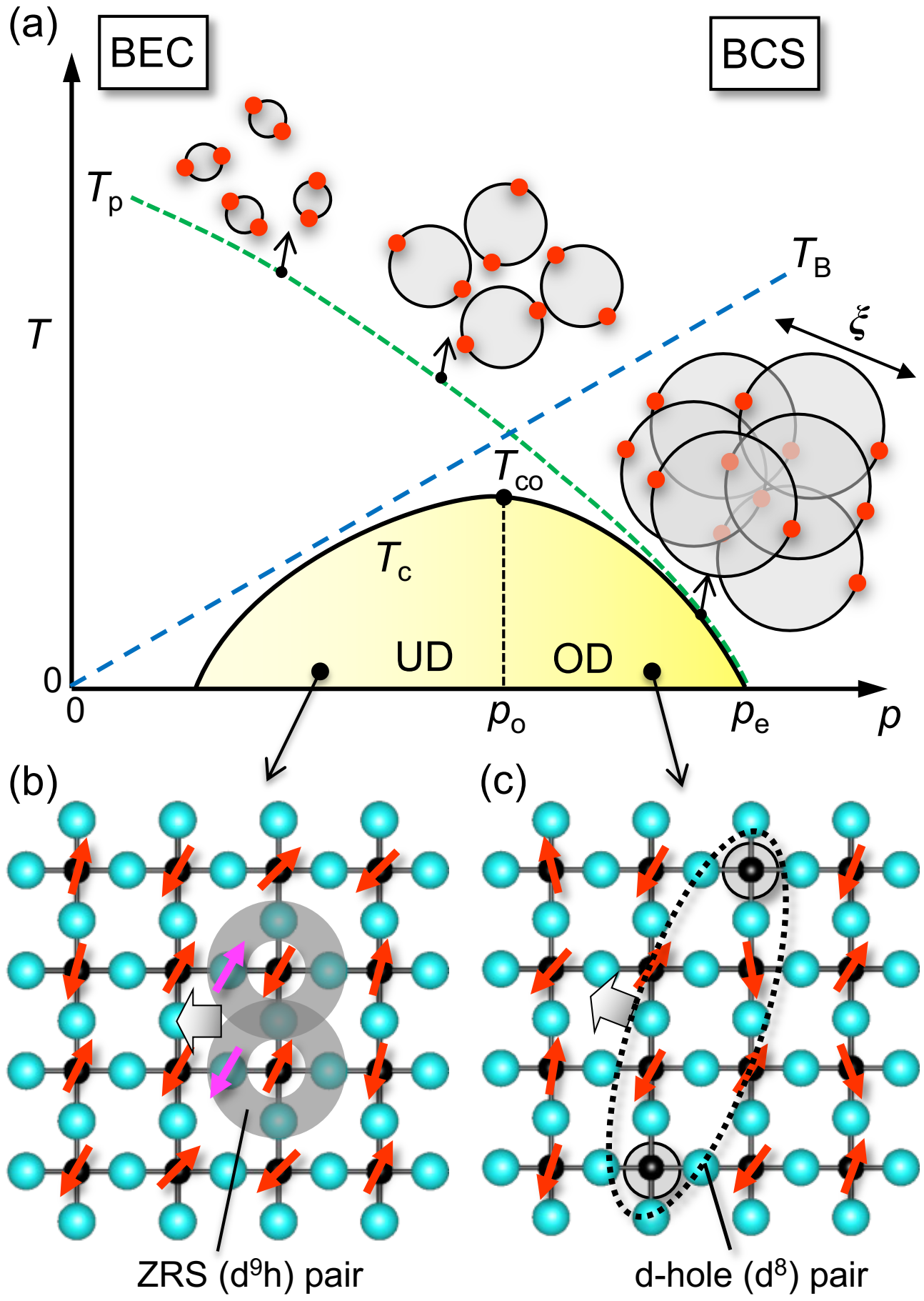

**Fig. 21.** (a) $T$–$p$ phase diagram and Cooper pairing for copper oxide superconductivity based on the BCS–BEC crossover in cold atom gas systems, as shown in Fig. 10. In the underdoped (UD) regime, to the left of the optimum hole concentration $p_o$, a large pairing interaction caused by a fluctuating but relatively rigid antiferromagnetic spin background keeps $T_p$ elevated. Preformed pairs form at $T_p$ during cooling, followed by BEC at $T_B$; a preformed pair is represented by a circle with two electrons (red balls); they are out of phase above $T_B$, as indicated by their random orientation, but in phase below it. BEC superconductivity occurs in real space below $T_c \sim T_B$. A small ZRS ($d^9h$) pair is expected just below $T_p$, as depicted in (b), but an actual Cooper pair below $T_c$ can be longer, measuring around 5–7 unit-cell length ($\xi = 2$–$3$ nm) in the plane. In the overdoped (OD) regime to the right of $p_o$, the antiferromagnetic spin background becomes weaker and diluted, making pairing interactions less effective and decreasing $T_p$. BCS superconductivity occurs in momentum space when larger d-hole ($d^8$) pairs, as illustrated in (c), form below $T_p$ and immediately overlap to one another to be in phase below $T_c \sim T_p$, producing Cooper pairs. Higher doping above the end $p_e$ suppresses superconductivity, leaving a normal metal state (Fermi liquid) with unpaired $d$ holes moving in a paramagnetic background.

When $p$ is increased further, $T_c$ deviates from the $T_B$ line, peaks at $T_{co}$ at $p_o$, and then begins to follow the dropping $T_p$ curve until it disappears at $p_e$. As the antiferromagnetic spin background decays with increasing $p$, reducing the effective attractive interaction for pairing, $T_p$ falls and becomes less than $T_B$ above around $p_o$. Thus, $T_c$ may scale with $T_p$. Neutron scattering experiments on OD La214 samples confirmed that $T_c$ is proportional to dynamic spin susceptibility, which is determined by $T_p$ [152]. In the UD regime, $T_c$ scales with $T_B$, so this relationship is not expected.

In contrast to a small ZRS pair in the UD regime, a weak attraction interaction in the OD regime generates a relatively large hole pair. When pairs form, their wavefunctions immediately overlap, resulting in phase coherence and superconductivity at the same time: this is unquestionably the picture of BCS superconductivity. The weak-coupling BCS superconductivity in the OD regime must be due to Cooper pairing in momentum space, as in the phonon mechanism shown in Fig. 7, rather than in real space, as shown in Fig. 19. The first hole induces an antiferromagnetically aligned Cu spin region in the paramagnetic background, which attracts the second hole before relaxing. This virtual process causes an effective weak pairing interaction between the two holes.

Because the attraction decreases with increasing $p$, the $T$–$p$ phase diagram of cuprate superconductivity can be interpreted as a left-to-right inversion of Fig. 10 for the BCS–BEC crossover [45]. Their distinction is that the particle number in the cold atom gas system is constant, whereas in cuprate superconductivity it varies. Consequently, a $T_c$ dome similar to that depicted in Fig. 21 appears, with rising $T_B$ determining $T_c$ in the UD regime, and dropping $T_p$ determining $T_c$ in the OD regime.

When a normal metal approaches, conducting carriers transform from ZRS, as imaged in real space, to d hole, also known as a band hole in momentum space. Above the $p_e$, independent band d holes form a non-superconducting Fermi liquid. According to the band diagram in Fig. 18, as $p$ increases and the Fermi level decreases, the Cu-d state's contribution to the carrier wavefunction grows. Holes will begin to occupy the $d_{LHB}$ state instead of the 2p state, reducing the spin component and destabilizing ZRS. Furthermore, as $p$ increases, the apical oxygen effect on the $CuO_2$ plane causes an increase in d–p hybridization, as covered in Section 4.5.1. As a result, rather than ZRS, conduction in the highly doped regime is governed by "d holes", which are strictly d–p holes with enhanced d components but a lower spin degree of freedom or localization tendency. ARPES experiments in the normal state show that holes in the low-doped regime behave like particles with a small Fermi surface in momentum space, whereas in the high-doped regime, they become extended band holes with a larger Fermi surface [53]. Doping should cause the Cooper pair character to shift from ZRS ($d^9h$) pairs (Fig. 21b) to d-hole ($d^8$) pairs (Fig. 21c).

### 4.3.3 Cooper pair size and shape

This section will look into the size and shape of a Cooper pair, as previously mentioned in Chapter 2.4, for copper oxide superconductivity. In the BEC regime, as illustrated in Fig. 21b, the pairs have strongly coupled extrema just below $T_p$; however, as temperature falls, their wavefunctions expand according to the de Broglie wavelength and overlap at $T_B = T_c$. The in-plane area per Cooper pair ($\xi^2$) is calculated as $(2/p)a^2$ ($a = 0.39$ nm); for $p = 0.08$, $\xi = 5a \sim 2$ nm. Experiments using optimally doped samples yielded $\xi$ values of 2–3 nm, consistent the BEC superconductivity. Thus, while the actual Cooper pair is not as small as depicted in Fig. 21b, it is still 1–2 orders of magnitude smaller than those found in other superconductors, indicating a strong pairing attraction. In contrast, BCS superconductivity in the OD regime results from larger Cooper pairs with weak coupling.

In terms of Cooper pair shape, the symmetry of the d$x^2$–$y^2$ orbital prevents pair formation in the [1 1 0] and [1 –1 0] directions, as shown in Fig. 19. The overlap integral between adjacent d$x^2$–$y^2$ orbitals arranged in phase vanishes along them. Thus, the superconducting gap in momentum space is maximally open along [1 0 0] and [0 1 0], but vanishes along [1 1 0] and [1 –1 0] [158]. As a result, the Cooper pair and superconducting gap have d$x^2$–$y^2$ wave symmetry [159], as depicted in Fig. 11b. Given the underlying antiferromagnetic interaction, the Cooper pair has a singlet spin component: because nearby Cu spins align up–down, the oxygen holes that form ZRSs with them have down–up spins, creating a singlet pair. The singlet spin channel is compatible with d-wave superconductivity.

The d-wave nature of Cooper pairs reflects the electrons' essentially repulsive interaction, as described in Section 2.4.3. Repulsive electron correlation leads to electron localization and antiferromagnetic spin order in the parent phase, as well as antiferromagnetic fluctuations when doped with holes. Remarkably, it can also cause a strong attraction for conducting oxygen holes in the antiferromagnetic spin background, resulting in high-$T_c$ superconductivity; many-body effects are fascinating! A dilute electron gas system should exhibit similar repulsion-induced pairing (Section 5.4.4, Fig. 45c).

4.4. Notable experimental findings regarding $T_c$'s material dependence

Let us consider the material dependence of $T_c$ using the superconducting mechanism described in Section 4.3. Even if all cuprate superconductors had the same shape of $T$–$p$ phase diagram, $T_{co}$ varies significantly depending on the material. Why this is the case is the most important question for solid state chemists to address. Here, we begin by enumerating the relevant four experimental findings: the dependence of $T_{co}$ on the number of stacking $CuO_2$ planes ($n$) within the conduction layer in Section 4.4.1; Uemura's plot derived from the µSR experiments in Section 4.4.2; the $T_c$–$p$ relationship in Section 4.4.3; and the $T_{co}$–$p_o$ relationship in Section 4.4.4. The first $n$ dependence of $T_{co}$ is well known, but it lacks an appropriate explanation. Uemura's plot is also regarded as important, albeit it has yet to be fully integrated into the existing mechanisms. Sections 4.4.3 and 4.4.4 demonstrate that the parabolic $T_c$–$p$ relationship found in La214, which is centered on $p = 0.16$, does not apply to most other compounds, even when $T_c$ is normalized. Moreover, we look for a true $T_c$–$p$ relationship. With these experimental foundations in mind, Section 4.5 will explore the factors that determine $T_c$.

4.4.1. Correlation between $T_{co}$ and the number of $CuO_2$ planes

The first notable chemical trend is that $T_{co}$ varies with $n$ in the same manner across members of each material system, with the highest value always occurring at C3, as shown in Fig. 22 [96, 117, 160]. In the Hg series with the highest $T_{co}$, $T_{co}$ increases from 97 K (C1) to 127 K (C2) and 135 K (C3) before decreasing to 127 K (C4), 110 K (C5), 107 K (N6), and 103 K (N7). Moreover, until $n = 16$, $T_{co}$ remains nearly constant at 105 K (not shown in the figure) [160]. $T_{co}$ in the Tl2 series with $Tl_2O_{2+\delta}$ in the B4 block layer varies as 90 K (C1), 110 K (C2), 125 K (C3), and 116 K (C4), while in the Tl1 series with the B3 $Ba_2TlO_{3-\delta}$ block layer, $T_{co}$ varies as 45 K (C1), 85 K (C2), 133.5 K (C3), 127 K (C4), and 115 K (C5). The highest $T_{co}$ in both series is found at C3. Furthermore, in Cu series such as Cu1212, $T_{co}$ ranges from 90 K (C2) to 119 K (C3), 105 K (C4), and 90 K (C5), with C3 having the highest $T_{co}$ once again. As a result, we must find a reasonable explanation for the shared $n$ dependence and the highest $T_{co}$ in C3. The alternative question, as mentioned at the end of Section 4.3.1, is why $T_{co}$ decreases in C2 while decreasing even more in C1 and approaching a slightly reduced value as $n$ exceeds five.

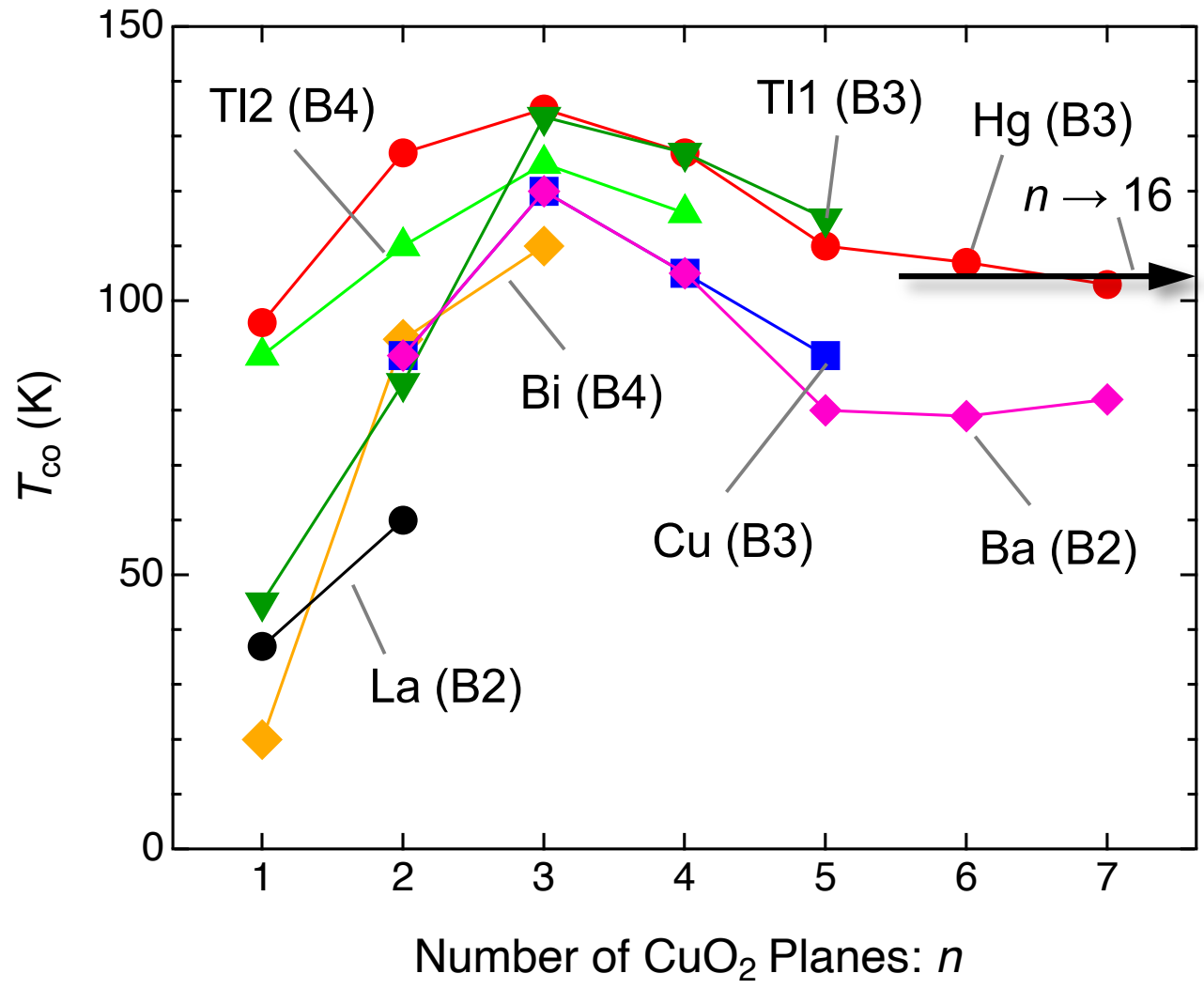

**Fig. 22.** $T_{co}$ versus $n$ plots for various compound series (Table 1) [96, 160, 161]. 'Hg', 'Bi', 'Cu', 'Ba', and 'La' refer to a group of compounds that typically contain Hg1201, Bi2201, Cu1212, Ba0212, and La214, respectively. 'Tl1' and 'Tl2' are Tl series with single (B3-NC) and double TlO sheets (B4-NC) in the block layers, such as Tl1201 and Tl2201, respectively. The $T_{co}$ for the Hg series remains nearly constant at around 105 K until a large number of $n = 16$ [160], as indicated by the arrow.

Another important feature of the chemical trend in Fig. 22 is that $T_{co}$ varies even for the same $n$ compounds. C1's $T_{co}$ varies greatly, with Hg, Tl2, La, and Bi series values of 97, 90, 39, and 25 K, respectively. There is a factor that reduces $T_{co}$, and its magnitude increases in this order. The difference is less for C2, and even less for C3, at 135, 125, 120, and 110 K for the Hg, Tl2, Cu, and Bi series, respectively. Then, as $n$ grows, similar differences remain. The material dependence of the reduction factor decreases with the order of C1, C2, and C3, but remains constant for larger $n$ values. Our goal is to figure out why C3 has the highest $T_{co}$ of any series, Hg1223 has the highest $T_{co}$ of any copper oxide superconductor, and C1 has a wide range of $T_{co}$.

Alternative physical explanations for observed chemical trends have been suggested. For example, three assumptions are made: Cooper pair quantum tunneling between stacking $CuO_2$ planes, an uneven hole concentration distribution per $CuO_2$ plane, and competition between superconductivity and secondary order [162]. Nonetheless, most chemical trends can be explained by more than two factors. Moreover, the story may not explain the scattering of $T_{co}$ in the same $n$ series. In this manuscript, we will use solid-state chemistry knowledge to better understand $T_{co}$'s $n$ dependence.

4.4.2. Uemura's plot

Uemura et al. conducted μSR experiments and discovered a significant correlation between $T_c$ and $p$ [163-165]. Figure 23 replicates Uemura's plot, with $T_c$ for various C1, C2, and C3 compounds on the vertical axis and zero-temperature extrapolation of the μSR relaxation rate on the horizontal axis. According to Uemura [163], the μSR relaxation rate is proportional to the superconducting carrier density ($n_s$) divided by effective carrier mass ($m^*$). The material dependence of $m^*$ may be negligible, as the conduction layers are composed of common $CuO_2$ planes with weak interplane couplings (Section 4.6.2) [166]. The horizontal axis must therefore scale in $n_s$. A similar relationship between $T_c$ and $n_s$ has been found in other superconductors, demonstrating its applicability [164].

Uemura's plot shows that $T_c$ values for C1, C2, and C3 increase in the low-doped regime along a common straight line beginning at $n_s$ = 0, but saturate and decline at different $n_s$ values in the high-doped regime. This implies that increased hole doping is responsible for the increase in $T_{co}$ from C1 to C3. In other words, there is a reason why $T_c$ deviates from the initial common slope after fewer $n_s$ in C1 and C2 than C3. This important finding can be explained by apical oxygen effects, as discussed in Section 4.5.1 [167], as well as the thickness of the superconducting layer (Section 4.7.2).

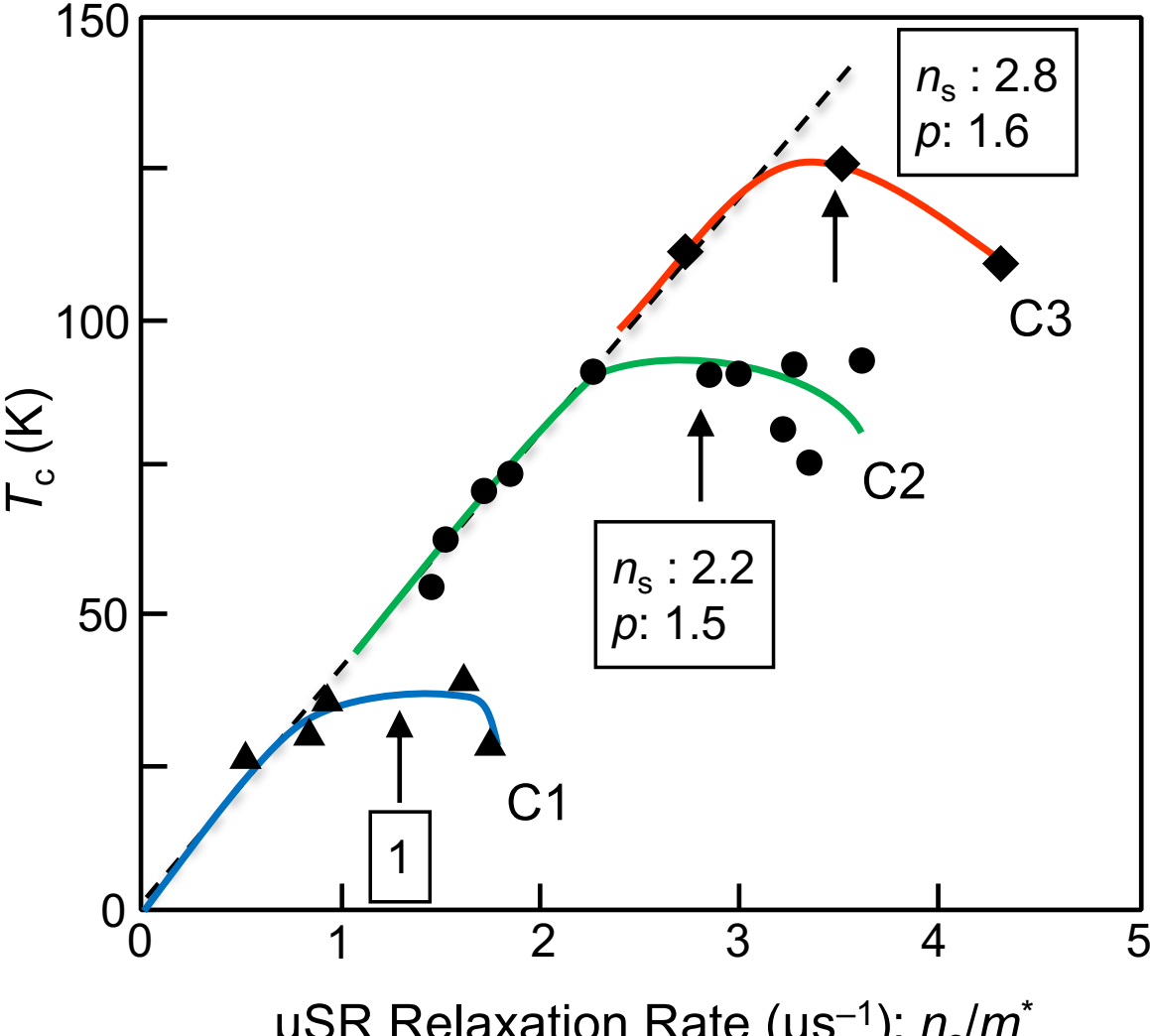

**Fig. 23.** Uemura's plot of the relationship between $T_c$ and μSR relaxation rates, extrapolated to zero temperatures. The latter scales to $n_s/m^*$, where $n_s$ and $m^*$ are superconducting carrier density per unit volume and effective carrier mass, respectively [163]. The arrows represent the estimated $T_c$ maximum positions for compounds C1 (triangles), C2 (circles), and C3 (diamonds). The boxes show how the $n_s$ values at the peak maximum differ from that of C1 for the same $m^*$. The corresponding $p$ values per Cu in the $CuO_2$ plane are also provided, calculated based on the crystal structures of the Hg series C1, C2, and C3 compounds, with uniform hole distributions across the $CuO_2$ planes.

The $n_s$ at the optimum doping yielding $T_{co}$ increases in the order of C1, C2, and C3, with C2 and C3 having $n_s$ values roughly 2.2 and 2.8 times greater than C1, respectively. It is worth noting that the number of carriers per $CuO_2$ plane ($p$) is not proportional to the number of carriers per unit volume ($n_s$), as the relative volume of $CuO_2$ layers increases with $n$. Given Hg series crystal structures and an even hole distribution across $CuO_2$ planes, the $p_o$ ratio is calculated using the $n_s$ ratio, which is 1, 1.5, and 1.6 for C1, C2, and C3, respectively. If the $p_o$ value for C1 is 0.16, then the $p_o$ values for C2 and C3 could be 0.24 and 0.26, respectively. As a result, both $p_o$ and $T_{co}$ increase with the number of $CuO_2$ planes.

4.4.3. $T_c$–$p$ relationship

Uemura's plot demonstrates that $p_o$ is not constant and varies with $n$. According to current consensus, however, all other systems have the same $p_o$ value of 0.16 as La214 [143, 168, 169]. Presland et al. proposed a parabolic $T_c$–$p$ equation as follows [143]:

$$T_c/T_{co} = 1 - 82.6(p - 0.16)^2. \qquad \text{Eq. 6}$$

This equation, developed by combining data from solid solutions containing variable alien metal atoms in the Bi and Tl systems with La214 data, is now widely accepted. Furthermore, in many cases where estimating the absolute value of $p$ is difficult, the equation has been used to calculate $p$ from experimental $T_c$ data. In some cases, the dependence of a physical quantity on hole concentration was studied using the $p$ values obtained.

It's unclear whether this relationship is universal. Presland's relative $p$ values for the Bi and Tl solid solutions were correct; however, their absolute values were calculated using the $T_c$–$p$ relationship for La214 rather than experimentally determined. Thus, there is insufficient experimental evidence to support the universality of Eq. 6. A previous study reported a non-parabolic relationship with a broad plateau at $0.12 < p < 0.25$, suggesting materials with $p_o > 0.20$ [167]. In this section, we will attempt to derive a reliable $T_c$–$p$ relationship for typical compounds.

Except for La214, almost all other cuprate superconductors have complicated structures, non-stoichiometry, and unexpected element substitutions (Section 4.1.4), making it difficult to estimate the formal charge based on chemical composition and calculate the hole concentration on the $CuO_2$ plane. As discussed in Section 4.4.5, several techniques for directly determining $p$ have been used, including chemical titrations, neutron diffraction-based crystal structure refinements, NMR, ARPES, Seebeck coefficient, and Hall coefficient measurements, but their reliability and applicability are limited. As a result, little is known about the $T_c$–$p$ relationship, which spans a wide range, including the top of the $T_c$ dome.

A sample's relative value of $p$ ($\Delta p$) can be accurately determined through systematic element substitution or changes in oxygen content. Shimakawa et al. carried out systematic oxygen extraction experiments on C1, C2, and C3 in the Tl2 system [170]. They annealed a sample in a reducing atmosphere at low temperatures to preserve its chemical compositions other than oxygen. Using as-grown samples as a baseline, a change in oxygen content was calculated by measuring sample weight loss during reduction. This enabled them to establish trustworthy $T_c$–$\Delta p$ relationships, with a single oxygen loss yielding two holes loss. Figure 24 reproduces their findings: The as-grown Tl2201 sample is not superconducting, but $T_c$ rises due to hole reduction and reaches saturation at 90 K with $\Delta p = -0.25$. The as-grown Tl2212 sample is a superconductor with a $T_c$ of 87 K that increases slightly with reduction. The as-grown Tl2223 is a superconductor with a $T_c$ of 115 K, which decreases slightly as $p$ decreases. Tl2201, Tl2212, and Tl2223 exhibit partially visible $T_c$ domes, with $\Delta p = 0$ at the right end, slightly right of the apex (OD), and slightly left of the apex (UD), respectively. Provided the common block layer's hole-donating ability, the shift in the $p$ range from OD to UD from C1 to C3 is reasonable.

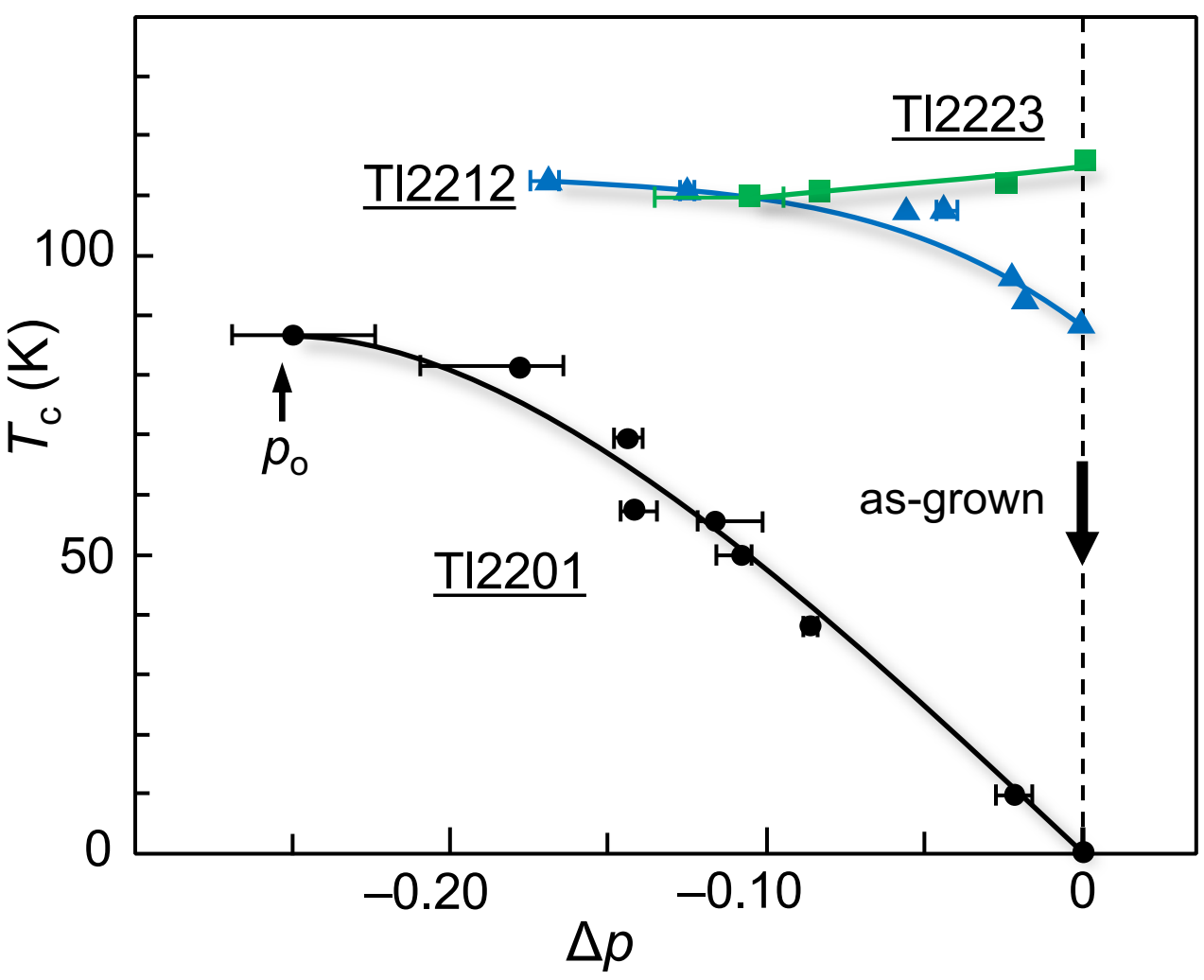

**Fig. 24.** $T_c$ variations with decreasing $p$ in the Tl2 series of compounds [170]. For each compound, $\Delta p$ represents the hole concentration in comparison to the as-grown sample prepared at 880–890 ºC in an oxygen atmosphere. To determine $p$ changes, oxygen loss was measured in weight during annealing at 350–600 ºC in an argon atmosphere. Tl2201, Tl2212, and Tl2223 have partially visible $T_c$ domes, with $\Delta p = 0$ at the right end, slightly right of the apex (OD), and slightly left of the apex (UD), respectively. Tl2201's half-$T_c$ dome is more than twice as large as La214's.

Figure 24 demonstrates that Tl2201 data in the OD regime, with $p_o$ at $\Delta p = -0.25$, spans almost the entire right half of the $T_c$ dome. This $\Delta p$ value is roughly twice as high as the corresponding value of 0.135 for the $T_c$ dome of La214. Equation 6's $T_c$–$p$ relationship is thus invalid, at least for Tl2201; if there is a general relationship, $p$ must be normalized. Figure 25a compares Tl2201 and La214, assuming that $\Delta p = 0$ corresponds to $p$ = 0.41 for Tl2201, based on NMR experiments: ($T_c$/K, $p$) = (72, 0.27), (42, 0.30), (0, 0.41) [171]. The $T_c$ dome of the Tl2201 appears to be larger than that of the La214, with higher $p$ and $T_c$ values.

Figure 25a depicts the $T_c$–$p$ relationship for additional compounds, where $p$ is determined using chemical titration, NMR, or ARPES experiments, which can yield relatively reliable absolute $p$ values, as described in Sections 4.4.5 and 4.6.2. Chemical titration reveals that Hg1212's $T_c$ dome is nearly parabolic, peaking at $p$ = 0.21 [129], while Bi2212's dome is asymmetrical, peaking slightly above 0.16 [126]; Bi's mixed valency effects are most likely responsible for the apparent extension to high doping levels. ARPES experiments on a Bi2212 crystal with varying $p$ via in-situ annealing revealed a $T_c$ parabola with a similar shape to La214, but with higher $T_{co}$ and a 0.02 shift toward high doping [172]. NMR Knight shift data indicate that the $T_c$–$p$ relationship peaks at $p$ = 0.21 for Y123 and its substituted systems [173], and at 0.23 for multilayered systems (Section 4.6.2) [166]. In contrast, a Bi2201 sample [(Bi, Pb)$_2$(Sr, La)$_2$CuO$_6$)] with a lower $T_{co}$ than the others has a narrower dome and a $p_o$ value of 0.12 [174]. As a result, La214's $T_c$ dome at $p_o$ = 0.16 is an exception, and $T_{co}$ rises with $p_o$, as predicted by Uemura's plot.

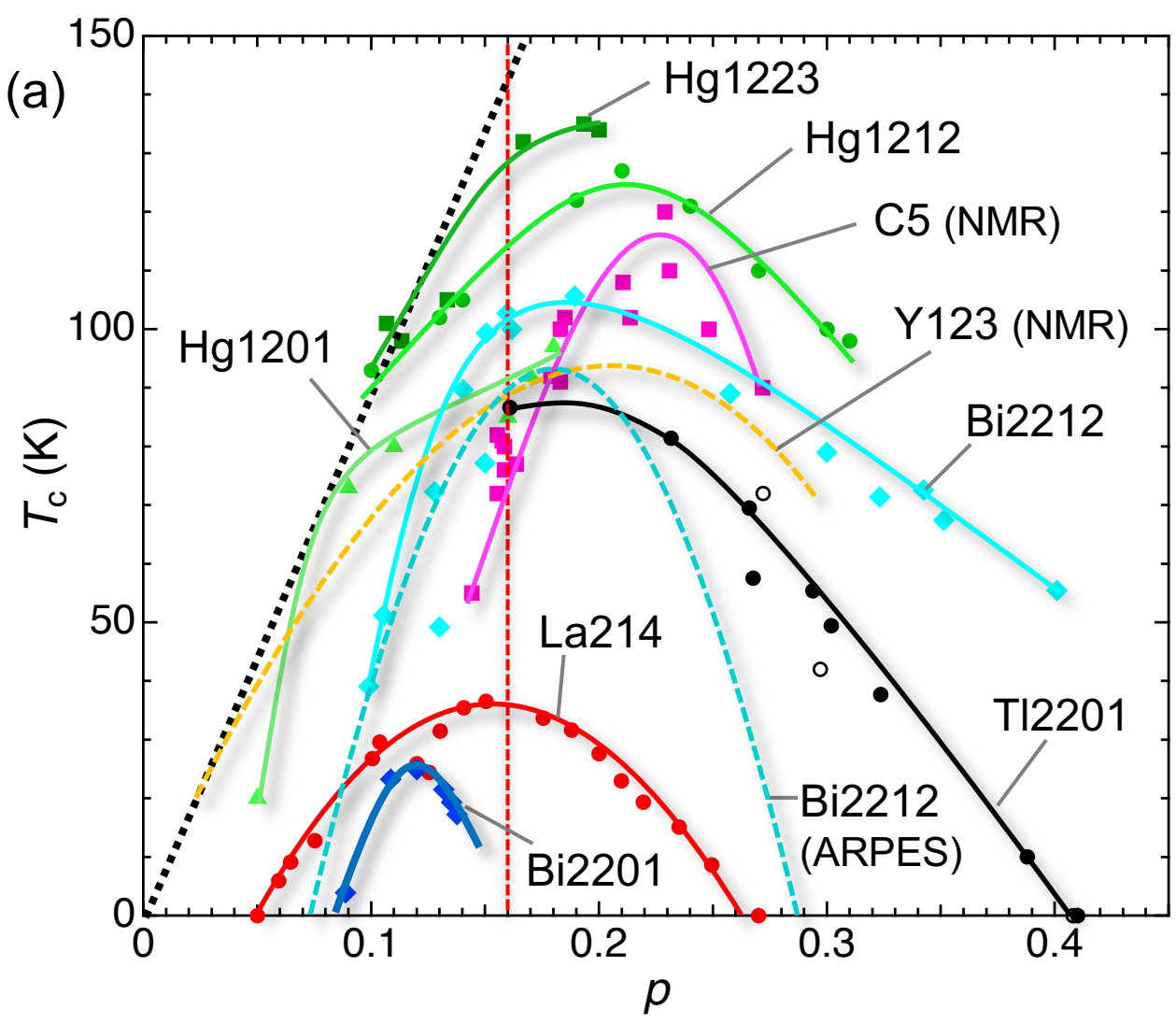

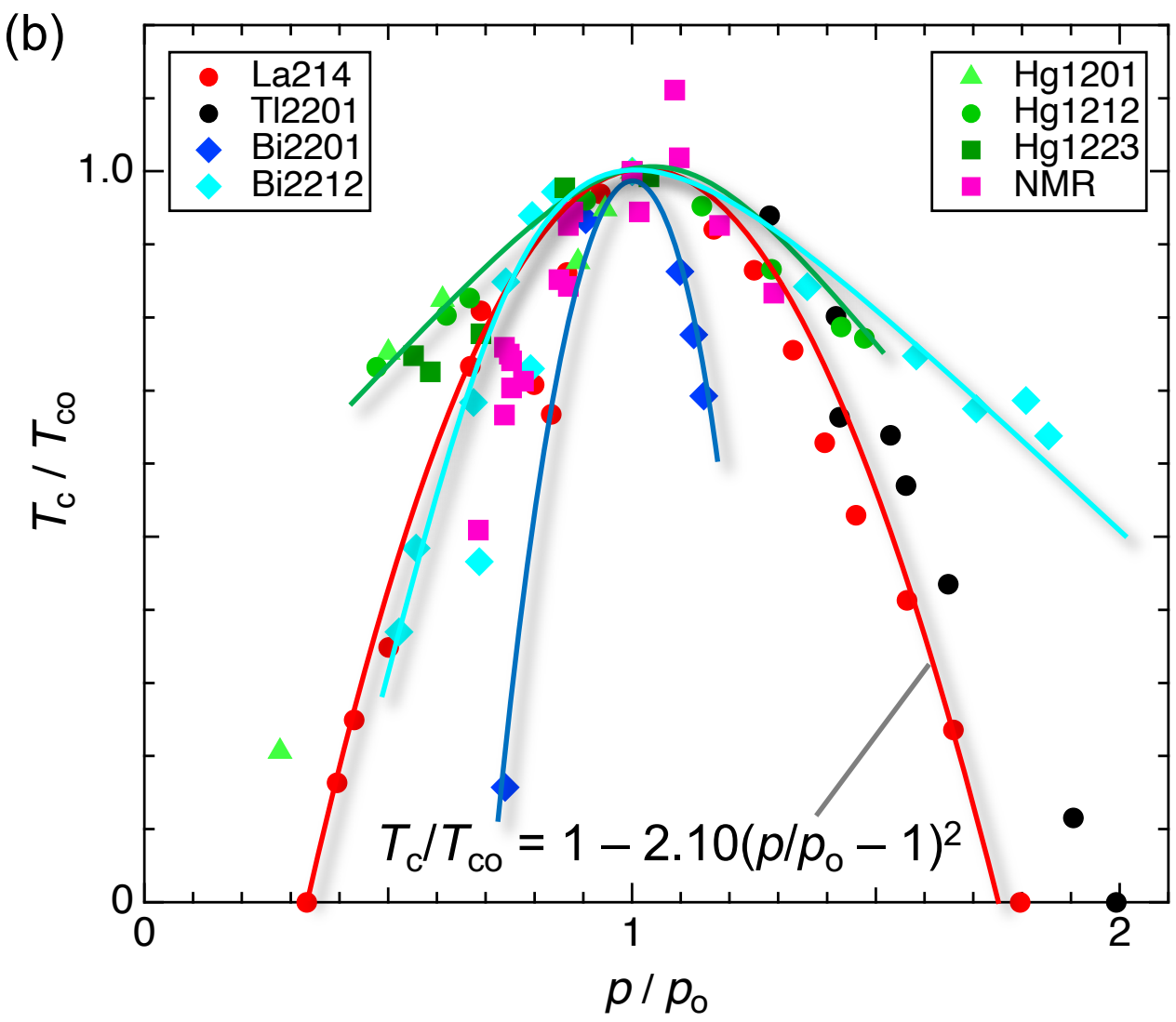

**Fig. 25.** (a) $T_c$ versus $p$ plot for selected compounds: La(Sr)214 with $p = x$ [140]; Bi2201 [174], Bi2212 [174], Hg1201 [175], Hg1212 [176], and Hg1223 [176] with $p$ determined by

chemical titration; Tl2201 [170], assuming that $\Delta p = 0$ in Fig. 24 corresponds to $p = 0.41$ based on NMR experiments that show $(T_c/\mathrm{K}, p) = (72, 0.27), (42, 0.30)$, and $(0, 0.41)$ (open circles) [171]. The 'C5 (NMR)' plot depicts the $T_c$–$p$ relationship derived from Cu NMR measurements for each $CuO_2$ plane in the C5 multilayered systems of the Hg and Ba series of compounds (Section 4.6.2; Fig. 34) [166]. The 'Y123 (NMR)' plot is also based on Cu NMR measurements, which selectively observe Cu in the $CuO_2$ plane [173]. The 'Bi2212 (ARPES)' plot uses the ARPES dataset [172]. The dotted line on the left side depicts a possible $T_B$ line with an 850 K slope (Fig. 37). (b) A normalized plot containing the majority of the data from (a). The red parabola fits the La214 data, disregarding the three points near the dip at around $p = 0.125$: $T_c/T_{co} = 1 - 2.10(p/p_o - 1)^2$, which corresponds to Presland's relation (Equation 6). Other $T_c$ dome curves are displayed for Bi2212, Hg1212, and Bi2201.

Figure 25b depicts the relationship between $T_c$ and $p$, normalized by $T_{co}$ and $p_o$ from Fig. 25a. The $T_c$ dome of La214 appears parabolic and follows this equation (which is the same as Presland's equation):

$$T_c/T_{co} = 1 - 2.04(p/p_o - 1)^2. \qquad \text{Eq. 7}$$

Some compounds happen to follow this relationship, but others do not. The low-doping region in the Hg series is wider, but Bi2201's dome is substantially smaller, demonstrating that superconductivity requires more holes to occur and vanishes with fewer holes. As a result, there is no universal relationship between $T_c/T_{co}$ and $p/p_o$, implying that additional material-dependent factors affecting $T_c$ should be explored. Given that the Hg series is relatively clean and Bi2201 is the dirtiest cuprate superconductor, their differences could be explained by randomness effects, as discussed in Section 4.5.2.

4.4.4. $T_{co}$–$p_o$ relationship

While there is limited information on the $T_c$–$p$ relationship, numerous experiments were performed on a single sample that was assumed to be representative of the system with the highest $T_c$. As a result, a wealth of information about the $T_{co}$–$p_o$ relationship has been compiled. However, we must be cautious because determining $p_o$ without first observing the $T_c$–$p$ relationship can result in significant errors. $T_c$ has little $p$-dependence near the $T_c$ dome apex, and shifting from $p_o$ to the UD or OD sides by the same amount produces comparable $T_c$s, so the $p_o$ value can vary significantly. In addition, in some studies, $p$ was calculated from $T_c$ using Eq. 6, rather than being determined experimentally. In this section, the author will gather as many dependable data points as possible in order to identify a potential chemical trend in $T_{co}$–$p_o$ relationship.

Figure 26 depicts the relationship between $T_{co}$ and $p_o$ for a variety of compounds. The aforementioned experimental uncertainty affects the determination of $p_o$; for Bi2212, it ranges from 0.17 to 0.27. Regardless of scatter, $T_{co}$ and $p_o$ in the Bi and Hg series tend to rise as $n$ increases. The remaining data also indicate that compounds with higher $T_{co}$ have larger $p_o$ values. Furthermore, $p_o$ clearly exceeds 0.20 at $T_{co}$ temperatures above 100 K. The observed trend is consistent with the Uemura plot prediction, with $p_o = 0.24$ and 0.26 for C2 and C3, respectively, assuming $p_o = 0.16$ for C1 (Fig. 23). We conclude that Presland's relationship is not universal, and $T_{co}$ and $p_o$ are positively correlated.

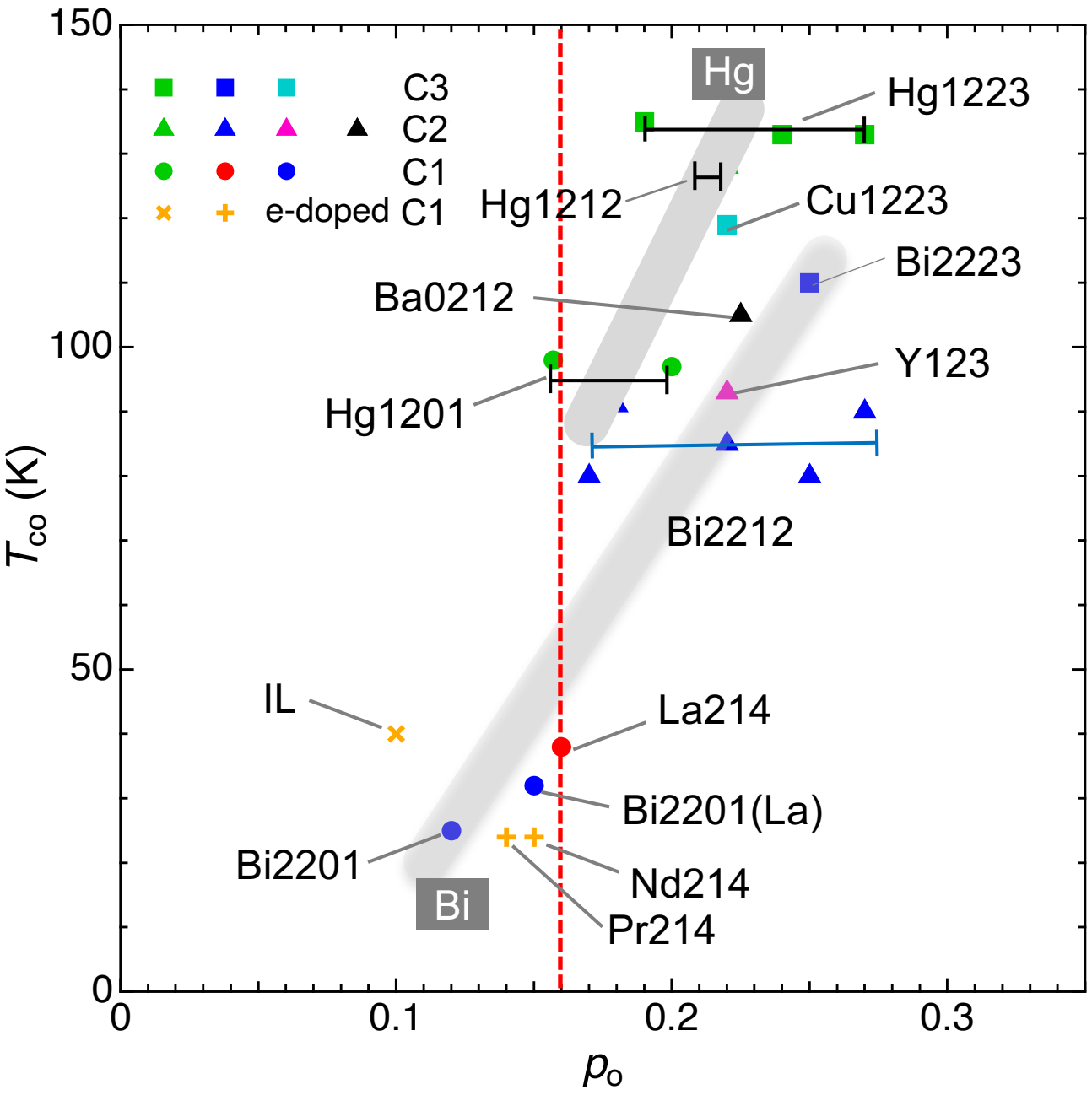

**Fig. 26.** $T_{co}$ versus $p_o$ plot. The data is presented in Table 1. The marks' shapes distinguish C1 (circle), C2 (triangle), and C3 (square), while the colors distinguish the material series. The plus and cross signs represent electron-doped C1 compounds. The error bar shows the variability in $p_o$ between studies. The thick grey lines highlight chemical trends in the Hg and Bi systems.

4.4.5. Experiments for $p$ estimation and additional information on the $T_{co}$–$p_o$ relationship

We examine the experimental techniques for $p$ estimation mentioned in the previous section, as well as other methods, and go over the $T_{co}$–$p_o$ relationship in more detail. Readers who are not interested can skip this section and move on to the next, which discusses the factors that determine the $T_c$ of cuprate superconductors.

Chemical titration (CT) using the oxidation–reduction reaction for Cu valence analysis is a reliable method for determining $p$ [174, 177]. Excess potassium iodide solution reacts with a single-phase copper oxide sample that has been resolved in acid to form precipitate copper iodite CuI and iodine $I_2$. Titrating the resulting iodine with a known concentration of sodium thiosulfate standard solution allows for precise determination of the copper ion's initial valence, $2 + p$. CT estimated the copper valence of a $La_{1.85}Sr_{0.15}CuO_4$ sample to be 2.140, which is close to the nominal $x$ value [178]. As previously mentioned, La-substituted Bi2201 with a low $T_{co}$ of 25 K has a CT-estimated $p_o$ of 0.12, which is significantly lower than 0.16 for La214 [126, 174]. Rao et al. found a chemical link between greater $p_o$ and higher $T_{co}$ [174]. CT was also performed on Bi2212, Bi2223, Hg1201, Hg1212, and Hg1223 (Table 1; Figs. 25-26). When another Cu atom exists outside the $CuO_2$ plane, as in Y123, CT is rendered ineffective because it can only determine average valence. In addition, caution should be exercised when estimating the valence of Cu in the presence of mixed aliovalent elements such as Bi and Tl, whose valences can vary with redox [177, 178].

Rietveld analysis of powder neutron diffraction data can determine overall chemical composition through structural refinements [97, 123, 179-181]. Unlike X-ray diffraction, neutron diffraction experiments provide precise information about the position and occupancy of light oxygen even in the presence of relatively heavy elements like Hg and Tl. Because copper oxides are multicomponent and require complex synthesis, powder samples, which are often of higher quality than single crystals, are commonly used for precise structural analysis. Powder neutron diffraction experiments were used to determine $p$, particularly in the Hg system (Table 1). Yamamoto et al. found a parabolic $T_c$–$p$ relationship in Hg systems using compositionally controlled sample synthesis, chemical analysis, and neutron diffraction experiments. The authors also discovered $p_o$ values ranging from 0.20 to 0.22 for C1, C2, and C3, with relatively high $T_c$ values [129, 176]. Furthermore, it was found that a typical Hg1201 sample with the highest $T_c$ = 98.0 K was identified as $Hg_{0.97}Ba_2CuO_{4.059}(CO_3)_{0.0088}$, yielding $p_o$ = 0.20 [129]. It should be noted that the substitution of carbon atoms or $CO_3$ molecules for Hg was unavoidable under actual preparation conditions, which complicated our understanding of the $T_c$–$p$ relationship. After accounting for carbon substitutions, the structural analysis yields $p$, which agrees with the CT value [182]. Experiments with this level of precision in determining a structure are uncommon.

Tokunaga and other NMR experimentalists employed $^{63}$Cu NMR measurements to determine $p$ with high accuracy. The Knight shift $K_s^{ab}$(RT) at 300 K measured in an in-plane magnetic field is proportional to $p$ [122, 166, 183-185]. Equation 8 provides a formula for calculating $p$ using observed $K_s^{ab}$(RT):

$$p = 0.502K_s^{ab}(\mathrm{RT}) + 0.0462. \qquad \text{Eq. 8}$$

In their review paper, Mukuda et al. used a modified relation [$p = 0.492K_s^{ab}(\mathrm{RT}) - 0.023$] based on Presland's relationship (Eq. 6) [166]. However, we use the previously established Eq. 8 [122]. Although this $p$ evaluation method is useful, it should be noted that Eq. 8 may not always be applicable. In the optimum and OD regimes, $K_s^{ab}$ does not change with temperature above $T_c$, and $K_s^{ab}$(RT) is regarded as a good indicator of the hole concentration, so Eq. 8 is correct. In contrast, in the lower-doping regime, owing to the formation of antiferromagnetic correlations and a pseudo-gap (Section 4.8.2.2), $K_s^{ab}$ is temperature dependent and decreases at low temperatures. The value at room temperature is thus used for convenience, but Eq. 8 has poor reliability at low doping levels [122].

Tokunaga et al. found $p_o$ values of 0.278 and 0.25 for C1 Tl2201 ($T_c$ = 80 K) and C2 Bi2212 ($T_c$ = 80 K), respectively, much higher than 0.16. The NMR spectrum's ability to distinguish between crystallographically distinct Cu sites enabled them to selectively evaluate $p$ on the $CuO_2$ plane of Y123, yielding $p_o$ = 0.22 (Fig. 26). Figure 25a displays the $T_c$–$p$ relationship for multilayer superconductors, which has a peak at $p_o$ = 0.23 based on NMR experiments (Section 4.6.2).

The size of the observed Fermi surface enables the ARPES experiments to estimate the exact number of electrons [172, 186]. In Bi2212, a systematic experiment was conducted to change the $p$ value using in-situ annealing in the measurement chamber. The resulting $T_c$ dome resembled Presland's parabola, but with a 0.02 shift towards the high-doping side (Fig. 25a) [172]. The APRES experiment is an excellent $p$ determination approach, as mentioned in Section 4.6.2.2 for multilayer systems, but because it requires a clean surface, it is only applicable to a limited number of compounds [53]. Furthermore, while high reliability is expected in the low-doped regime, where small hole pockets exist, this is not true in the high-doping region, where the Fermi surface expands and some parts are missing (Fermi arc). As a result, ARPES is a highly reliable method for determining $p$ in the low-doped regime, while NMR is in the high-doped regime.

The Hall coefficient $R_H$ and the Seebeck coefficient can be used to estimate $p$ [140, 187-191]. These transport parameters, however, are not always reliable estimates of $p$ and frequently show significant temperature dependency due to other causes. In the UD regime, $p$ can be reasonably calculated using $R_H$, but not in the OD regime. The positive $p$ from the $R_H$ in La214, for example, equals $x$ in the UD regime, making it an accurate measure of $p$; however, in the OD region, it deviates greatly and even reverses sign [140, 187]. A Bi2212 UD sample gives a plausible $p$ value based on $R_H$ [126]. Moreover, Bi2201, with an increased $T_{co}$ of 32 K due to La substitution in $Bi_2Sr_{2-x}La_xCuO_y$ ($x$ = 0.4), produces $p_o$ = 0.15 from $R_H$, which is similar to the La214' value with comparable $T_{co}$ [192]. In contrast, in the Tl2201 OD sample discussed previously, the change in $p$ estimated from $R_H$ is inconsistent with $\Delta p$ from changes in oxygen content [188]. $R_H$ provides a good $p$ estimate in the UD regime because of the particle hole character, but not in the OD regime because of the band hole character, as discussed in Section 4.3.2. As a result, $R_H$ data is insufficient to derive a complete $T_c$–$p$ relationship.

Determining $p$ thus offers a number of experimental obstacles; however, by carefully assessing the applicability and reliability of each method, the chemical trends between $T_c$ and $p$ can be identified. In the following section, we explore the factors that influence the chemical trends of $T_{co}$ and $p_o$ in order to provide a consistent explanation for the remarkable experimental findings summarized in Section 4.4. This is the most important issue a solid state chemist should know when developing a material.

### 4.5. What determines $T_{co}$?

To gain a better understanding of the material dependence of $T_{co}$ and $p_o$, we will look at two factors: the role of apical oxygen and its impact on ZRS stability in Section 4.5.1, which is important in the OD regime, and the randomness effect in Section 4.5.2, which is critical in the UD regime. Many researchers have already discussed these issues, but we will address them together to provide a plausible explanation for the chemical trends in $T_{co}$ and $p_o$ described in Section 4.4. Furthermore, the thickness of the superconducting layer is a third factor that affects $T_{co}$. This effect is associated with the stability of the 3D superconducting order against hole doping and serves similarly as the apical oxygen effect. This point is addressed, along with the implications for multilayer systems in Section 4.7.2, which covers the ideal electronic phase diagram for C*n*.

4.5.1. Role of apical oxygen

As depicted in Fig. 21, $T_c$ should rise initially with hole doping along $T_B$, which is proportional to $p$ and shared by all compounds. Then, additional doping reduces pair attraction and thus $T_p$, causing $T_c$ to shift downward across $p_o$. More holes lead to higher $T_c$ values, as demonstrated by Uemura's plot (Fig. 23) and the $T_{co}$–$p_o$ relationship (Fig. 26). These findings reveal that $T_p$ decreases more slowly with doping in a higher-$T_{co}$ compound. In addition, the fact that the maximum $T_{co}$ is always found in C3 in the $T_{co}$–$n$ relationship (Fig. 22) implies that the $T_p$ line declines more slowly as $n$ approaches three. Ohta, Tohyama, and Maekawa's discussion on the role of apical oxygen is relevant in this context [193]. They discovered that the greater the difference between the electrostatic potentials for holes at the $O_a$ and $O_p$ sites in different compounds, the higher the $T_{co}$. In other words, the more the hole favors the $O_p$ site, the higher the $T_{co}$; conversely, the stronger $O_a$'s influence on the $CuO_2$ plane's electronic state, the lower the $p_o$ and $T_{co}$.

4.5.1.1. Material dependence of the apical oxygen effect

As depicted in Fig. 15, the Cu atom of the $CuO_2$ plane has two $O_a$ in the upper and lower block layers of C1, and one $O_a$ in the adjacent block layer in C2. C3 has one $O_a$ for the outer $CuO_2$ plane (OP) but none for the inner $CuO_2$ plane (IP). Consequently, $O_a$'s influence over the $CuO_2$ plane should be reduced from C1 to C3. If apical oxygen enhances $T_p$ suppression and thus reduces $T_{co}$, higher $T_{co}$ is naturally achieved in the order of C1, C2, and C3, which are less affected by $O_a$, as depicted schematically in Fig. 27.

Even in the same C*n* compounds, the apical oxygen effect should depend on the distance between Cu and $O_a$ [$d$(Cu–$O_a$)]. Tl2201 and Hg1201 from the C1 family have significantly longer $d$(Cu–$O_a$) (2.722 Å and 2.79 Å, respectively [128, 194]) than La214 (2.40 Å) [195] (Fig. 28), indicating a weaker $O_a$ effect. Thus, despite C1, they have higher $T_{co}$; the $T_c$ domes of both compounds may be expanded to resemble the C2 or C3 domes in Fig. 27. This scatter in $d$(Cu–$O_a$) could partly explain the large scatter in $T_{co}$ for C1 in Fig. 22. The same scatter in $d$(Cu–$O_a$) is ineffective for C2 and has no effect on C3's IP, resulting in the fewest scatters in $T_{co}$. Hg1223 has the largest $d$(Cu–$O_a$) of 2.82 Å [123, 196].

The apical oxygen effect has been discussed in $Sr_2CuO_2F_{2+\delta}$ (F214; $T_c$ = 46 K) [197] and $Ca_{2-x}Na_xCuO_2Cl_2$ (Cl214; $T_c$ = 26 K) [115, 116, 198] in the C1 system, with F and Cl ions located at the octahedral apex sites, respectively. Because the two ionic apical atoms are univalent anions, the electrostatic potential on the $CuO_2$ plane should be lower than that of the divalent oxide ion in La214. They should therefore have a higher $T_{co}$, as does F214. However, Cl214 has a lower $T_{co}$. Thus, understanding the entire compound requires more than just the apical anion effects. The randomness effect, which must be pronounced in the low-doped regime where these compounds are found, may be the primary cause of the discrepancy, as discussed in Section 4.5.2. In fact, STM observations revealed significant inhomogeneity in a Cl214 sample [199].

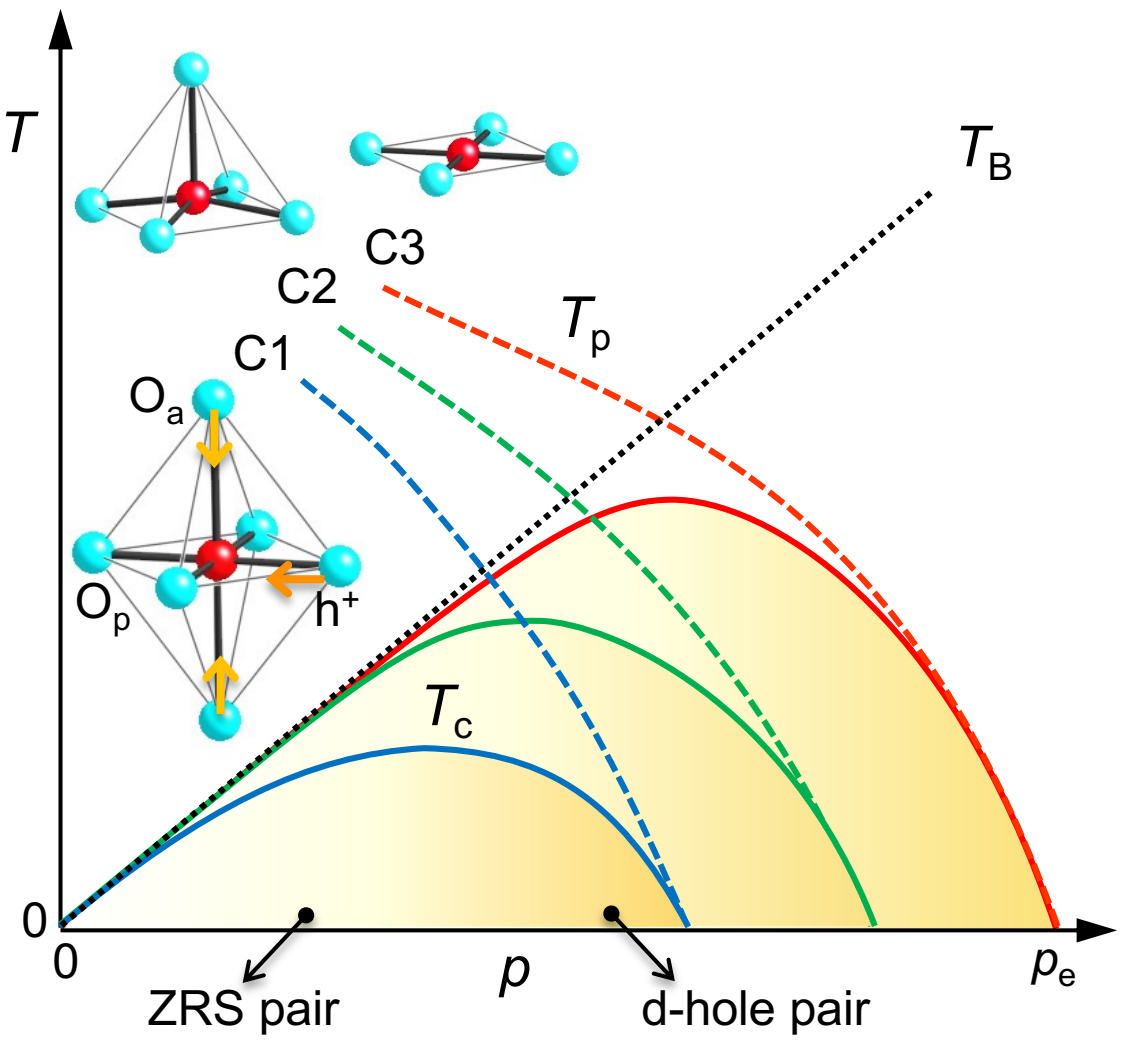

**Fig. 27.** Schematic representation of the $T_c$–$p$ relationship for the C1, C2, and C3 compounds in the absence of randomness effects, demonstrating the apical oxygen effect. The $T_B$ line, proportional to $p$, represents the BEC temperature in two dimensions. The $T_p$ curve, which generates hole pairs, shifts to the right as the apical oxygen effect decreases from C1 to C3. The Cu–O octahedron in the inset illustrates what happens as $p$ increases, particularly above $p_o$: as $O_a$ approaches the $CuO_2$ plane, a hole at $O_p$ ($h^+$) moves to Cu, causing Cooper pairs' characters to change from ZRS to d-hole pairs. The change occurs at higher $p$ levels in the order of C1, C2, and C3, resulting in a higher $T_{co}$ at a larger $p_o$.

4.5.1.2. Doping dependence of the apical oxygen effect

Let's look at how doping influences the apical oxygen effect. Figure 28 shows the doping dependence of distances from Cu to $O_p$ and $O_a$ [$d$(Cu–$O_p$) and $d$(Cu–$O_a$)] for three C1 compounds: La214 [195, 200], Tl2201 [128], and Hg1201 [194]. All of the data are based on neutron diffraction experiments conducted on powder samples with systematic composition variations and are deemed reliable; however, comparable data for C2 and C3 may be missing. In Tl and Hg systems with unknown $p$ values, the occupancy $g$($O_{ex}$) of the excess oxygen site $O_{ex}$ is used to substitute for relative $p$ values; $p = 2g(O_{ex})$ if no other sites had vacancies and no substitution by different valence ions occurred.

Doping reduces $d$(Cu–$O_a$) in all three compounds, resulting in increased $O_a$ effects. The increase in positively charged holes draws negatively charged $O_a$ to the $CuO_2$ plane. La214 shows that as $x$ increases to 0.38, $d$(Cu–$O_a$) and $d$(Cu–$O_p$) decrease monotonically by 1.3% and 1.1%, respectively. Interestingly, despite similar decreases, their doping dependences differ significantly. As $x$ increases, $d$(Cu–$O_p$) drops quickly, followed by a gradual decrease, whereas $d$(Cu–$O_a$) decreases slowly at first, then rapidly. These transitions appear to happen throughout the $p_o$. In Tl2201, $g$ = 0.005 surpasses the $p_o$, followed by superconductivity loss at 0.028. The difference in $g$ corresponds to $\Delta p$ = 0.046. In the narrow OD region, $d$(Cu–$O_p$) remains almost constant, while $d$(Cu–$O_a$) drops as $T_c$ lowers. Hg1201 has $g$ values ranging from 0.04 to 0.23, equivalent to $\Delta p$ = 0.38 and covering almost the entire $T_c$ dome.

Both $d$(Cu–$O_p$) and $d$(Cu–$O_a$) decline monotonically, but the latter drops significantly above $p_o$. Consequently, it is clear that for any C1 compound, $O_a$ quickly approaches Cu when hole doping exceeds $p_o$. These experimental data indicate that the decrease in $T_c$ is triggered by the sudden approach of $O_a$ to Cu, or vice versa.

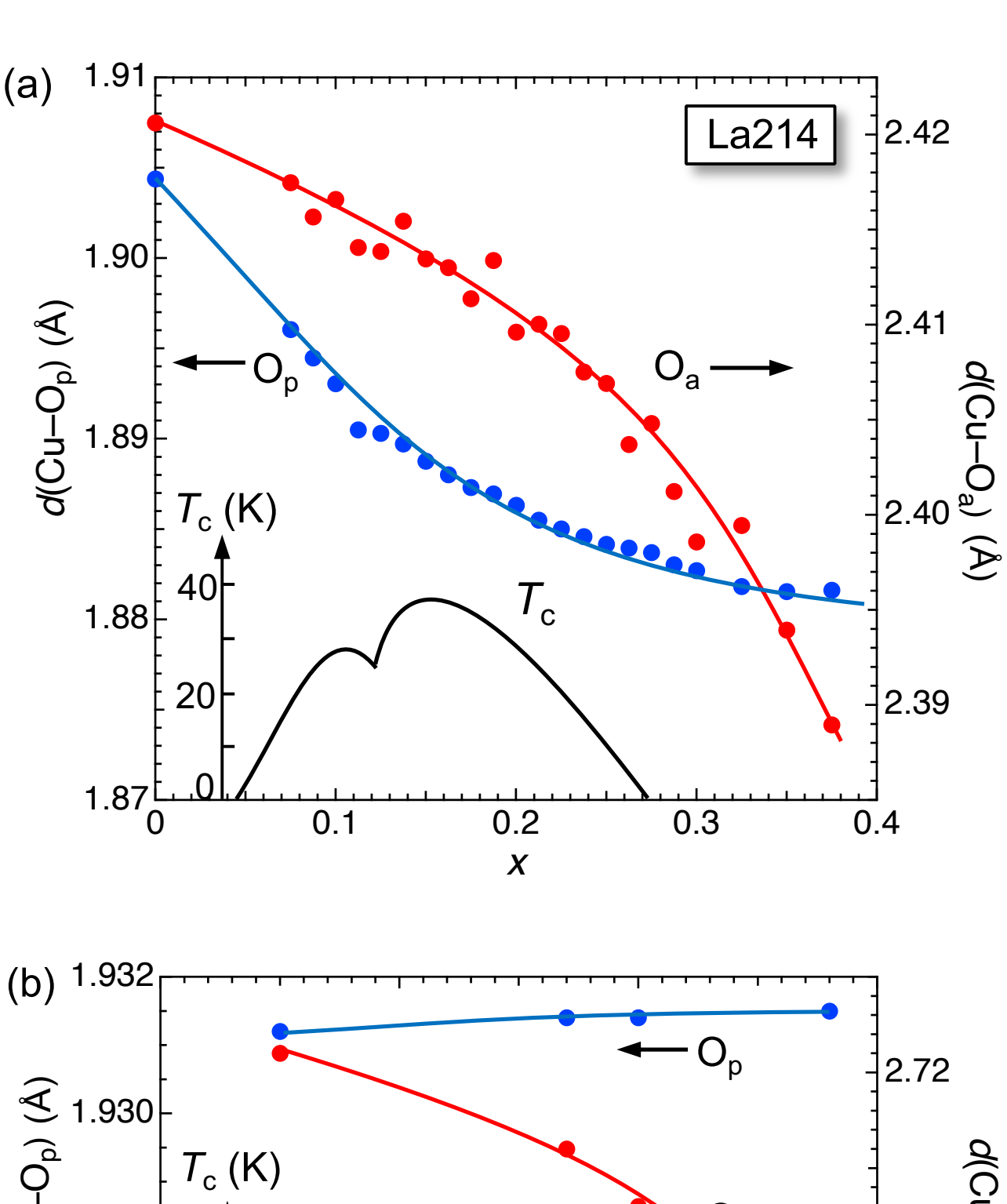

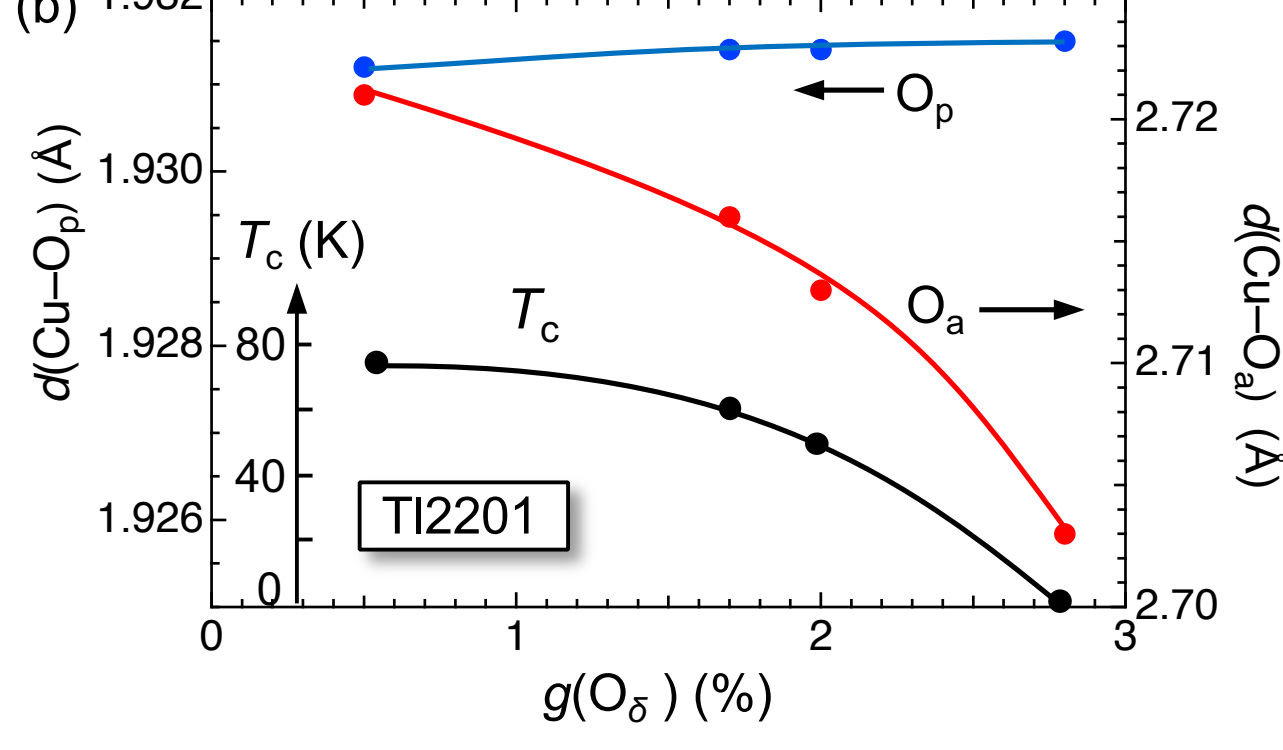

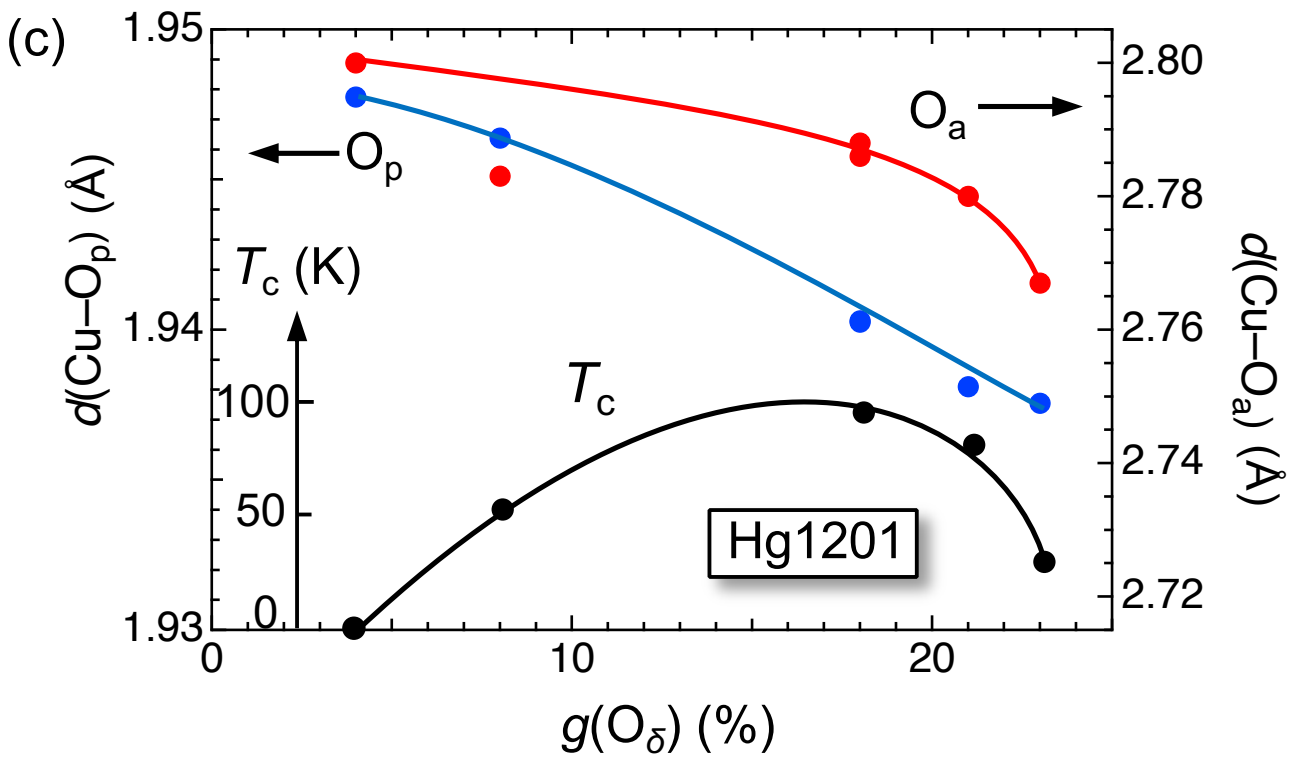

**Fig. 28.** Doping dependence of the Cu–O distances, $d$(Cu–$O_p$) (left axis) and $d$(Cu–$O_a$) (right axis), as determined by powder neutron diffraction experiments for (a) La214 [195, 200], (b) Tl2201 [128], and (c) Hg1201 [194]. In Tl2201 and Hg1201, the occupancy at the excess oxygen site $O_\delta$ [$g(O_\delta)$] replaces $p$, which may scale with $2p$. Each figure's lower inset depicts the corresponding $T_c$ variation.

4.5.1.3. ZRS destabilization caused by the apical oxygen effect

Apical oxygen's effect on $T_c$ is thought to be linked to the ZRS's stability [193]. When $O_a$ is located away from the $CuO_2$ plane, the hole $h^+$ is stable on $O_p$, as depicted in the inset of Fig. 27, forming a ZRS in the low-doped regime. However, as $p$ increases, the negatively charged ionic $O_a$ approaches Cu, attracting holes from $O_p$ to Cu sites. According to the energy diagram in Fig. 18, as electrostatic repulsion from $O_a$ increases, the $d_{LHB}$ band rises while the 2p band falls. As d–p hybridization progresses and the d orbital contribution to the hole state grows, the electronic state transitions from p-hole to d-hole-like. In other words, a ZRS with an oxygen p-hole in the $d^9h$ electron configuration converts to a d-hole in the $d^8$ configuration, which lacks an unpaired electron spin in the d orbital. In Fig. 28, $O_a$ approaches Cu faster above $p_o$ because it is drawn to d-holes, which are closer than p-holes. Zheng et al. [201] and Rybicki et al. [202] conducted NQR (nuclear quadrupole resonance) and NMR experiments that supported the transition from ZRS to d-hole.

Using the explanation provided above, consider Fig. 27's ideal phase diagram, which disregards the randomness effect discussed in the following section. The initial doping raises $T_c$ along the $T_B$ line. In the case of C1 containing two $O_a$, the strong apical oxygen effect induces a transition from ZRSs to d holes at a few holes. When the d hole annihilates the Cu spin, the source of pair attraction is quickly eliminated, leaving a weaker antiferromagnetic background than with ZRS. $T_p$ is thus expected to fall rapidly, while $T_c$ remains below the $T_p$ line. In contrast, the transition in C2 and C3 OPs, which both have a single $O_a$, occurs at a higher $p$, shifting the $T_p$ line to the highly doped side while increasing $T_{co}$.

Because C3 IP lacks $O_a$, the apical oxygen effect must be ineffective in suppressing $T_c$, allowing it to rise further along the $T_B$ line. Nonetheless, as mentioned in Section 4.3 and illustrated in Fig. 21, an excessive increase in the number of holes on the $CuO_2$ plane lowers the Fermi level and increases d–p hybridization, eventually leading to the transition from ZRSs to d holes. In addition, dilution may reduce the effective magnetic interaction $J_{eff}$. As a result, $T_p$ and $T_c$ should begin to decline in a similar manner to the $O_a$ effect. It should be noted that the apical oxygen effect causes this final transition at lower hole concentrations, thereby limiting $T_{co}$. Then, in the OD regime, Cooper pairing of d holes (rather) takes place. Additional doping reduces attraction, prevents Cooper pairs from forming, and restores the normal Fermi liquid state, which allows multiple d holes to move independently. C3's $T_p$ line extends to the highest doping regime, resulting in many hole pairs, which accounts for its high $T_{co}$.

To summarize the apical oxygen effect, differences in $T_{co}$ and $p_o$ between materials, as well as $T_c$ in the highly doped regime, are attributed to variations in the stability of the ZRS, or the antiferromagnetic spin background that generates the attractive force. This factor is critically dependent on the crystal structure of the apical oxygen contribution, resulting in higher $T_{co}$ and $p_o$ from C1 to C3. It is important to remember that when the same fundamental pairing mechanism is employed (Fig. 19), both ZRS and d-hole pairs can superconduct as Cooper pairs. The ZRS and d-hole differ in that the former simply masks the Cu spin while maintaining a strong antiferromagnetic background and an attraction that is less affected by higher hole concentrations, whereas the latter quenches the Cu spin, causing the antiferromagnetic background to decay more quickly. The apical oxygen effect is vital for achieving high $T_c$ because $T_c$ should be directly proportional to $p$ without it.

4.5.2. Randomness issue

Another important factor in determining $T_c$ is the effect of randomness on superconductivity in the $CuO_2$ plane, which results from chemical modifications to block layers. It is challenging to incorporate into theoretical models and Hamiltonians in a meaningful way because it is not a fundamental property of materials and varies by sample. However, as mentioned in Section 3.3, randomness cannot be completely avoided in any real material, so its impact must be considered when analyzing material properties, especially in low dimensions. In a theoretical model, a Mott insulator is rendered metallic with a single carrier doping, while an actual metal requires a certain amount of doping [28]. The parent phase of cuprate superconductors has strong Coulomb interactions and is close to an electron-localized state, where the randomness effect is likely to be most noticeable [203].

Several studies have examined the relationship between $T_c$ and randomness caused by chemical modifications to the block layers. In their study of La-site substitution effects in La214 for various elements, Attfield et al. discovered that $T_c$ decreased proportionally to a randomness parameter determined by the size mismatch between La and the substituted atoms [87]. Eisaki, Uchida, and colleagues carried out comparable tests on the La214, Bi2201, and Bi2212 systems and found that reducing randomness caused by elemental substitutions greatly raised $T_c$ [88, 89]. Thus, randomness in the block layers has a significant impact on $T_c$. It is worth noting that these randomness effects occur near optimum doping rather than at the previously mentioned insulator–metal boundary.

The story will proceed as follows: after introducing the experimental findings on randomness (Section 4.5.2.1), the second section will employ a simple model to explain how randomness disrupts the electronic state of the $CuO_2$ plane (4.5.2.2). The model is then applied to investigate an insulator–metal transition at low doping levels (4.5.2.3) and how it influences the $T_c$ dome (4.5.2.4). Finally, we will discuss material dependence (4.5.2.5). The randomness effect in cuprate superconductors is associated with both the conventional pair breaking effects caused by impurity scattering, which occur in most superconductors, and a reduction in the effective movable hole number due to carrier trapping, the latter of which may be critical in explaining the $T_c$–$p$ relationship.

4.5.2.1. Experiments on randomness

Electronic inhomogeneities have been detected in the low-doped regime using various measurements. Scanning tunneling microscopy (STM) directly observed superconducting regions with a diameter of approximately 3 nm embedded in the non-superconducting matrix in UD Bi2212 [204-206], as well as comparable inhomogeneities in Cl214 [199]. Cu NQR experiments on La214 identified two types of Cu sites: those near substitutional sites and those farther away [207, 208]. In addition, UD samples demonstrated the coexistence of antiferromagnetism and superconductivity [209]. Even in relatively clean Hg1201 with high $T_c$, $^{63}$Cu NMR could detect some inhomogeneities [210].

The linewidth of Cu NQR spectra provides a good indicator of randomness. Cu NQR signal frequency is determined by the magnitude of the electric-field gradient (EFG) at the copper nucleus, which detects even minor differences in local structure. The experimental spectrum represents the sum of signals from all copper nuclei in the sample. Its linewidth is an effective indicator of inhomogeneity because it widens to reflect the magnitude difference in EFG experienced by each copper nucleus. The EFG distribution in La214 is determined by two factors: the variable distance between Cu atoms in the $CuO_2$ plane and nearby substituted Sr atoms in the block layer, as well as the uneven distribution of holes within the $CuO_2$ plane caused by randomly distributed Sr. The latter is thought to be the predominant factor, with hole shading patchy at sizes of about 3 nm [208, 211]. This size matches the scale of hole shading observed in STM experiments on Bi2212 [204, 205]. Therefore, regardless of the system, the $CuO_2$ plane exhibits a similar, few nm-scale inhomogeneity in the UD regime.

Magnetic susceptibility, a bulk probe, revealed a lower superconducting volume fraction in the UD regime and nearly 100% fraction at or above the optimal doping level in several compounds [116, 140, 188]. These experimental results demonstrate that UD samples exhibit partial superconductivity. Furthermore, in the UD regime, other competing orders emerge due to randomness effects, as discussed in Section 4.8.2. One obvious example is the spin glass (SG) located between AFI and SC in La214 (Fig. 20), which freezes spins so that they are randomly oriented rather than antiparallel.

4.5.2.2. Randomness caused by chemical modifications to the block layer

Because of the ionic nature of the block layers above and/or below the $CuO_2$ plane, substitution atoms, such as $Sr^{2+}$ replacing $La^{3+}$ in La214, add a significant electrostatic impurity potential to the $CuO_2$ plane. The distribution of replaced atoms is commonly regarded as "random". The entropy term in Gibbs free energy predicts that at sufficiently high synthesis temperatures, an entropically favorable random configuration will be achieved. The random configuration is then quenched and kept below ambient temperature. There is a common misconception that a random distribution always produces a uniform distribution. This is incorrect because it depends on the coverage area of the physical quantity of interest. Large-area probes detect uniform distributions; small-area sensitive probes do not. The coherence length ($\xi$) of a Cooper pair is the characteristic length of superconductivity (Section 2.4.1). Cuprates have shorter coherence lengths and lower carrier mobility than other superconductors, so averaging effects may be less effective. It is noted that the short coherence length allows superconductivity to be maintained even in small spaces, potentially stabilizing a microscopic mixture of superconductivity and other orders.

Figure 29 illustrates schematics of random substitution atom arrangements, assuming La214. A random number generator is used to analyze a 20 × 20 lattice plane ($CuO_2$) with 1%, 2.5%, and 5% substitution atoms in each upper and lower block layers. For the LaO–$CuO_2$–LaO stacking unit, these correspond to $p = 0.02$ (AFI termination), 0.05 (SC edge), and 0.10 (SC), respectively (Fig. 20). Apparently, the random distribution in Fig. 29 does not always imply a uniform distribution. To be more specific, at 2.5% (Fig. 29b), substitution atoms are concentrated in the lower left and sparse in the upper right. This type of inhomogeneity occurs regardless of the random number used; it is always present.

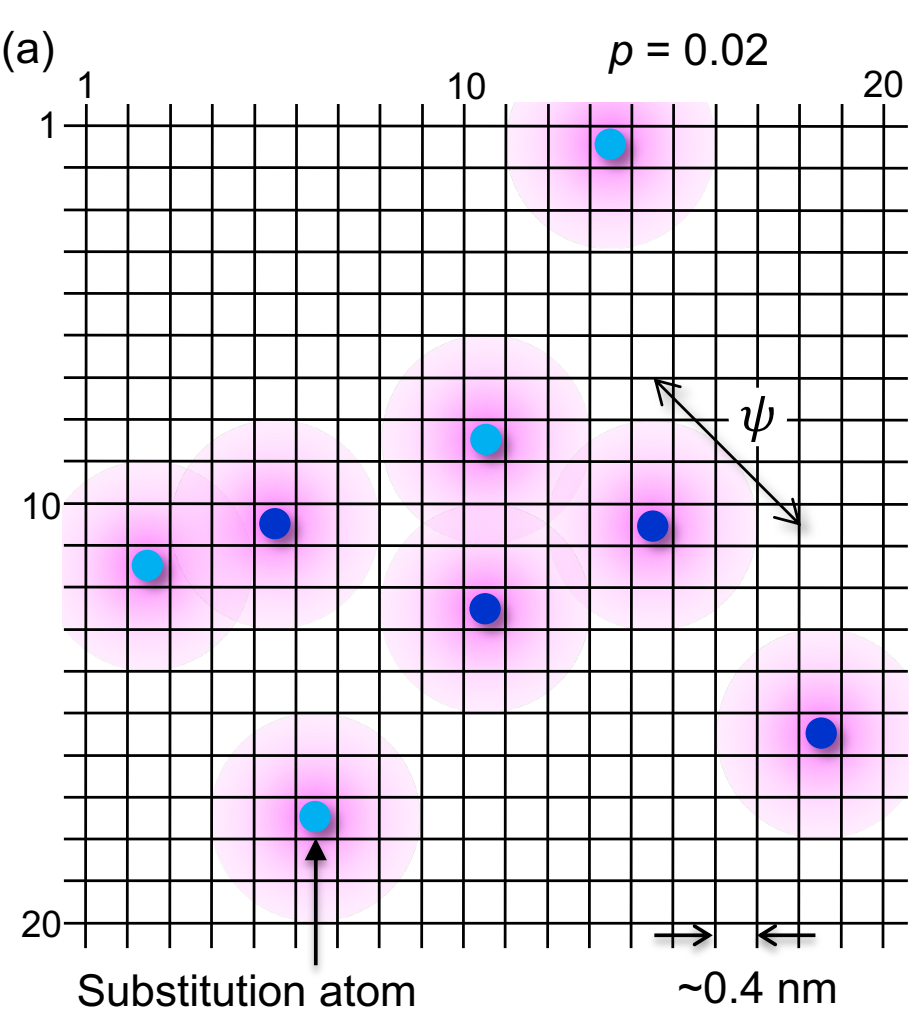

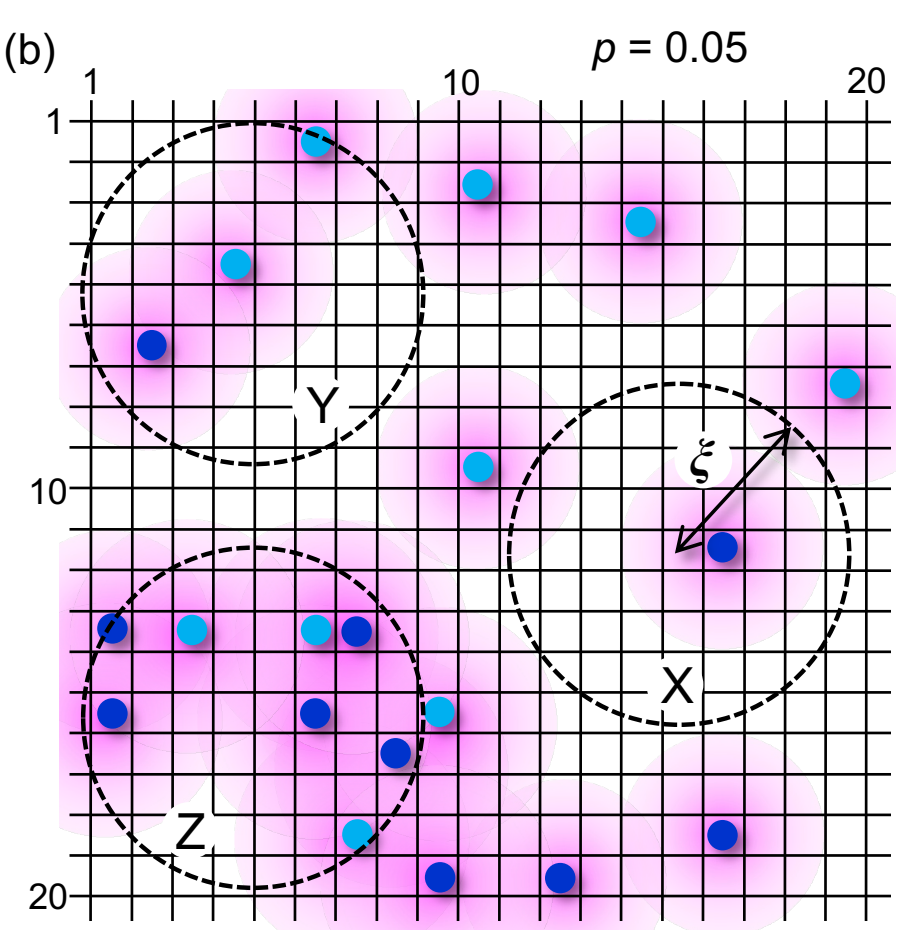

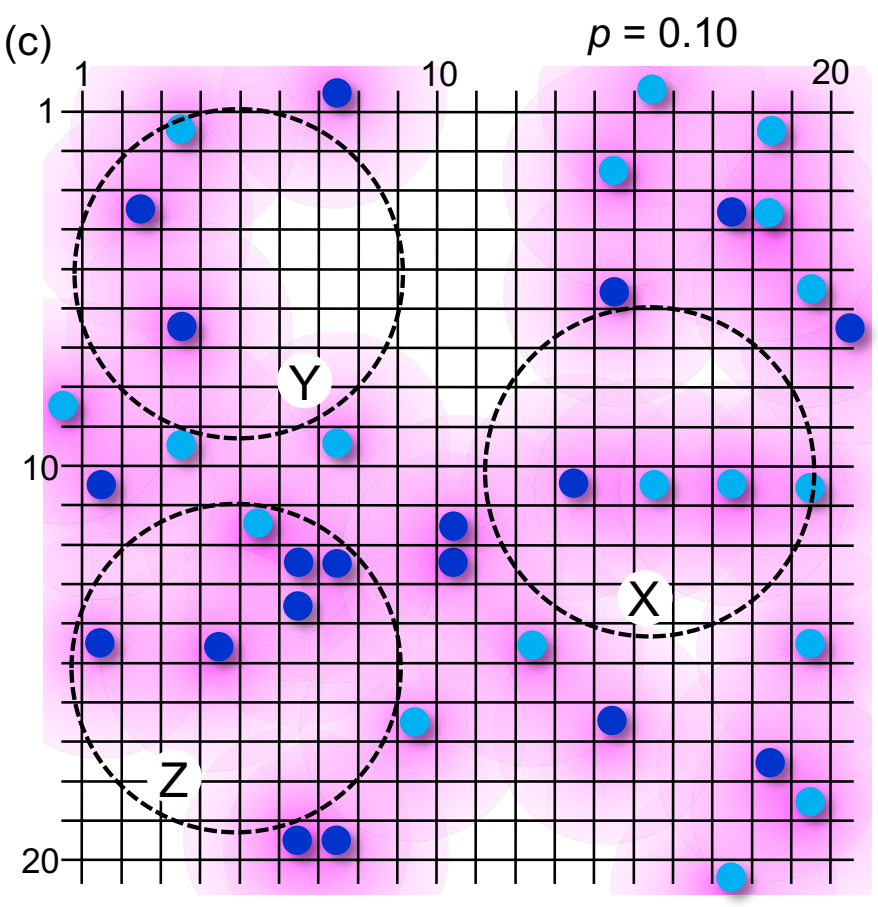

**Fig. 29.** Cartoons illustrating how random chemical substitution in the block layers causes uneven distributions of substitution atoms, resulting in inhomogeneous electronic states in the $CuO_2$ plane. The La214 stacking unit, (La, Sr)O–$CuO_2$–(La, Sr)O, randomly arranges Sr atoms in block layers above (sky blue balls) and below (blue balls) a 20 × 20 square $CuO_2$ sheet. 1.0%, 2.5%, and 5% Sr substitutions are assumed in (a), (b), and (c), yielding $p$ values of 0.02 (AFI termination), 0.05 (SC edge), and 0.10 (SC), respectively (Fig. 20); there are 8, 20, and 40 substitution atoms near the $CuO_2$ plane, with some overlapping. The magenta circle depicts the spread of a hole wavefunction $\psi$ around a substitution atom, which may correspond to the localization length in Anderson localization. The broken circle displays the area covered by a 2 nm superconducting coherence length ($\xi$). At $p$ = 0.02 in (a), the substitution atom is sparse. At $p$ = 0.05 in (b), the substitution atom is sparse in circle X, medium in circle Y, and dense in circle Z, illustrating an uneven distribution. These three typical areas can support antiferromagnetic insulators, competing secondary orders, and superconducting states, respectively. The difference in (c) at $p$ = 0.10 is less significant, suggesting a more uniform distribution. Extended hole wavefunctions eventually overlap, creating a uniform metallic and superconducting state.

4.5.2.3. Insulator-to-metal transition at low doping

The parent phase of copper oxide superconductors is a Mott insulator with an electron correlation-induced gap (Fig. 18). $V_2O_3$ typically undergoes a Mott transition with temperature [27, 28]. However, in most other materials, such as $VO_2$, both structural instability and electron correlation are important in driving the metal–insulator transition (MIT) [27, 28]. The author coined the term 'molecular orbital crystal (MOC)' to describe many crystals that form dimers or larger clusters during MITs or at any temperature [212], with the goal of gaining chemical bonding energy through molecule formation rather than Fermi liquid instability. Thus, true Mott transitions are uncommon. Non-doped copper oxides lacked a high-temperature metallic phase; however, if one existed, a Mott transition could occur at temperatures above the melting point of around 1300 K. Some argue that doping in copper oxides induces a filling-control Mott transition [28]. Nevertheless, it is important to remember that doping is always accompanied by randomness. A filling-control Mott transition is difficult to achieve in any chemically modified compound; however, it may be attainable with clean doping techniques such as electric double layer (EDL) field effects [213].

Anderson demonstrated that a one-electron wavefunction spread across a crystal can be localized in a random potential field [214]. A small number of doped carriers in a disordered semiconductor may become trapped in a random field of dopants, impeding their movement. Then, as their number grows and their energy exceeds the "mobility edge" threshold, they contribute to metallic conductivity. Mott dubbed this type of transition from insulator to metal the Anderson transition [27]. Mott's textbook describes the filling-control transition to metal in copper oxides as an Anderson transition.

SCES, like cuprate superconductors, requires more doping for metallization than conventional semiconductors. Doping in La214 removes AFI from the parent phase at 2% hole concentration (Fig. 20) [27], resulting in an Anderson transition rather than a filling-control Mott transition. The temperature dependence of electrical conductivity at high temperatures in the AFI regime follows the variable-range hopping mechanism expected for randomness electron localization [140]. Carriers are localized around impurities and can hop between wavefunctions, spreading with a "localization length". Apparently, in this regime, all doped holes eventually stop at zero temperature, implying that no mobile holes remain. The Anderson transition occurs when localized wavefunctions overlap and their localization lengths diverge.

The critical hole concentration of 2% in La214 suggests that the wavefunction broadening $\psi$ of electrons trapped in the random potentials of substituted atoms is not so large: in Fig. 29a at $p$ = 0.02, wavefunctions with $\psi$ = 2 nm do not overlap each other, and no in-plane connected conduction path occurs. High temperatures only permit hopping conduction between wavefunctions. As a result, AFI remains stable. It is important to remember that $p$ is a nominal quantity, and the actual number of movable holes ($p^*$) is zero! In contrast, NMR and ARPES experiments on clean IPs of multilayer systems, as described in Section 4.6.2, reveal that antiferromagnetic metal (AFM) rather than AFI appears next to SC. The AFI in the parent phase appears to survive at low hole doping levels, possibly due to randomness hole trapping. A clean $CuO_2$ plane would probably transition to AFM with minimal hole doping (Fig. 38a).

The uneven distribution of substituted atoms becomes more pronounced as doping increases to $p$ = 0.05. Figure 29b depicts a calculated distribution pattern of substitution atoms, with dashed circles with a radius of $\xi$ = 2 nm indicating Cooper pair size. The number of substituted atoms in the $\xi$ circle differs between regions, with fewer in region X and more in region Z. In the former, holes perceive the bare impurity's potential for shrinkage and do not overlap, resulting in a nonconducting state that must be an AFI, as shown in Fig. 29a. In the latter case, the wavefunctions overlap enough to allow holes to move, producing an AFM or SC phase. If a third metastable state competes with them, it will appear in region Y, which contains a medium number of substitution atoms.

Figure 29c shows no significant difference between $\xi$ circles at $p$ = 0.10, compared to Fig. 29b at $p$ = 0.05. In addition to the high concentration, increased screening expands the hole wavefunction, allowing it to overlap and form a percolating conduction path through the crystal. As the doping level increases, the system approaches a uniform metallic state, with the inhomogeneous substitution atom distribution averaged out and all Cu atoms in comparable environments. Thus, with a few nm of coherence length and a similar localization length, we can capture La214's low-doped region. Furthermore, Fig. 29b ($p$ = 0.05) replicates the inhomogeneity observed using STM probes [204-206]. We emphasize that the current insulator-to-metal transition is not a filling-control Mott transition, but rather an Anderson transition with randomness.

4.5.2.4. Hole trapping and the parabolic $T_c$ dome

In the AFI regime of the La214 phase diagram, all holes are trapped by impurity potential and unable to move, so no holes contribute to electrical conduction, even though the nominal $p$ is finite. When more holes are added, the electronic state becomes inhomogeneous, as depicted in Fig. 29b. Some may become trapped near impurities, while others can contribute to metallic conductivity and superconductivity [87]. Even in superconducting regimes with additional doping, some holes may remain trapped, particularly in highly random compounds. In fact, $R_H$ predicts a reasonable $p$ value for Bi2212 but a lower $p$ value than expected for Bi2201 due to increased disorder [126]. In contrast, Fujita et al. found that all holes contribute to conduction in the nearly-optimally hole-doped states of La214 and La-substituted, relatively clean Bi2201 ($T_c$ = 35 K) when studying the relationship between nominal and actual movable hole amounts [89]. As a result, as hole concentration increases, hole trapping becomes less effective. When more holes are added, even those that were previously trapped at low doping levels can become free and contribute to superconductivity; however, this may not occur in dirty systems. One must consider randomness-induced hole trapping especially in the UD regime.

The preceding discussion strongly implies that the $T_c$ dome's parabolic shape is due to hole trapping by randomness at low doping levels, rather than intrinsic. In the low-doped regime, the mobile hole concentration $p^*$ is less than $p$; the smaller $p$, the greater the difference. Drawing $T_B{}^*$, which is proportional to $p^*$, as a function of $p$ yields a complex curve, as illustrated schematically in Fig. 30: it starts out at zero, gradually increases, and then rapidly rises as it approaches the metallic and superconducting phases. This is just a demonstration; using $p^*$ as the abscissa should result in a clear proportional relationship. Because $T_p$ is high enough in the low-doped regime and $T_c$ is parallel to $T_B{}^*$, the complex $T_B{}^*$ line can form a parabolic shape on the left side of the $T_c$ dome. Uemura's plot demonstrates a linear relationship between $T_c$ and superconducting carrier density $n_s$, which is due to the μSR experiment's focus on moving holes [163]. It is also worth noting that the $T_c$ dome of Hg1201 with relatively low disorder in Fig. 27c rises linearly when compared to La214 in Fig. 27a, and the dome is obviously asymmetric. This is also true for clean multilayer systems (Fig. 34). To fully comprehend the experimental results, it is critical to take into account the randomness effect. Again, the parabolic $T_c$ dome is just an artifact, regardless of its shape or vertex location.

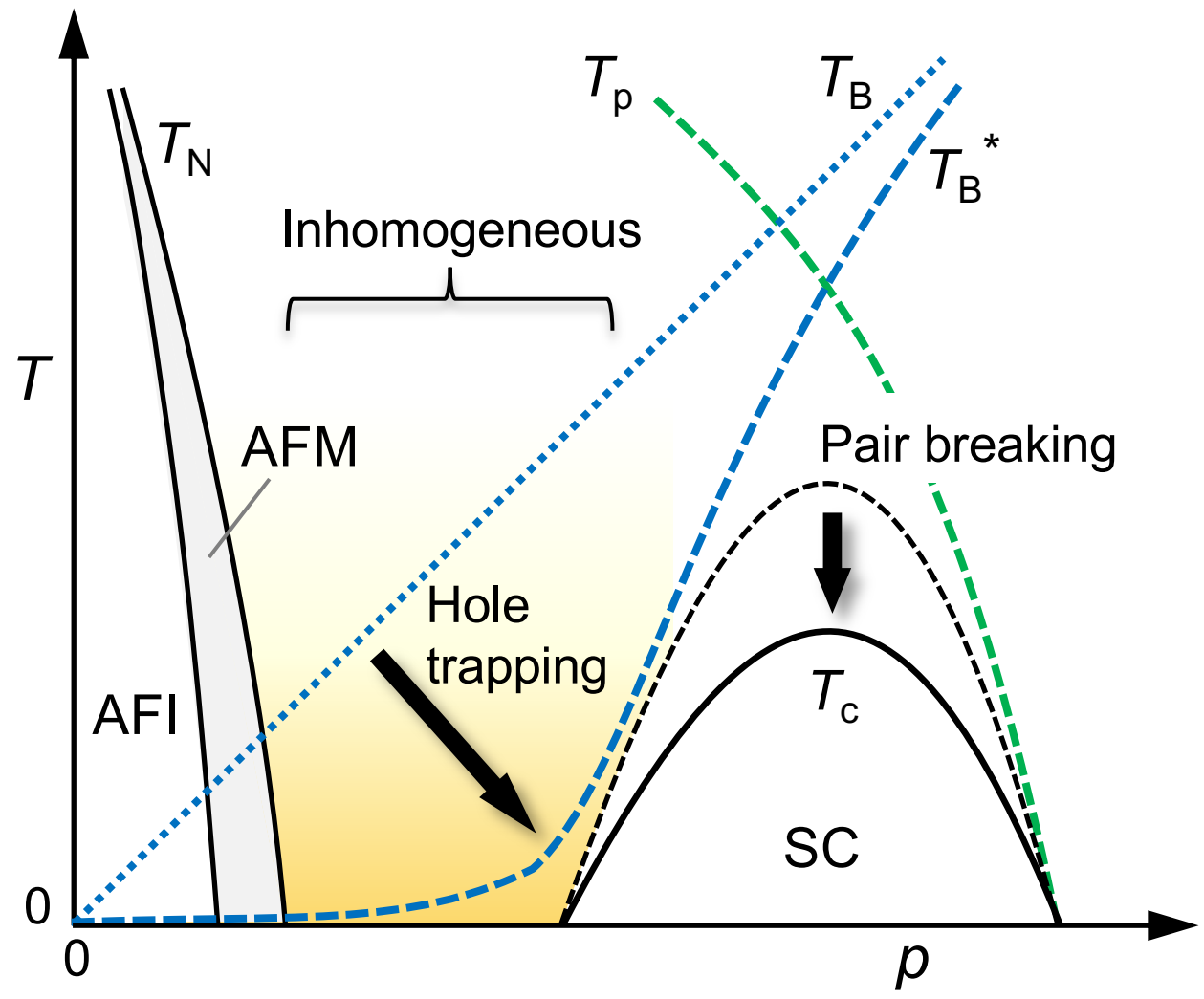

**Fig. 30.** Schematic phase diagram demonstrating how randomness alters its appearance. Nominal $p$ does not equal $p^*$ which represents the actual mobile hole concentration; $p^*$ is less than $p$, especially at lower doping levels, due to increased hole entrapment caused by random potential from the block layer. When plotted against $p$, $T_B^*$, which is proportional to $p^*$, appears to be zero at first before rapidly rising with $p$, approaching the $T_B$ line. AFI survives as $p$ increases initially, even if $p$ is finite, because $p^* = 0$ when $T = 0$. After AFI is replaced by AFM but before SC appears, a window with an inhomogeneous hole distribution emerges, as depicted in Fig. 29b, in which various secondary phases or phenomena like "phase separation" may occur. Then, $T_c$ develops along the $T_B^*$ line, eventually resembling a parabolic shape as observed in La214 (Fig. 20). Furthermore, randomness effects cause conventional pair breaking effects to lower the $T_c$ dome top, as indicated by the vertical thick arrow.

The typical impurity effects in superconductors cause $T_c$ suppression due to pair breaking. In copper oxide superconductors, it does occur when Zn replaces Cu in the $CuO_2$ plane; the $T_c$ in La214 ($x = 0.15$) decreases by 20% and 50% with 1% and 2% Zn substitutions, respectively, and disappears with 4% substitution [215, 216]. Thus, the presence of Zn atoms in the conduction layer causes strong scattering, which breaks Cooper pairs and lowers $T_c$. In contrast, the previously mentioned randomness effect, which causes partial hole trapping, is weak and indirect. Nevertheless, many studies have shown that when randomness is reduced via improved sample synthesis methods or systematic elemental substitutions, $T_c$ rises significantly, even at around the optimum doping level, where hole trapping is minor [87-89]. Therefore, pair breaking must be effective at determining $T_c$ in cuprates. As depicted by the vertical line in Fig. 30, pair breaking by randomness results in a significantly lower $T_c$ dome top than an ideal crossover curve below $T_B^*$, which is determined by hole trapping, and $T_p$, which decreases as the pairing interaction decreases with increasing $p$.

4.5.2.5. Material dependence of randomness effects

Any cuprate superconductor will experience randomness effects, but the strength varies depending on the material. Eisaki et al. provided an overview of the randomness effects induced by chemical modification patterns in different crystal structures [88]. The randomness effect varies according to the type of block layers, the number and location of alien atoms, and their distance from the $CuO_2$ plane. It, like the apical oxygen effect, is determined by the number of stacking $CuO_2$ planes in the conduction layer. As depicted schematically in Fig. 31, the $CuO_2$ plane in C1 is significantly influenced by the impurity potential of the block layers immediately above and below. In C2, only one block layer affects the OP, whereas in C3, the IP sandwiched between the OPs is less affected because it is isolated from the block layers and protected by the OP carriers. As $n$ increases, the initial $T_c$ curve should shift toward the low-doping side, as depicted in Fig. 31d.

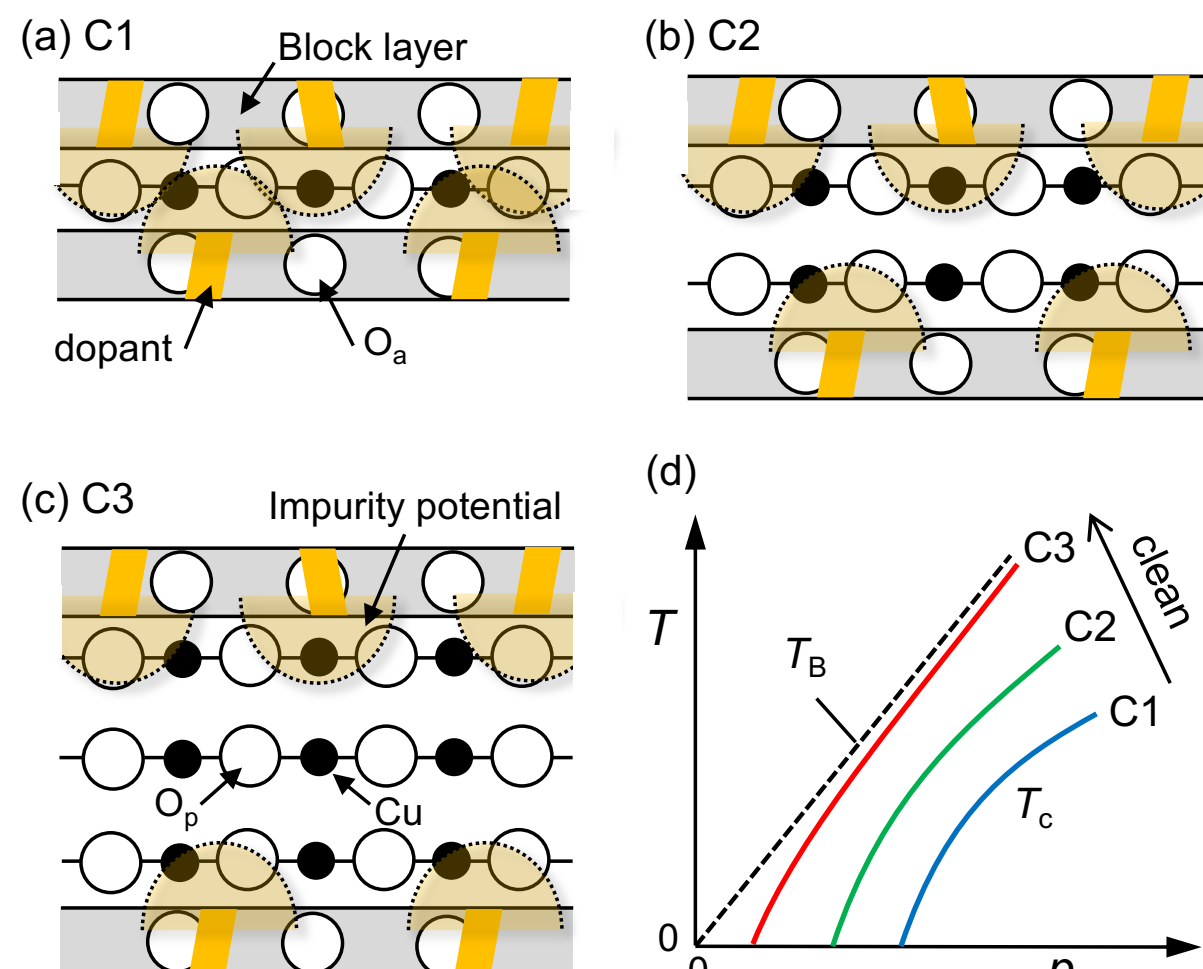

**Fig. 31.** Schematic drawings of how substituents or excess oxygen in block layers affect the $CuO_2$ plane for (a) C1, (b) C2, and (c) C3 compounds. The dotted half dome represents the random potential generated by them. (d) $T_c$ curve evolution in the low-doping regime, where $CuO_2$ planes become clean from C1 to C3. As randomness decreases, the initial $T_c$ curve may shift to the left and eventually converge to the $p$-proportional $T_B$ line in the ideal case.

Randomness has a particularly strong influence from the beginning of doping until superconductivity appears: inhomogeneous electronic states between the AFI/AFM and SC phases are likely to occur, leading to a variety of secondary and complicated phenomena such as the appearance of SG and metastable competing phases (Section 4.8.2), as well as electronic phase separation. As discussed in Section 4.3.1, the $CuO_2$ plane has a parameter that is susceptible to electronic phase separation, and randomness may exacerbate this tendency. It should be noted, however, that electronic phase separation refers to the spontaneous formation and contact of domains with different hole concentrations and properties across a sharp boundary. In the presence of randomness, the electronic phase separation is indistinguishable from an inhomogeneous distribution with a blurred boundary, as depicted in Fig. 29b.

With the exception of NMR and STM, most physical measurements produce average results, so be cautious in low-doping experiments, which are subject to disorder. The interpretation of $p$-dependent physical quantities, such as the $T$–$p$ phase diagram, necessitates extra care. To better understand the superconducting mechanism, materials with relatively uniform electronic states due to low randomness effects should be studied. In this regard, La214, which has been regarded as the standard, is clearly an inappropriate system.

4.6. Understanding of the material dependence

4.6.1. C1 to C3

Let us examine $T_{co}$'s material dependence, as shown in Figs. 22–26 using the previously mentioned apical oxygen and randomness effects. First, we will go over C1 through C3. Because the single $CuO_2$ plane has two apical oxygen atoms and is sandwiched between two block layers (Figs. 15 and 31), the C1 compound has a lower $T_{co}$ at a lower $p_o$ due to the greater influence of the two effects. The large scatter in $T_{co}$ between 7 and 90 K, even for the same C1, as shown in $T_{co}$ vs $n$ in Fig. 22, is caused by magnitude variances between compounds. C1-B3 Hg1201 and C1-B4 Tl2201 have longer apical oxygen-Cu distances ($d$(Cu–$O_a$) = 2.7 and 2.8 Å, respectively), indicating higher $T_{co}$ values than C1-B2 La214 with $d$(Cu–$O_a$) = 2.4 Å. Furthermore, their block layers with excess oxygen in the middle are thicker than La214's block layer with elemental substitutions, indicating a smaller randomness effect (Fig. 17, section 4.5.1.1). On the other hand, C1-B4 Bi2201 has much lower $T_{co}$ values of 7–25 K due to prominent disorder factors other than chemical modification for hole doping, as mentioned in the previous section. As a result, Bi2201's $T_c$ dome shrinks further and enters the left side, as shown in Fig. 25a.

The $CuO_2$ plane in C2 has only one apical oxygen, which explains the higher $T_{co}$. $T_{co}$ ranges between 90 and 125 K due to differences in both apical oxygen and randomness effects, as does C1; $T_{co}$ is higher when a cleaner block layer is farther away from the $CuO_2$ plane. $T_{co}$ in C3 has low material dependence because the two effects are less pronounced at the OP and the block layer has little influence on the common IP. The weakest chemical bond to the apical oxygen in all OPs occurs between $O_a$ and Cu in C3 Hg1223, where $O_a$ is covalently bonded to Hg to form a $HgO_2$ dumbbell, as depicted in Fig. 15c ($d$(Cu–$O_a$) = 2.82 Å [123, 196]). Therefore, even in OP, apical oxygen is unlikely to effectively suppress $T_{co}$. Furthermore, the three $CuO_2$ planes become superconducting at the same $T_c$ [166], which may aid in the development of a stable superconducting order resistant to hole doping, as will be mentioned in Section 4.6.3. The characteristics of multilayer systems with $n$ greater than three will be addressed in the following section.

The $T_{co}$–$n$ relationship (Fig. 22) reveals a notable distinction between B3 Tl1 and B4 Tl2 within the same Tl system. In the former, C1 and C2 have lower $T_{co}$, indicating a greater contribution from randomness, most likely because B3's excess oxygen site is closer to the conduction layer than B4 (Fig. 17). Nevertheless, Tl1 has a higher $T_{co}$ than Tl2 in C3 and C4. B3's higher hole-supplying capacity (Section 4.1.3) may be more important in this case than the randomness effect because it is necessary to dope enough holes into the multiple $CuO_2$ planes. As demonstrated by this comparison, in the large $n$ case, the bock layer's hole-supplying capacity must also be considered.

The material dependence of $T_{co}$ has been explained in a variety of ways. For example, Kivelson and Fradkin argued that $T_B$, not $T_p$, is material-dependent [217]. Stronger interplane interactions cause a steeper $T_B$ line slope (why?), which leads to a higher $T_{co}$. However, this scenario does not necessitate a higher $T_{co}$ at a larger $p_o$, as observed in the current manuscript. While many mechanisms emphasize the importance of interplane interactions in the superconducting mechanism, we believe that the essence of cuprate superconductivity lies solely in the 2D $CuO_2$ plane, which determines the $T_B$ line slope, with interplane interactions needed only to stabilize 3D long-range orders by suppressing 2D fluctuations. We believe that many of the other hypothesis proposed thus far would make it difficult to explain the chemical trends described here without introducing contradictions.

4.6.2. Multilayer systems

This section examines the chemical trend of $T_{co}$ in multilayer systems with $n > 3$, which provides valuable experimental data for understanding the fundamentals of copper oxide superconductivity. They have ($n$ – 2) IPs that lack apical oxygen and are protected from randomness in the block layer by outer OPs, resulting in clean $CuO_2$ planes with fewer $T_c$ suppression effects. The $T_{co}$–$n$ diagram in Fig. 22 demonstrates that for Hg multilayer compounds with $n > 3$, $T_{co}$ decreases slightly with $n$ before becoming almost constant at 105 K for $7 < n < 17$ [160]. This is due to the uneven distribution of holes across the multiple $CuO_2$ planes [166]. The $T_{co}$ scatter is independent of $n$ in multilayer systems, as superconductivity occurs in OPs under nearly identical environments.

4.6.2.1. Hole distributions across the $CuO_2$ planes

We employ a simple electrostatic potential model to improve the outlook for discussion. Figure 32 depicts model calculations of possible hole distributions across the $CuO_2$ planes of the Hg system's C1, C2, C3, and C5. Because the block layer is ionic and electronic states near the Fermi level are restricted to Cu and O states, the total hole supply ($p_B$) from a single block layer should be doped on any of the $n$ $CuO_2$ planes. Furthermore, we assume that the hole distribution is determined by the Coulomb potential of a block layer's negative charge of magnitude (–$p_B$) after the hole supply, which is simply inversely proportional to its stacking distance from the block layer center ($d$). The potential at a $CuO_2$ plane at a distance $d$ is calculated as $A/d$ using crystal structure data, with $A$ determined solely by $n$ and $p_B$ (Fig. 32 footnote). Previous studies estimated hole distribution using similar but distinct models, taking into account the electrostatic potential of the apical oxygen [183], Madelung energy [218], and similarities to graphite intercalation [219].

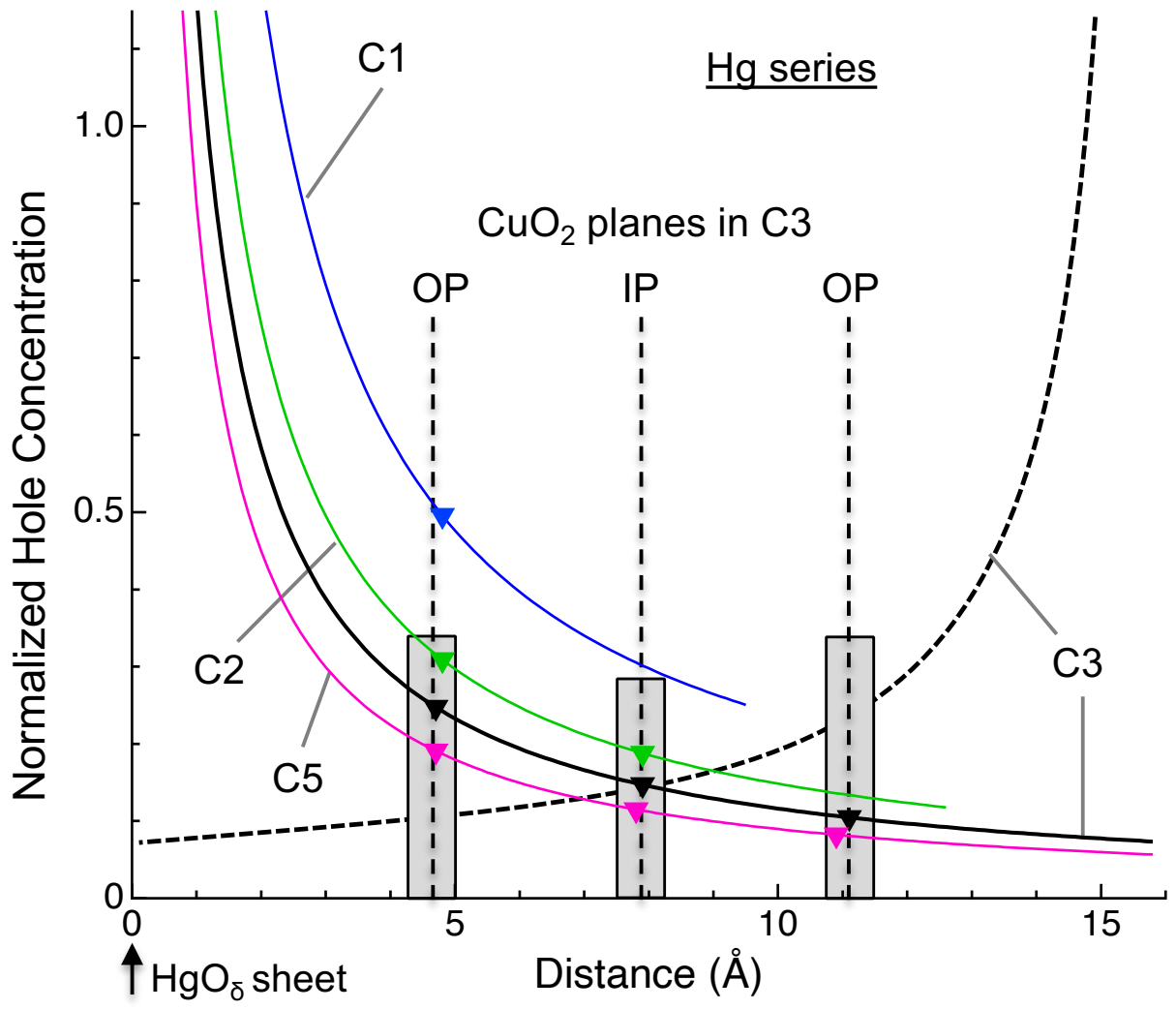

**Fig. 32.** Calculated hole concentration, normalized to $p_B$, for each $CuO_2$ plane of various Hg compounds. The horizontal axis represents the distance $d$ along the $c$ axis from the block layer center at the $HgO_\delta$ sheet in the B3-NC block layer (Fig. 17d) for C1 (blue line) [194], C2 (green line) [181], C3 (black line) [123], and C5 (magenta line) [220]. Each curve is calculated using the formula $A/d$, with $A$ set so that the sum of $p$ values across all planes equals $p_B$: $A$ = 2.3822 (C1), 1.4856 (C2), 1.1637 (C3), and 0.8983 (C5). Each curve's inverse triangles represent $CuO_2$ plane positions. In C3, the broken black curve represents the counter block layer's contribution, while the height of rectangles at the OPs and IP positions represents the combined $p$ values provided by the two block layers ($0.35p_B$ and $0.30p_B$, respectively).

A single $CuO_2$ plane in C1 is supplied with $p_B/2$ holes from the upper and lower block layers, ensuring that $p = p_B$. In C2, each of the two $CuO_2$ planes is doped with $p_B/2$, which is the sum of $(A/d_1)p_B = 0.31p_B$ from the adjacent block layer and $[A/(d_1+ d_2)]p_B = 0.19p_B$ from the counter block layer; $d_1$ is the distance from the block layer center to the OP (4.76 Å), and $d_2$ is the distance between the two OPs (3.14 Å) (or, the OP–IP distance in the case of C3). Similarly, for C3, the distribution of holes is $0.35p_B$ in the OP and $0.30p_B$ in the IP, with only minor variations in $p$. However, the difference becomes more pronounced as $n$ exceeds 3. In terms of proximity to the block layer, C5 has an OP at $0.24p_B$, $IP_1$ at $0.18p_B$, and $IP_0$ at $0.16p_B$. As a result, the hole distribution is highly uneven, with more holes in the OP than in IPs. The difference grows with $n$, and OPs contain the vast majority of holes. Because the commonly discussed $p$ is the average value for all $CuO_2$ planes, caution should be exercised when discussing the $T_c$–$p$ relationship or other quantities' $p$ dependence for $n > 3$.

Tokunaga, Kotegawa, Mukuda, Shimizu, and colleagues conducted Cu NMR experiments to identify Cu atoms in OPs and IPs. They determined $p$ for each $CuO_2$ plane by comparing it to the Knight shift $K_s^{ab}$(RT) measured at room temperature in an in-plane magnetic field (Equation 8, Section 4.4.5) [122, 166, 183-185]. Remarkably, they discovered significant variations in $p$ and electronic states across the planes. NMR experiments on a C3 Hg1223 sample with $T_c$ = 133 K, close to its $T_{co}$, revealed 0.252 and 0.207 holes in the OP and IP, respectively (Fig. 33a), for a total of $p_B$ = 0.711 [122]. Using this $p_B$, the electrostatic potential calculations above yield distribution values of 0.25 and 0.21, which, despite the model's simplicity, are clearly consistent with the experimental results. This agreement implies that electrostatic potential is the sole determinant of hole division. If excess oxygen is the only source of hole generation, then $\delta$ = 0.36, which is comparable to values of 0.29 [221], 0.41 [124], and 0.44 [123], as determined by neutron diffraction experiments on nearly optimally doped samples. NMR experiments also revealed that another doping level in the C3 UD sample ($T_c$ = 115K) was $p$(OP) = 0.196, $p$(IP) = 0.182, and $p_B$ = 0.574 [122]. As a result, regardless of the total amount of doping, a nearly uniform hole distribution is achieved; thus, the uneven distribution of holes is considered a minor issue when discussing the $T_c$–$p$ relationship for C3.

The OP in the $T_c$ = 133 K sample has a $p$ value of 0.25, indicating that it is close to $p_o$ and produces a high $T_{co}$. With $p$ = 0.21, the IP's $T_c$ should be lower, possibly around 120 K. However, site-selective NMR experiments reveal that both planes transition around 133 K. It is likely that as the OP transitions, the IP, which already has well-developed superconducting correlations, will become superconducting due to proximity effects [1, 8]. Furthermore, because the OP may experience apical oxygen and randomness effects from the block layers while not in the IP, the difference in $T_{co}$ may be reduced. As a result, the three $CuO_2$ planes can combine to form a thick superconducting layer with the same $T_c$. This increased thickness may stabilize the superconducting long-range order against hole doping, thereby preserving the high $T_c$. As discussed in the following section, IPs in systems above C3 have significantly lower $T_c$ or are in AFM states with lower doping, which prevents them from coupling with OPs and forming a thick superconducting layer like C3.

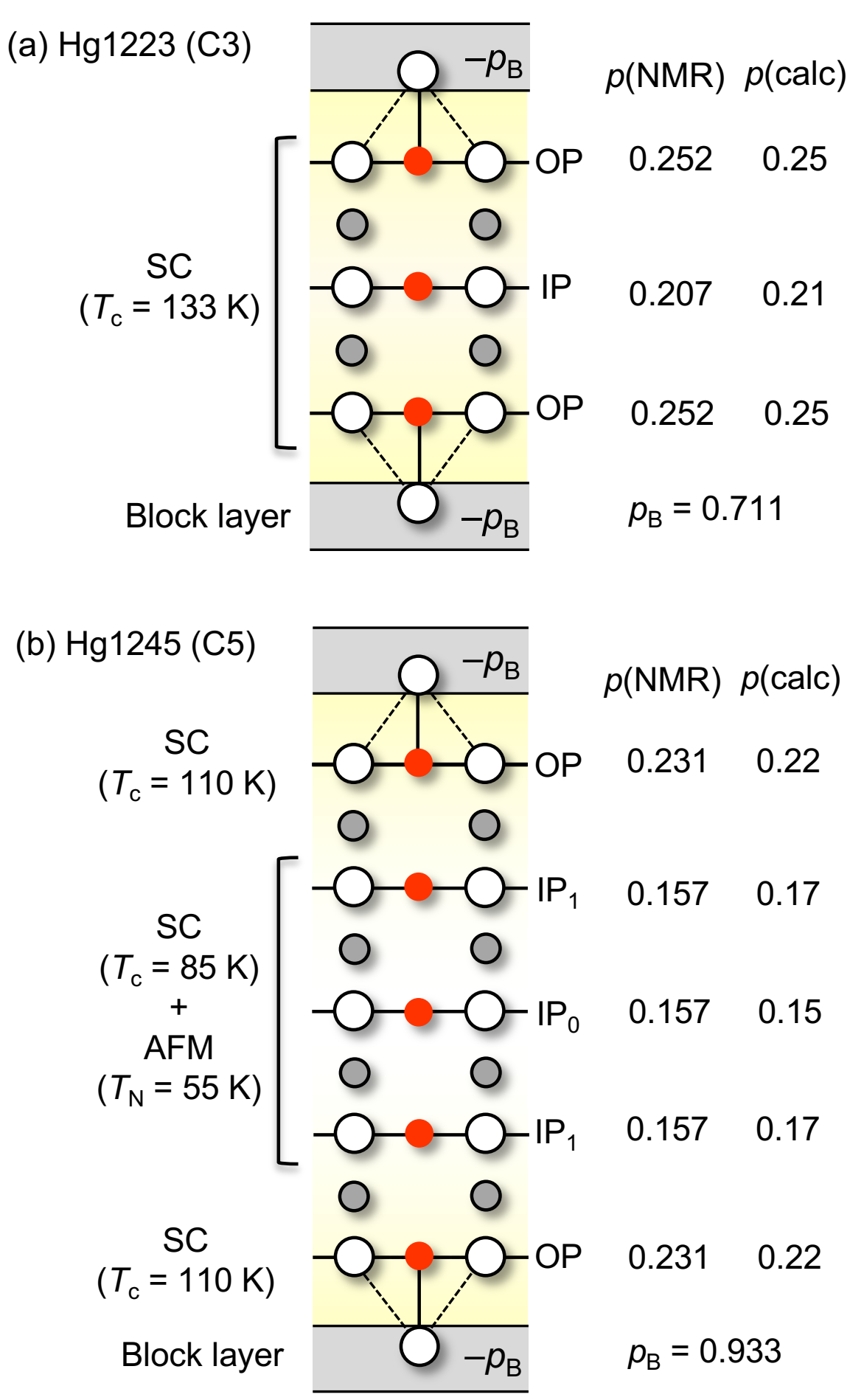

**Fig. 33.** Distribution of holes across the $CuO_2$ planes of (a) Hg1223 and (b) Hg1245, as revealed by NMR experiments on samples with $T_c$ = 133 K [122] and 110 K [166], respectively. The total hole supply ($p_B$) from the block layer with $-p_B$ charge is determined by adding the NMR values, $p$(NMR), and then used in the electrostatic potential calculation. In both cases, the calculated hole distribution, $p$(calc), is consistent with $p$(NMR). The yellow shading illustrates the nearly even and uneven distributions of holes in C3 and C5, respectively. The NMR experiments also determined the electronic states of each plane, as shown on the left: simultaneous superconducting transitions at 133 K in the OPs and IP in Hg1223; a superconducting transition at 110 K in the OP in Hg1245; and a superconducting transition at 85 K, followed by a transition to AFM at 55 K in Hg1245's Ips ($IP_1$ and $IP_0$).

4.6.2.2. Uneven hole distributions in C5

The distribution of holes is relatively uniform up to C3, but it becomes noticeably uneven in C5 Hg1245. The Cu NMR experiments on a Hg1245 sample with $T_c$ = 110 K yielded $p$ values of 0.231 for the OP and 0.157 for the IPs (Fig. 33b) [166, 222], indicating that the OP is nearly optimally doped while the IPs are clearly in the UD regime. It should be noted that two IPs ($IP_1$ and $IP_0$) are not resolved, in contrast to the ARPES observations described later. The Hg1245 sample has a $p_B$ value of 0.93, which is significantly greater than the Hg1223 value of 0.711. Using this $p_B$ value in the electrostatic potential model calculations yields estimated $p$ values of 0.22 for OP, 0.17 for outer $IP_1$, and 0.15 for inner $IP_0$. Consequently, the NMR and calculated results are in good agreement again. NMR experiments on C4 Ba0234 and C5 Ba0245 revealed similar uneven hole distributions [166].

Figure 34 depicts a united phase diagram that includes all of the relationships between transition temperatures and hole concentrations estimated for the OP and IPs of seven C5 samples (Hg1245, Tl1245, and Cu1245) with varying levels of doping using site-selective Cu NMR experiments [166, 184]. This seamless integration of data from different systems and samples strongly suggests that hole concentration is the sole determinant of electronic states in these clean systems. The $T_N$ data comes from IPs, while the $T_c$ data comes from both OP and IPs.

The low-doped IPs only show an AFM transition, with $T_N$ rapidly decreasing as $p$ increases in Fig. 34. Above $p$ = 0.15, the variation appears to weaken, and above 0.17, the transition is lost. Superconducting and AFM transitions have been observed in IPs with $p$ values ranging from 0.15 to 0.17. For example, in IPs with $p$ = 0.17, SC occurs at 90 K and AFM at 45 K. This could be an intrinsic two-step transition on the phase diagram, but the nearly vertical $T_N$ phase line in Fig. 34 indicates otherwise. It could be due to in-plane inhomogeneity in hole distribution caused by weak randomness, which worsens near the phase boundary (Fig. 30). Alternatively, as the calculation indicates, there could be a difference in hole concentration between $IP_1$ and $IP_0$. Although NMR did not detect the difference, APRES found, as mentioned in the next paragraph, that $IP_1$ has more holes than $IP_0$, resulting in superconductivity and antiferromagnetism, respectively. In either case, keep in mind that the $p$ value in this range is only an average. In contrast, the other $p$ values in Fig. 34 represent the actual hole concentration in each $CuO_2$ plane. The appearance of two phases in the interfacial region is simply an illusion caused by using the average $p$ in a single phase diagram.

Only superconductivity can be observed in OP with sufficient hole doping from 0.15 to 0.28. The OP's $T_c$ dome has a peak at $p_o$ = 0.23 and $T_{co}$ = 110 K, which is smoothly connected to the IPs' $T_c$. Interestingly, the observed $T_c$ dome is obviously asymmetric, with a shallow linear variation pointing to the origin on the left and a sharp drop on the right. This $T_c$ dome's shape must reflect clean superconductivity in separate $CuO_2$ planes with minimal interplane couplings.

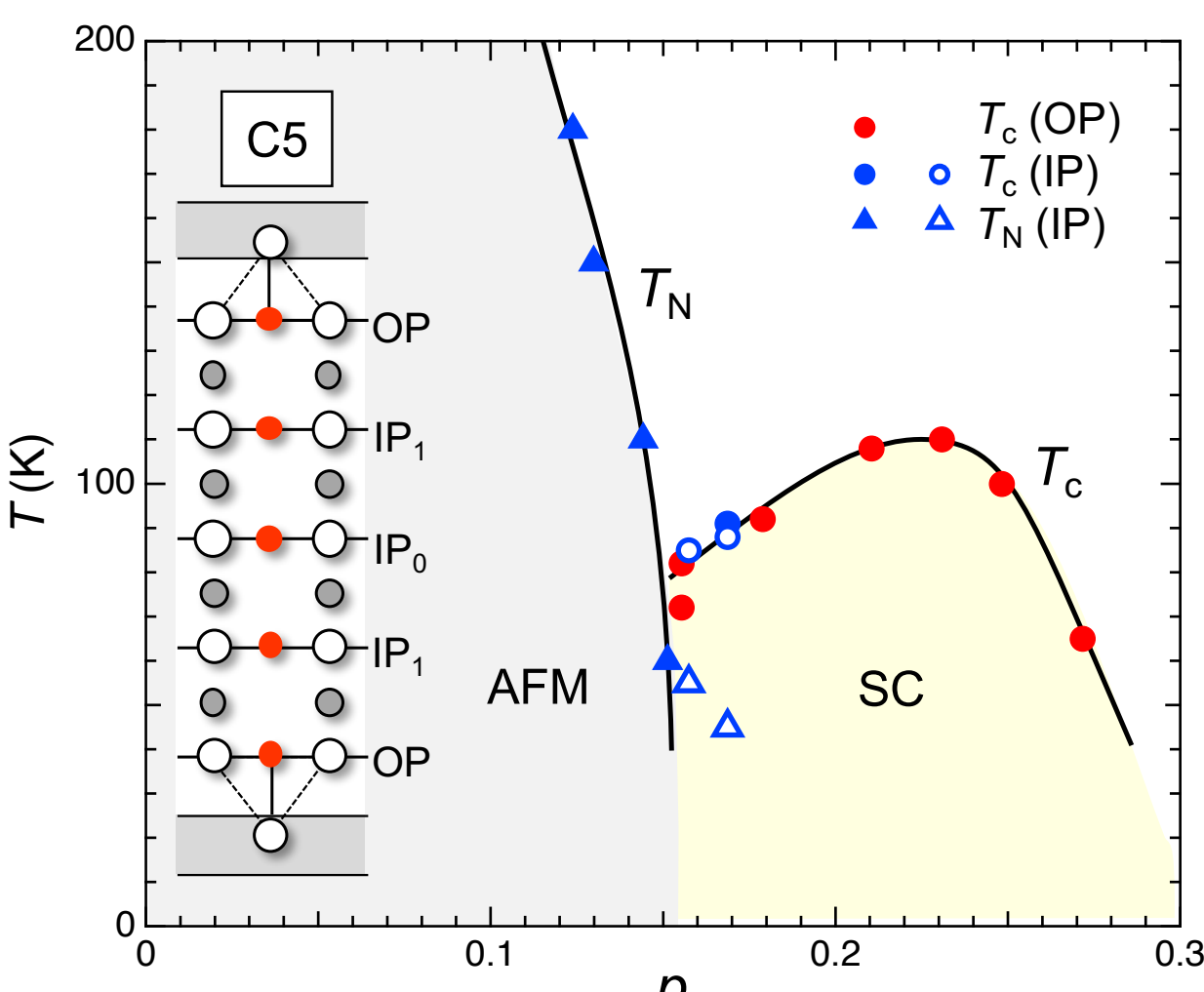

**Fig. 34.** Phase diagram derived from NMR experiments on seven C5 samples (Hg, Ba, and Cu systems) with varying doping levels [166]. The experiments distinguished between OP and IPs (not $IP_1$ and $IP_0$) and determined their hole concentration and ordering temperatures: $T_N$ for antiferromagnetic metal (AFM) and $T_c$ for superconductivity (SC). The data is combined into a single phase diagram as a function of $p$ in each plane. The red and blue circles represent the OP's and IPs' $T_c$ values, respectively, and the triangles represent the IPs' $T_N$ value. The open circle and triangle at $p$ = 0.157 and 0.169 represent the SC and subsequent AFM transitions, respectively. They may occur in more doped $IP_1$ and less doped $IP_0$, respectively (see text). It's worth noting that their $p$ values are the averages for $IP_1$ and $IP_0$. Because of the thick conduction layer, the AFM order extends to the hole range, reaching a higher $p$ value (0.15) than in C1 La214 (0.02). The $T_c$ dome determined for OP and IP ($IP_1$) appears to be asymmetric, with nearly linear expansion to the origin on the left and a relatively rapid drop on the right.

Kunisada and coworkers performed ARPES and quantum oscillation experiments on a C5 Ba0245 [$Ba_2Ca_4Cu_5O_{10}(O_{1-y}F_y)_2$] sample in the UD regime ($T_c$ = 65 K) [186], revealing IPs and OP differences that are consistent with NMR findings [166]. The used Ba compounds have slightly lower $T_{co}$s than the Hg compounds studied by NMR, with $T_{co}$ values of 120 K for $n$ = 3 and 80 K for $n$ > 5 [117] (Fig. 22). This could be due to increased randomness in their block layer; holes are generated by substituting F for O in the thin B2-NC $Ba_2O_2$ block layer.

ARPES discovered two small Fermi surface hole pockets of varying sizes around the Brillouin zone point ($\pi/2$, $\pi/2$) and an atypical Fermi arc (partially missing ring) surrounding them [53]. The presence of a Fermi surface is strong evidence of a metallic state. The former hole pockets correspond to metal states in $IP_0$ and $IP_1$, while the latter represents the OP. The expanding Fermi surfaces indicate that $p$ increases in the order $IP_0$, $IP_1$, and OP, which is consistent with the electrostatic potential model. $IP_0$ and $IP_1$ have $p$ values of 0.02 and 0.045 based on hole pocket size, respectively, which are supported by bulk measurements using the de Haas–van Alphen (dHvA) effect, indicating that they are highly reliable; however, no similar estimate could be made for the Fermi arc for OP.

In addition, temperature-dependent ARPES experiments discovered superconducting gaps at $IP_1$ and OP but not at $IP_0$. $IP_0$ remains an antiferromagnetic metal, with $p$ = 0.02. This clearly demonstrates that AFM is adjacent to SC rather than AFI. Remarkably, the superconducting gap size in $IP_1$ is larger than in OP, indicating that $IP_1$ has a higher $T_c$ despite having a lower hole concentration. This contradicts the $T_c$–$p$ relationship, but it could be explained by increased randomness at OP, which could also account for the origin of the Fermi arc, as well as apical oxygen effects at OP. As a result, instead of OP as in Hg1245, $IP_1$ with $p$ = 0.045 exhibits UD superconductivity at $T_c$ = 65 K, which governs the superconductivity in the Ba0245 sample.

#### 4.6.2.3. Key factors affecting $T_{co}$ in multilayer systems

The NMR results for Hg1245 show that the OP is superconducting at $T_c$ = 110 K and $p$ = 0.23, whereas the three IPs with significantly lower average hole concentrations ($p$ = 0.16) have a lower $T_c$ of 85 K. Unlike in Hg1223, where all three $CuO_2$ planes are superconducting simultaneously (Fig.

35a), the IPs remain normal metal as the OPs transition (Fig. 35b); superconducting fluctuations in the IPs have not developed sufficiently to render superconductivity via the proximity effect. When cooled further, $IP_1$ and $IP_0$ exhibit superconductivity and antiferromagnetism, respectively. This must be true for larger $n$-systems. As a result, the bulk $T_c$ of multilayer systems is entirely determined by their individual OPs (Ba0245 in ARPES is an exception due to increased disorder).

In a weakly 2D system, weak interplane coupling $J'$ determines the critical temperature of a 3D LRO. In contrast, in a highly 2D system with significantly lower $J'$ than strong in-plane coupling $J$, a 3D LRO emerges at a higher critical temperature than $J'$, as the in-plane correlation grows and begins to diverge when cooled to the temperature [223, 224]. For example, Figure 35c shows that when a quasi-2D antiferromagnetic spin system is ordered, the in-plane spin orientations nearly align over a magnetic correlation length $\xi$(AF) that scales with $J$ and grows rapidly with cooling. Weak interplane interactions eventually produce large effective coupling ($\xi$•$J'$) between $\xi$ areas, resulting in 3D ordering. Even minor interactions between planes can have a significant cumulative effect as $\xi$ diverges. As a result, the critical temperature is determined by the properties of a single plane (the magnitudes of $J$ and 2D fluctuations), not their coupling.

In multilayer systems, nearly optimally doped OPs have a strong superconducting correlation and are coupled via thick IP layers with fewer holes and low correlation ($J_{IP}$) and a block layer ($J_{BL}$), resulting in a quasi-2D systems (Fig. 35b). Regardless of separation, the temperature at which the superconducting correlation develops within a single OP plane determines the $T_c$ of multilayer superconductivity, much like the antiferromagnetic order in the quasi-2D system in Fig. 35c. This clarifies why $T_c$ converges with large $n$. The $T_{co}$ scatter remains constant as $n$ increases because the $T_{co}$ is commonly determined by nearly identical OPs.

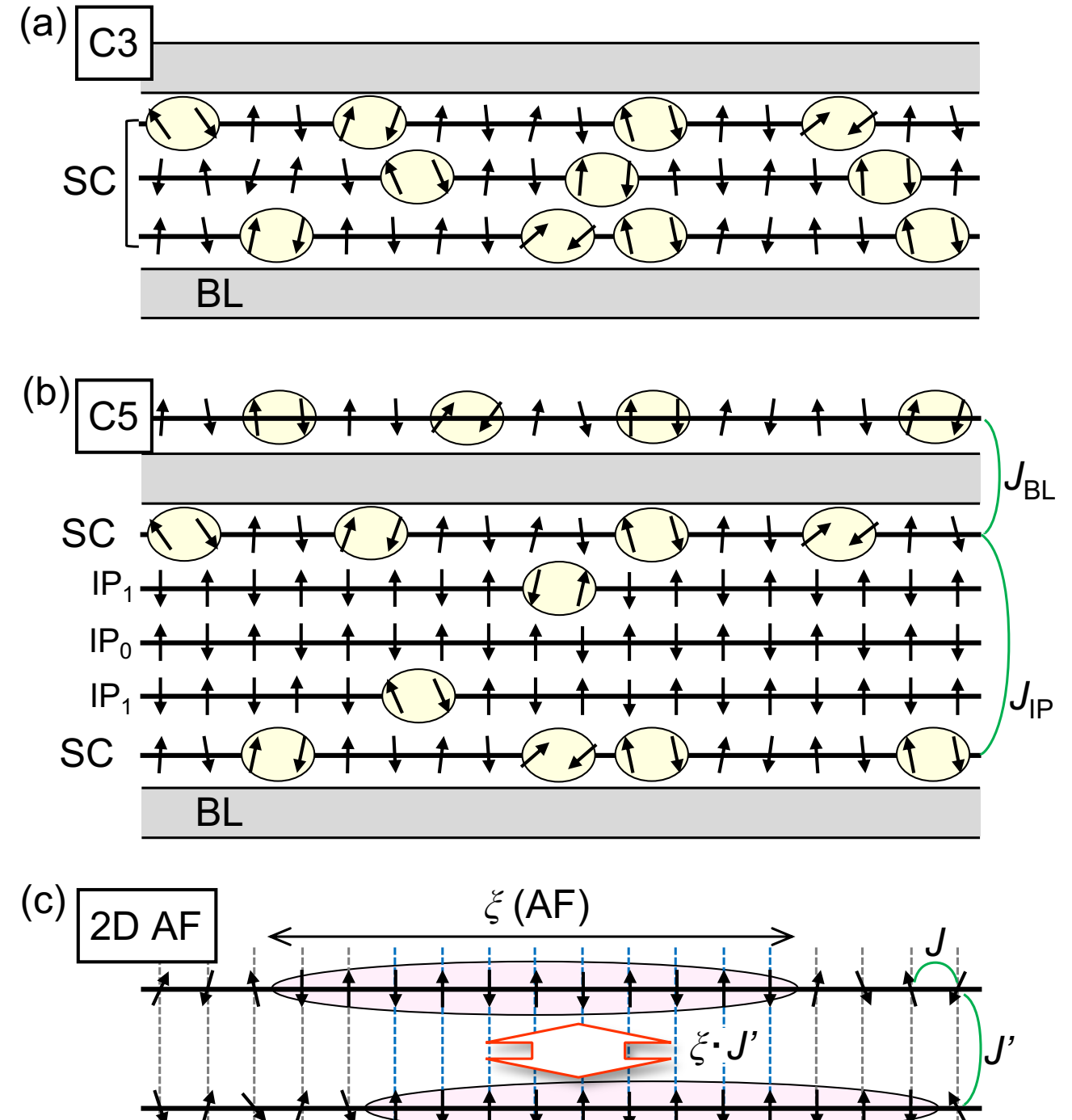

**Fig. 35.** Schematic representations of the electronic states at optimum doping for (a) C3 and (b) C5, as shown in Fig. 33. A pair of encircled arrows represents a BEC Copper pair made up of ZRSs, while other arrows depict Cu spins that are almost antiferromagnetically ordered (straight up and down) or fluctuating (inclined). At $T_c$ = 133 K in (a), three $CuO_2$ planes form a thick superconducting layer. When C5 is cooled to $T_c$ = 110 K in (b), the OP superconducts, but three IPs with fewer holes remain in the paramagnetic metal state, separating the superconducting OPs. $J_{IP}$ and $J_{BL}$ are couplings between them via IPs and a block layer, respectively. When cooled further, $IP_1$ becomes superconducting below 85 K, while $IP_0$ transitions to an AFM at 55 K (Fig. 33). (c) A quasi-2D antiferromagnet with a large in-plane coupling $J$ and a negligible interplane coupling $J'$. When cooled to a critical temperature that scales with $J$ and is reduced by 2D fluctuations, a plane's magnetic correlation diverges, resulting in elongated coherence length $\xi$(AF). Minor $J'$ interactions can result in significant coupling ($\xi$•$J'$) between nearly ordered spins within $\xi$, leading to 3D long-range order at the critical temperature.

In a simple electrostatic model, the $p$ of each $CuO_2$ plane should decrease monotonically with increasing $n$. Figure 36 displays the calculated hole distributions up to C11 for a simple structure ($d_1$ = 4.7 Å, $d_2$ = 3.2 Å, $p_B$ = 1). In C9, for example, $p$ in the OP is calculated to be 0.17 with $A$ = 0.680, a significant decrease from C5's 0.23, and $T_c$ above 100 K is not expected. The $p$ values in each of the seven IPs are much lower, but they are large overall, resulting in a lower $p$ in the OP. Because Coulomb interactions are long-range, increasing $n$ only slightly reduces the potential near the IPs' center. Therefore, the observed convergence of $T_{co}$ to 105 K for large $n$ (Fig. 22) [160] is incomprehensible. A constant $T_{co}$ is unlikely to result from an unlimitedly increasing $p_B$ that accidentally compensates for the $p$ decrease in the OP.

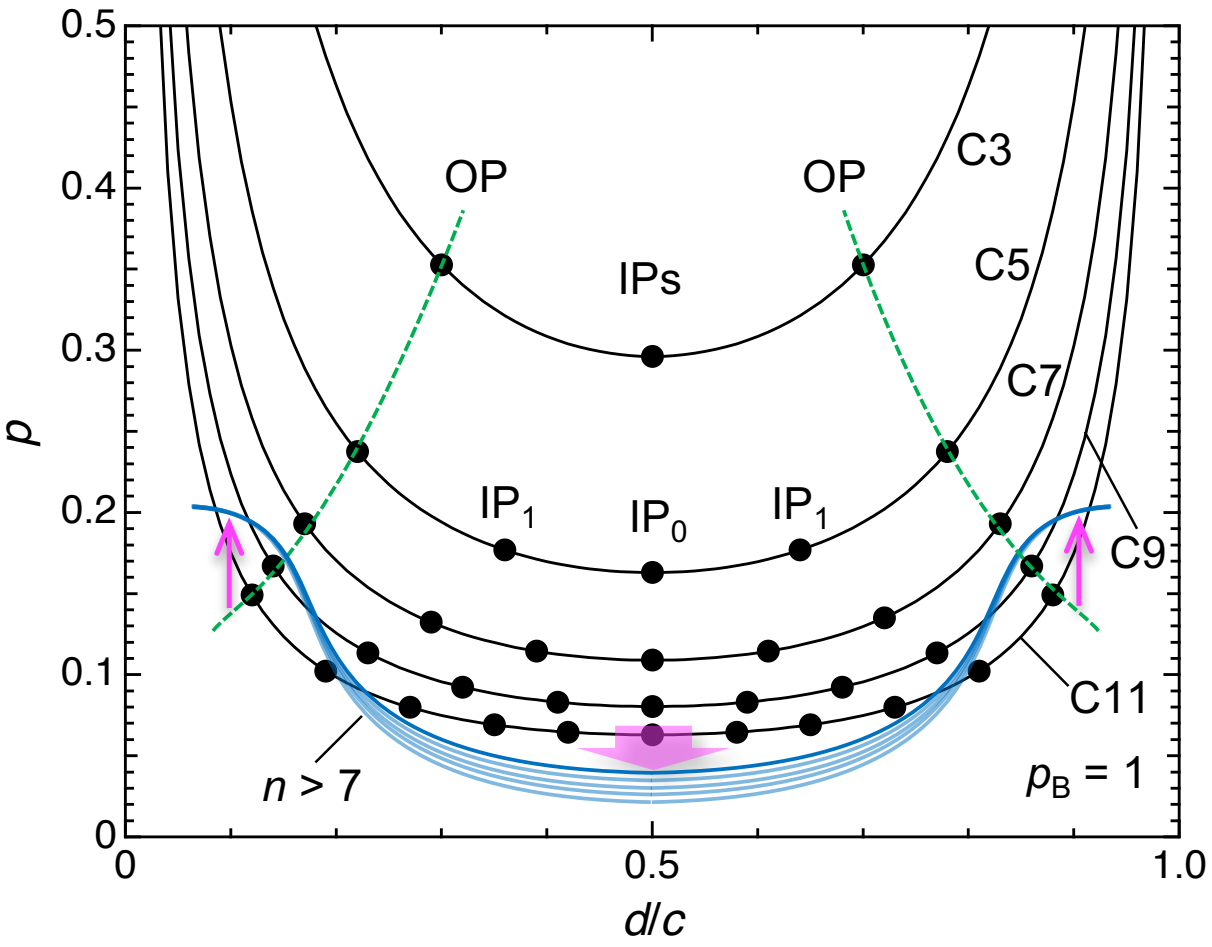

**Fig. 36.** Calculated hole distributions over OPs and IPs for C3, C5, C7, C9, and C11, with $p_B$ = 1 and a simple structural model with stacking distances of $d_1$ = 4.7 Å (between the block layer center and OP) and $d_2$ = 3.2 Å [between OP (IP) and IP]. The circle on each curve represents the plane's position, which is normalized by the $c$-axis length. The calculation predicts that $p$ at OP will continue to decrease as $n$ increases, as shown by the green dotted line, which contradicts the observed convergence of $T_{co}$ to 105 K in Hg compounds (Fig. 22). For large $n$ cases, the blue lines depict the most likely hole distributions across

the planes as a result of modifications caused by carrier screening effects near the block layer. The magenta arrows represent potential changes after corrections that reduce holes in IPs while keeping the OP's $p$ constant at around 0.2, resulting in a constant $T_{co}$ value.

It is plausible that the simple electrostatic model breaks down as $n$ increases. For $n > 5$, the shielding effects of holes in the OP and adjacent IPs must reduce long-range Coulomb interactions, lowering the electrostatic potential at the inner IPs. As a result, holes struggle to enter those IPs, raising the $p$ in the OP. Modified hole distributions, as schematically depicted by a set of blue curves in Fig. 36, are therefore expected. However, a simple electrostatic potential model with an exponential dumping factor for the screening effect cannot produce these curves; a more sophisticated model is needed. In the large-$n$ limit, only a few $CuO_2$ planes above and below the block layer are doped with enough holes to be superconducting, separated by a thick nonsuperconducting and most likely AFM spacer layer made up of many rarely doped IPs (Fig. 35b). If the OP in these separate superconducting layers is always doped with approximately 0.20 holes, the $T_c$ will remain at 105 K.

In $n \geq 4$, an increase in $p_B$ can produce $T_c$ values comparable to 135 K in Hg1223. However, the 2D nature of isolated OP dominates $T_{co}$ suppression (Section 4.7.1), so this is unlikely. Nevertheless, it is important to remember that any block layer's ability to supply holes is constrained by structural capacity. As stated in Section 4.1.3, the B3-NC block layer of Hg compounds can have $p_B$ values as high as one, while the B2-NC and B4-NC block layers have lower limits below 0.6. A low $p_B$ limit results in a lower $T_{co}$ in the large $n$ limit, as illustrated in Fig. 22 for the Ba and Cu systems.

4.6.3. $T_c$–$p$ relationship in the relatively clean $CuO_2$ plane

Figure 37 summarizes the relationship between $T_c$ and $p$ based on the NMR and ARPES experiments. NMR experiments show that the $T_c$ domes of Hg C1, C2, and C3 have similar shapes and gradually expand into the high-doped regime, with $T_{co}$ observed at $p_o = 0.16$, 0.21, and 0.25 [166, 225], which agrees with Uemura's plot in Fig. 23. In contrast, the $T_c$ dome formed by the C5's OP in Fig. 34 has an intermediate $T_{co}$ between C1 and C2 in Fig. 37, and its $T_c$ curve rapidly decreases and approaches C1 as the doping level exceeds $p_o$. This is thought to be because the superconducting order is controlled by a single $CuO_2$ plane (OP) like C1, rather than multiple coupled planes like C2 and C3, which causes it to become unstable rapidly with doping, similar to the AFM order discussed in Section 4.7.1. In contrast, C3 has a high $T_{co}$ due to the superconducting order's resistance to doping in integrated triple $CuO_2$ planes with reduced 2D fluctuations. Therefore, the thickness of the superconducting layer is one of the most important parameters for achieving high $T_{co}$.

The initial $T_c$ rise of the low-doped Hg system has a slope of approximately 850 K (Fig. 37), which is most likely a guide to the universal $T_B$ line. When a line with the same slope is drawn on the $T_c$–$p$ phase diagram for various materials (Fig. 25a), it appears to be the steepest of all systems. However, even in the relatively clean Hg system, randomness-induced hole trapping cannot be ignored, resulting in a lower slope. In contrast, the $T_c$–$p$ relationship from ARPES experiments has a much larger slope: the slope for $IP_1$ ($T_c$ = 65 K, $p$ = 0.045) in the C5 Ba system exceeds 1400 K. Given ARPES's superior reliability over NMR in the low-doped regime, as well as cleaner IPs in ARPES than in OP in NMR, the actual $T_B$ line must have a steep slope similar to ARPES. This important point will be addressed further in Section 6.2, which will focus on the strategy for increasing $T_c$.

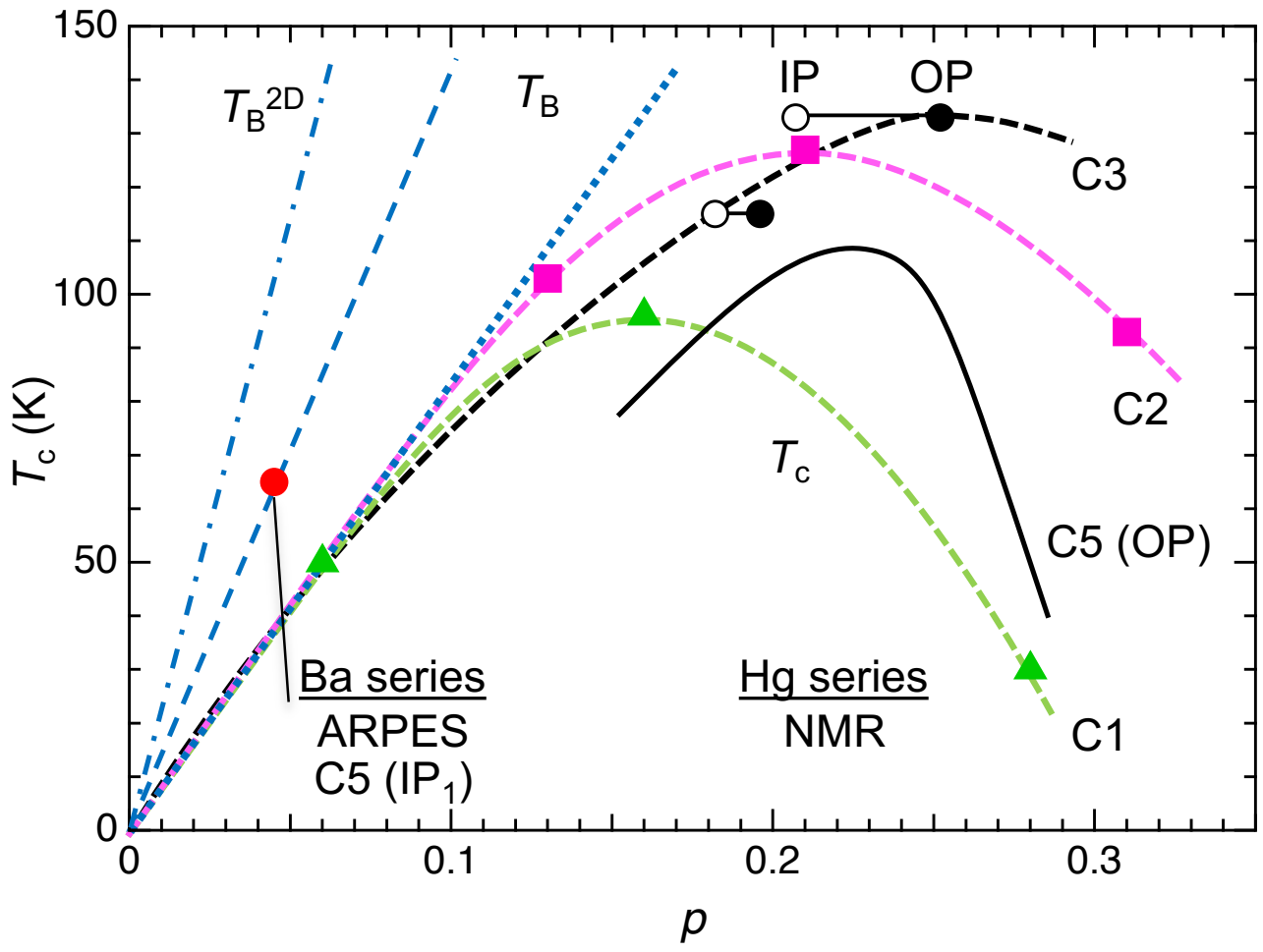

**Fig. 37.** $T_c$–$p$ plot based on ARPES data for the C5 Ba compound [186] and NMR data for the Hg series of C1 and C2 [225], as well as C3 and C5 (OP) (Fig. 34) [166]. The NMR data points show an initial rise of 850 K (blue dotted line), while the ARPES data point for C5 Ba $IP_1$ yields a larger slope of 1400 K (blue dashed line). The initial slope may become steeper as it approaches the clean limit, which corresponds to the ideal $T_B^{2D}$ line predicted for 2D BEC superconductivity at a slope of 2300 K (blue dot-dash line) [226, 227].

4.6.4. The ultimate copper oxide superconductor

The experimental findings and discussion presented in this section offer a plausible explanation for the material dependence of the $T_{co}$–$n$ and $T_c$–$p$ relationships. After sorting through the various materials, we discovered that Hg1223 possesses a distinct advantage: it has an outer $CuO_2$ plane that is least susceptible to the effects of apical oxygen and randomness from the block layers with high hole donation capacity, as well as an ideal single inner $CuO_2$ plane that is further protected from randomness by the OP. Furthermore, the OP and IP are nearly optimally and slightly less doped with holes, respectively, allowing them to exhibit superconductivity concurrently, resulting in thick superconducting layers that stabilize the 3D LRO while remaining resistant to hole doping. When $n$ increases, only a pair of OPs separated by a thick nonsuperconducting spacer IP layer contribute to 2D superconductivity, which is less resistant to hole doping. Consequently, Hg1223 has the highest $T_c$ value of 135 K. It is not an exaggeration to call Hg1223 the ultimate copper oxide superconductor in this regard. If the maximum expected $T_c$ value is 150 K, as mentioned in Section 4.3.1, there is still a 10% decrease, implying that it may be possible to increase the $T_c$ using a method other than pressure.

4.7. Expected electronic states in an ideal $CuO_2$ plane

4.7.1. The appearance of AFM instead of AFI

ARPES and dHvA experiments on the Ba0245 sample ($T_c$ = 65 K) revealed that $IP_0$ with $p$ = 0.02 was a metal with a small Fermi pocket [186], while NMR experiments [166], which are sensitive to magnetism, revealed that the Hg1245 sample ($T_c$ = 110 K) had IPs at $p$ = 0.157 on average, exhibiting antiferromagnetism at $T_N$ = 55 K (Fig. 33b). The IPs had a magnetic moment of 0.10$\mu_B$ ($\mu_B$ is the Bohr magneton), far lower than the parent phase's 0.4$\mu_B$ in La214 ($T_N$ = 325 K) [136]. This is consistent with reduced magnetic moments caused by 2D fluctuations in conventional itinerant magnets [147]. AFM is thus a natural state for clean IPs in multilayer superconductors, as demonstrated by comparable results for Cu- and Ba-based multilayer superconductors [166, 228].

When the randomness effect is minimal, AFI transforms to AFM with doping of less than 2%, most likely much less, and the superconducting phase's left neighbor should be AFM, as shown in Fig. 38a, rather than AFI, as in La214's phase diagram (Fig. 20). While AFI is a simple antiferromagnetic order of localized Cu spins, AFM is thought to be a metallic state in which oxygen holes (ZRSs) move through a forest of Cu spins while maintaining their antiferromagnetic long-range order. The coexistence of charge and spin degrees of freedom is a hallmark of SCES.

The NMR experiments on multilayer systems reveal that the hole concentrations required for the AFM phase to appear vary significantly with $n$ [166]. Figure 34 shows that the AFM phase is extended to $p$ = 0.15 for C5 Hg1245, while the critical concentrations for C4, C3, and C2 are lowered to $p$ = 0.12, 0.11, and 0.09, respectively. Moreover, the AFI phase disappears at $p$ = 0.02 in C1 La214. As $n$ increases, AFM survives up to larger $p$ values. When multiple $CuO_2$ planes with few holes and high antiferromagnetic correlations are stacked and strongly coupled by interplane magnetic interactions, they form a single antiferromagnetic layer that thickens as $n$ increases (refer to Fig. 35b with fewer holes). Weakening 2D magnetic fluctuations enhances antiferromagnetic correlations inside the layer. Then, a minimal interlayer coupling across the block layer ($J_{BL}$), which is unaffected by $n$, results in a robust 3D antiferromagnetic LRO that is resistant to hole doping and expands to high doping levels. It was suggested that enhanced three-dimensionality helped to stabilize AFM [166]. However, caution is advised because increasing three dimensionality in quasi-2D magnets frequently implies increased interlayer coupling, which is not the case here. We think that as $n$ increases, the antiferromagnetic correlations in each layer become more resistant to doping. The growing AFM phase with $n$ obscures the initial $T_c$ rise, causing $T_c$ to appear suddenly, as shown in Fig. 34.

#### 4.7.2. The intrinsic electronic phase diagram

The NMR and ARPES experiments on multilayer systems revealed the characteristics of a clean $CuO_2$ plane, which are summarized as an electronic phase diagram for ideal copper oxide superconductivity in Fig. 38a. It depicts the expected C3 phase diagram, in which the conduction layer is made up of three $CuO_2$ planes without apical oxygens (even in OPs), connected by a minimal interlayer interaction across clean block layers, and evenly doped with holes. The horizontal axis represents mobile hole concentration $p^*$ instead of nominal $p$. The AFI phase (Mott insulator), which is only present near $p^*$ = 0, immediately gives way to the AFM phase; one hole may be sufficient to render metallicity. In contrast, AFM's $T_N$ gradually decreases until it vanishes at a critical hole concentration. Based on NMR results [166], the value for ideal C3 must be smaller than 0.15 for thicker C5 compounds in Fig. 34 and slightly larger than 0.08 for actual C3, taking into account hole trapping loss (~0.1 in Fig. 38a).

The AFM phase generates an extended region with significant antiferromagnetic spin fluctuations on the right. The attractive force produced by the antiferromagnetic spin background causes ZRS pairs to form below $T_p$, where fluctuations are highly developed. As hole doping progresses, the AFM's 3D order weakens, and when $T_N$ falls below $T_B$, which rises proportionally to $p^*$ (assuming an 850 K slope), the ZRS pairs share phases at $T_B$ and transform into Cooper pairs at $T_c$. As a result, the BEC superconducting phase appears rather than the AFM; however, as discussed in Section 4.3.1, if $t_{eff}$ is small or becomes negligible near $p^*$ = 0, a miscibility gap will open between AFM and SC, which is inaccessible by experiments on actual disordered materials.

As $p^*$ increases, the $T_p$ line decreases due to the loss of attraction caused by the collapse of the antiferromagnetic spin background, and $T_c$ peaks at an optimum hole concentration of around 0.25 (Figs. 25, 26, and 37). BCS superconductivity with d-hole pairing occurs in the OD regime, as opposed to ZRS pairing in the UD regime. Superconductivity disappears at $p_e$, which is unknown but believed to be around 0.4 based on Figs. 25 and 37.

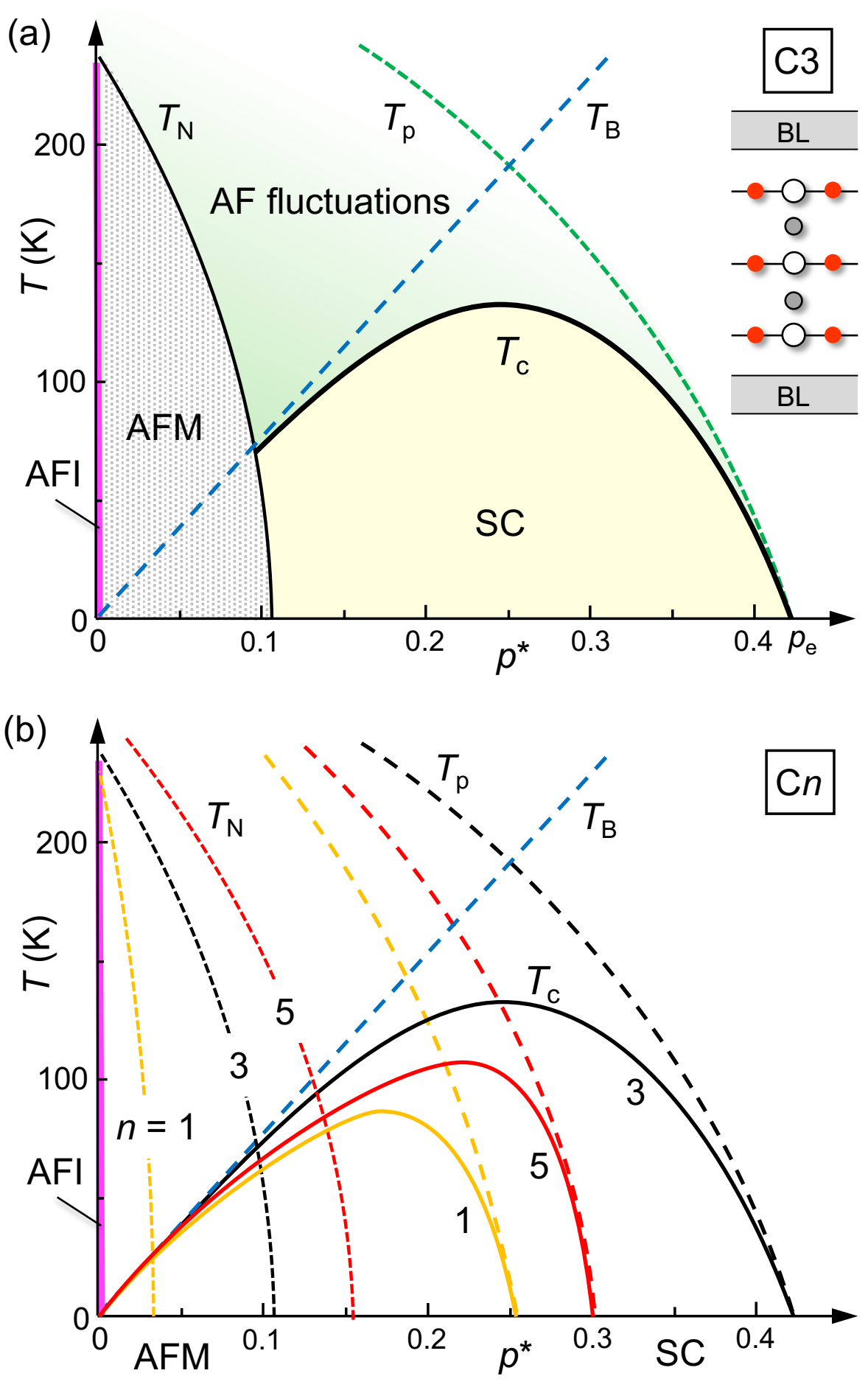

**Fig. 38.** Hypothetical phase diagram for ideal copper oxide superconductivity. The horizontal axis represents the mobile hole concentration ($p^*$), not the nominal $p$. (a) assumes C3 with conduction layers made up of three $CuO_2$ planes (no apical oxygen even in OP), evenly doped with holes, and coupled via minimal interlayer interactions across a clean block layer (BL)

to maintain 3D long-range order. As $p^*$ increases, the antiferromagnetic insulator (AFI) vanishes immediately with one hole or after phase separation due to small $t_{eff}$, the antiferromagnetic metal (AFM) disappears around 0.1, and the superconducting (SC) phases emerge, peaking at $p^* = 0.25$ and disappearing around $p_e = 0.4$. The green area to the right of the $T_N$ line represents an antiferromagnetic fluctuation region that causes ZRS or d-hole pairing below $T_p$, followed by Cooper pairing to BEC superconductivity below $T_c \sim T_B$ in the UD regime and BCS superconductivity below $T_c \sim T_p$ in the OD regime. All copper oxide superconductors share the same $T_B$ line, which is assumed to have an 850 K slope based on Fig. 37 or a steeper slope in the clean limit, but the $T_p$ curve varies by material. $n$ alters the phase diagram in (b). Because of the robust 3D order in thick conduction layers, the AFM region simply expands with $n$. The SC dome, on the other hand, reaches its maximum at $n = 3$ and shrinks at $n = 5$ as $n$ increases, because only isolated OPs superconduct. No more changes occur for $n > 5$. It should be noted that for $n$ greater than three and with uneven hole division, $p^*$ represents hole concentration in IPs for AFM and OP for SC.

Unlike C3 in Fig. 38a, doping rapidly destabilizes long-range order in C2 and C1, which have thinner conduction layers, and both the AFM and SC phases shrink toward the low-doping side, as seen in Fig. 38b. C1's SC phase may disappear at $p_e \sim 0.25$, as observed in La214 (Fig. 20) and Hg2201 (Fig. 28c). $p_e$ appears to be independent of block layer type. On the other hand, if the hole distribution was even, with enough holes at C4 or higher, both phases would expand into the high-doped regime, resulting in higher $T_{co}$. In reality, the hole concentration differs significantly between the OP and IPs. In C5 of Fig. 38b, which is drawn according to Fig. 34, LRO in thick AFM layers is doping-resistant, surviving until $p = 0.15$, while separated thin OPs allow 2D-like superconductivity (Fig. 35b), resulting in a smaller $T_c$ dome, lower $T_{co}$, and a rapid drop in $T_c$ on the right side of the dome, vanishing at 0.3. Further increases in $n$ have little effect on the SC dome and only a minor expansion of AFM, which is insufficient to cover the SC dome. The thickness of the superconducting layer thus governs the $T_{co}$–$n$ relationship, as shown in Fig. 22.

In Section 4.6.1, we discussed the chemical trend of $T_{co}$ from C1 to C3 in terms of the apical oxygen effect and the randomness effect; however, the former's influence is nearly identical to the superconducting layer thickness effect, which is highlighted here. The apical oxygen effect is clearly important because it contributes to the variation in $T_{co}$ at C1. However, explaining the decrease in $T_{co}$ and its convergence to a constant value at C5 and above using only the two characters of the block layers is difficult, necessitating the thickness effect of the superconducting layer. As a result, it is believed that all three of these effects play an important role in $T_{co}$'s chemical trend.

Although even hole distribution over $CuO_2$ planes cannot be achieved in C4 or higher, supplying additional holes to the IP may result in a thicker superconducting layer than C3, potentially rasing $T_{co}$. For example, if a monovalent metal (such as $Na^+$) was substituted into the Ca site sandwiched between the IPs of C5 Hg1245 and the three IPs underwent a superconducting transition simultaneously as the OP, a higher $T_{co}$ with a higher $p_o$ could be obtained; however, $T_{co}$ may decrease because the randomness effect caused by Ca-site substitution is dominant. While conventional thermodynamic synthesis methods may make it difficult to develop such materials, thin film fabrication using layer-by-layer growth is thought to be feasible.

In actual phase diagrams, the $T_B$ line is shared by all materials, while the $T_p$ line and $T_c$ dome are affected by block layer properties. At high doping levels, the apical oxygen effect destabilizes the ZRS, pushing $T_p$ to the low-doping side (Fig. 27) and shrinking the right side of the $T_c$ dome. At low doping levels, the randomness effect becomes critical near the AFM/SC boundary, resulting in inhomogeneity or two-phase coexistence. In addition, spin glass and other disorder-induced phases will form in the gap, reducing the left edge of the $T_c$ dome (Fig. 41). When the horizontal axis is assumed to be nominal $p$, the AFI or other insulating phases appear to spread to the high-doping side due to conduction carrier loss caused by random hole trapping, further reducing the left side of the $T_c$ dome (Fig. 30). As a result, when compared to the ideal shape in Fig. 38a, the measured $T_c$ dome shrinks to the lower left and resembles a parabolic shape (Fig. 25).

4.7.3. Remarks on the superconductivity mechanism

Soon after the discovery of cuprate superconductors, Baskaran and Anderson proposed climbing the $T_c$ dome from the left. They developed the resonating valence bond (RVB) theory [229, 230]. When the crossover temperature to the RVB state ($T_{RVB}$) is used instead of $T_p$, the electronic phase diagram in Fig. 38a matches RVB superconductivity predictions [139, 231]. The RVB theory is beyond the author's understanding, so he cannot argue it in this paper. In contrast, climbing the $T_c$ dome from the right relies on the spin fluctuation mechanism, which predicts an AFM phase to the left of the SC phase rather than an AFI phase [148]. Because of the intrinsic phase diagram in Fig. 38a, the right-handed approach may provide a better approximation than climbing from the left. It does not, however, cover the entire doping regime because, contrary to what spin fluctuation theory assumes, BEC superconductivity in the low-doped regime is more closely related to real-space pairing than momentum space. As previously stated, ascending the $T_c$ dome from either side simply changes the starting point for the approximation; the truth is always somewhere in the middle. Anyway, the attraction mechanism derived from the antiferromagnetic interaction of copper spins is shared by both. The BCS–BEC crossover concept is applicable to all high-temperature superconductivity, regardless of Cooper pair attraction source.

In general, the quantum critical point (QCP) scenario, which will be covered in Chapter 5.1, can account for the vast majority of superconductivity phase diagrams [62, 232]. Because of the increased fluctuations used for Cooper pairing, a $T_c$ dome is expected to form around a QCP, at which point the order associated with superconductivity is lost (Fig. 42). Cuprate superconductivity has been extensively investigated under the QCP scenario, assuming QCP at $p_o$ [102, 233]. However, the neighboring AFM phase vanishes at lower doping levels than $p_o$. Although large fluctuations are expected to enhance pairing interaction at that QCP, the $T_c$ remains low due to the low boson concentration and $T_B$. The maximum $T_c$ is thus achieved at $p_o$ through a trade-off between increasing $T_B$ and decreasing $T_p$. The QCP scenario applies to cuprate superconductivity, but the resulting phase diagram (Fig. 38a) differs significantly from the general one (Fig. 42). It should be noted that the QCP scenario considers fluctuations in BCS superconductivity rather than carrier numbers, which are critical in BEC superconductivity. There are claims that alternative "hidden" orders with QCP at $p_o$ exist, resulting in the $T_c$ dome under the QCP scenario (see Section 4.8.2.1). Nevertheless, it is difficult to believe that the observed variety of material-dependent extra orders contributes to the common phase diagram of cuprate superconductivity.

### 4.8. Additional aspects of copper oxide superconductivity

This section will cover electron-doped systems and competing orders, both of which are important aspects of copper oxide superconductivity. Despite extensive debate, both seem to be contentious. The author will provide insights into the issues at hand, assisting readers to better understand copper oxide superconductors.

#### 4.8.1. Electron-doped superconductors

Copper oxides, such as $Nd_{2-x}CeCuO_4$ (Nd214) and $Sr_{1-x}Nd_xCuO_2$ [IL(Nd)], achieve superconductivity by doping electrons rather than holes [234, 235]. For more information, read the review [103, 236]. Nd214 is a C1-B2-CF containing a $CaF_2$-type $Nd_2O_2$ block layer (see Figs. 17c and 39). When $Ce^{4+}$ replaces $Nd^{3+}$, the system becomes superconducting because electrons generated under charge-neutral conditions flow into the $CuO_2$ plane. It is common to compare the hole-doped C1-B2-NC La214 to the electron-doped system, emphasizing the difference in superconducting properties (Fig. 13).

##### 4.8.1.1. Electron–hole symmetry

There are a few notable distinctions between electron-doped Nd214 and hole-doped La214 in the same C1-B2 (Figs. 13 and 39) [235]. The former $T_{co}$ of 24 K is significantly lower than the latter $T_{co}$ of 39 K. In the former, $T_c$ simply decreases with doping, with no peak after the AFI phase survives electron doping up to $x = 0.14$, which is much higher than the La214's doping level of 0.02. Many previous books and reviews emphasized the presence of a wide AFI region as a distinguishing feature compared to hole-doped systems [103]. The observed electron–hole asymmetry is thought to represent either two different superconductivity mechanisms in the same $CuO_2$ plane or, in an extended $t$–$J$ model with different electron transfer parameters for electrons and holes or additional distant transfer integrals only in the former.

Other electron-doped superconductors behave differently than Nd214. In Fig. 39, electron-doped C1-B1 $Sr_{1-x}La_xCuO_2$ [IL(La)] shows superconductivity at a low doping level of 0.05, followed by a $T_c$ dome with a higher $T_{co}$ of 40 K above 0.1 [237, 238]. In addition, C1-B2-CF $Pr_{2-x}Ce_xCuO_{4-\delta}$ (Pr214), which crystallizes in the same structure as Nd214, was rendered superconducting at $x = 0.04$–$0.17$ by extracting excess oxygen in a reducing preparation condition [239]. A systematic study on $(Pr, La)_{2-x}Ce_xCuO_{4-\delta}$ found a $T_c$ dome with an electron concentration ($n_e$) of 0.09–0.20 [240-242]. Therefore, the myth that electron doping makes the AFI phase more stable than hole doping must be dispelled; in reality, the two appear to be quite similar.

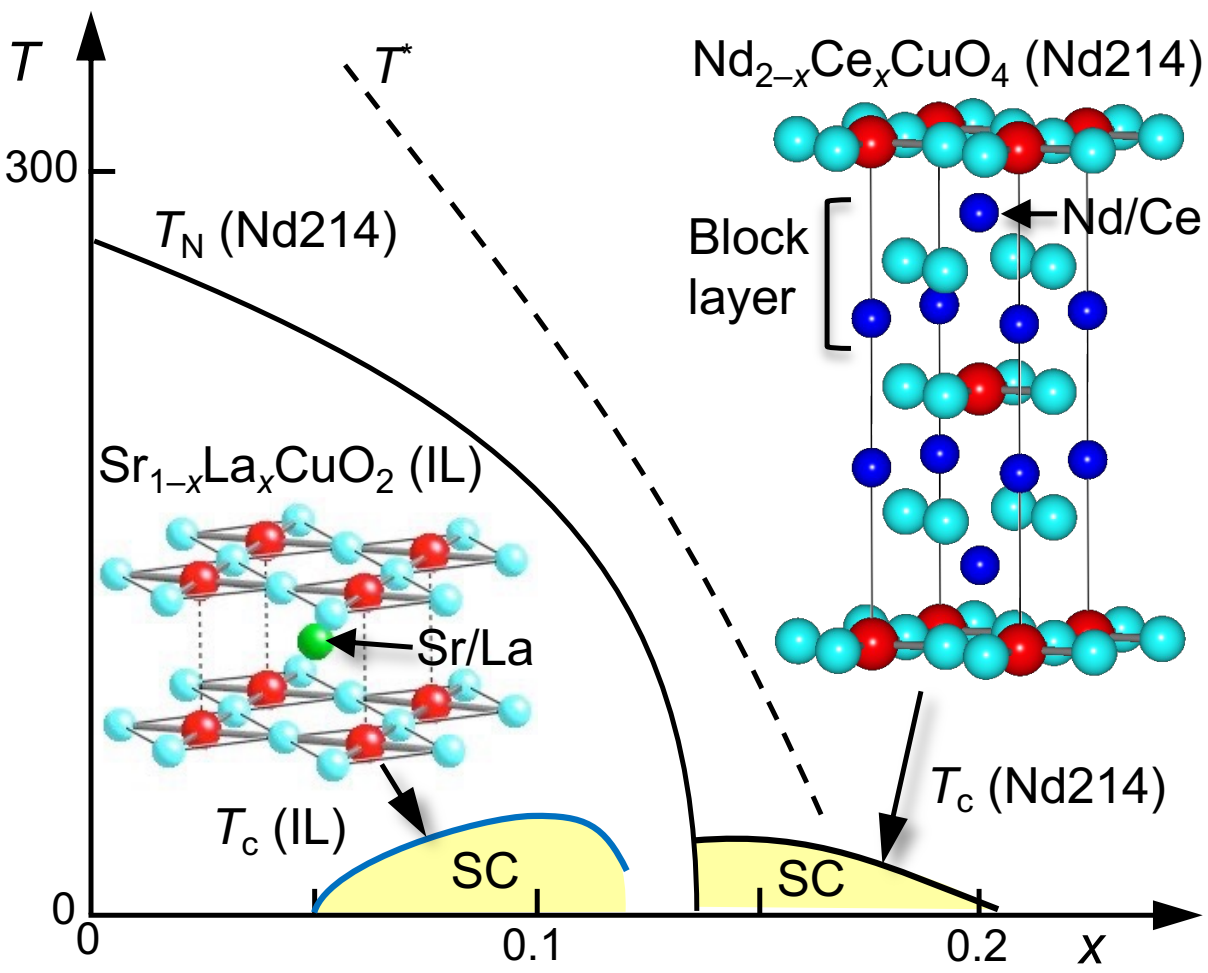

**Fig. 39.** Phase diagram for electron-doped copper oxide superconductors. For C1-B2-CF Nd214, an AFI phase exists up to $x = 0.14$, followed by a portion of the $T_c$ dome with $T_{co} = 24$ K [235]. In contrast, C1-B1 IL(La) exhibits a $T_c$ dome with a higher $T_{co}$ of 40 K at a lower doping range of $x = 0.05$–$0.12$ [237, 238], which is similar to hole-doped La214.

##### 4.8.1.2. Randomness effects in electron-doped systems

The electron-doped system, like the hole-doped system discussed in Section 4.5.2, should experience a randomness effect that lowers $T_c$. The randomness effect caused by the Ce substitution may explain Nd214's low $T_{co}$ and apparent wide AFI range. The crystal structures of the block layers in Nd214 and La214 are $CaF_2$-type and NaCl-type, respectively (Fig. 17). In the former, where the oxygen-free Nd sheet is adjacent to the apical oxygen-free $CuO_2$ plane (Fig. 39), Ce substitution for Nd must result in a higher random potential; in La214, structural relaxation and local polarization of oxygen atoms in the La(Sr)O sheet may weaken the impurity potential.

Pr214's Cu NMR spectrum consists of two components with small and large linewidths [243, 244]. The sharp component is thought to represent an AFI region with few Ce atoms and all electrons localized, while the broad component represents a SC region with more Ce atoms and moving electrons. Thus, the superconductivity occurs in highly disordered regions. This robust randomness effect must be the source of the low $T_{co}$. In addition, strong randomness expands the apparent AFI region on the phase diagram by trapping more electron carriers, reducing the effective number of mobile electrons. The antiferromagnetic region can also expand in the presence of thick magnetic layers, as discussed in Section 4.7.1, but this is unlikely for Nd214 and La214, both of which only have one $CuO_2$ plane.

IL(Nd) has a higher $T_{co}$ of 43 K than Nd214, indicating either increased stability of LRO due to interplane couplings or a relatively weak randomness effect, despite the structural feature that every $CuO_2$ plane in IL(Nd) is exposed to substitutions. In addition, IL(La) has a $T_c$ dome around $x = 0.05$–$0.12$ (Fig. 39) [237], resulting in a narrower AFI region due to weak electron trapping by randomness. When considering the randomness effect, there is no discernible difference between electron- and hole-doped superconductors.

4.8.1.3. Cooper pairing of electrons

Electrons added to the $CuO_2$ plane do not produce ZRS, as observed in hole-doped systems. The newly added electron pairs with an existing electron in the Cu $3dx^2–y^2$ state, forming a $d^{10}$ electron configuration (Fig. 18) [103, 238]. This spinless state is analogous to a d-hole, which contains empty $3dx^2–y^2$ states ($d^8$) under heavy hole doping. Both are voids in the spin-1/2 square lattice known as "d-holes". As a result, the superconductivity mechanisms in hole and electron doping should differ, albeit not fundamentally. Although it has been claimed that metallic states or even superconductivity can be achieved in parent C1-B2-CF materials without electron doping [245, 246], this claim has yet to be experimentally verified. It should be noted that the phase diagrams of clean $CuO_2$ planes (Figs. 38 and 40) demonstrate that even minor doping (most likely electron injection caused by oxygen vacancies in Nd214) converts AFI to AFM.

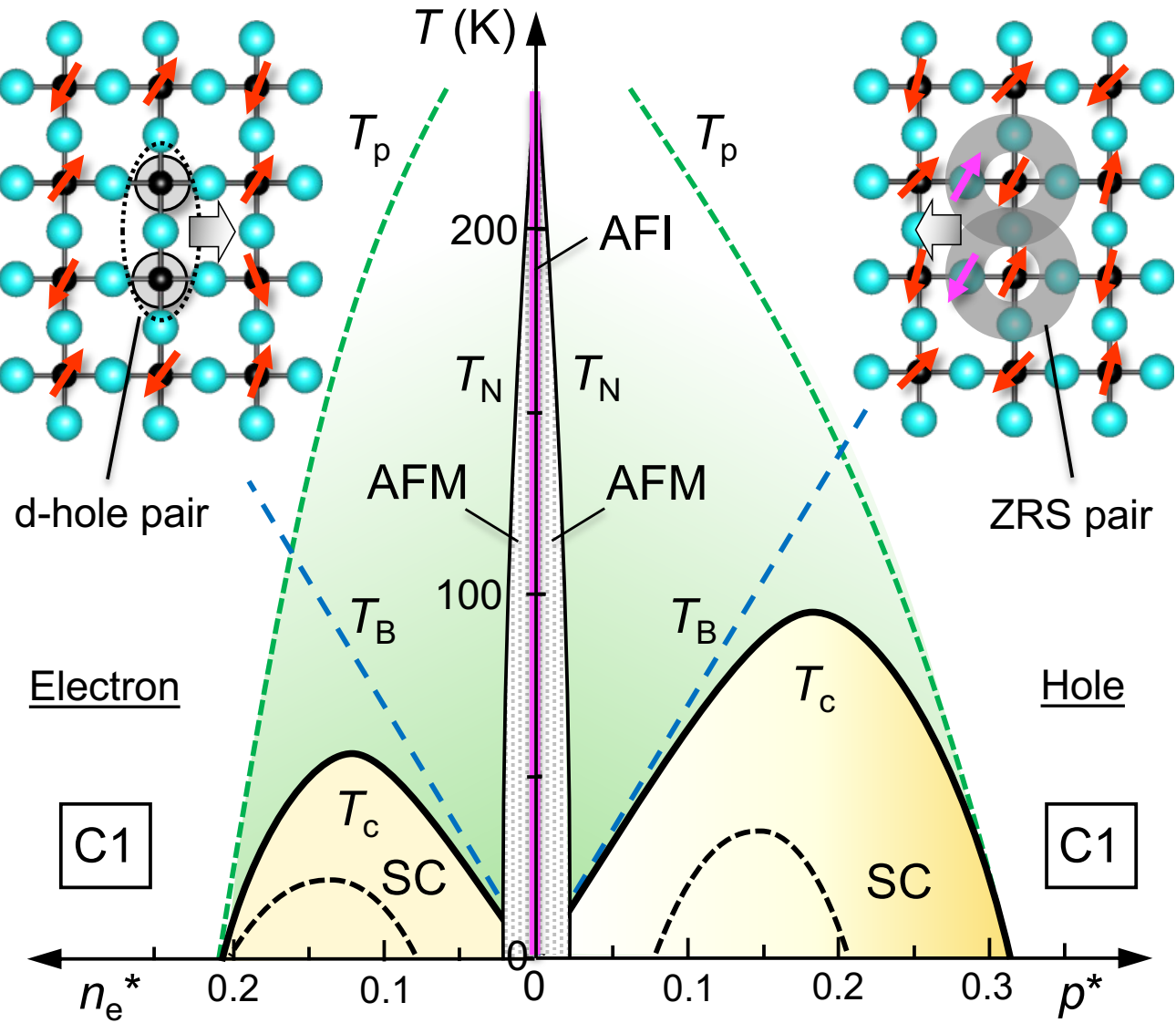

**Fig. 40.** Combined ideal phase diagrams for hole- and electron-doped C1 compounds. Their $T_c$s commonly rise first along $T_B$ lines following AFI/AFM suppression with doping. The $T_B$ lines share the same slope and are proportional to the number of mobile electrons ($n_e^*$) on the left and mobile holes ($p^*$) on the right. As previously mentioned for hole doping, the $T_p$ line on the right gradually decreases as ZRSs increase, then rapidly when transitioning to d-holes ($d^8$). In contrast, on the left for electron doping, the $T_p$ line decreases faster because the antiferromagnetic spin background is simply diluted by d-holes ($d^{10}$) as $n_e^*$ increases, causing $T_c$ to fall faster. At low doping levels, random carrier trapping shrinks the apparent $T_c$ dome of Nd214 to the bottom left (broken curves), similar to hole doping in La214 and Bi2201. When randomness effects are properly taken into account, nearly complete electron–hole symmetry arises. This ideal phase diagram will be compared to the experimental one in Fig. 13, which has been widely used in previous textbooks and should be replaced.

The formation of ZRS pairs at the beginning, or spinless d-hole pairs, distinguishes between hole and electron doping. According to the Cooper pairing image in Fig. 19, the formation of either pair leads to a $J$ energy gain over two freely moving carriers, enabling BEC superconductivity. Figure 40 illustrates a comparison of hypothetical phase diagrams. For hole doping on the right, a ZRS crossover to a d-hole ($d^8$) in the OD regime reduces $T_p$ faster than in the UD regime (Fig. 27). Superconductivity emerges with d-hole ($d^{10}$) pairing at the start of electron doping on the left, while the antiferromagnetic spin background is simply diluted, reducing effective attraction. Thus, $T_p$ drops more quickly, resulting in lower $T_{co}$ at lower optimum doping levels. It is worth noting that the slope of the $T_B$ line, which is proportional to pair concentration, is identical for hole and electron doping; however, electron doping does not allow for comparable $T_{co}$ to hole doping due to the faster drop in $T_p$. Despite minor differences between Cooper pair entities, the underlying attraction mechanism, which is driven by the antiferromagnetic spin background, remains the same. As previously stated, when the randomness effect is considered, the phase diagrams of hole- and electron-doped superconductors must be comparable, as depicted in Fig. 40. Hence, the electron–hole symmetry is almost completely preserved.

4.8.1.4. Solid state chemistry of electron-doped superconductors

There are a limited number of electron-doped cuprate superconductors. This is due to the difficulty of matching the block layer's in-plane size to that of the $CuO_2$ plane. Hole doping shortens the Cu–$O_p$ distance and shrinks the $CuO_2$ plane, whereas electron doping has the opposite effect. Several small block layers can be lattice-matched to the shrunken $CuO_2$ plane, but there are fewer large block layers: the $CaF_2$-type $Nd_2O_2$ layer of Nd214 is significantly larger in-plane ($a$ = 0.395 nm) than the NaCl-type $La_2O_2$ layer (0.381 nm) of La214 (Fig. 17), and the block layer in C1-B1 ILs is entirely Sr atoms; size-matching constraints do not apply.

C2 and other higher members are difficult to prepare in electron-doped cuprate crystal chemistry because their structures require the use of small metals such as Ca, Sr, and Y to form thick conducting layers. For example, when making "$Nd_{2-x}Ce_xCaCu_2O_6$" of C2-B2-CF with a $Nd_2O_2$ block layer, analogous to hole-doped C2-B2-NC $La_{2-x}Sr_xCaCu_2O_6$, it is difficult to avoid Ca partially replacing Nd because they have similar ionic radii and prefer similar crystallographic environments; unless there is a significant difference in site potentials, two atoms will always mix to gain mixing entropy at high synthesis temperatures. As a result of this mixing, the block layer transforms into a (Nd, Ce, $Ca)_2O_2$ layer with a smaller average metal radius, preferring the NaCl-type structure over the $CaF_2$-type structure and shrinking to prevent electron doping; C2-B2-NC forms rather than C2-B2-CF, shrinking $CuO_2$ planes that accept only holes; and nature frequently behaves unexpectedly. In addition, while using large Ba instead of Ca may produce a B2-CF block layer, the $BaCuO_2$ unit, unlike $SrCuO_2$, is unable to form an oxygen-deficient perovskite layer and thus does not function as a conducting layer.

Electron-doped C1 compounds with less disorder will have a higher $T_{co}$. Clean doping using the EDL technique [213] could achieve high $T_c$ electron-doped superconductivity in C1-B1 $SrCuO_2$ or a stretched $CuO_2$ plane at the thin film interface under epitaxial strain; however, the maximum number of carriers that can be doped by the EDL method is more than one order of magnitude smaller than the amount required to produce high $T_c$ [213]. If clean doping is possible, the current C1 $T_c$ dome (broken line) on the left side of Fig. 40 will expand to the ideal one (solid line), resulting in a higher $T_{co}$. Nevertheless, it is unlikely to achieve a higher $T_{co}$ than hole doping because electron doping inevitably causes a faster decrease in $T_p$, due to d-hole pairing rather than ZRS pairing.

4.8.2. Other orders and pseudogap phenomenon

The "general" $T$–$p$ phase diagram of cuprate superconductivity was compiled using the vast amount of experimental data accumulated through the La214 study, along with additional results from other materials [247]. Keimer's 2015 review includes a phase diagram with multiple competing phases [102]. However, the author believes that the other states and associated fluctuations, in addition to the antiferromagnetic insulating/metallic, superconducting, and normal metallic phases in Fig. 38a, are not intrinsic; rather, they may be metastable orders hidden in the $CuO_2$ plane or related to a structural or electronic instability inherent in each material that would not manifest in the absence of disorder. This level of complexity is common in SCES, which pits multiple ground states against one another. In this section, we will briefly review the other competing orders and talk about the pseudogap phenomenon.

4.8.2.1. Competing metastable orders

Figure 41 depicts the presence of metastable orders between the AFI/AFM and SC phases in a phase diagram for dirty superconductors [248]. Even in copper oxides with strong electron correlations, electron–phonon interactions cannot be overlooked [249]. As illustrated in Fig. 3, the two perturbations lead to different types of ordering depending on their strength, resolving the inherent Fermi liquid instability. When both have a comparable impact, even minor variations in energy balance caused by temperature, pressure, doping, and material parameter differences, as well as randomness, can cause a shift in ground state or create a novel order that incorporates both instabilities. The electron correlation-induced antiferromagnetic and superconducting phases of copper oxides can be replaced by CDW-like phases with electron–phonon interactions causing lattice distortions. A typical example is found in La214: the "1/8 conundrum" refers to the suppression of superconductivity at $x = 0.125$ (the dip in the $T_c$ dome shown in Fig. 20). This issue is associated with a CDW phase known as "stripe order" [250, 251], which is only stabilized at the composition through electron–phonon interactions. The stripe order is defined by a charge and spin wave with a period of $4a$ in the $CuO_2$ plane. SCES distinguishes itself from ordinary CDWs by allowing spins to be ordered simultaneously.

Bi2212's checkerboard CDW [252] and stripe fluctuations are additional examples of competing orders [253-255]. Furthermore, CDW-like electronic orders have been observed in UD Y123 [256] when superconductivity is suppressed in high magnetic fields [257, 258]. Torque magnetometry detected a nematic phase transition in Y123 at the pseudogap temperature $T^*$ [259]. Micro-X-ray diffraction imaging reveals inhomogeneous electronic states called "CDW puddles" [260]. The observed metastable phase diversity is the result of complex structural and electronic instability. Common electron correlations favor antiferromagnetic or superconducting orders, while material-dependent electron–phonon instabilities favor a competing order with a distinctive lattice distortion.

Many of these metastable states were observed in an inhomogeneous state caused by hole trapping: weak screening by a few holes results in complete hole trapping, yielding an AFI phase in region X with few dopants (Fig. 29b), metastable phases in intermediate region Y, and superconductivity in region Z with more holes (Section 4.5.2.2). It should be noted, however, that the few-nanometer-scale mixture results from randomness rather than true electronic phase separation, which should have a clear domain boundary. As a result, in Fig. 41's practical phase diagram, these three regions appear to coexist in the transitional regime of hole concentration. It is important to remember that this is simply a ruse caused by using the nominal average hole concentration along the abscissa axis.

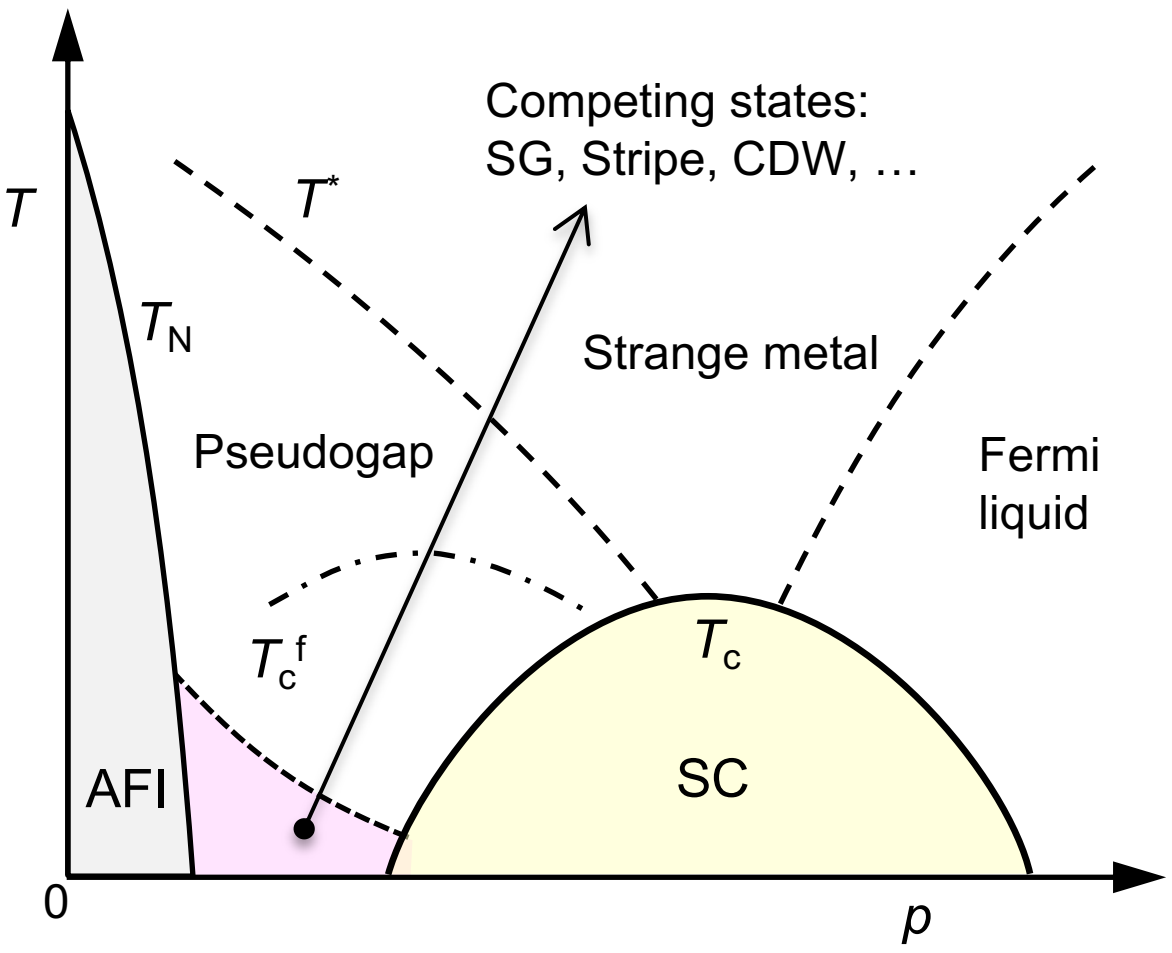

**Fig. 41.** Schematic phase diagram for dirty copper oxide superconductors [102, 254]. A few nanometer-scale mixtures between the antiferromagnetic insulator (AFI) (AFM is usually hidden) and the superconducting $T_c$ dome (SC) can appear, along with competing states such as spin glass (SG), stripe order, and various CDW phases accompanied by lattice–spin order. The two broken lines represent crossover temperatures: the pseudogap temperature $T^*$ on the left side of the $T_c$ dome and the one between strange metal, which has $T$-linear resistivity across a wide temperature range, and Fermi liquid on the right. The dash-dotted line represents $T_c^f$, which is the temperature at which superconducting fluctuations develop with existing pairs starting to share their wavefunction phase.

The author disagrees with those who claim that the origins of superconductivity are inextricably linked to these competing orders and their QCPs at $p_o$ [217, 254, 261]. As stated in Section 4.7.3, the true QCP lies at a lower doping level, with vanishing AFM. The competing orders exhibit notable material dependence at different energy scales, so the observed general chemical trends cannot be explained solely by the common hole concentration. We believe that the Cooper pairing mechanism, which involves antiferromagnetic fluctuations in the $CuO_2$ plane and the general BCS–BEC crossover concept, can explain cuprate superconductivity without relying on such a "exotic" mechanism.

4.8.2.2. Pseudogap phenomenon

The pseudogap phenomenon refers to anomalies observed at $T^*$ above $T_c$ in various physical quantities in the low-doped regime (Fig. 41). They are classified as intrinsic $T_p$ events, in which preformed pairs are created prior to BEC superconductivity [45, 157, 262]. Examples include the $T^*$ lines in the phase diagrams of La214 and Nd214 (Figs. 13 and 20), the formation of spin gaps in Cu NMR in Y123 and other compounds [263, 264], pseudogaps in ARPES in Bi2201 and Bi2212 [265, 266], and enhanced Nernst signal regions [267]. The significant scatter in observed temperatures could be attributed to the fact that $T_p$ is a crossover temperature rather than a phase transition; the temperature at which the anomaly manifests itself will most likely vary depending on the type of experimental probe used and the observation time window. However, some anomalies appear to be linked to competing orders on energy scales comparable to superconductivity [157, 266]. In any case, the presence of $T_p$ in experiments is well documented.

$T_p$ is the temperature at which pairs form, not the temperature ($T_c^f$) at which superconducting fluctuations develop. $T_c^f$ is the temperature at which already existing pairs begin to share their wavefunction phase as they approach the BEC-superconducting transition at $T_c$, and it is expected to be substantially lower than $T_p$. At $T_p$, phenomena such as a decrease in magnetic susceptibility due to singlet pair formation can be noticed [263, 264], while at $T_c^f$, gap-like features in the energy spectrum that lead to the superconducting gap are observed [266]. It is crucial to distinguish these characteristic crossover temperatures for interpreting what actually happens upon cooling in the electronic state.

4.9. Final remarks on copper oxide superconductivity

Although numerous cuprate superconductors have been discovered and studied in each area of interest, the first La214 has garnered significant attention and is still regarded as the "standard material" for three reasons. First, the hole concentration is simply equal to the substituent composition $x$, which is easily determined experimentally. Other systems complicate estimating the absolute value of $p$ due to difficulties in determining oxygen content, non-stoichiometric compositions, uneven hole distribution, and the resulting complex hole supply mechanisms. Second, in La214, the hole concentration can be varied continuously over a wide range up to 0.4, effectively covering the entire electronic phase diagram, whereas other systems, such as the Tl system shown in Fig. 24, can only cover a fraction. Third, and perhaps most importantly, the ease of sample synthesis has allowed many researchers to participate in and carry out a wide range of experiments with their skills. Particularly important were neutron scattering experiments with $x$ ($p$) controlled samples, which require large single crystals and provide critical information on magnetism and spin fluctuations that are directly related to the superconductivity mechanism [153].

The cuprate superconductivity mechanism remains unresolved 40 years after its discovery, in part because we have yet to identify the variable effects of randomness in real materials. The secondary effects mentioned by Vojta in the theory review must be randomness effects [104]. Unfortunately for the cuprate superconductivity study, the standard La214 proved to be one of the unclean systems. Furthermore, two important experiments, STM and ARPES, were carried out on Bi2212, a system with a superior crystal surface than the others but greater randomness than the Hg system. This critical fact has received insufficient attention even in recent years. Uemura's plot is an excellent example of successfully avoiding this annoying issue [163]. The author primarily relies on recent NMR and ARPES experiments on multilayer systems with clean IPs [166, 186]. These experiments successfully distinguished $CuO_2$ planes with variable disorder effects and drew significant conclusions from clean IPs. These dependable data will be critical for gaining a thorough understanding of cuprate superconductivity, and findings from other dirty, low $T_{co}$ systems must be carefully examined.

The author believes the fundamental $T$–$p$ phase diagram is as simple as the one depicted in Fig. 38a. SCES must have a similarly straightforward picture of superconductivity (Fig. 19), just as BCS theory reduced conventional superconductivity from a complex many-body problem to a two-body problem, providing a clear description of Cooper pairing via phonon (Fig. 7). It is difficult to accept that no comprehensive mechanism exists, as previously stated [104]. Removing the branches and leaves of a natural phenomenon often reveals that it is governed by a simple organizing principle.

There is no doubt that Hg1223 is the current best copper oxide superconductor. It is more difficult to prepare a $CuO_2$ plane that is less susceptible to apical oxygen and randomness effects, regardless of the other block layers used, and optimally dope a thick conduction layer composed of three of them than in Hg1223. This means that copper oxides with higher $T_c$ values should not be expected. To raise $T_{co}$, a trick is required to counteract the drop in $T_p$ while increasing $p_o$ in Fig. 38a. Chapter 6 will look at whether it is possible to expand the $T_p$ region.

Some argue that the $T_c$ race began over a century ago with the element Hg and ended with Hg1223, which also contained the element. We developed a scenario by collecting the $T_c$'s material dependence in the synthesized copper oxide superconductors and explaining why Hg1223 has the highest $T_c$. The author wonders if a critical factor is overlooked. He'd like to believe that there are still undiscovered routes to increase the $T_c$ of copper oxides, even slightly, by exposing flaws in current arguments.

In Chapter 4, the author distills everything he knows about copper oxide superconductors into his own personal image, which he has honed over many years of discussing difficult physics topics. The author does not understand how so many experimental findings and theoretical considerations can be combined. Rather, it would be fortunate if this chapter provided an opportunity for researchers to reflect on previous findings and reconsider their interpretations and implications. Readers seeking a more in-depth or advanced understanding of the physics should consult other excellent commentaries and reviews; however, they should be aware that the physics described in those works is extremely difficult for solid state chemists to grasp.

## 5. Various superconductivity mechanisms and related superconductors

This chapter provides a brief overview of superconducting mechanisms and the materials that are expected to meet their criteria (albeit some are questionable). This is a comprehensive list because each item is tough to describe in depth. Table 2 lists typical superconductors other than cuprates. In addition to each $T_c$, possible relevant orders, fluctuations, or glues are provided, along with their respective ordering temperatures or energies.

**Table 3.** Classification of various superconductors based on the Cooper pairing mechanisms, whose characteristics are summarized in Tables 1 and 2. Some of the compounds listed may be relevant to the mechanisms, while others are uncertain.

| Degrees of freedom | Relevant order and fluctuations | Potentially related superconductors |
|---|---|---|
| Phonon | Normal phonons | Al, Pb, $MgB_2$, C(B), $H_3S$ |
| | Structural instability | $Nb_3Sn$, $V_3Si$, β-$KOs_2O_6$ |
| | CDW | $NbSe_3$, 1T-$TaS_2$, $Cu_xTiSe_2$, $IrTe_2$, $LuPt_2In$, $CsV_3Sb_5$ |
| Spin | Antiferromagnetic order | Copper oxides (Table 1), $CeCu_2Si_2$, $CeIn_3$, $CeRhIn_5$, $UPt_3$, $(TMTSF)_2PF_6$, CrAs, MnP, LaFeAs(O, F), LaFeAs(O, H), (Ba, K)$Fe_2As_2$ |
| | Ferromagnetic order | $UGe_2$, $UTe_2$, URhGe, UCoGe |
| | Spin liquid | $\kappa$-(BEDT–TTF)$_2$Cu$_2$(CN)$_3$, (Sr, Ca)$_{14}$Cu$_{24}$O$_{41}$ |
| Charge | Charge order | (BEDT-TTF)$_2$I$_3$ |
| | Valence fluctuations | $CeCu_2$(Si, Ge)$_2$, β-$YbAlB_4$ |
| | Valence skipping | $BaBi_{1-x}Pb_xO_3$, $Ba_{1-x}K_xBiO_3$, (Pb, Tl)Te |
| | Dilute electron gas | $Li_xZrNCl$, $Li_xHfNCl$ |
| | Exciton insulator | $Ta_2NiSe_5$, NaAlGe |
| Orbital | | α-FeSe |
| Multipole | | $PrOs_4Sb_{12}$, α-$Cd_2Re_2O_7$, $La_2IOs_2$ |
| Unidentified | | $Sr_2RuO_4$, $BaTi_2Sb_2O$, $Na_{0.35}CoO_2$•1.3$H_2O$, $La_2PrNi_2O_7$ |

The degree of freedom used as a glue to generate Cooper pairs, as well as the type of order and fluctuation, all influence the superconducting mechanism. Table 3 classifies typical superconductors according to three characteristics: degree of freedom (major classification), order type (intermediate classification), and material type (minor classification). In all cases, high $T_c$ can be achieved if the glue employed provides a strong attraction force. The BCS mechanism takes advantage of the vibrational degrees of freedom of the surrounding atoms. The electron's fundamental degrees of freedom, which are spin, charge, and orbital, can be used to make more glues [268].

Superconductivity, which uses spin ordering and fluctuations as a glue, has been developed in a variety of materials, with copper oxides leading the way in terms of $T_c$. Charge degree of freedom has long been assumed to generate a glue for high $T_c$ [269, 270]. Coulomb interactions often have high energy, around eV (~12,000 K). In terms of orbital degree of freedom, the degeneracy associated with p-orbitals in typical element compounds and d-orbitals in transition metal compounds can persist at low temperatures, causing fluctuations to act as a glue.

### 5.1. Order, fluctuation, and superconductivity: the quantum critical point (QCP) scenario

We'll begin by talking about the general origin of Cooper pair attraction, which involves various degrees of freedom. Superconductors with high $T_c$ or unusual mechanisms, such as cuprates, commonly have one or more ordered phases in their neighbor [271]. Cooper pair-forming interactions destabilize electronic systems, suppressing the normal-metallic phase and stabilizing superconductivity. When the interaction becomes too strong, a parent long-range order (LRO) phase of the degree of freedom or a quantum-mechanical state will emerge instead of a superconducting phase. Figure 3 illustrates that when electron–phonon interactions become too strong, structural transitions or CDW orders occur. Moreover, when magnetic interactions become too strong with electron correlations, a variety of magnetic orders or quantum-mechanically entangled states emerge, such as spin liquids. If these states are suppressed by adjusting a control parameter such as carrier number or pressure, superconductivity with the flavor of the original states should appear (Fig. 42). Of course, electron–phonon interactions occur in all crystals; however, we will focus on the case in which another distinct interaction acts as the primary glue.

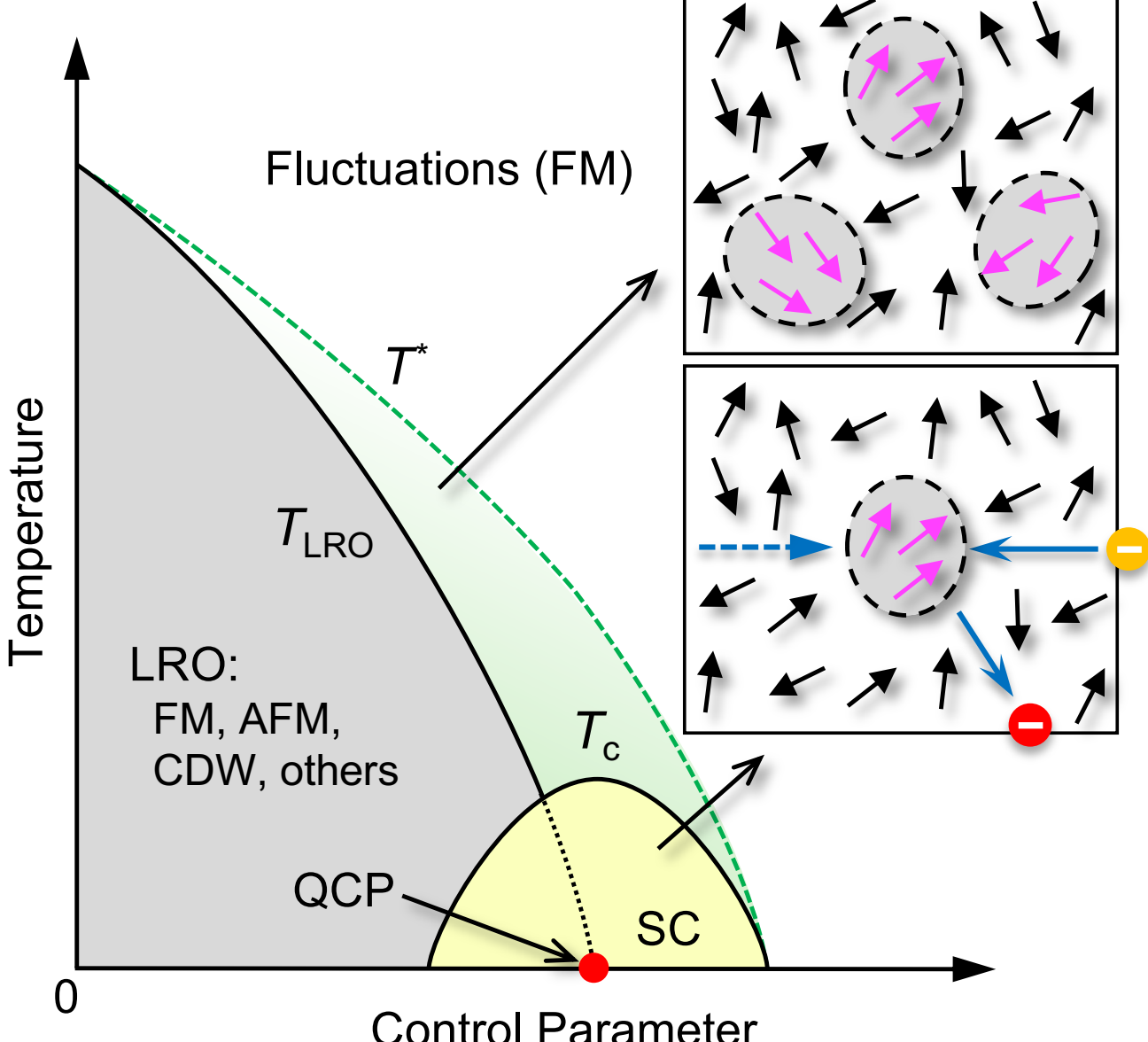

**Fig. 42.** General phase diagram for superconductivity derived from a relevant long-range order (LRO). LROs in ferromagnetic metal (FM), antiferromagnetic metal (AFM), charge-density wave (CDW) insulators, and others can be suppressed by increasing a control parameter, such as carrier number or pressure, with the ordering temperature $T_{LRO}$ vanishing at a quantum critical point (QCP). Above $T_{LRO}$, there is a crossover temperature $T^*$, below which a short-range order

emerges. Fluctuations within the temperature window can cause electron pairing, which results in Cooper pairing below $T_c$. Superconductivity occurs near the QCP, where the dome's $T_c$ is highest due to the most intense fluctuations. The top inset cartoon depicts ferromagnetic fluctuations that cause spatially and temporally variable regions of nearly parallel spin alignment in a matrix of randomly oriented spins. The bottom inset illustrates how Cooper pairing works with ferromagnetic fluctuations: the first electron (red ball) creates a ferromagnetically spin-aligned region that immediately attracts the second electron (orange ball) before disappearing. Note that the QCP scenario assumes variable fluctuations in BCS superconductivity rather than BEC type.

Pair attraction is typically caused by a fluctuation present above the LRO's critical temperature ($T_{LRO}$) but below a crossover temperature $T^*$, where the fluctuation begins. Thermal fluctuations during a general second-order phase transition leads to the formation and dissipation of local regions with non-zero order parameters in both time and space. As the temperature approaches $T_{LRO}$, the ordered region rapidly expands, eventually spreading throughout the crystal and resulting in a phase transition to an LRO at $T_{LRO}$. In the case of ferromagnetic ordering, for example, spins are oriented randomly with zero net magnetization, which is the order parameter, in the high-temperature paramagnetic state; however, before all spins align ferromagnetically below $T_{LRO}$, local regions of nearly ferromagnetically aligned spins with a finite total magnetization appear in the temperature window between $T^*$ and $T_{LRO}$, as illustrated in the inset of Fig. 42.

These fluctuating regions may cause electron or hole pairings. Similar to the phonon mechanism depicted in Fig. 7, the first electron will create a local region with some spins oriented ferromagnetically, and the second electron, which favors ferromagnetic interactions, will be drawn to this region before relaxing and vanishing (bottom inset in Fig. 42). Because of this virtual process, ferromagnetic fluctuations may induce an attraction between two electrons. Similarly, in antiferromagnetic and alternative interactions, fluctuations corresponding to an associated order generate attraction force.

Temperature changes are not the only route for controlling fluctuations. Quantum fluctuations caused by zero-point oscillations can exist at absolute zero if the order corresponds to quantum mechanical degrees of freedom like spin [272]. The ordered phase experiences second-order suppression as the quantum fluctuation grows, resulting in $T_{LRO} = 0$ at the QCP [232]. Because it occurs at absolute zero, it is regarded as a transition caused by the divergence of quantum fluctuations rather than thermal fluctuations. Pressure, for example, often broadens bandwidth by shortening interatomic distances, resulting in lower electron correlations, weaker magnetism, and thus a lower magnetic transition temperature. The fluctuation regime extends both to the right and above the phase transition line, as shown in Fig. 42. As a result, the QCP experiences the greatest fluctuation. If these fluctuations serve as Cooper pair glue, a superconducting dome with the highest $T_c$ at QCP will emerge.

Achieving high $T_c$ requires a relatively robust LRO; an overly stable order is difficult to suppress, and the resulting fluctuations are insufficient to function as an efficient Cooper pair glue; in contrast, a fragile LRO with a weaker interaction can only produce low $T_c$. The location and shape of the $T_c$ dome in the QCP superconductivity phase diagram differs according to the type of fluctuation and system [62, 272]. It is worth noting that the general phase diagram in Fig. 42 depicts the link between $T_c$ and fluctuations in BCS superconductivity rather than BEC superconductivity. As previously stated in Section 4.7.3 for cuprate superconductivity with variable carrier density, the $T_c$ dome forms at much higher doping levels than the AFM's QCP, where $T_c$ is limited by low carrier density.

5.2. Phonon superconductivity enhanced by structural instability

Ordinary phonon superconductivity has a relatively high $T_c$ when fluctuations at the QCP suppress a nearby ordered phase with a specific structural distortion. This is due to soft phonons. They lose energy as the system approaches the boundary, increasing electron–phonon interactions and causing $T_c$ to rise despite phonon energy loss. The high $T_c$ of A15-type compounds, such as $Nb_3Sn$ ($T_c$ = 18.1 K) and $V_3Si$ ($T_c$ = 17.1 K), is thought to be due to increased electron–phonon interactions caused by structural instability associated with the martensite transformation, which is a cubic–tetragonal structural transition [273].

Many superconductors have parent phases with varying CDW orders, which occur when electronic instability in low-dimensional systems is removed via crystal structure deformation, resulting in charge-modulated superstructures. Transition metal chalcogenides are a typical example. $NbSe_3$, a quasi-1D conductor, undergoes a CDW transition at 59 K, which is suppressed at 0.7 GPa pressure, and eventually transforms into a superconductor at 2.5 K [76]. In quasi-2D dichalcogenides, pressure suppresses the CDW phase in 1T-$TaS_2$, resulting in 5 K superconductivity at 5 GPa [274]. In $Cu_xTiSe_2$, electron doping with Cu intercalation suppresses the parent's CDW order ($T_{CDW}$ = 220 K) and causes superconductivity with the maximum $T_c$ of 4.2 K at $x$ = 0.08 (Fig. 43a) [275]. $IrTe_2$ has a CDW transition at 250 K, but with 3.5% Pt-for-Ir substitution, it achieves superconductivity at 3.1 K, surpassing the QCP [276]. Pressure suppresses the CDW (or CO) phase in β-$Na_{0.33}V_2O_5$ below 135 K, resulting in superconductivity with $T_c$ = 2.3 K at $P$ = 8 GPa [277]. Intermetallic compounds provide additional examples: in $LuPt_2In$, which has a CDW transition temperature of 90 K, replacing 60% of the Pt with Pd yields a QCP with a significant increase in $T_c$ up to 1.10 K [278].

$AV_3Sb_5$ (A = K, Rb, Cs), a relatively new superconductor discovered in 2019, is a fascinating system with a kagome lattice composed of V-atom triangles with shared corners [279, 280]. The Cs compound exhibits CDW order at $T_{CDW}$ = 94 K and superconductivity at $T_c$ = 2.5 K. Because the $T_c$ values of the Rb and Cs analogues with lighter alkaline metal elements are lower (around 0.9 K), a simple electron–phonon model may be insufficient to explain the chemical trend in $T_c$. CDW instability competes with superconductivity and most likely results in a chiral state with the complex pair wavefunction of $dx^2–y^2 + idxy$ [281]. As pressure increases, $T_c$ initially decreases, followed by the formation of a second $T_c$ dome [282]. CDW order fluctuations and topological features in electronic structures can contribute to complex superconducting mechanisms [280].

Similar double-$T_c$ domed superconductivity phase diagrams were observed for $CeCu_2Ge_2$ (Fig. 43b) [283, 284] and $LaFeAs(O_{1-x}H_x)$ (Fig. 43c) [285] as a function of pressure and H content, respectively. These are assumed to be composite phase diagrams, with the phase diagram from Fig. 42 inverted horizontally and joined to the original on the right. The superconducting double dome has two distinct parent phases on its left and right sides. This phase diagram is produced by two distinct Cooper pairing mechanisms in the same system, which are converted according to the control parameter. If one is a standard phonon mechanism, the other is most likely an exotic mechanism with a different degree of freedom. Because phonon superconductivity can occur in any compound, the presence of another pairing source naturally results in a double-$T_c$ domed phase diagram.

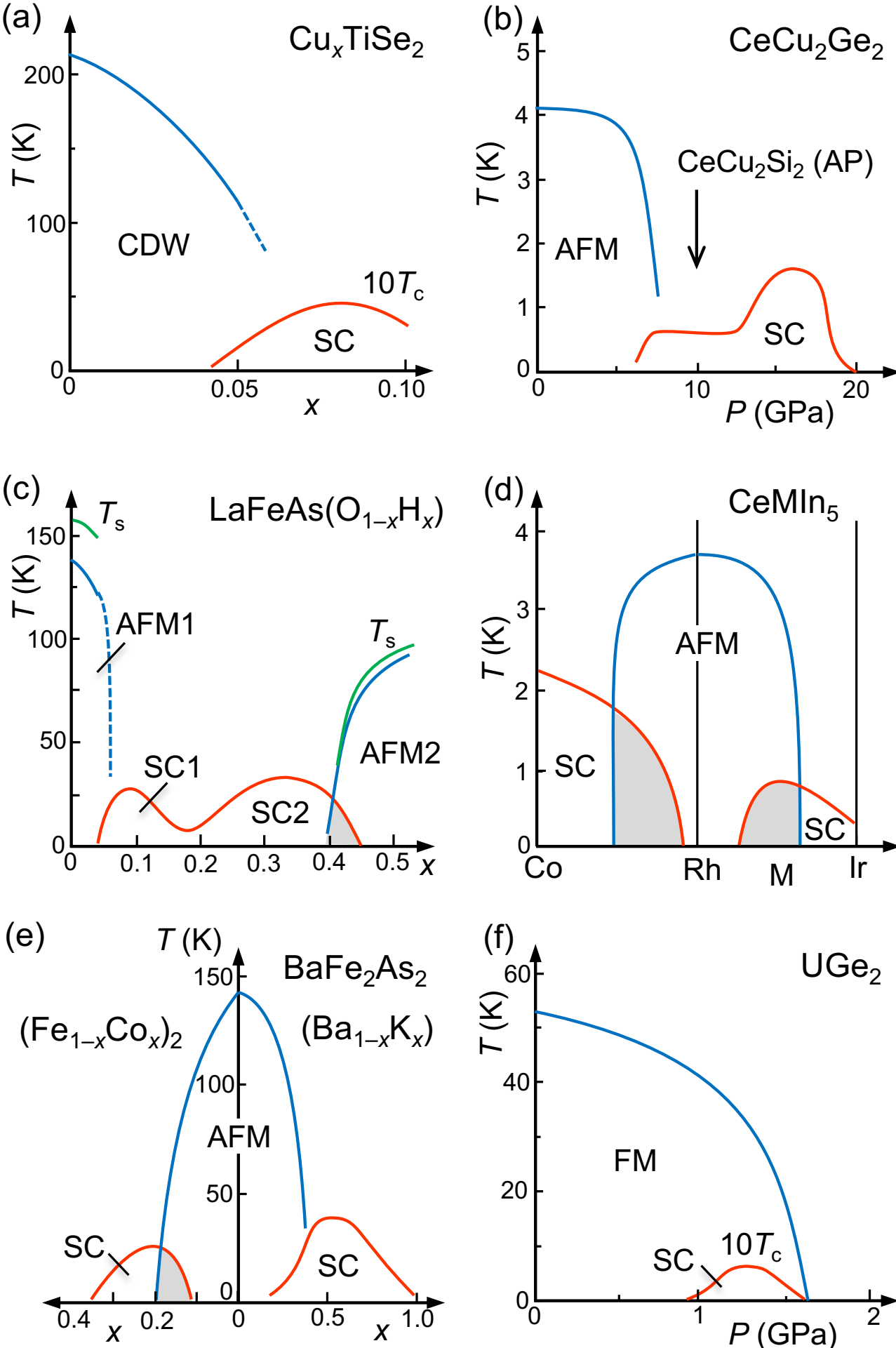

**Fig. 43.** Collection of phase diagrams for various types of superconductors. (a) Cu intercalation suppresses the CDW phase of $TiSe_2$, resulting in superconductivity at 4.2 K [275]. The figure shows the $T_c$ values multiplied by ten. (b) $P$–$T$ phase diagram for $CeCu_2Ge_2$ [286, 287]. The first SC next to AFM at ambient pressure is caused by antiferromagnetic spin fluctuations, while the second $T_c$ dome may be due to charge fluctuations [284]. Because the smaller Si adds chemical pressure, the $CeCu_2Si_2$ at ambient pressure (AP) is approximately located at 10 GPa in the Ge compound [288]. (c) Double $T_c$ domes form in $LaFeAsO_{1-x}H_x$ when antiferromagnetic metal phases of LaFeAsO (AFM1) and $LaFeAsO_{0.5}H_{0.5}$ (AFM2) are doped with electrons and holes, respectively [285]. $T_s$ is slightly higher than $T_N$ and represents the tetragonal–orthorhombic structural transition temperature, which can cause orbital fluctuations in addition to antiferromagnetic ones. (d) $T_c$ domes appear in the $CeMIn_5$ (M = Co, Rh, Ir) solid solutions around the AFM $CeRhIn_5$ on both sides of $CeCoIn_5$ and $CeIrIn_5$ [289]. (e) A typical phase diagram of Fe-based superconductors, starting from AFM $BaFe_2As_2$ [62, 290]: hole doping in $(Ba_{1-x}K_x)Fe_2As_2$ on the right and electron doping in $Ba(Fe_{1-x}Co_x)_2As_2$ on the left cause superconductivity at $T_c$ = 38 K and 22 K, respectively. The original figures used SDW instead of AFM. The shaded area around the SC–AFM border in (c), (d), and (e) represents a phase mixture; however, the author believes it is the result of elemental substitution-induced randomness rather than intrinsic. In the $BaFe_2As_2$ systems in (e), only the Co-for-Fe substitution causes a mixture, but the K-for-Ba does not, which could be attributed to the former's greater randomness. (f) Ferromagnetic spin fluctuations induce SC in $UGe_2$ [291]. The $T_c$ values are multiplied by ten.

### 5.3. Superconductivity utilizing the spin degree of freedom

#### 5.3.1. Antiferromagnetic spin fluctuations

In the QCP scenario, many superconductors employ antiferromagnetic spin fluctuations [271]. Numerous studies have been carried out on heavy-fermion superconductors containing heavy elements such as Ce and U [77, 292, 293]. The f electrons of heavy elements, which are highly concentrated around the nucleus and thus tend to localize, couple with one another via the RKKY interaction, which is mediated by the more expanded s and p electrons of counter light elements. The accompanying magnetic LRO at low temperatures below $T_N$ varies in character depending on the hybridization magnitude between the two: when it is small, the f electrons are almost localized and behave as large magnetic moments, resulting in a conventional magnetic LRO; when it is large, spin density wave (SDW) magnetism is produced, with the magnitude of the magnetic moment reduced and modulating as a wave. As the f–conduction electron hybridization further increases, another anti-RKKY interaction emerges: the Kondo effect, which occurs when conduction electrons obscure the localized spins and obliterate the magnetic moment. As the intensity of Kondo screening increases, the magnetic order disappears at the QCP, leaving a nonmagnetic metallic state (Doniach phase diagram [294]). The itinerant electrons appear unusually heavy because they drag f electrons that tend to localize, and their effective mass can be 1000 times that of free electrons, hence the term heavy fermion. Antiferromagnetic spin fluctuation-induced superconductivity is common in heavy fermion compounds near the QCP [293].

$CeCu_2Si_2$, the first heavy fermion superconductor discovered in 1979 [295], is a nonmagnetic, itinerant metal that undergoes superconductivity at 0.7 K. Under negative chemical pressure (volume expansion) caused by the Ge-for-Si substitution, an antiferromagnetic metal phase appears next to the superconducting phase at temperatures below 0.8–2 K [77]. $CeCu_2Ge_2$ shows $T_c$ = 0.6 K superconductivity following pressure-induced suppression of $T_N$ = 4 K AFM (Fig. 43b) [286, 287]. These findings suggest that antiferromagnetic fluctuations play an important role in superconductivity. More interestingly, each displays another $T_c$ dome at higher pressures, which could be related to charge-fluctuation superconductivity, as discussed in Section 5.4.2 [284].

$CeIn_3$ is an AFM with a $T_N$ of 10.2 K at ambient pressure. As pressure increases, $T_N$ falls until it reaches a QCP at $P$ = 2.65 GPa, when a superconducting dome appears with a maximum $T_c$ of 0.19 K [296]. $CeRhIn_5$ transitions from an AFM ($T_N$ = 3.8 K) at ambient pressure to a superconducting phase ($T_c$ = 2.1 K) at around 1.7 GPa [297]. An intriguing phase diagram for the related $CeMIn_5$ (M = Co, Rh, Ir) solid solutions has been obtained, revealing an antiferromagnetic phase extending around $CeRhIn_5$, as well as two separate superconducting phases appearing on both sides of the Co and Ir compounds (Fig. 43d) [77, 289]. As mentioned in Section 2.4.3, $CeCoIn_5$ has Cooper pairs with d$x^2$–$y^2$ waves and must be an antiferromagnetic fluctuation superconductor based on electron correlations [50]. Therefore, it appears to be simply AFM QCP superconductivity. However, the phase diagram in Fig. 43d can be viewed differently. That is, all materials can exhibit antiferromagnetic fluctuation superconductivity, with $T_c$ decreasing toward the right; however, for some reason, antiferromagnetic correlation becomes prevalent in Rh compounds, replacing the competing SC order. When it comes to uranium compounds, $UPt_3$ transitions from an AFM ($T_N$ = 5 K) to a superconductor ($T_c$ = 0.54 K).

Although these heavy-fermion superconductors are attracting from a physics viewpoint, their $T_c$s are low due to weak antiferromagnetic interactions that act as a glue. Unlike d electrons, f electrons concentrate near the nucleus, and magnetic interactions between f electrons are weakly mediated by the other s/p conduction electrons. As a result, both $T_N$ and $T_c$ are low, rendering them unsuitable for high-temperature superconductivity. However, their low energy scale and temperature range are advantageous for exploring various quantum phenomena, making them important in physics field.

Other antiferromagnetic spin-fluctuation superconductors include molecular conductors and d-electron compounds. At 12 K, the quasi-1D molecular conductor $(TMTSF)_2PF_6$ enters an SDW phase, which is suppressed by pressure. At 0.9 GPa, a superconducting phase with $T_c$ = 1.2 K appears [298, 299]. In the latter, CrAs becomes an AFM below 265 K at ambient pressure while showing superconductivity at $T_c$ = 2.2 K at 0.7 GPa [300, 301]. MnP exhibits a ferromagnetic order at 290 K at ambient pressure, but an antiferromagnetic order with a helical spin arrangement above $P$ = 2 GPa and 1 K superconductivity near QCP at 8 GPa [302, 303]. Despite the high $T_N$ values due to strong antiferromagnetic interactions, the much lower $T_c$ values are most likely the result of weak coupling between antiferromagnetic fluctuations and conducting electrons.

Several superconductors have been discovered in iron-based compounds since the first publication in 2006 [304]. They have relatively high $T_c$ values and display distinctive material dependency of $T_c$, similar to cuprate superconductors. The parent phases of LaFeAsO and $BaFe_2As_2$ are AFMs with $T_N$ values of 150 and 135 K, respectively [62, 63, 290, 305]. In the former, electron doping with $F^-$ and $H^-$, which replace $O^{2-}$ in $LaFeAs(O_{1-x}F_x)$ and $LaFeAs(O_{1-x}H_x)$, leads to superconductivity at $T_c$ = 26 and 36 K, respectively, after reducing AFMs [285, 306, 307]. Another superconducting dome (SC2) appears in $LaFeAs(O_{1-x}H_x)$, this time with an AFM2 phase distinguished by a different spin arrangement on the high doping side about $x$ = 0.5, in contrast to AFM1 at $x$ = 0 and nearby SC1 in Fig. 43c [285]. On the other hand, hole doping in $BaFe_2As_2$ suppresses AFM (SDW) order in $(Ba_{1-x}K_x)Fe_2As_2$, resulting in superconductivity at $T_c$ = 38 K at $x$ = 0.5 (Fig. 43e) [308]. Furthermore, electron doping in $Ba(Fe_{1-x}Co_x)_2As_2$ yields $T_c$ = 22 K superconductivity at $x$ = 0.2 [309]. Superconductivity, like copper oxide superconductors, originates from electron or hole doping in the parent phase. The key distinction from copper oxides is that the parent phase is AFM rather than AFI, which is thought to result from weaker electron correlations.

Although antiferromagnetic fluctuations have a role in the superconducting mechanism of iron-based compounds [62, 305], the emphasis has been on orbital fluctuations, which are uncommon and arise from degenerate orbitals or bands [63, 310, 311]. Unlike the simple d$x^2$–$y^2$ orbital-derived single band in copper oxides, the Fe atom's d$xy$, d$yz$, and d$zx$ orbitals in the $d^6$ electron configuration are close to the Fermi level, resulting in a complex electronic state. Because the last two orbitals are degenerate in the tetragonal crystal structure, an electronic instability can arise to remove the degeneracy. The transition to a low-temperature orthorhombic phase occurs by spontaneously breaking the fourfold rotation symmetry, known as the nematic transition, which is triggered by an electronic rather than a structural instability. The nematic transition temperature is denoted by $T_s$ in Fig. 43c, which is located immediately above $T_N$. Fluctuations in orbital degrees of freedom could potentially contribute to superconducting mechanisms.

α-FeSe exhibits superconductivity at 8 K under normal conditions [312, 313]. However, no antiferromagnetic phase was found nearby, suggesting that nematic fluctuations were predominant [314, 315]. Nevertheless, antiferromagnetic fluctuations are significant in increasing $T_c$ because antiferromagnetic phases are constantly close to superconductivity in higher $T_c$ iron-based superconductors. It is plausible that orbital and antiferromagnetic spin fluctuations provide weak and strong glues, respectively. Otherwise, they combine to create novel fluctuations. This duality could be linked to the observed complexity, such as double-$T_c$ domed superconductivity phase diagrams in $LaFeAs(O_{1-x}H_x)$ [285]. It is noted that both mechanisms cause d-wave pairing, allowing them to occur concurrently [316]; antiferromagnetic and phonon mechanisms cannot coexist, because they cause incompatible d-wave and s-wave pairing.

$SmFeAs(O_{1-x}F_x)$ has the highest $T_c$ of any iron-based superconductor at 55 K [317, 318]. Using Eq. 5 and a reduction factor of 0.1, the glue's energy is calculated to be approximately 550 K. An electronic structure calculation predicts that the maximum antiferromagnetic interaction occurs at 550 K in LaFeAsO [319], while a neutron scattering experiment on $BaFe_2As_2$ yielded a comparable result of 700 K [320]. In this context, achieving higher $T_c$ values in iron-based superconductors may be challenging, but additional research will elucidate the mechanism, comprehend the $T_c$'s material dependence, and lead to a novel approach toward higher $T_c$.

5.3.2. Ferromagnetic spin fluctuations

Only a few uranium compounds demonstrate ferromagnetic fluctuation-mediated superconductivity [77, 321]. When magnetic rare earth elements replace the Pb and Er sites in $PbMo_6S_8$ [322] and $ErRh_4B_4$ [323], both superconductivity and ferromagnetism occur. Coexistence of the two phenomena has also been observed in $(Ce_{1-x}Gd_x)Ru_2$ [324] and $GdSr_2RuCu_2O_8$ (C2-B3-PV: Ru replacing Cu in the Y123 block layer) [325, 326]. In the former compound, the localized Gd 4f electrons show ferromagnetism, while the Ru sheets show superconductivity. In the latter, itinerant d electrons on the $RuO_2$ sheets cause ferromagnetism, while superconductivity is caused by the $CuO_2$ planes. Thus, superconductivity and ferromagnetism occur in separate locations in these compounds, and ferromagnetic spin fluctuations may not directly mediate superconductivity.

In contrast, $UGe_2$, discovered in 2000, converts to a ferromagnetic metal below 52 K, which is suppressed by pressure and disappears at 1.6 GPa [291]. The *T–P* phase diagram depicts a superconducting dome with a maximum $T_c$ of 0.8 K in the ferromagnetic metal phase, just prior to the QCP (Fig. 43f). $UTe_2$ undergoes a superconducting transition at 1.6 K and has no neighboring ferromagnetic phases. However, this transition is most likely caused by ferromagnetic fluctuations associated with a hidden ferromagnetic order [321, 327]. Other URhGe ($T_c$ = 0.25 K) [328] and UCoGe ($T_c$ = 0.8 K) [329] have been shown to be superconducting at ambient pressure. Itinerant U 5f electrons play a leading role in these U-based compounds, causing ferromagnetic spin fluctuations. The $T_c$ values in ferromagnetic fluctuation superconductivity are extremely low, as are those in other f-electron systems, reflecting the glue's low energy scale caused by weak magnetic couplings between f electrons. Despite their low $T_c$, uranium compounds may contain intriguing novel physics, as have long been studied in solid state physics [65].

Unlike antiferromagnetism, ferromagnetic superconductivity requires that the pair spins align in the same direction. As a result, rather than a spin singlet, superconductivity is achieved via a spin triplet channel, similar to $^3He$ superfluidity. One distinguishing feature is that the magnetic field stabilizes rather than suppresses superconductivity, similar to FFLO superconductivity (Section 2.4.3). A 2 T magnetic field in URhGe suppresses superconductivity, but a $T_c$ dome reappears at 9–13.5 T, with a peak at 0.4 K [330]. This reentrant superconductivity is characteristic of spin-triplet superconductivity caused by ferromagnetic fluctuations.

5.3.3. Spin liquid states

A phase adjacent to a superconducting phase is not always LRO. In general, antiferromagnetically interacting spin systems favors simple spin up–down Néel ordering. However, when spins are arranged in triangular lattice points, antiferromagnetic interactions between adjacent spins cannot be satisfied simultaneously. This geometrical frustration inhibits simple spin ordering, resulting in LROs with complex spin orientations or exotic many-body states at low temperatures [80]. Even at absolute zero, spin-1/2 systems with large quantum fluctuations have unordered spins that are entangled in a liquid-like quantum mechanical state known as the "spin liquid".

A link between spin liquid and superconductivity has been proposed in a few compounds. In molecular compounds, the quasi-2D material κ-$(BEDT\text{-}TTF)_2Cu_2(CN)_3$ is antiferromagnetic with a relatively large *J* of 250 K. However, because of geometrical frustration in the triangular lattice, spins do not order even at 32 mK and can remain in a spin liquid [331]. Superconductivity is observed at $T_c$ = 4 K and 0.4 GPa pressure [332], but the relationship between the superconductivity and spin liquid remains unclear.

The resonating valence bond (RVB) state is a simple representation of the spin liquid, as described in the cuprate superconductivity mechanism [333]. It is a dynamic state composed of superimposed spin singlet pairs. The formation of spin singlet pairs indicates the presence of an energy gap. To break the spin liquid state, one of the singlet pairs' spins must be flipped to form parallel triplets, which requires energy injection greater than the gap (spin gap). The spin gap widens as the pair's size shrinks (attractive force increases). The introduction of two holes into a spin liquid is thought to replace one spin singlet pair, resulting in a BEC Cooper pair swimming in a sea of spin singlets and exotic superconductivity. The spin gap is expected to directly transform into a superconducting gap via doping, with large spin gaps resulting in high $T_c$.

The ground state of the spin-1/2 kagome lattice is considered to be a spin liquid [80], and doping is expected to induce superconductivity. However, the spin singlet pairs that form there are quite large, with a negligible spin gap (long-range RVB state). As a result, the superconducting gap narrows, and even if superconductivity is achieved, the $T_c$ must be relatively low. The superconductivity of $AV_3Sb_5$, as mentioned in Section 5.2, was primarily driven by CDW instability rather than V's kagome lattice magnetism.

The spin ladder is a typical system under investigation as a potential spin liquid-based superconductor, beginning with a large spin gap [156, 334]. It consists of two spin-1/2 antiferromagnetic chains arranged in a ladder-like configuration, with the ground state representing a superposition of adjacent singlet pairs (short-range RVB state), as depicted in Fig. 44a. Similar to the frustrated spin system, its unique spin arrangement produces a spin liquid rather than Néel order. The resulting spin gap shares the same order as the antiferromagnetic interaction *J*. The spin ladder model is supported by the discovery of an expected spin gap in $SrCu_2O_3$ with a $Cu_2O_3$ plane, as depicted in Fig. 44d [335, 336].

Let's consider what happens when we add holes to the spin ladder [156]. When one hole is introduced, three *J* bonds are lost; when two holes are placed next to each other, only five *J* bonds are lost instead of six, yielding a one *J* energy gain (Figs. 44b and 44c). The produced pair can move up the ladder to become a Cooper pair, resulting in BEC superconductivity; this Dagotto's reasoning is analogous to the picture of the pairing on the $CuO_2$ plane in Fig. 19. The starting positions differ between them, with AFI for the $CuO_2$ plane and spin liquid for the spin ladder. However, one may argue that the $CuO_2$ plane after doping below $T_p$ resembles the RVB spin liquid state [337]. Spin-liquid-induced superconductivity describes a straightforward scenario in which a spin pair is replaced by a hole pair (this simplicity inspired the author's research in the spin ladder field).

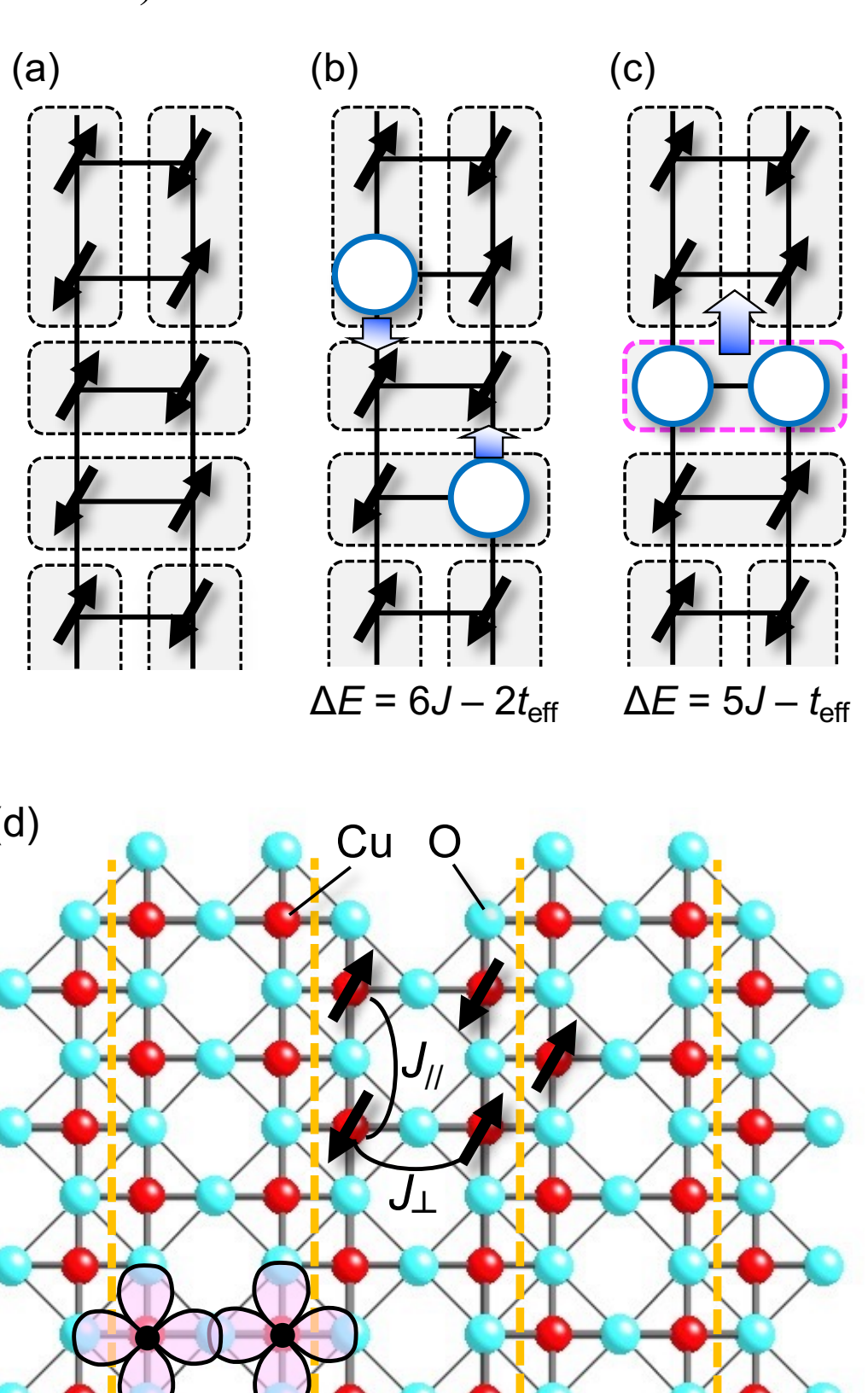

**Fig. 44.** Superconductivity on the spin ladder [156]. (a) A snapshot of singlet spin pairs captures the spin liquid (short-range RVB) state of a spin-1/2 ladder. Pairs form primarily on neighboring spins, resulting in a large spin gap on the order of the antiferromagnetic interaction $J$. (b) Two separate holes lose six $J$ bonds and gain kinetic energy of two $t_{eff}$. (c) When a hole pair forms nearby, the loss is reduced to five $J$ bonds, resulting in a $J$ energy gain and one $t_{eff}$ loss. At temperatures below $T_c$, the resulting hole pair is assumed to be a BEC Cooper pair, yielding spin liquid-induced superconductivity. (d) The spin ladder in the $Cu_2O_3$ plane is found in $SrCu_2O_3$ [336] and (Sr, Ca)$_{14}Cu_{24}O_{41}$ [156, 338]. The antiferromagnetic interactions parallel and perpendicular to the ladder were estimated at $J_{//}$ = 2000 K and $J_{\perp}$ = 1000 K, respectively. When an antiferromagnetic correlation develops along the ladder leg at temperatures much lower than $J_{//}$, the magnetic coupling between ladders in the zigzag Cu arrangement at the interface is effectively cancelled out, resulting in the plane being divided into a series of independent spin ladders. Thus, the quasi-1D spin ladder emerges from the 2D $Cu_2O_3$ plane as a result of dimensional reduction by frustration [339, 340].

This mechanism is believed to be responsible for the superconductivity found in the spin ladder copper oxide (Sr, Ca)$_{14}Cu_{24}O_{41}$ [156, 338]. The compound shares the same $Cu_2O_3$ surface as $SrCu_2O_3$ (Fig. 44d). The $Cu_2O_3$ surface is viewed as a reconstruction of the cuprate superconductors' $CuO_2$ plane, which has been divided into two-legged ladders that connect after being shifted halfway in the leg direction. Despite the obvious two-dimensionality in the crystal structural, the magnetic properties of the $Cu_2O_3$ surface exhibit 1D spin ladder behavior. At temperatures sufficiently below $J_{//}$ = 2000 K, antiferromagnetic correlations develop along the ladder's leg direction. Then, the interface zigzag coupling effectively cancels out the magnetic interactions between two ladders, dividing the $Cu_2O_3$ surface into a series of 1D spin ladders [335]. This dimensional reduction by frustration [339, 340] yields a spin system with a dimension lower than a crystal lattice, which occurs in several frustrated spin lattices.

The $Cu_2O_3$ surface of (Sr, Ca)$_{14}Cu_{24}O_{41}$ has a spin liquid ground state and an expected spin gap, indicating a spin ladder system [156]. Uehara, Akimitsu, and colleagues discovered that pressure induces superconductivity at 12 K, while only hole doping does not [338]. Despite extensive research, it is still unclear whether the aforementioned spin-ladder superconductivity has been realized. Prior to superconductivity, electronic anisotropy decreases with increasing hole concentration and pressure [156]. As previously mentioned, a spin ladder forms on the $Cu_2O_3$ surface when the antiferromagnetic correlations along it have developed long enough. However, because holes disrupt the antiferromagnetic chain and shorten the magnetic correlation length, doping must cause the loss of one-dimensionality and the ladder feature. Ladder superconductivity may be difficult to achieve on a $Cu_2O_3$ surface. The superconductivity of (Sr, Ca)$_{14}Cu_{24}O_{41}$ could be attributed to a specific magnetic fluctuation on the 2D $Cu_2O_3$ surface with a weak 1D feature; otherwise, it is caused by phonons or unknown glues.

Another candidate compound with the spin ladder feature prior to hole doping is $LaCuO_{2.5}$; however, no superconductivity was observed after hole doping metallization [341]. One possible explanation is that the strong randomness effect in 1D systems may suppress superconductivity; even a single defect disrupts the conduction path and cannot be bypassed, unlike in 2D or 3D systems. It is difficult to find an appropriate 1D material in practice, despite the fact that 1D systems have high $J$ and simple theoretical modelling.

Compounds for spin ladders with more than two legs are available [342]. $Sr_2Cu_3O_5$ has three legs because the $CuO_2$ plane is divided into strips with three chains rather than two in $SrCu_2O_3$ [335]; those with more legs were identified as crystalline defects, indicating the possibility of preparing a homologous series of $Sr_nCu_{n+1}O_{2n-1}$ compounds via improved synthesis. Interestingly, the spin ladder's ground state exhibits an even-odd-number effect: an odd-number ladder has no gap, whereas an even-number ladder has a spin gap that shrinks as the leg number increases [155]. Even-leg ladders, including four-leg ladders, are predicted to demonstrate spin-ladder superconductivity [343]. The smaller spin gap may cause a lower $T_c$ than the two-leg ladder; however, reduced randomness effects mitigate the reduction in $T_c$. The $CuO_2$ plane with high $T_c$ is the end member of increasing the leg number. It is not surprising that as we approach the $CuO_2$ plane, actual $T_c$ exceeds that of the two-leg ladder compound. In conclusion, no clear evidence of spin ladder superconductivity has been found thus far. However, its simple BEC superconducting mechanism is appealing, and further material research in cuprates and other systems is encouraged.

The Haldane chain is a 1D spin-gap system made up of atoms like $Ni^{2+}$ ($3d^8$) with spin quantum number 1 [344]. If a Haldane chain compound is doped to induce metallization, it is possible to achieve novel superconductivity by replacing a Haldane gap with a superconducting gap, similar to spin ladder superconductivity. Verification in Haldane materials, such as $Y_2BaNiO_5$ [345] and other organic compounds, is awaited.

### 5.4. Superconductivity utilizing the charge degree of freedom

#### 5.4.1. Charge fluctuations

Charge fluctuations have long been considered a novel superconductivity glue. Coulomb interactions can cause charge fluctuations with energies greater than 10,000 K [270]. More than 60 years ago, Little proposed a pairing mechanism using side branch polarization in a 1D conduction path (Fig. 45a) [269]. It was thought that a material design could successfully decorate a conducting molecular chain with easily polarizable molecules as side branches, but no such compound has been discovered. Ginzburg investigated 2D models containing a semiconductor–metal interface or a metal layer sandwiched between semiconductor layers, as illustrated in Fig. 45b. Cooper pairs can be mediated by holes on the semiconductor side of the interface, which are formed by conducting electrons on the metal layer [270]. He also considered that superconductivity resulted from instability caused by excitons, or electron–hole pairs. To the author's knowledge, no evidence has been discovered that provides compelling experimental support for these intriguing models. It would be fascinating to revisit these issues in light of recent advances in cuprates and subsequently discovered superconductors.

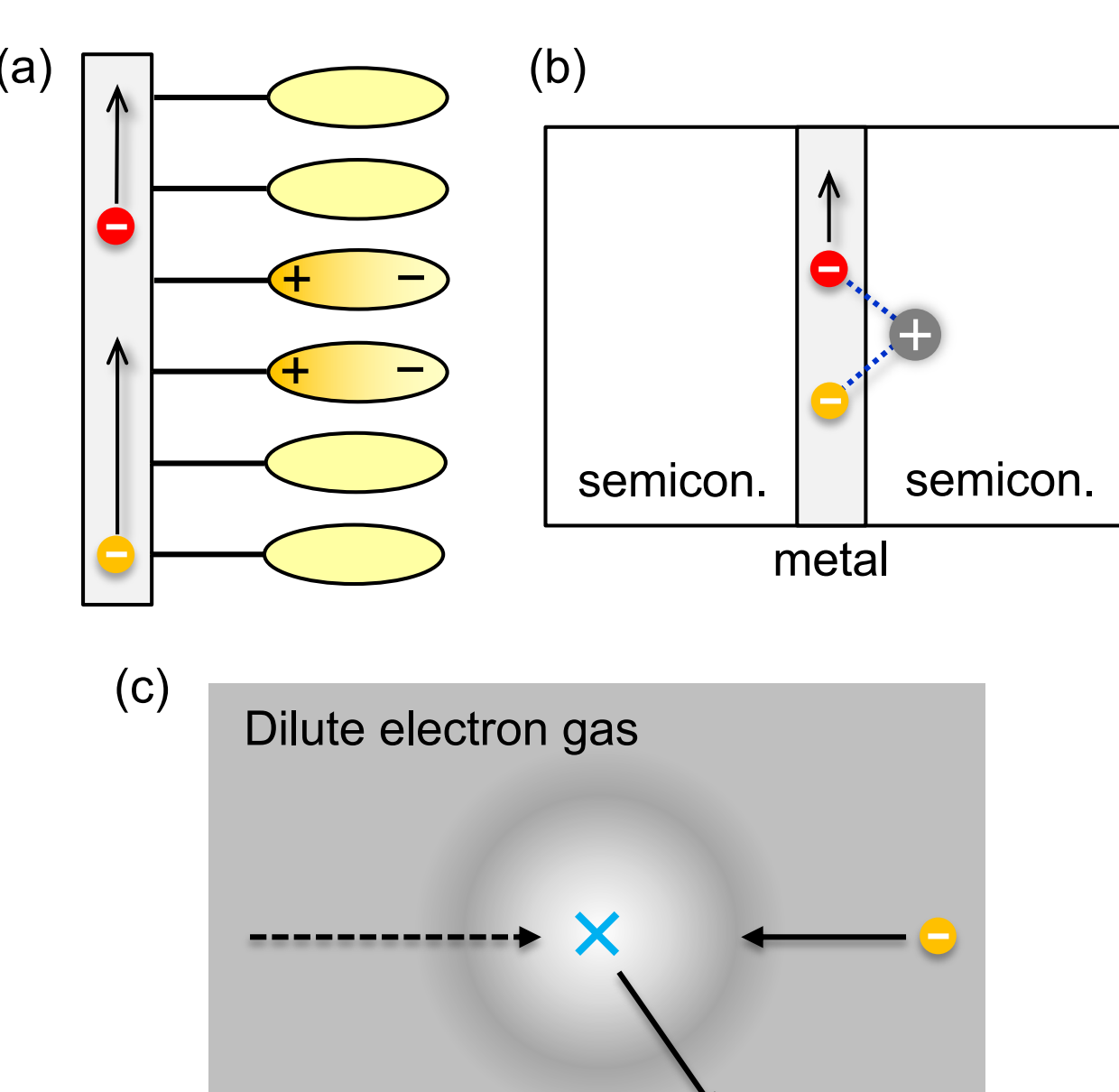

**Fig. 45.** (a) Little's superconductivity model based on charge fluctuations [269]. The polarization of molecules attached to a 1D conduction path causes two electrons to pair: the first electron (red ball) polarizes the side molecules while passing, then the second electron (orange ball) is drawn to the induced plus charge. Cooper pairing occurs when a polar region is formed and absorbed. (b) Ginzburg's 2D model of a metal–semiconductor interface for the charge fluctuation mechanism [270]. Similar to Little's model, a hole created in a semiconductor near the interface can cause two electrons in the metal layer to couple. (c) A pairing mechanism for dilute electron gases and excitonic insulators. The shading represents the distribution of electrons or excitons (electron–hole pairs). In the dilute electron gas case, near the image's center, the first electron scatters by pushing surrounding electrons away via unscreened Coulomb repulsion, creating a region of lower carrier concentration that attracts the second electron and causes an effective coupling between them. When it comes to excitons, the first electron breaks them and scatters, leaving a region with fewer excitons to attract the second electron, resulting in virtual coupling. These simplified interpretations are similar to those given for the phonon mechanism in Fig. 7.

The quasi-2D molecular conductor α-(BEDT-TTF)$_2$I$_3$ undergoes CDW-like charge ordering at 135 K, resulting in an insulator. Superconductivity was observed at 7 K with a uniaxial pressure of 0.2 GPa [331, 346]. These phenomena are thought to be caused by charge fluctuations resulting from charge order suppression [347]. Nevertheless, because molecular solids are soft and easily distorted under pressure, one must consider enhanced electron–phonon interactions and randomness effects, which can overwhelm the phenomenon.

#### 5.4.2. Valence fluctuations

Valence fluctuations, a type of charge fluctuation, have been observed in f-electron compounds containing ions with two stable valences, such as $Eu^{2+}$–$Eu^{3+}$, $Yb^{2+}$–$Yb^{3+}$, and $Ce^{3+}$–$Ce^{4+}$ [77]. The suppression of the antiferromagnetic metallic phase in $CeCu_2Ge_2$ (Fig. 43b) causes the first superconducting phase to appear with increasing pressure, which is associated with antiferromagnetic fluctuations. In contrast, the second superconducting dome at higher pressures is thought to result from Ce valence fluctuations [284]. β-$YbAlB_4$ exhibits superconductivity at a low $T_c$ of 80 mK, possibly due to Yb valence fluctuations [20, 348].

#### 5.4.3. Valence skipping

Valence skipping is a common charge fluctuation phenomenon found in Bi and Tl-containing compounds [14]. Bi (Tl) can form cations with valences of 3 (1) or 5 (3), which represent the electronic states $6s^2$ and $6s^0$, respectively. Valence skipping, rather than valence fluctuation, refers to the absence of the unstable $6s^1$ state. The $6s^2$ state consists of two s-electrons, and superconductivity can occur when these extreme electron pairs move around the crystal and Bose–Einstein condense into Cooper pairs.

In $BaBiO_3$ with a perovskite structure, Bi has the formal valence $Bi^{4+}$ but divides into the more stable $Bi^{3+}$ and $Bi^{5+}$. This charge disproportionation couples strongly with the lattice, producing alternating packing of large and small $BiO_6$ octahedra for $Bi^{3+}$ and $Bi^{5+}$, respectively. As a result, $6s^2$ electron pairs become trapped on the large $Bi^{3+}O_6$ octahedra. When '$Bi^{4+}$' is partially replaced by $Pb^{4+}$ with no 6s electrons in $BaBi_{1-x}Pb_xO_3$, a hole forms, allowing the 6s electrons to hop between sites, structural distortion disappears, and a metallic state with $T_c$ = 13 K superconductivity appears [14]. Alternatively, $Ba_{1-x}K_xBiO_3$ introduces holes by substituting K for Ba and achieves superconductivity with a higher $T_c$ of 30 K without introducing disorder into the Bi–O conduction path [86,

349].

Charge fluctuations caused by valence skipping may contribute to the superconductivity mechanism in these compounds. However, doping softens and reduces the energy of phonons that produce large and small octahedra, also known as breathing phonons. The resulting structural instability must amplify electron–phonon interactions, which could be the primary cause of superconductivity. The high $T_c$ values come from the large coupling constant $\lambda$, which compensates for the low energy of phonons. As illustrated in this example, as an electronic instability grows, so does the phononic instability that it couples with. It is frequently difficult to distinguish between the two contributions; it is always a chicken and egg problem [29].

PbTe, a narrow-gap semiconductor, exhibits superconductivity with $T_c < 1.5$ K when Pb is replaced by 0.5–1.5% Tl [350]. Charge fluctuations caused by Tl's valence skipping have been attributed to its appearance. Interestingly, cooling increases the normal state's electrical resistance, followed by a metallic decrease, which could be due to charge screening by the Kondo effect; as mentioned in Section 5.3.1, the Kondo effect can mask a localized magnetic moment and a charge in this case. Thus, the charge degree of freedom has a significant influence on the electronic state. Nevertheless, there is no conclusive evidence that charge fluctuations induce superconductivity in this or any other valence-skipping compound.

5.4.4. Dilute electron gas

Charge fluctuations in dilute electron gas have long been thought to cause superconductivity [351], with theoretical $T_c$ values exceeding 200 K [352]. Coulomb interactions between electrons are usually screened by nearby conduction electrons, resulting in a weak and perturbative effect, except at half band filling, such as in cuprates; however, when carrier density is extremely low, the screening is rendered ineffective, and the Coulomb interaction is expected to have a critical destiny. When an electron-deficient region forms in a uniform dilute electron gas, electrons separate more due to enhanced Coulomb repulsion, resulting in a sparse region with lower density and a dense surrounding region. As a result, the dilute electron gas tends to separate into two regions: electron-sparse and dense, characterized by strong and weak Coulomb repulsions, respectively. This charge disproportionation can vary in space and time, causing charge fluctuations that could be the source of the attractive force for Cooper pairing. As depicted in Fig. 45c, the first electron's Coulomb repulsion creates a sparse electron region into which the second electron is drawn after the first scatters. It is worth noting that the many-body effect can generate an attractive force even via the Coulomb interaction, which usually causes repulsion between two bodies [352].

$Li_xZrNCl$ and $Li_xHfNCl$ are promising quasi-2D materials with charge-fluctuation-induced superconductivity in a dilute electron gas [226, 353-356]. As shown in Fig. 46, Zr (or Hf) and N atoms form a honeycomb lattice sheet, and two of them stack to form a conduction layer with a wide band consisting of Zr 4d and N 2p orbitals. ZrNCl layers are constructed by stacking Cl sheets above and below, and they crystallize via van der Waals forces. Intercalated $Li^+$ ions between the layers donate electrons to the conduction layer, converting the parent band insulator into a superconductor. As $Li^+$ ions or electrons are reduced toward the parent phase, the $T_c$ rises from 11.5 K ($x = 0.13$) to 15.2 K ($x = 0.06$) [355]. Then, $T_c$ peaks at 19.0 K ($x = 0.011$) and gradually decreases, resulting in a $T_c$ dome in Fig. 46 [226]. Furthermore, at higher temperatures of $T^*$, a pseudogap opens, with $T^*$ approaching $T_c$ as $x$ increases, indicating preformed pair formation.

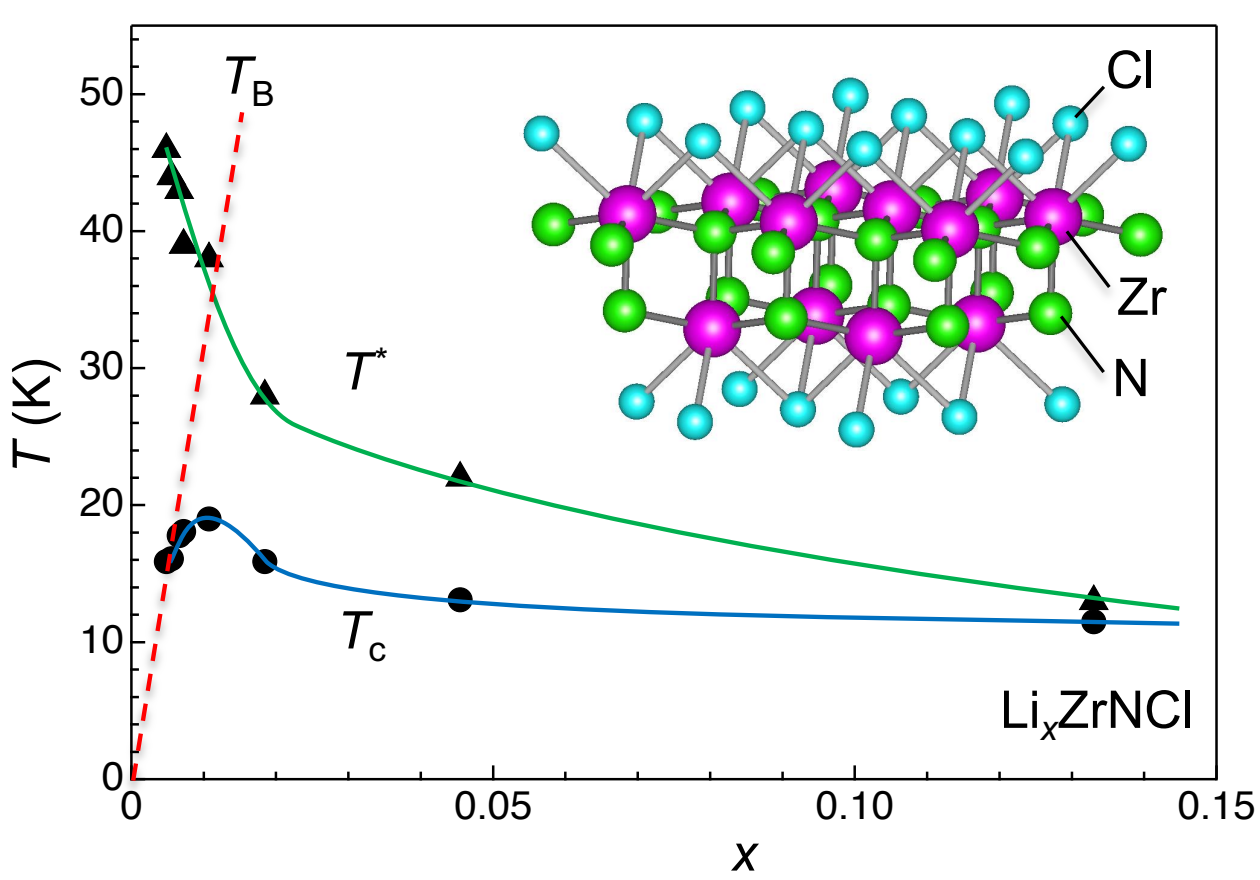

**Fig. 46.** Phase diagram of $Li_xZrNCl$, an electron-doped superconductor studied using the EDL technique [226], which may exhibit charge-fluctuation-induced 2D superconductivity in a dilute electron gas system. $T^*$ denotes the pseudogap-opening temperature, and the BEC temperature ($T_B$) has a slope of 3300 K. The inset depicts the crystal structure of pristine ZrNCl, which is made up of double honeycomb ZrN sheets, where transport occurs, and Cl sheets above and below. Electron carriers are generated by $Li^+$ ions intercalated between the ZrNCl layers, so $x$ simply equals the electron carrier density which agrees with those calculated from Hall measurements.

The observed decrease in $T_c$ toward the parent phase, combined with the presence of $T^*$ ($T_p$), supports 2D BEC superconductivity. The $T_B$ slope is approximately 3300 K from $T_c$ = 15.9 K at the smallest $x$ of 0.0048, which is nearly four times larger than 850 K in cuprates (Fig. 37). Cuprates have greater randomness effects, which may explain the difference. $Li_xZrNCl$ has fewer randomness effects in the weakly correlated electron system based on expanded Zr 4d and N 2p orbitals and a chemical modification that lowers the impurity potential. Cleaner compounds exhibit higher levels of ideal BEC superconductivity, which transitions to BCS superconductivity as doping increases [226].

Superconductivity in $Li_xZrNCl$ appears to be caused by charge fluctuations acting as a glue in a dilute electron gas, rather than magnetism. Although charge fluctuations may have a larger energy scale than $J$ in cuprates, the observed $T^*$ in Fig. 45 is much lower than $T_p$ in cuprates at comparable doping levels. However, $T^*$ is likely to be significantly higher near zero doping. If this is the case, doping causes $T_p$ to rapidly drop as screening effects increase, weakening charge fluctuations. As a result, $T_c$ reaches its peak at only 1% doping. Employing a "trick" to suppress the drop in $T^*$, $T_c$ could be dramatically raised, surpassing cuprates at 4.5% doping and 300 K at 9%. The implications of designing high $T_c$ materials will be explored further in Chapter 6.

5.4.5. Excitonic instability

Electron–hole interaction can be another source of Fermi liquid instability, albeit it is not mentioned in Fig. 3. It can create electron–hole pairs (excitons) in a conventional semiconductor when valence electrons are excited by photons. Although excitons are not popular in itinerant electron systems, they are considered critical in a semiconductor with a narrow band gap or a semimetal with a shallow band overlap. They have a BEC, similar to Cooper pairs in the superconducting state, which results in a 'excitonic insulator' [357]. However, unlike Cooper pairs, excitons have no net charge and thus cannot carry electrical current or become superconducting even in their BEC state; therefore, unlike superconductivity, their presence is extremely difficult to verify experimentally and distinguish from typical CDW states. It is not surprising that a superconductor induced by excitonic fluctuations exists.

Superconductivity may appear near the excitonic insulator due to exciton density fluctuations. When an excitonic insulator is doped with an extra electron, it is expected to destabilize because the added electron repels the exciton's electron while attracting its hole, causing the exciton to split up, as is also true for hole doping. Similarly to the dilute electron gas in Fig. 45c, the first electron depletes an exciton in the background, leaving a sparse region into which the second electron is drawn, resulting in an effective attraction between them. Thus, excitonic superconductivity can occur when density fluctuations in the excitonic insulator serve as a Cooper pairing glue.

Material candidates for excitonic insulators include 1T-$TiSe_2$ [358, 359], $Ta_2NiSe_5$ [360, 361], ZrSiS [362], and NaAlGe [363]. If superconductivity appears in these materials as a result of suppressing parent states, it could be attributed to excitonic fluctuations. $Ta_2NiSe_5$, an insulator at ambient pressure, exhibits superconductivity at $T_c$ = 1.2 K at 8 GPa following pressure-induced metallization [364]. However, the superconducting phase has a different crystal structure than the excitonic phase at ambient pressure, making it difficult to apply a simple QCP scenario. In NaAlGe, hole doping with Zn-for-Al substitution suppresses the 100 K pseudogap, most likely due to excitonic instability, resulting in superconductivity at $T_c$ = 2 K [365]. However, it is unclear whether this superconductivity stems from excitonic fluctuations. To demonstrate the possibility of excitonic superconductivity, further search for new materials is encouraged.

5.5. Orbital and multipole fluctuations, and related superconductors

Orbital degrees of freedom correspond to p-orbital degeneracy in typical elements, while transition metals have d-orbital degeneracy. Their associated fluctuations may produce superconductivity [310, 366]. In most compounds, however, the crystal field, which is a local field generated by the surrounding ligands, can remove the degeneracy of the p and d orbitals [367]. Furthermore, orbital degrees of freedom are always combined with crystal structure distortion, resulting in spontaneous degeneracy loss; the Jahn–Teller effect is a common example. $LiNiO_2$ [368, 369] and $FeCr_2S_4$ [370] are believed to be exceptional materials with "liquid orbitals", but they are difficult to inject carriers into while remaining insulators. Although α-FeSe may exhibit superconductivity due to orbital fluctuations [310, 314], as mentioned in Section 5.3.1, research is still ongoing with unknown details. It is preferable to seek out other materials with surviving orbital degeneracy, which can couple with electronic properties.

The final perturbative interaction described in Fig. 3 that causes electronic instability is the spin–orbit (SO) interaction. In relatively heavy elements, SO interaction combines spin and orbital degrees of freedom. The resulting multipoles with a higher degree of freedom, such as quadrupoles and octupoles, can be used as glues [366]. The physics of multipoles has been extensively studied in f-electron compounds. $PrOs_4Sb_{12}$, a skutterudite compound, exhibits superconductivity at 1.85 K [371]. When a magnetic field suppresses superconductivity, an electric quadrupole (one of Pr's f-electron multipoles) appears in an ordered phase, suggesting that quadrupole fluctuations are responsible for superconductivity [372].

Superconductivity and multipole fluctuations were also investigated in 5d electron compounds with SO interaction smaller than f electrons but larger than 3d electrons [373]. The author studied the 5d pyrochlore oxide α-$Cd_2Re_2O_7$ for years [43]. It achieves superconductivity at 0.97 K (Fig. 9) after losing spatial inversion symmetry via a phase transition at 200 K to an electric-toroidal-quadrupole order, an odd-parity itinerant multipole [25, 43, 374]. Cooper pairing with a mix of s-wave and p-wave characters is theoretically expected [375], but experiments show that s-wave pairing is predominant (the theory predicts the mixing of two types but not their ratio, as mentioned in Section 2.4.3). In addition, the superconductor $La_2IOs_2$, discovered in 2023 [376], may contain 5d multipoles, which play an important role in the superconductivity mechanism with a relatively high $T_c$ of 12 K.

Although the energy scale of the aforementioned orbitals and multipole fluctuations is unknown, sufficient energy fluctuation that strongly interacts with electrons could result in a high $T_c$. However, because additional electron–phonon interactions might significantly couple, achieving pure orbital/multipole fluctuation superconductivity may prove difficult (in any case, increasing $T_c$ would be beneficial).

When reviewing current materials, high-energy electron-origin glues are frequently relaxed by electron–phonon interactions, giving the impression that the two are competitive. In Section 5.3.1 for α-FeSe, two attraction sources can compete or assist each other by producing distinct or identical pairing symmetry. Antiferromagnetic fluctuations and phonons are incompatible because they prefer d-wave and s-wave Cooper pairing, while the former and orbital fluctuations can cooperate at d-wave superconductivity. Is there any combination that strengthens one another? It would be interesting to see if their combination creates a powerful glue or acts as a new type of fluctuation, leading to an unknown superconductivity mechanism.

5.6 Other transition metal compound superconductors

Aside from the material classification based on the aforementioned superconducting mechanisms, this section introduces other notable transition metal compound superconductors that are caused by self-evident phonon mechanisms or are difficult to classify using current knowledge.

Many exotic superconductivities that compete with magnetism due to strong electronic correlations have been discovered in 3d transition metal compounds, with copper oxides leading the way. The transition metal element possesses a $s^2d^z$ electron configuration, loses $s^2$ electrons in solids, and has $(z - v + 2)$ d electrons, depending on its valence $v$. Figure 47 depicts a superconductor map with horizontal axis $z$ and vertical axis $v$ (rather than the previous mechanism-based classification, it would be helpful for solid state chemists). Cu, for example, has a $s^2d^9$ configuration, and $Cu^{2+}$ contains nine d electrons, giving rise to an antiferromagnetic order in the parent Mott insulator, as well as electron-doped and hole-doped superconductivity appearing above and below this phase. Furthermore, Fe-based compounds containing $Fe^{2+}$ ($d^6$) are either AFM or SC, with doping induced superconductivity above and below, as shown in Fig. 47. It should be noted that the actual valence of heavy anions like Sb and Bi can differ significantly from their formal one (for example, $Sb^{3+}$ and $Bi^{3+}$) due to strong covalency, which influences the TM's valence estimation.

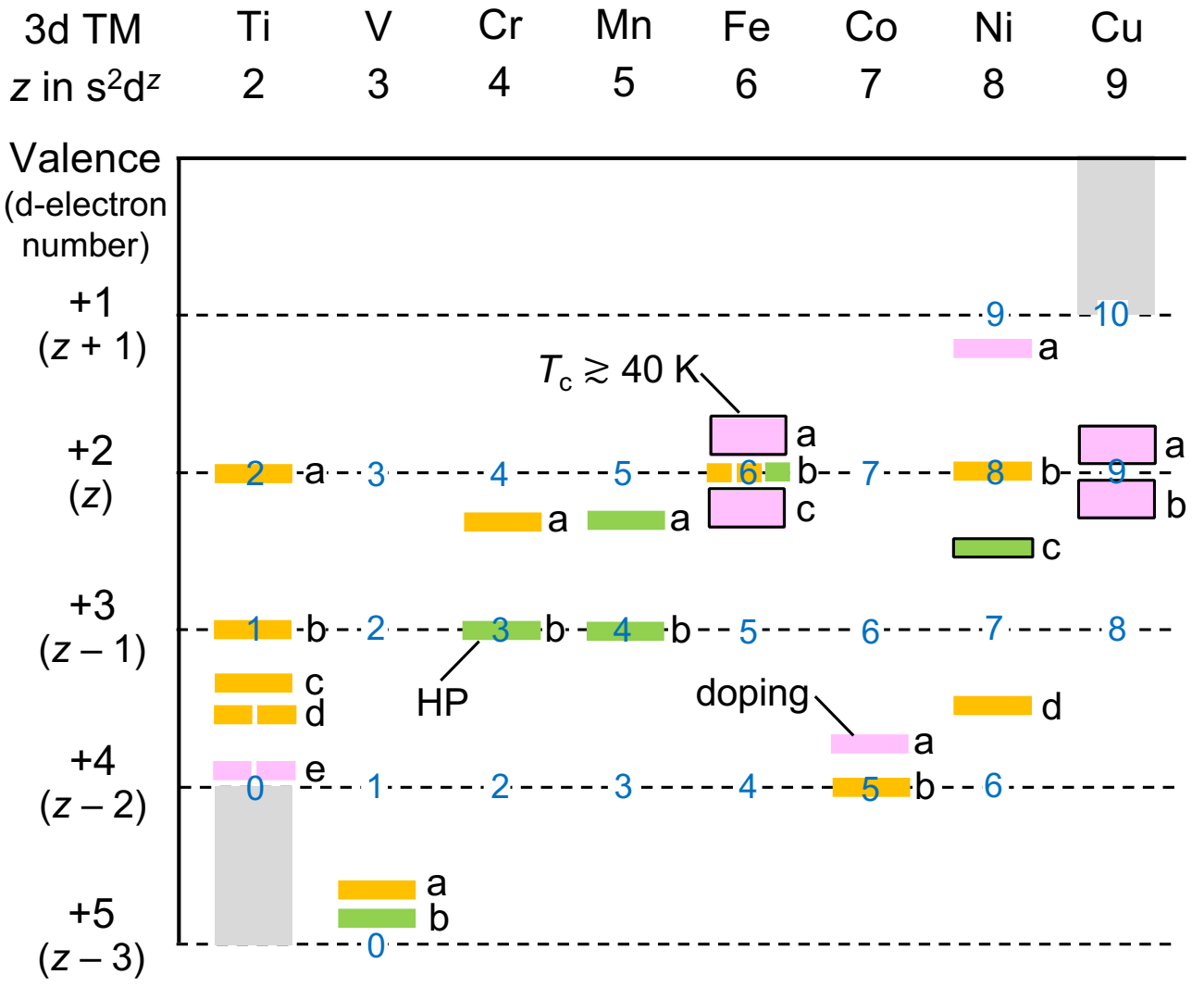

**Fig. 47.** Superconductor map with 3d transition metal (TM) elements as the key ingredient. The 3d TM has a $4s^23d^z$ electron configuration that acquires +2 valence after losing $4s^2$ electrons in a solid. The d-electron number varies with valence $(z - v + 2)$, as shown on the left of the figure, with the actual number indicated in blue on the horizontal broken line for each element. The bars represent superconductors found at ambient conditions (orange), under high pressures (green), and induced by intentional carrier doping (lavender). They are a (TiO), b ($BaTi_2Sb_2O$), c ($\alpha$-$Ti_3O_5$), d ($LiTi_2O_4$, $Ti_4O_7$), and e ($SrTiO_{3-\delta}$, $Cu_xTiSe_2$) for Ti; a ($CsV_3Sb_5$) and b ($\beta$-$Na_{0.33}V_2O_5$) for V; a ($K_2Cr_3As_3$) and b (CrAs) for Cr; a ($KMn_6Bi_5$) and b (MnP) for Mn; a ($Ba_{1-x}K_xFe_2As_2$), b (LaOFeP, $\alpha$-FeSe, $BaFe_2S_3$), and c ($SmFeAsO_{1-x}F_x$) for Fe; a ($Na_{0.35}CoO_2$•$1.3H_2O$) and b ($Na_2CoSe_2O$) for Co; a ($Nd_{0.8}Sr_{0.2}NiO_2$), b ($SrNi_2P_2$), c ($La_2PrNi_2O_7$), and d ($YNi_2B_2C$) for Ni; a (Nd214) and b (La214) typically for Cu. Table 2 details the compounds. The high-temperature superconductors with $T_c$ near or higher than 40 K are marked by square frames, which appear in late TM compounds with enhanced electron correlation and moderate antiferromagnetic order.

Many superconductors have been discovered in Ti compounds with few d electrons. For instance, $SrTiO_{3-\delta}$, a perovskite, is a superconductor ($T_c$ = 1.2 K) obtained by doping a band insulator of $Ti^{4+}$ ($d^0$; $z$ = 2, $v$ = 4) with electrons [18]. $BaTi_2Sb_2O$ is a quasi-2D system with a square $Ti^{3+}$ ($d^1$) lattice, similar to copper oxides, but with a much lower $T_c$ of 1.2 K [377]. In the V series, $\beta$-$Na_{0.33}V_2O_5$ [277] and $CsV_3Sb_5$ [279] are superconductors with $V^{4.835+}$ ($d^{0.165}$) and $V^{4.66+}$ ($d^{0.33}$) configurations, respectively.

Cr and Mn prefer $d^3$ and $d^5$ electron configurations, respectively, leading to the Hund's rule-based selection of high-spin states with large spin quantum numbers, strong magnetism, and a tendency to suppress superconductivity. Pressure broadens bandwidth and weakens electronic correlations, suppressing magnetic order and leading to superconductivity (CrAs [300] and MnP [302]). $Fe^{3+}$ ($d^5$) has even more magnetism, while in $Fe^{2+}$ ($d^6$), the Hund coupling and crystal field splitting are comparable, resulting in a weaker magnetic state. As a result, in iron-based superconductors, an antiferromagnetic metal of sufficient strength appears around the $d^6$ state, while superconductivity with relatively high $T_c$ values occurs above and below this state.

Co-based superconductors are found near $Co^{4+}$ ($d^5$). Takada et al., experts in soft chemistry, discovered that superconductivity emerged at 4 K in $Na_{0.35}CoO_2$•$1.3H_2O$, composed of $CoO_2$ layers formed by $Co^{3.65+}$ ($d^{5.35}$) atoms arranged in a triangular lattice [378, 379]. It occurs when $Na_xCoO_2$, an expected thermoelectric material [380], is intercalated with water molecules to enhance two-dimensionality. The superconducting mechanism remains unclear, but an exotic f-wave spin triplet state has been proposed theoretically [51]. It is noted that in strongly correlated electron systems where orbital and lattice compatibility is crucial, square-lattice systems like copper oxides favor d-wave superconductivity, while triangular-based lattices favor f-wave superconductivity. $Na_2CoSe_2O$, found in 2024 [381], has a layered structure composed of $Na_2O$ and $CoSe_2$ layers. It exhibits superconductivity at $T_c$ = 5.4 K in the $CoSe_2$ layer, which contains a triangular sheet of $Co^{4+}$.

Nickel oxide superconductors are a new member gaining popularity. Although many unknowns remain, they are likely to play an important role in future superconductivity research. In 2019, a thin film of $Nd_{0.8}Sr_{0.2}NiO_2$ with a C1-B1 structure demonstrated a superconducting transition at 12 K [382]. The composition indicates that $Ni^+$ ($d^9$) is 20% doped with holes, yielding an electronic state comparable to copper oxides. In 2023, another nickelate '$La_3Ni_2O_7$' was found to be a superconductor with a relatively high $T_c$ of 70–80 K at high pressures of 20 GPa or higher [383]. This new superconductor is classified as Ruddlesden–Popper (RP) series oxides, with a C2-B2-NC structure based on the copper oxide scheme. However, the used single crystal samples contained additional RP oxides, $La_2NiO_4$ (C1-B2-NC) and $La_4Ni_3O_{10}$ (C3-B2-NC), as stacking faults. Furthermore, alternate stacking of the two produced a superstructure (C1-B2–C3-B2) with the same composition as $La_3Ni_2O_7$ [384]. In 2024, $La_2PrNi_2O_7$ was discovered to have a more uniform C2-B2-NC structure and to be a bulk superconductor with a $T_c$ of 75 K at 20 GPa [17].

$La_2PrNi_2O_7$ has a formal valence of $Ni^{2.5+}$ ($d^{7.5}$), with 1.5

electrons distributed in the d$x^2$–$y^2$ and d3$z^2$–$r^2$ orbitals, which appears to be significantly different from that of the copper oxides (Fig. 18). However, the bonding between the two $NiO_2$ planes significantly separated the d3$z^2$–$r^2$ band, leaving the unsplit d$x^2$–$y^2$ band in the gap. As a result, the higher d3$z^2$–$r^2$ antibonding band remains unoccupied, and the electronic state near the Fermi level is expected to be dominated by the half-filled d$x^2$–$y^2$ band [385]. If this is the case, superconductivity may occur in the same way that copper oxides do. It is interesting to see what types of Cooper pairs are responsible for superconductivity, which will be clarified by future research. The pressure required to achieve superconductivity is currently too high. It is desirable to clarify the significance of the pressure effect and develop nickel oxide superconductors that can be tested at normal or low pressure.

The most notable 4d transition metal compound is $Sr_2RuO_4$, which is structurally identical to La214 and has a C1-B2-NC structure. This compound exhibits 2D conduction within the $RuO_2$ plane, replacing $Cu^{2+}$ with $Ru^{4+}$ ($4d^4$ in the Fe series in Fig. 47). In 1994, Maeno et al. reported superconductivity at 0.93 K [114], which was identified as a spin-triplet p-wave. However, recent research suggests a more complex Cooper pair symmetry [386]. Furthermore, the $T_c$ of high-quality single crystals increased to 1.5 K and was found to rise to 3.5 K under uniaxial pressure. There is no magnetic state in the vicinity, and because it is not a simple s wave, the phonon mechanism is ruled out. Its origin is still being investigated after 30 years.

The second half of 3d TMs contains high-temperature superconductors with $T_c$ values greater than ~40 K, as represented by square frames in Fig. 47. One probable explanation is the high electron correlation and moderate antiferromagnetic order. Strong electron correlations are most effective in a half-filled single band, which is difficult to achieve in early TMs with multiple bands made up of degenerate orbitals. This is one of the causes of the low $T_c$ in $BaTi_2Sb_2O$. Magnetism, on the other hand, appears to be excessively strong in the TM map's center. As a result, these areas lack a sufficient supply of strong glues. However, in some cases, the excessive magnetism at the middle can be weakened.

Another important consideration is the energy difference between the TM d and counter anion p levels. The d band has higher energy than the oxygen p band at Ti, but it decreases as nuclear charge increases to the right, eventually overlapping around Cu [28]. When heavier counter anions are used, the balance appears to shift to the left. Fe-based superconductors must meet this condition by combining Fe and As at comparable d and p levels; important combinations include $Cu^{2+}$ and O and $Fe^{2+}$ and As. Furthermore, $Ni^{2.5+}$ has slightly lower d level than $Ni^{2+}$, which put it closer to the O 2p level. As a result, there is a strong coupling between the d electrons, which are primarily responsible for magnetism, and the p electrons, which dominate conduction, resulting in a strong glue that promote high-temperature superconductivity, as typically observed for ZRS in copper oxide superconductors.

The material search field will change based on the type of glue. In contrast to magnetism-induced superconductivity, charge fluctuation superconductivity requires a broad band with weak electron correlations, such as electron doping into the $Zr^{4+}$ ($4d^0$) and N bands in $Li_xZrNCl$. Thus, 4d and 5d electron systems are more likely to be targeted than 3d. Heavy elements are also required for enhanced orbital and multipole fluctuations.

As previously mentioned, a strategy based on superconducting mechanisms may prove advantageous. However, if one becomes overly focused on the mechanism (physics), chances of encountering unexpected materials decrease. When considering material exploration strategies, chemists may find it useful to let their imaginations run wild with Fig. 47. They could look for materials in empty windows or in the neighborhoods of known compounds. Furthermore, they could consider material designs that eliminate the factors that contribute the reduction of $T_c$ in known superconductors or that construct novel platforms with the same electron configuration. Based on the information presented above, relying on rapidly evolving generative AI may yield unexpected materials when combined with massive amounts of online data (albeit the author is skeptical). It will be difficult to overcome the advantages of copper oxides (Chapter 4), which include strong electron correlations (rare at half band filling), large magnetic interactions as well as appropriately stable magnetic order, and weak lattice coupling, but the author believes it is not impossible.

## 6. Where is the road to a room-temperature superconductor?

In Chapter 5, the author classified superconductors according to the type of Cooper pairing glue. The prospect of achieving a high $T_c$ using these glues will be discussed below. Given that Equation 3 [$T_c = \omega_0 \exp(-1/\lambda)$] can formulate $T_c$, the key issues are the elementary excitation's energy, $\omega_0$, which generates the glue, and the exponential term (reduction factor), which includes $\lambda$, determined by the coupling between the excitation and electron. In addition, using the general BCS–BEC crossover diagram, we will discuss a strategy for a room-temperature superconductor based on our findings from copper oxide superconductivity research. A personalized approach to locating superconducting materials will also be addressed.

### 6.1. Glue types and energy

First, we'll review the glue types covered in Chapter 5 and their energy scales in preparation for high $T_c$. The phonon energy scale is roughly equivalent to the Debye temperature, which is usually around 400 K for most materials. A 10% reduction factor yields the highest value of $T_c$ of 40 K; in fact, the highest $T_c$ for phonon systems at ambient pressure is 39 K for $MgB_2$ [387]. $T_c$ does not increase because increasing $\omega_0$ ($\omega_{ph}$) in Eq. 3 decreases $\lambda$ ($V$). Higher energy phonons interact less with electrons, leading to lower $\lambda$ and limited $T_c$. Boron-doped diamond [C(B)] has a $T_c$ of only 4–7 K [74, 75], which is far from diamond's high $\omega_{ph}$ of 2250 K (Table 2). This means that weak electron–phonon interactions preclude the use of high phonon energies. In contrast, the phonon energy that contributes to superconductivity in $Nb_3Sn$ is low (176 K), but a relatively high $T_c$ = 23.2 K is obtained due to a reduction factor of 0.13 greater than 0.1; the low energy phonons, i.e., slow ionic motion, can interact strongly with electrons. However, excessive $\lambda$ can disrupt crystal structures, distort lattices, and lower DOS at the Fermi level, resulting in reduced electron–phonon interactions. The trade-off between $\omega_{ph}$ and $\lambda$ sets the upper limit for $T_c$ in phono-mediated superconductivity.

Superconductivity above $T_c$ = 200 K in $H_3S$ and $LaH_{10}$ at ultrahigh pressures above 100 GPa [16, 388, 389] may be due to the high energy phonons of light hydrogen atoms present at high density. High pressure can reduce structural instability and maintains large $\lambda$ values, potentially mimicking theoretical superconductivity for solid hydrogen [73]. The discovery of these ultrahigh-pressure stable phases demonstrated that room-temperature superconductivity was possible. However, because these phases cannot be quenched or grasped in hands under normal conditions, they are difficult to classify as "materials" in solid state chemistry. Their superconducting mechanism is also thought to be a conventional phonon mechanism, which is unlikely to lead to the discovery of a new glue.

Superconductivity, which exists alongside spin ordering and uses its fluctuations as glue, has been observed in a variety of materials, most notably cuprates. Copper oxides are expected to have the highest antiferromagnetic fluctuations of any known 2D magnetic material, with $J$ at 1500 K and $T_c$ at 150 K. As a result, it may be challenging to achieve even higher $T_c$ from antiferromagnetic interactions in limited materials unless the reduction factor is raised above 0.1; conversely, if there is a trick to increase the reduction factor, the world will change; this issue will be discussed in the following section. However, 1D cuprates have a large $J$ up to 3000 K in the chain direction [390], so a $T_c$ of 300 K is not surprising. Nevertheless, no doping-induced metallization was observed in typical quasi-1D cuprates $Sr_2CuO_3$ or $SrCuO_2$ with Cu–O chains. Additionally, spin ladder compounds with $J \sim 2000$ K along the leg have a $T_c$ as low as 12 K [338]. Unfortunately, no known 1D materials can efficiently harness their high $J$ for superconductivity. One possible explanation is that the randomness effect is more problematic in 1D structures. Future material searches should focus on quasi-1D cuprates and other superconductors that are less susceptible to the randomness effect. Other reasons include excessive 1D fluctuations, which destabilize any order, including superconductivity. If this is the case, then highly 1D systems are not targets for high $T_c$, while weakly 1D systems may be advantageous.

Exotic superconductivity mechanisms are expected in ferromagnetic fluctuations. However, high $T_c$ is unlikely because their energy scale is much smaller than that of antiferromagnetism. The only known ferromagnetic superconductors are uranium compounds with $T_c$ below 1 K. There are ferromagnets with high magnetic transition temperatures, but they are made up of classical spins with large spin quantum numbers, making it difficult to suppress their order and obtain the required fluctuations. Thus, they are unsuitable for use as efficient superconducting glue. In contrast, in systems with multiple competing magnetic interactions or geometrical frustration, magnetic order with complex spin arrangements occurs, potentially leading to superconductivity, such as seen in MnP and CrAs. Nonetheless, the strength of magnetic interactions will fundamentally limit $T_c$. Similarly, $J$ governs $T_c$ in superconductivity that starts with a spin liquid, such as a spin ladder.

Despite their high energy, superconductors induced by charge fluctuations have low $T_c$s. As demonstrated in phonon superconductivity, maintaining a strong coupling between excitation and electron becomes more challenging as excitation energy increases. The $T_c$ of superconductors based on fluctuations caused by orbital and multipole degrees of freedom with unknown energy scales is similarly low. High-temperature superconductivity requires the development of a glue with sufficient energy and reduction factor.

6.2. Improving the reduction factor: insights from copper oxide superconductivity

A high-energy fluctuation and its strong coupling with electrons can serve as an effective glue, resulting in tightly bound Cooper pairs and even room-temperature superconductivity. However, there are no general guidelines for achieving both at the same time; the situation will differ greatly depending on the type of glue used and the system's specifics. If Eq. 5 [$T_c = J\exp(-1/\lambda)$] is applicable to cuprate superconductivity, a small increase in $\lambda$ for the same $J$ will significantly raise $T_c$. Multiplying $\lambda$ by 1.4 yields a reduction factor of 0.2, enabling $T_c$ to reach 300 K.

The above discussion of the BCS equation does not explicitly consider the variable DOS in $\lambda$ [$= N(E_F)V$]. When electrons are doped from scratch in a 2D band, $T_c$ is unaffected because the DOS is finite from the beginning and remains constant throughout the doping process (Fig. 12). In contrast, the $T_c$ of 2D BEC superconductivity in cuprates and dilute electron gas systems varies with carrier density, setting it apart from BCS superconductivity. It is extremely low, especially at the beginning of doping, in comparison to high values in BCS superconductivity. Thus, the $T_c$ cannot be discussed using the BCS formulation based on momentum space pairing. Nevertheless, mapping it to the BCS equation provides useful insight, requiring a reduction factor that is significantly dependent on carrier density. In this section, we will consider a potential route to high-$T_c$ by generalizing what we've learned about copper oxide superconductors.

Let us begin by comparing the ideal phase diagram of cuprate superconductivity (Fig. 38a) in Fig. 48a, after omitting the AFI and AFM phases for simplicity, to the left-right inversion of a BCS–BEC crossover image (Fig. 10) in Fig. 48b, and then consider how to raise $T_c$. In the low-doped regime of the phase diagram, the pairing interaction caused by antiferromagnetic fluctuations is sufficient to achieve BEC superconductivity, as shown in the left part of Fig. 48. If the pairing interaction is independent of $n_s$ ($p$), then $T_c = T_B = (h^2/m)(n_s/2)$. The vertical arrow in Fig. 48b indicates that $T_c$ is proportional to $n_s$ at the maximum interaction ($\omega_0$). Copper oxides exhibit a rightward tilt (arrow 0) as $p$ increases, indicating that the antiferromagnetic background weakens and the pairing interaction decreases. Importantly, in this UD regime, the attraction is strong enough with high $T_p$ that $T_c$ is insensitive to changes in the pairing interaction and grows almost linearly with $p$. In other words, because ZRS pairs have small boson sizes, pair density is the sole determinant of BEC temperature. As $p$ increases, the attraction weakens and the $T_p$ line falls, forming a $T_c$ dome. $T_c$ is then determined by pair size rather than pair density in the BCS regime shown on the right side of Fig. 48.

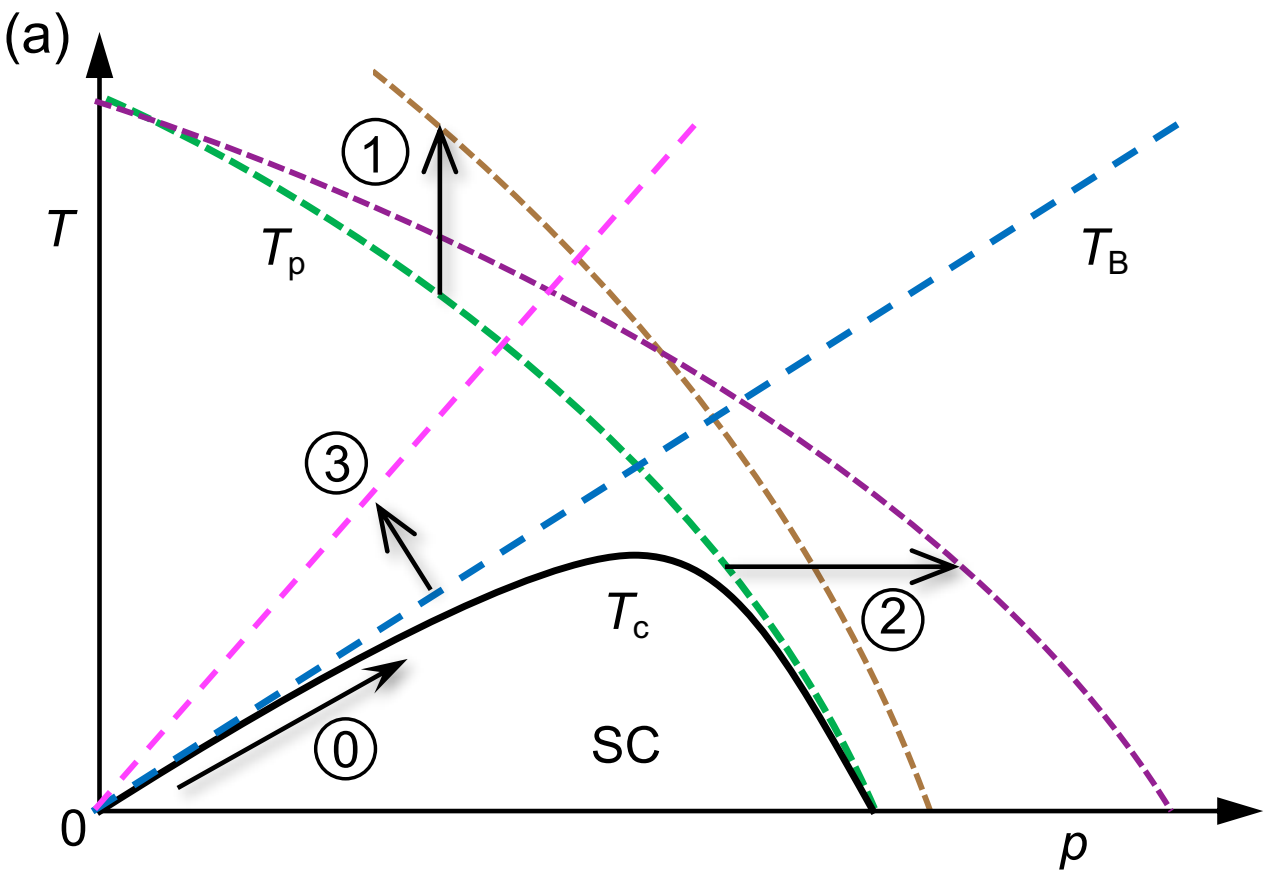

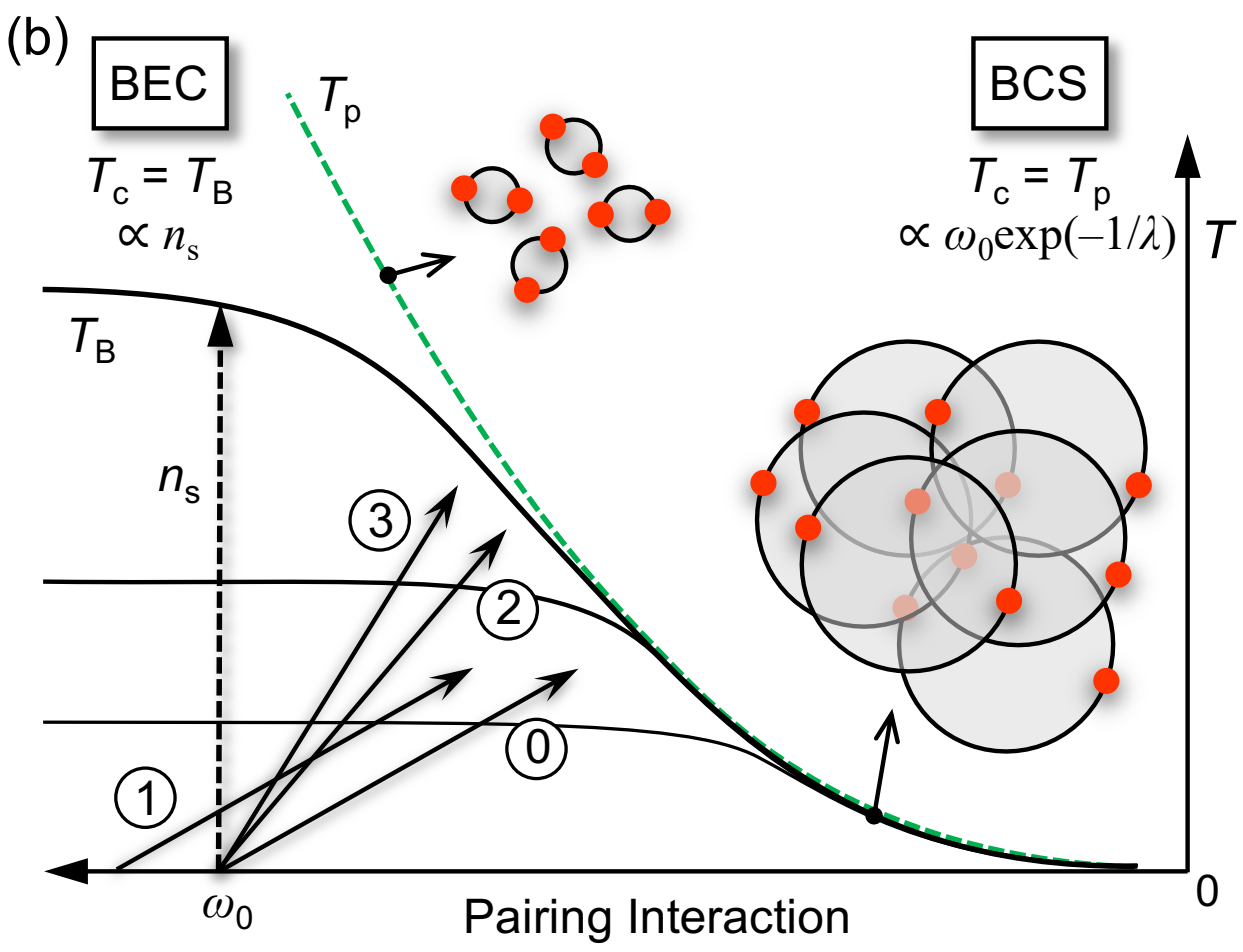

**Fig. 48.** (a) Ideal phase diagram of copper oxide superconductivity (Fig. 38a), with the AFI and AFM phases omitted for clarity. (b) The BCS–BEC crossover image, which is a left-right inversion of Fig. 10. $T_c$ nearly equals $T_B$ in the BEC regime on the left due to a strong pairing interaction, where $T_p$ is much higher than $T_B$. The vertical broken arrow at the maximum pairing interaction $\omega_0$ in (b) indicates that $T_c$ ($T_B$) increases proportionally to carrier density $n_s$. The actual $T_c$ in copper oxide superconductors rises along the inclined arrow 0, as the pairing interaction decreases with increasing $p$. The $T_p$ line falls as the BCS regime approaches, and $T_c$ is suppressed below $T_p$, resulting in a $T_c$ dome in (a). To increase the maximum $T_c$ ($T_{co}$), move the $T_p$ line upward in (a) (operation 1), which corresponds to an increase in $J$ and moving arrow 0 to the left (arrow 1) in (b). Alternatively, move the $T_p$ line to the right in (a) (operation 2) and increase the slope of arrow 0 to arrow 2 in (b). This means that the pairing interaction becomes less reduced as $p$ increases. The third option is to increase the $T_B$ line's slope following operation 3 in (a) and arrow 3 in (b). The left-wing BEC regime can achieve high $T_c$ values, and the majority of high-$T_c$ superconductors are found in the BCS–BEC crossover regime. Large pairing interactions and high carrier density could be combined to achieve room-temperature superconductivity in or near the BEC regime.

According to the preceding discussion, we consider three approaches for increasing $T_{co}$ at the top of the $T_c$ dome. The first step is to increase the original pairing interaction in the parent phase by moving the $T_p$ line upward in Fig. 48a (operation 1) and shifting arrow 0 to the left in Fig. 48b (arrow 1). In cuprate superconductors, high pressure increased $J$ and thus $T_{co}$ from 135 K to 153 K [32]. Taking into account the reduction caused by doping, the parent phase's $T_p$ should be significantly higher than room temperature to obtain room-temperature superconductivity at optimum doping. It is critical to use a larger interaction with associated fluctuations as the pairing glue.

To mitigate the reduction in pairing attraction caused by hole doping, the second strategy involves shifting the $T_p$ line to the right in Fig. 48a (operation 2) and increasing the slope of arrow 0 in Fig. 48b (arrow 2). In other words, boost $T_{co}$ by doping more holes while keeping pairing attraction above a certain threshold. This requires that the attraction mechanism be insensitive and resistant to doping. As discussed in Section 4.8, in the case of cuprate superconductivity, doping destroys the antiferromagnetic spin background, destabilizing the ZRS and lowering $T_p$; $T_{co}$ is reached at 25% doping in C3 (Fig. 38a); however, as discussed in Section 4.7.2, if a thick evenly- doped superconducting layer is achieved for $n > 3$, $T_p$ will shift to the right, giving a higher $T_{co}$ at a larger $p_o$. The background that creates pairing attraction may respond differently depending on the nature of the underlying interaction and the carrier features (for example, ZRS or d-holes in Section 4.8.1.3). It is desirable to develop an appealing mechanism that is resistant to doping. As discussed in Section 6.4, one option is to spatially separate the conduction and attraction origins.

The third option for raising $T_{co}$ is to increase the $T_B$ line's slope in Fig. 48a (operation 3); similarly to the second case, this results in a shift from arrow 0 to arrow 3 in Fig. 48b. The BCS–BEC crossover scenario for a conventional parabolic band in 2D Fermi gas [226, 227] limits the $T_B$ to 1/8 of the Fermi temperature $T_F$, yielding a 2300 K slope. Figure 37 predicts a lower $T_B$ slope for cuprate superconductivity at 850 or 1400 K; however, this deviation may be reduced in the clean and low-doping limits. As demonstrated in Section 5.4.4 for the dilute electron gas compound $Li_xZrNCl$, 1.1% (0.48%) electron doping yields 19.0 (15.9) K $T_c$, implying that the $T_B$ curve has a slope of 1730 (3300) K, which may be close to the theoretical value within the experimental error or indicates a larger slope due to a lighter carrier mass. The low $T_c$, despite the steep slope, must be due to the sensitivity of background charge fluctuations to doping, specifically a rapid drop in $T_p$. Again, reducing the $T_p$ drop is critical for increasing $T_{co}$.

Although it is unclear whether the $T_B$ slope can be increased further, a system with a particularly light electron mass and broad bandwidth would have a high $T_F$ even at low doping levels, resulting in a significant $T_B$ slope. In contrast, the $T_B$ of a 3D electron system is 0.218 times the Fermi energy [44, 48]. Given that the Fermi energy can exceed 10,000 K, a $T_c$ of greater than 2,000 K is not surprising, assuming the attraction is strong enough to enter the BEC regime; however, we wonder if obtaining an effective glue will be more difficult in 3D due to less intense fluctuations than in 2D.

### 6.3. High-$T_c$ superconductivity should occur in the BCS–BEC crossover regime

The BCS–BEC crossover picture applies to the general superconductivity mechanism, regardless of pairing mechanisms [44, 164, 226, 391, 392]. High $T_c$ is only achievable in or near the BEC regime, not in the BCS regime. If the pairing interaction is strong enough to produce BEC superconductivity, increasing Cooper pair density simply leads to a higher $T_c$. The highest $T_c$ in superconductors with monotonically decreasing pairing interaction caused by carrier doping, such as cuprates and dilute electron gas systems, occurs naturally in the BCS–BEC crossover regime, where an increase in Cooper pair density is balanced by a decrease in pairing interaction (unfortunately, nature always prefers a trade-off). In general, as the number of carriers increases, screening effects reduce the majority of many-body interactions, such as electron–phonon and Coulomb. To achieve a high $T_c$, a specific pairing interaction is needed, which is difficult to suppress via screening. What is it?

In a typical QCP superconductivity (Fig. 42), which is caused by pressure suppression of the ordered phase rather than carrier doping, the large number of carriers already present in the metallic parent phase can be used to form highly concentrated Cooper pairs. Consider the appearance of pressure-induced QCP superconductivity following the suppression of antiferromagnetic order. Pressure reduces interatomic distances, broadens bandwidth, and weakens antiferromagnetic interactions. As a result, pressure suppresses magnetic order, enhancing magnetic fluctuations and raising the $T_c$ initially. When the bandwidth is increased to exceed the QCP by applying additional pressure, magnetic fluctuations decrease and $T_c$ falls. Assuming a constant number of carriers under pressure, the $T_c$–pressure relationship shown in Fig. 42 is translated as follows in Fig. 48b: Magnetic order appears on the right side of the phase diagram; as pressure increases, it is replaced by superconducting order, with $T_c$ moving to the left (increasing pairing interaction) along a specific $n_s$ constant curve (e.g., the top thick black curve), but returning to the right (decreasing pairing interaction) beyond QCP, causing $T_c$ to decrease. Thus, the $T_c$ dome as a function of pressure is created by climbing the hill in Fig. 47b from the right, stopping halfway up, and returning to the starting point. The current low $T_c$ values in pressure-induced superconductivity indicate that the BEC regime has not been reached. Unfortunately, many carriers are useless for high $T_c$ due to the difficulty of meeting the BEC requirements. Similarly, as mentioned in the section on doing-induced superconductivity, it is necessary to find a glue that works even at high electron density.

In the case of pressure-induced superconductivity adjacent to an insulating phase, such as CDW, a normal metal at a higher pressure is shown on the right side of the phase diagram in Fig. 48b. As the pressure decreases and the system approaches the insulating phase, the attraction force rises due to relevant fluctuations, stabilizing superconductivity and increasing $T_c$. However, the $T_c$ line eventually ends because the superconducting phase does not continue to the insulating phase. A high $T_c$ is possible if the attraction force just before the insulating phase is strong enough to reach the BEC regime. As a result, discovering pairing attraction forces that reach the BEC region is critical for achieving a high $T_c$ in pressure-induced superconductivity while maintaining sufficient carrier density.

## 6.4. Potential setups towards high-$T_c$ superconductivity

High-$T_c$ superconductivity necessitates both strong pairing interactions and a high Cooper pair density. BEC superconductivity at high doping levels is difficult to achieve because pairing interaction and carrier density are often correlated. However, their divorce can be maintained by keeping them somewhat independent in specific models. For example, in Little's model (Fig. 45a), the ionic charge fluctuation degrees of freedom of the side-chain molecules may overcome the constraint, resulting in a high $T_c$ due to the abundance of Cooper pairs in the BEC regime. Nevertheless, it is unclear whether the molecule's charge degrees of freedom and conduction electrons in the 1D path can be sufficiently coupled in such a two-component system.

In terms of dimensionality, 3D systems are unsuitable for obtaining high-$T_c$ superconductors because the long-range order competing with superconductivity may be too stable, and the resulting fluctuations are too weak. Although 1D systems are expected to have large interactions and DOS, the fluctuations are far too large to achieve any 3D order, including superconductivity. In addition, actual 1D compounds suffer from randomness effects. To achieve a high $T_c$, a 2D system with adequate fluctuation, a larger DOS than 3D, and a lower randomness effect than 1D is preferable. As a result, it is natural to seek high-temperature superconductivity in quasi-2D materials.

Two-dimensional systems can also be useful in structural chemistry. The crystal structure's high flexibility allows for the design of a diverse range of materials. While there are few ways to pack common atomic clusters like octahedra and tetrahedra into three dimensions with counter atoms, stacking atomic cluster-based layers is straightforward. For example, the rigid and small $SiO_4$ tetrahedron is only stable in layered structures at ambient pressures [109].

Figure 49 depicts two testing beds. The first approach assumes a crystal made up of excitation layers that generate pairing interactions and conducting layers with high carrier density (both containing a few atom-thick sheets) (Fig. 49a). Similar to the Ginzburg model in Fig. 45b, high-$T_c$ superconducting 3D LRO is not surprising if the fluctuations caused by high-energy excitations of the spin, charge, and orbital degrees of freedom in the excitation layer generate sufficient pairing interactions for dense electrons in the conduction layer. In other words, the charge-reserving block layer in cuprate superconductors serves as an excitation layer, producing dense Cooper pairs in conduction layers that connect them.

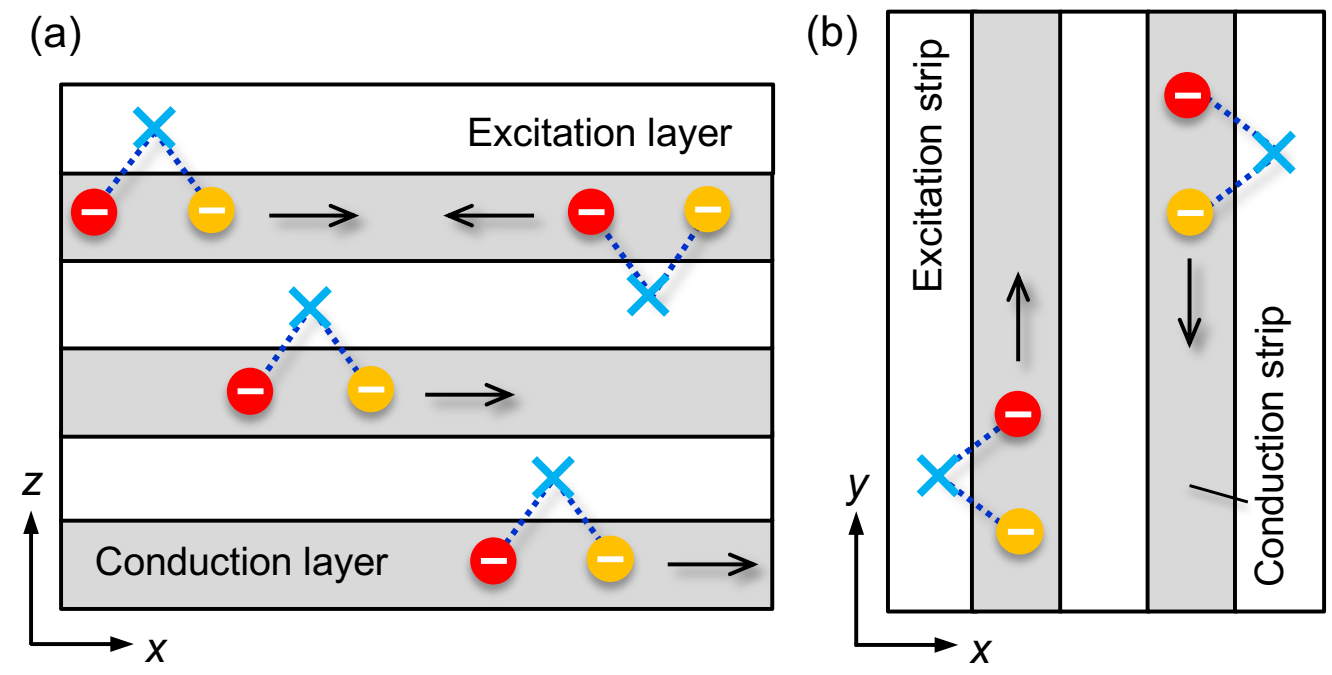

**Fig. 49.** Cartoon illustrating possible quasi-2D crystal models

for high-$T_c$ superconductors. (a) Conducting layers with high light carrier density alternate along the *z* axis with excitation layers with specific pairing sources (blue cross), such as charge and spin fluctuations, resulting in dense and tiny Cooper pairs in the conduction layer. (b) A quasi-2D crystal's component layer, consisting of conduction and excitation strips. The former can contain carbon chains, other atom chains, or linked TM-ligand octahedra, while the latter has 1D spin chains, ladders, and so on. Topological insulators with conducting surface and edge states can serve as the conduction layer in (a) and strip in (b), respectively.

Let us consider two scenarios. The first model consists of an excitation layer with a very low carrier density or a narrow band gap and a conducting layer with a high carrier density at the matched Fermi level. Strong pairing in the conducting layer can be caused by charge fluctuations (Section 5.4.4) or exciton fluctuations (Section 5.4.5) in the excitation layer; superconductivity can also be observed in the excitation layer, but the $T_c$ may be extremely low due to low carrier density. To develop such a model, the original chemical potential or work function must be considered when fine-tuning the electronic state between the two layers. Even if perfect matching is difficult, their contact may cause a charge transfer, balancing the electron potential and, hopefully, producing the situation described above. Recent advances in data science will make this possible.

The second model incorporates a spin network with a specific geometry and strong magnetic fluctuations in the excitation layer, in addition to the conduction layer, which already has sufficient carrier density. It may be able to defeat cuprate superconductivity due to the independence of the pairing source and carrier density, as well as the lack of randomness effects. $AV_3Sb_5$ from Section 5.2 [279, 280] may be a relevant example. However, in $AV_3Sb_5$, CDW instability takes precedence over purely electronic instability, which appears to limit $T_c$. Coupling with low-energy excitations like phonons can eliminate high-energy electronic instability, which could have served as an efficient glue. Avoiding lattice instability is generally necessary for achieving high $T_c$.

The second testing bed, depicted in Fig. 49b, is a quasi-2D system consisting of layers with conduction and excitation strips aligned alternately, combining the benefits of 1D and 2D. The conduction strip may contain carbon chains, other atom chains, or TM-ligand octahedral chains, whereas the excitation strip may consist of 1D spin chains, ladders, or a variety of other structures that employ Cooper pairing tricks. If both are properly coupled, BEC superconductivity can be achieved in the conduction strip. Furthermore, the stacking of these layers allows for orientation flexibility; rotating the stacking by 90 degrees may yield interesting results. Conventional thermodynamic synthesis methods will not be sufficient to produce such materials, necessitating the use of novel "brute force" techniques.

The surface and edge states of topological insulators (TIs) can be used as conduction routes in the two testing beds depicted in Fig. 49. The discovery of metallic conduction states on their crystal surfaces and edges has increased interest in TIs [393]. When TI is used as the conduction layer in Fig. 49a, conduction carriers form at the interface with the excitation layer and can be converted to Cooper pairs. In Fig. 49b, superconductivity may be induced near the edge of the TI strip. It is unclear whether TI's surface/edge states can attain enough Cooper pair concentration to produce high $T_c$, but it is worthwhile to test these material designs. An unusual superconducting state was observed at the interface of a TI (HgTe) and a superconductor (Nb) [394].

As with Little's model, it is unclear whether space-separated excitations and carriers are intimately coupled in these two cases. Copper oxide superconductivity allows conducting oxygen holes (ZRSs) and strongly antiferromagnetically interacting copper spins that generate pairing interactions to coexist in a delicate balance in a single $CuO_2$ plane, resulting in exceptionally high $T_c$ values—a natural miracle! When the author considers the feasibility of his previously developed room-temperature superconductivity fantasy, he is left to admire nature's magnificent setting for cuprate superconductivity. Despite the extremely high copper oxide barrier, the author believes it is not impossible. Chemists believe that we'll never know until we try.

6.5. New material strategy: look for superconductors with high $T_c$ or exotic properties; otherwise, anything interesting

The author's limited ability and poor imagination make it difficult to respond to questions about unknown glue and attraction mechanisms that lead to even higher $T_c$, as well as whether there are ways to avoid the previously mentioned dilemma of obtaining both strong pairing interactions and a high Cooper pair density. The author apologizes to those who have come this far with patience and high expectations, but he wishes to irresponsibly delegate the answers to future generations of researchers. However, in terms of material exploration, we would avoid putting too much emphasis on high $T_c$ because it will make every effort tough. Instead, we should unwind and have fun while searching for new materials.

Solid state chemists should work on projects that are relevant to their interests in synthetic routes, material novelty, and crystal structure beauty (the author's motivation). As a result, even if the $T_c$ is low, they may discover intriguing superconductors with previously unknown mechanisms and properties (for example, the rattling superconductor in Fig. 9 was completely unexpected; it was discovered by chance by a student attempting to synthesize another compound). The discovery of a new superconductor, such as copper oxides, will greatly advance materials science, delighting solid state physicists while broadening the playground for solid state chemists. If they're lucky, they might even discover a room-temperature superconductor (optimism is the most important strategy in life and science).

6.6. New direction: single-atomic layers derived from van der Waals crystals

In this paper, we considered the search for superconducting materials while assuming bulk crystals. One of the most recent major trends in materials science is the formation of single atomic sheets by exfoliating layered compounds that are weakly stacked via van der Waals interactions [82, 395-397]. Monolayer superconductors are generally disadvantageous in terms of high transition temperatures and large current transport; however, there is great potential for using superconductivity (zero resistance) and Josephson junctions in nanoscale devices, and research will be encouraged.

Pure two dimensions lack long-range order. Instead, the Berezinskii–Kosterlitz–Thouless (BKT) transition takes place at $T_{BKT}$ [398, 399]. As previously mentioned in the section on cuprate superconductivity, at temperatures where superconducting correlations develop well within the 2D plane, 3D long-range order emerges in the bulk crystal with minimal interplane interactions (Fig. 35b). At a comparable temperature, a single sheet undergoes a BKT transition, resulting in a zero-resistance state similar to superconductivity. It is not surprising given that electron pairs have already formed and the superconducting correlation is well established within the plane. The essence of the BKT transition is thought to be a topological quasi-long-range order accompanied by the pairing of vortices (vortex currents in the case of superconductivity) and anti-vortices (vortices in the opposite direction) (this is too complicated for the author to explain in a clear manner). A single one-unit-cell thick Y123 film sandwiched between non-superconducting Pr123 layers demonstrated zero resistance at 30 K, indicating a BKT transition [400].

2D superconductivity has previously been investigated using thin films of bulk materials, but due to disorder, they become insulators with decreasing thickness and fail to achieve superconductivity [83]. However, single atomic sheets extracted from van der Waals crystals retain a high crystallinity, leading to the discovery of new physical phenomena [396, 397]; van der Waals crystals are prone to stacking faults, rather making it difficult to produce high-quality crystals in bulk. Graphene, a single carbon atomic sheet derived from graphite, shows unusually high electron mobility due to Dirac electrons with zero effective mass, making it a promising candidate for novel electronic devices [82, 395]. Interestingly, twisting and stacking two graphene sheets (twisted bilayer graphene: TBG) produces a moiré structure with a Mott insulator and superconductivity ($T_c$ = $T_{BKT}$ = 1.7 K), depending on the twist angle [401].

In a single atomic layer of $NbSe_2$, the CDW transition temperature (not for a phase transition) rises from 33.5 K in bulk to 145 K, while $T_c$ ($T_{BKT}$) decreases from 7.2 K to 3 K [402]. A half-unit-cell layer (only one C2-B4) peeled off from a Bi2212 crystal yields a bulk-comparable $T_c$ of 88 K [403]. The reason for a single FeSe atomic layer's significantly higher zero-resistance temperature of 23 K [404] or ~100 K [405] compared to the bulk $T_c$ of 8 K remains unknown. In addition, field-effect-doping superconductivity was achieved using an EDL transistor in $SrTiO_3$ ($T_c$ = 0.4 K) [213], $MoS_2$ ($T_c$ = 11 K) [406], and ZrNCl ($T_c$ = 19 K) [226]. As demonstrated in the examples above, 2D superconductivity has been discovered in a variety of systems, and further advancements are expected in the future.

6.7. Room-temperature superconductors and their applications

Although the term "room-temperature superconductivity" sounds appealing, $T_c$ of 300 K, a typical room temperature, is insufficient for practical room temperature applications because the critical current density and critical magnetic field of superconductivity just below $T_c$ are so small that zero resistance is difficult to maintain; thus, $T_c$ of at least 400 K is most likely required. However, it is still uncertain whether the superconductor will function properly at room temperature [407]. Magnetic vortices form above the lower critical field in type II superconductors used for applications. A magnetic vortex passing through a superconductor generates voltage even below $T_c$, breaking the zero-resistance state [9, 408]. To address this issue, materials must contain a carefully designed array of superconductively weak components, such as defects or impurities, that pin the magnetic vortex [407, 408]. At higher temperatures, however, thermal fluctuation activates magnetic vortex motion, suppressing the zero-resistance state, which poses a significant challenge in applications. Actual material properties are frequently determined by extrinsic modifications intended to achieve the desired applications, rather than bulk properties.

Low-temperature superconductors such as NbTi ($T_c$ = 9.8 K) and $Nb_3Sn$ ($T_c$ = 18.1 K) took decades to mature into viable materials for liquid helium temperature applications. Superconducting magnets used in medical MRIs and magnetic levitation trains must be extremely stable and reliable, and they can now be used in real-world applications that require liquid helium or refrigeration cooling. Y123 tapes and Bi-based wires are currently being developed to operate at liquid nitrogen temperatures [409]. $MgB_2$ ($T_c$ = 39 K) is also used to produce cryogen-free magnets [410]. Thus, compounds with the properties required for superconducting applications are already available. It will be critical to improve their dependability and integrate them into practical applications. Nevertheless, the discovery of new superconductors will almost certainly result in advances in materials science, which we are extremely excited about.

## 7. Final remarks

The author attempted to explain superconductivity to solid state chemists in an understandable manner using chemist-friendly images, but what impression does the reader get? He summarized the characteristics of the copper oxide superconductors with the highest $T_c$ and discussed how to comprehend the material dependence of $T_c$ using a simplified and slightly biased narrative. Other superconducting mechanisms and materials for implementing them were also addressed. Regardless of the specifics of Cooper pairing interactions, the essence of superconductivity can be captured in a single image: electron/hole pair formation and BEC. If the author's intended audience grasps this concept, he is pleased. Furthermore, if the allure of the structure–property relationship in superconductors inspires young researchers to develop new compounds, this paper will have accomplished its goal. Once the barrier to superconductivity research is removed, chemists' ears will gradually adjust to the difficult physics, as the author discovered.

The discovery of novel compounds has consistently led to significant jumps in materials science. When a solid state chemist finds an unexpected compound without regard for physics, and solid state physicists conduct extensive research on the compound, materials science advances significantly as a result of both wheels turning in tandem. Furthermore, engineering researchers join the growing field, and it matures into applied research that enables the "impossible" (in contrast to technology, someone once said that natural science is the study of making the "unknown" known, i.e., revealing the hidden mystery itself before solving). The greater the material's importance, the more serendipitously it was discovered. Innovative materials do not emerge through deliberate efforts; instead, they appear in unexpected places; no one expected jerry fish research to result in a 2008 Nobel Prize in Chemistry [411].

Future superconductors are unlikely to follow the extended line of concepts presented in this manuscript. Perhaps it is too late to say, but the readers should put everything aside, reset their minds, and approach the subject with a fresh perspective. Rather than focusing solely on superconductors, compounds can be discovered by "playing" with different private material interests. The author encourages young people to take on the challenge because older people, like the author, who have gained too much knowledge from their minor successes and hardened their brains, are incapable of undertaking such an endeavor. If someone reads this review, conducts their own research, and discovers superconductivity at room temperature in a copper oxide or another compound in ten, twenty, or one hundred years, the author will be overjoyed. He is optimistic that the day will come when we can hold a room-temperature superconductor in our hands, as illustrated in Fig. 50.

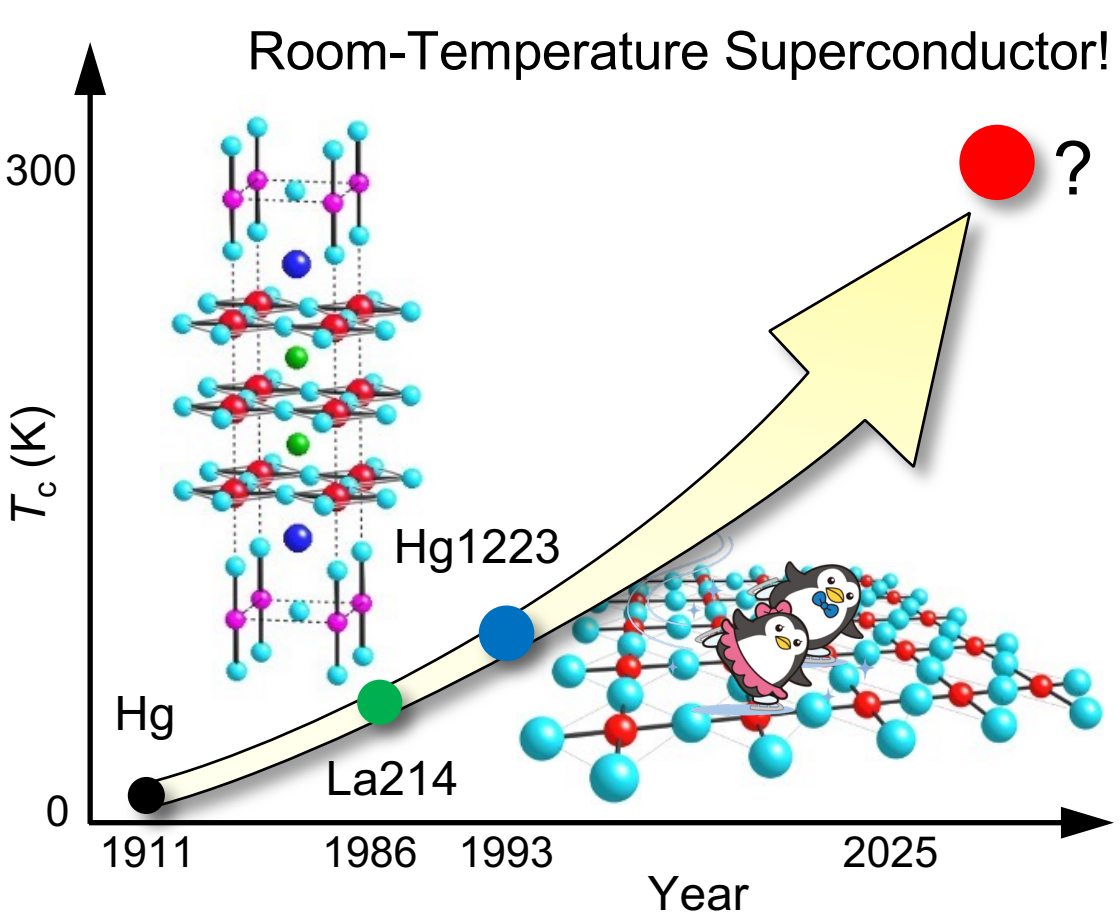

**Fig. 50.** Room-temperature superconductivity can be achieved in a clean $CuO_2$ plane doped with a higher number of holes with minimal loss of the AF spin background, or in an entirely unknown platform with an efficient pairing interaction and a sufficiently high carrier density to enter the BEC superconductivity regime.

**Acknowledgements**

The author wishes to thank H. Mukuda, Y. Okamoto, C. Michioka, Y. Koike, and K. Poeppelmeir for their thorough reviews and valuable feedback. He also thanks M. Ogata, K. Tsunetsugu, M. Takigawa, K. Ishida, T. Yagi, J. Yamaura, and H. Takagi for their helpful suggestions. Many NMR and ARPES researchers' experimental findings have proven particularly useful in discussions about copper oxide superconductivity. The review paper by H. Mukuda et al. [166] was especially helpful in constructing the present story and understanding $T_c$'s material dependence.

More than two decades ago, the author considered writing a review of the solid state chemistry and physics of copper oxide superconductivity. After the initial superconductivity fever passed, the research was divided into numerous complex physics subjects. The author then moved on to other material systems and research topics. He intends to retire in March 2026, but he hopes to leave something meaningful for future generations. He began writing this manuscript five years ago, intrigued by the benefits of having a solid state chemist write such a review. Because of the lengthy blank, recalling previous experimental results and arguments took longer than expected, and the task is still incomplete. Many important results and papers were undoubtedly overlooked due to the author's inability to do so, and he sincerely apologizes to all parties involved.

As the author organized the experimental results and prepared the manuscript, the plot took a turn that defied his initial expectations. As he contemplated how to organize his thoughts into a coherent article, his doubts grew, and more thinking was required to make everything go together. He eventually experienced a sense of clarity. During the conceptualization phase, the author realized that a solid state chemist-friendly introduction to superconductivity in general was necessary before covering high-temperature superconductivity, thus it was incorporated and expanded to its current form. Although some may question the utility of the resulting lengthy text, we hope it will be useful to solid state chemists interested in superconductors. This manuscript is freely available in both English and Japanese from the ISSP Note Collection on the ISSP website [412].

The author improved his English expression by using Quillbot, an internet-based English-to-English paraphrasing software [413]. This software is an excellent resource for non-native authors who want to improve their English writing because it not only checks grammar but also recommends precise, sophisticated, and readable expressions. He strongly recommends it to any non-native speaker who plans to write an English paper in the future.

While working on this manuscript in July 2024, we learned of the unfortunate passing of Professor Maurice Rice, a long-time leader in a variety of physics, including copper oxide superconductivity. We pray for his soul, as he created the Zhang–Rice singlet, which is responsible for cuprate superconductivity.

Finally, I'd like to thank everyone who encouraged me to write this manuscript, especially my wife Kiyomi and our two children. I spent a good deal of time polishing this paper at the sushi bar 'Hama' on the University of Tokyo's Kashiwa campus (the occasional casual phrasing is not due to drinking). I'm very grateful to Hama's owner and wife.

**Table 1.** Copper oxide superconductors.

| Compound | Ideal Composition | $n$ | Block Layer | $m$ | St. Type | $T_{co}$ (K) | $p_o$ | Comments | Reference |
|---|---|---|---|---|---|---|---|---|---|
| IL | $Sr_{1-x}Nd_xCuO_2$ | 1 | $Sr_{1-x}Nd_x$ | 1 | C1-B1 | 43 | $x = 0.14$ | e-doping | [112] |
| IL | $Sr_{1-x}La_xCuO_2$ | 1 | $Sr_{1-x}La_x$ | 1 | C1-B1 | 40 | $x = 0.10$ | e-doping; thin film | [237] |
| La(Sr)214 | $La_{2-x}Sr_xCuO_4$ | 1 | $La_{2-x}Sr_xO_2$ | 2 | C1-B2-NC | 36<br>39<br>36 | <br>0.15<br>0.15–0.24 | $x = 0.20$; T structure<br>$T_c$ dome<br>Synthesis at HP $O_2$ | [414]<br>[140]<br>[145] |
| La(Ba)214 | $La_{2-x}Ba_xCuO_4$ | 1 | $La_{2-x}Ba_xO_2$ | 2 | C1-B2-NC | 25 | 0.15 | | [415] |
| La214 | $La_2CuO_{4+\delta}$ | 1 | $La_2O_{2+\delta}$ | 2 | C1-B2-NC | 38 | | High-oxygen pressure synthesis | [119] |
| (Nd-Ce-Sr)214 | $(Nd, Ce, Sr)_2CuO_{4-\delta}$ | 1 | $(Nd, Ce, Sr)_2O_2$ | 2 | C1-B2-(NC–CF) | 28 | | T* structure; alternating stack of NC and CF | [416] |
| La2126 | $La_{2-x}Sr_xCaCu_2O_6$ | 2 | $La_{2-x}Sr_xO_2$ | 2 | C2-B2-NC | 60 | | | [417] |
| F214 | $Sr_2CuO_2F_{2+\delta}$ | 1 | $Sr_2F_{2+\delta}$ | 2 | C1-B2-NC | 46 | 0.2–0.3 | $p_o$ from the nominal composition | [197] |
| Cl214 | $Ca_{2-x}Na_xCuO_2Cl_2$ | 1 | $Ca_{2-x}Na_xCl_2$ | 2 | C1-B2-NC | 26 | | | [115, 116] |
| Ba0212 | $Ba_2CaCu_2O_4(O_{1-y}F_y)_2$ | 2 | $Ba_2(O_{1-y}F_y)_2$ | 2 | C2-B2-NC | 90<br>105 | <br>0.225 | HP synthesis<br>NMR | [117]<br>[418] |
| Ba0223 | $Ba_2Ca_2Cu_3O_6(O_{1-y}F_y)_2$ | 3 | $Ba_2(O_{1-y}F_y)_2$ | 2 | C3-B2-NC | 120 | | HP synthesis | [117] |
| Ba0234 | $Ba_2Ca_3Cu_4O_8(O_{1-y}F_y)_2$ | 4 | $Ba_2(O_{1-y}F_y)_2$ | 2 | C4-B2-NC | 105 | | HP synthesis | [117] |
| Ba0245 | $Ba_2Ca_4Cu_5O_{10}(O_{1-y}F_y)_2$ | 5 | $Ba_2(O_{1-y}F_y)_2$ | 2 | C5-B2-NC | 90 | | HP synthesis | [117] |
| Sr0212 | $Sr_2CaCu_2O_4(O_{1-y}F_y)_2$ | 2 | $Sr_2(O_{1-y}F_y)_2$ | 2 | C2-B2-NC | 99 | | HP synthesis | [118] |
| Sr0223 | $Sr_2Ca_2Cu_3O_6(O_{1-y}F_y)_2$ | 3 | $Sr_2(O_{1-y}F_y)_2$ | 2 | C3-B2-NC | 111 | | HP synthesis | [118] |
| Nd214 | $Nd_{2-x}Ce_xCuO_4$ | 1 | $Nd_{2-x}Ce_xO_2$ | 2 | C1-B2-CF | 24 | $x = 0.15$ | T' structure; e-doping | [234, 235] |
| Pr214 | $Pr_{2-x}Ce_xCuO_4$ | 1 | $Pr_{2-x}Ce_xO_2$ | 2 | C1-B2-CF | 22<br>20<br>24 | $x = 0.10$<br><br>0.14 | T' structure; e-doping<br><br>NMR | [234]<br>[239]<br>[243, 244] |
| Y123 | $YBa_2Cu_3O_{7-\delta}$ | 2 | $Ba_2CuO_{3-\delta}$ | 3 | C2-B3-PV | 93<br><br><br>30 | <br>0.22<br>0.25<br> | <br>NMR<br>Even $p$ for the two Cu sites<br>One-unit-cell thick film | [125]<br>[122]<br>[419]<br>[400] |
| Gd123(Ru) | $GdSr_2RuCu_2O_8$ | 2 | $Sr_2RuO_4$ | 3 | C2-B3-PV | 16 | | Ferromagnetic order in the $RuO_2$ sheets below 133 K | [325, 326] |
| Y124 | $YBa_2Cu_4O_8$ | 2 | $Ba_2Cu_2O_4$ | 4 | C2-B4-PV | 82.5 | | Block layer with double Cu-O chains | [420-422] |
| Y123.5 | $YBa_2Cu_{3.5}O_{8-\delta}$ | 2 | $Ba_2CuO_{3-\delta}/Ba_2Cu_2O_4$ | 3 / 4 | C2-(B3–B4) | 95 | | Alternating block layers of Y123 and Y124 | [423] |
| Hg1201 | $HgBa_2CuO_{4+\delta}$ | 1 | $Ba_2HgO_{2+\delta}$ | 3 | C1-B3-NC | 97.0<br>97<br>95<br>95<br>98 | 0.20<br>0.18<br>0.18<br>0.18<br>0.16 | $Hg_{0.97}Ba_2CuO_{4.059}(CO_3)_{0.0088}$<br>CT[a]<br>ND[b]<br>ND<br>*S*[c] | [129]<br>[175]<br>[424]<br>[194]<br>[190] |
| Hg1212 | $HgBa_2CaCu_2O_{6+\delta}$ | 2 | $Ba_2HgO_{2+\delta}$ | 3 | C2-B3-NC | 127<br>128 | 0.21<br>0.22 | CT<br>ND | [176]<br>[181] |
| Hg1223 | $HgBa_2Ca_2Cu_3O_{8+\delta}$ | 3 | $Ba_2HgO_{2+\delta}$ | 3 | C3-B3-NC | 135<br>133<br>133 | 0.19<br>0.27<br>0.252, 0.207 | CT<br>ND<br>$p$(OP), $p$(IP); NMR | [176]<br>[123]<br>[122] |
| Hg1234 | $HgBa_2Ca_3Cu_4O_{10+\delta}$ | 4 | $Ba_2HgO_{2+\delta}$ | 3 | C4-B3-NC | 127<br>123 | <br>0.222, 0.157 | <br>$p$(OP), $p$(IP); NMR | [425]<br>[122] |
| Hg1245 | $HgBa_2Ca_4Cu_5O_{12+\delta}$ | 5 | $Ba_2HgO_{2+\delta}$ | 3 | C5-B3-NC | 110 | 0.23 | $p$(OP); NMR | [166] |
| Hg1256 | $HgBa_2Ca_5Cu_6O_{14+\delta}$ | 6 | $Ba_2HgO_{2+\delta}$ | 3 | C6-B3-NC | 107 | | | [425] |
| Tl1201 | $TlBa_2CuO_{5-\delta}$ | 1 | $Ba_2TlO_{3-\delta}$ | 3 | C1-B3-NC | 45 | | $TlBa_{2-x}La_xCuO_{5-\delta}$ | [426] |
| Tl1212 | $TlBa_2CaCu_2O_{7-\delta}$ | 2 | $Ba_2TlO_{3-\delta}$ | 3 | C2-B3-NC | 65–85 | | | [427] |
| Tl1212 | $TlSr_2CaCu_2O_{7-\delta}$ | 2 | $Sr_2TlO_{3-\delta}$ | 3 | C2-B3-NC | 85<br>68 | | Lu-for-Ca substitution | [428]<br>[429] |
| Tl1223 | $TlBa_2Ca_2Cu_3O_{9-\delta}$ | 3 | $Ba_2TlO_{3-\delta}$ | 3 | C3-B3-NC | 133.5<br>132 | | <br>ND | [430]<br>[431] |
| Tl1234 | $TlBa_2Ca_3Cu_4O_{11-\delta}$ | 4 | $Ba_2TlO_{3-\delta}$ | 3 | C4-B3-NC | 122<br>127 | | | [432]<br>[430] |
| Cu1212 | $CuBa_2CaCu_2O_{6+\delta}$ | 2 | $Ba_2CuO_{2+\delta}$ | 3 | C2-B3-NC | 90 | | | |
| Cu1223 | $CuBa_2Ca_2Cu_3O_{8+\delta}$ | 3 | $Ba_2CuO_{2+\delta}$ | 3 | C3-B3-NC | 119 | 0.22 | Average $p$; NMR | [183] |
| Cu1234 | $CuBa_2Ca_3Cu_4O_{10+\delta}$ | 4 | $Ba_2CuO_{2+\delta}$ | 3 | C4-B3-NC | 105<br>117 | <br>0.313, 0.192 | <br>$p$(OP), $p$(IP); NMR | [433]<br>[122] |
| Cu1245 | $CuBa_2Ca_4Cu_5O_{12+\delta}$ | 5 | $Ba_2CuO_{2+\delta}$ | 3 | C5-B3-NC | 90 | | | [433] |
| Pb1212 | $PbSr_2YCu_2O_{7-\delta}$ | 2 | $Sr_2PbO_{3-\delta}$ | 3 | C2-B3-NC | 52 | | $(Pb, Cu)Sr_2(Y, Ca)Cu_2O_{7-\delta}$ | [127] |
| Sr0201-$CO_3$ | $Sr_2CuO_2CO_3$ | 1 | $Sr_2CO_3$ | 3 | C1-B3-NC | ~40 | | $(Ba_{1-x}Sr_x)_2CuO_2(CuO_\delta)_{0.1}(CO_3)_{0.9}$ ($x = 0.4–0.65$) | [130] |
| Bi2201 | $Bi_2Sr_2CuO_{6+\delta}$ | 1 | $Sr_2Bi_2O_{4+\delta}$ | 4 | C1-B4-NC | 7<br>15<br>25<br>25<br>32 | <br>0.13<br>0.12<br>0.12<br>0.15 | $Bi_{2+x}Sr_{2-x}CuO_{6+\delta}$<br>CT; $Bi_2Sr_2CuO_{6+\delta}$<br>CT; $BiPbSr_{2-x}La_xCuO_{6+\delta}$<br>CT; $Bi_2Sr_{2-x}La_xCuO_{6+\delta}$<br>$R_H$[d]; $Bi_2Sr_{2-x}La_xCuO_{6+\delta}$ ($x = 0.4$) | [434, 435]<br>[436]<br>[174]<br>[126]<br>[192] |
| Bi2212 | $Bi_2Sr_2CaCu_2O_{8+\delta}$ | 2 | $Sr_2Bi_2O_{4+\delta}$ | 4 | C2-B4-NC | 80<br>85<br>80 | <br>0.26<br>0.17 | <br>$R_H$; $Bi_2Sr_2Ca_{1-x}Lu_xCu_2O_y$<br>CT; $Bi_2Sr_{1.8}(Ca_{1-x}Y_x)_{1.2}Cu_2O_y$ | [437]<br>[438]<br>[126] |
| | | | | | | 85<br>80<br>91<br>88 | 0.22<br>0.25<br>0.18 | CT; $BiPbSr_2Ca_{1-x}Y_xCu_2O_y$<br>NMR<br>ARPES<br>Monolayer | [174]<br>[122]<br>[172]<br>[403] |
| Bi2223 | $Bi_2Sr_2Ca_2Cu_3O_{10+\delta}$ | 3 | $Sr_2Bi_2O_{4+\delta}$ | 4 | C3-B4-NC | 105<br>110 | <br>0.25 | <br>CT | [437]<br>[174] |
| Tl2201 | $Tl_2Ba_2CuO_{6+\delta}$ | 1 | $Ba_2Tl_2O_{4+\delta}$ | 4 | C1-B4-NC | 90<br>87<br>80 | 0.1/0.2<br>~0.25<br>0.28 | $\delta \sim 0$, 5% Cu-for-Tl sub.<br>$\Delta p = -0.25$<br>NMR; overdoped | [128]<br>[170]<br>[122] |
| Tl2212 | $Tl_2Ba_2CaCu_2O_{8+\delta}$ | 2 | $Ba_2Tl_2O_{4+\delta}$ | 4 | C2-B4-NC | 110 | | ND; $\delta = 0.3$ | [439] |
| Tl2223 | $Tl_2Ba_2Ca_2Cu_3O_{10+\delta}$ | 3 | $Ba_2Tl_2O_{4+\delta}$ | 4 | C3-B4-NC | 125 | | | [427, 440-442] |
| Tl2234 | $Tl_2Ba_2Ca_3Cu_4O_{12+\delta}$ | 4 | $Ba_2Tl_2O_{4+\delta}$ | 4 | C4-B4-NC | 116 | | $Tl_{2-x}Ba_2Ca_{3+x}Cu_4O_{12+\delta}$ | [443, 444] |
| Pb2213 | $Pb_2Sr_2YCu_3O_{8+\delta}$ | 2 | $Sr_2Pb_2CuO_{4+\delta}$ | 5 | C2-B5-NC | 68 | | $Pb_2Sr_2Y_{0.5}Ca_{0.5}Cu_3O_8$ with the SrO–PbO–$CuO_\delta$–PbO–SrO block layer | [113] |

[a]Chemical titration; [b]Hall coefficient; [c]Neutron diffraction; [d]Seebeck coefficient.

**Table 2.** Typical superconductors other than cuprates.

| Compound | $T_c$ (K) | Related order (fluctuation) or possible glue | $T_{LRO}$ or $\hbar\omega_0$ (K) | Comment | Reference |
|---|---|---|---|---|---|
| Elements | | | | | |
| Al | 1.2 | Phonon | 296[a] | Weak-coupling BCS type | [445] |
| Pb | 7.2 | Phonon | 56[a] | Strong-coupling BCS type | [445] |
| Nb | 9.2 | Phonon | 150[a] | Strong-coupling BCS type | [445] |
| Li | 0.0004 | Phonon | 344[b] | Weak-coupling BCS type | [70] |
| Bi | 0.00053 | Phonon | 120[b] | Semimetal with a low carrier density | [22] |
| a-Bi | 6.1 | Phonon | | Amorphous prepared by quenching | [68] |
| Ca | 29 | Phonon | 229[b] | $P$ = 125 GPa | [71] |
| $O_2$ | 0.6 | Phonon | | $P$ = 125 GPa | [72] |
| Nb–Ti | 9.8 | Phonon | | Alloy for commercial superconducting magnets | |
| | | | | | |
| Carbon-based | | | | | |
| $KC_8$ | 0.55 | Phonon | 235[b] | K-intercalated graphite | [446] |
| $K_3C_{60}$ | 19.5 | Phonon | | Fulleride; intramolecular $H_g$ phonons | [447] |
| $Cs_3C_{60}$ | 35 | Phonon | | Fulleride | [448] |
| C(B) | 4–7 | Phonon | 2250[b] | Boron-doped diamond; High-pressure synthesis or thin films | [74, 75] |
| TBG | 1.7 | | | Twisted bilayer graphene; adjacent to a Mott-like insulator | [401] |
| $YNi_2B_2C$ | 12 | Phonon | | $Ni^{3.5+}$ ($3d^{6.5}$) | [449] |
| $LuNi_2B_2C$ | 16.6 | Phonon | | $Ni^{3.5+}$ ($3d^{6.5}$) | [450] |
| | | | | | |
| Intermetallics | | | | | |
| $Nb_3Sn$ | 18.1 | Phonon | 124[a] | Strong-coupling BCS type; Martensite transformation at 43 K | [445] |
| $V_3Si$ | 17.1 | Phonon | 245[a] | Strong-coupling BCS type; Martensite transformation at 21 K | [445] |
| $Nb_3Ge$ | 23.2 | Phonon | 176[a] | Strong-coupling BCS type | [445] |
| $MgB_2$ | 39 | High-energy B phonons | 700[a] | Two superconducting gaps | [61, 387] |
| $ErRh_4B_4$ | 8.7 | | | $T_c$ = 11.8 K for $LuRh_4B_4$ | [323] |
| $LuPt_2In$ | 1.10 | CDW | 480 | QCP at 60% Pd-for-Pt substitution | [278] |
| $Au_{64}Ge_{22}Yb_{14}$ | 0.68 | | | Tsai-type crystalline approximants of quasicrystals | [69] |
| NaAlGe | 2.8 | | 100[c] | Zn-for-Al substitution | [365] |
| | | | | | |
| f-electron systems | | | | | |
| $CeCu_2Si_2$ | 0.7 | AFM | 0.8–2 | SDW stabilized in $CeCu_2(Si, Ge)_2$ | [295] |
| $CeCu_2Si_2$ | 2.5 | Valence fluctuations | | $P$ = 4 GPa | [284] |
| $CeCu_2Ge_2$ | 0.6, 1.5 | AFM, Valence fluctuations | 4 | $P$ = 8 GPa, $P$ = 16 GPa | [286, 287] |
| $CeIn_3$ | 0.19 | AFM | 10.2 | $P$ = 2.65 GPa | [296] |
| $CeRhIn_5$ | 2.1 | AFM | 3.8 | $P$ = 1.7 GPa | [297] |
| $CeCoIn_5$ | 2.3 | AFM | | | [50, 451] |
| $UPt_3$ | 0.54 | AFM | 5 | $P$ = 2.5 GPa | [452] |
| $UBe_{13}$ | 0.85 | | | | [453] |
| $UGe_2$ | 0.8 | FM | 52 | $P$ = 1.6 GPa | [291] |
| $UTe_2$ | 1.6 | F fluctuation | | | [65, 321, 327] |
| URhGe | 0.25 | FM | 9.5 | | [328, 330] |
| UCoGe | 0.8 | FM | 2.5 | | [329] |
| β-$YbAlB_4$ | 0.080 | Valence fluctuation? | | | [20] |
| $PrOs_4Sb_{12}$ | 1.85 | AF quandupole order | 1.3 | $B$ = 4–14 T | [371, 372] |
| $PuCoGa_5$ | 18.5 | AF spin fluctuations | | | [454] |
| | | | | | |
| Oxides | | | | | |
| TiO | 2.3 | Phonon | | NaCl structure; $Ti^{2+}$ ($3d^2$) | [455] |
| $BaTi_2Sb_2O$ | 1.2 | CDW/SDW | 50 | $Ti^{3+}$ ($3d^1$); a square lattice of Ti | [377] |
| α-$Ti_3O_5$ | 7.1 | Phonon | | Magnéli phase; $Ti^{3.3+}$ ($3d^{0.7}$); thin film | [456] |
| $Ti_4O_7$ | 3.5 | Phonon | | Magnéli phase; $Ti^{3.5+}$ ($3d^{0.5}$); thin film | [456] |
| $LiTi_2O_4$ | 13.7 | Phonon | 630[b] | $Li_{1+x}Ti_{2-x}O_4$; near $Ti^{3.5+}$ ($3d^{0.5}$) | [457] |
| $SrTiO_{3-\delta}$ | 0.25 | Phonon | | Perovskite structure; near $Ti^{4+}$ ($3d^0$)<br>$T_c$ = 0.4 K by EDL doping | [18]<br>[213] |
| β-$Na_{0.33}V_2O_5$ | 8 | CDW | 135 | $P$ = 8 GPa; $V^{4.66+}$ ($3d^{0.33}$) | [277] |
| $Na_{0.35}CoO_2 \cdot 1.3H_2O$ | 4 | | | 2D SC; $Co^{3.65+}$ ($3d^{5.35}$) | [378, 379] |
| $Nd_{0.8}Sr_{0.2}NiO_2$ | 12 | Mott insulator | 200 | Thin film; $Ni^{1.2+}$ ($3d^{8.8}$) | [382] |
| $La_2PrNi_2O_7$ | 75 | | | $P$ = 20 GPa; $Ni^{2.5+}$ ($3d^{7.5}$); C2-B2-NC type structure; Orthorhombic-to-tetragonal transition at 11 GPa | [383] [17] |
| $Ba(Pb_{1-x}Bi_x)O_3$ | 13 | Breathing phonon / Valence fluctuation | 195[b] | BPBO | [14] |
| $(Ba_{1-x}K_x)BiO_3$ | 30 | Breathing phonon / Valence fluctuation | 210[b] | BKBO; $x$ = 0.4 | [86, 349] |
| $Sr_2RuO_4$ | 1.0 | | | $Ru^{4+}$ ($4d^4$) | [114, 386] |
| $K_xWO_3$ | 1.7 | | 242[b] | | [458, 459] |
| (Sr, Ca)$_{14}$Cu$_{24}$O$_{41}$ | 12 | | | $P$ = 3 GPa; spin ladder | [338] |
| α-$Cd_2Re_2O_7$ | 0.97 | Electric troidal quadrupole order | 200 | Noncentrosymmetric SC; spin–orbit-coupled metal | [43] |
| β-$KOs_2O_6$ | 9.6 | Rattling phonons | 57[a] | Rattling-induced SC | [41] |
| 12CaO・7$Al_2O_3$ | 0.2 | Phonon | | | [21] |
| | | | | | |
| Iron-based compounds | | | | | |
| LaOFeP | 5 | | | $Fe^{2+}$ ($3d^6$) | [304] |
| LaFeAs($O_{1-x}F_x$) | 26 | AFM | 150 | $x$ = 0.1–0.2; $T_c$ = 43 K at $P$ = 4 GPa | [306, 307] |
| SmFeAs($O_{1-x}F_x$) | 55 | AFM | 130 | $x$ ~ 0.2 | [317, 318] |
| LaFeAs($O_{1-x}H_x$) | 36 | AFM | 160 | Double $T_c$ domes at $x$ = 0.1 and 0.4 | [285] |
| ($Ba_{1-x}K_x$)$Fe_2As_2$ | 38 | AFM | 135 | $x$ = 0.5; hole doping | [308] |
| Ba($Fe_{1-x}Co_x$)$_2As_2$ | 22 | AFM | 135 | $x$ = 0.2; electron doping | [309] |
| α-FeSe | 8 | Orbital fluctusation? | | $Fe^{2+}$ ($3d^6$)<br>$T_c$ = 27 K at $P$ = 1.5 GPa<br>$T_{BKT}$ = 23 or 100 K in a one-unit-cell thick film on a $SrTiO_3$ substrate | [312]<br>[313]<br>[404, 405] |
| ($Ca_{1-x}La_x$)$FeAs_2$ | 34 | | | $x$ = 0.10<br>$T_c$ = 47 K in ($Ca_{1-x}La_x$)Fe($As_{1-y}Sb_y$)$_2$ | [460]<br>[461] |
| | | | | | |
| Organics | | | | | |
| (SN)$_x$ | 0.26 | | | Quasi-1D SC | [19] |
| $(TMTSF)_2PF_6$ | 1.2 | AFM | 12 | $P$ = 0.9 GPa | [298, 299] |
| (BEDT-TTF)$_2$Cu(SCN)$_2$ | 10.4 | | | | [462] |
| κ-(BEDT-TTF)$_2$Cu$_2$(CN)$_3$ | 4 | Spin liquid | 250[d] | $P$ = 0.4 GPa | [332] |
| α-(BEDT-TTF)$_2$I$_3$ | 7 | Charge order | 135 | Uniaxial pressure of 0.2 GPa | [346] |
| | | | | | |
| Chalcogenides | | | | | |
| $PbMo_6S_8$ | 15 | Phonon | 140[a] | Strong-coupling BCS type | [49, 322] |
| 1T-$TaS_2$ | 5 | CDW | 350 | $P$ = 5 GPa | [274] |
| $LaOBiS_2$ | 10 | | | HP synthesis | [463] |
| $Bi_4O_4S_3$ | 4.5 | | | | [464] |
| $BaFe_2S_3$ | 14 | | | $P$ = 11 GPa; $Fe^{2+}$ ($3d^6$) | [465] |
| $MoS_2$ | 11 | | | EDL doping | [406] |
| $NbSe_2$ | 7.2 | CDW | 33.5<br>145 | Bulk<br>$T_c$ = 3 K in a monolayer | [466]<br>[402] |
| $NbSe_3$ | 2.5 | CDW | 59 | $P$ = 0.7 GPa; quasi-1D | [76] |
| $Cu_xTiSe_2$ | 4.2 | CDW | 220 | $x$ = 0.08; $Ti^{3.92+}$ ($3d^{0.08}$) | [275] |
| $Ta_2NiSe_5$ | 1.2 | Excitonic fluctuation? | 328 | $P$ = 8 GPa | [364] |
| $Cu_xBi_2Se_3$ | 3.8 | | | $0.12 \leq x \leq 0.15$ | [467] |
| ($Pb_{1-x}Tl_x$)Te | 1.5 | Valence fluctuation? | | Tl-for-Pb substitution; $x$ = 0.015 | [350] |
| $IrTe_2$ | 3.1 | CDW | 250 | QCP at 3.5% Pt-for-Ir substitution | [276] |
| $WTe_2$ | 7 | | | $P$ = 17 GPa<br>$T_{BKT}$ ~ 0.5 K in a monolayer | [468, 469]<br>[470] |
| $MoTe_2$ | 8.2 | | | $T_c$ = 0.10 K at AP and 8.2 K at $P$ = 11.7 GPa | [471] |
| $Sc_6FeTe_2$ | 4.7 | | | | [472] |
| | | | | | |
| Pnictides | | | | | |
| MnP | 1 | FM/Helical AFM | 290 | $P$ = 8 GPa; $Mn^{3+}$ ($3d^4$) | [302, 303] |
| $SrNi_2P_2$ | 1.4 | | | | [473] |
| CrAs | 2.2 | Helical AFM | 265 | $P$ = 0.7 GPa; $Cr^{3+}$ ($3d^3$) | [300, 301] |
| $K_2Cr_3As_3$ | 6.1 | | | $Cr^{2.33+}$ ($3d^{3.66}$); Quasi-1D; Strong electron correlations | [474, 475] |
| $CsV_3Sb_5$ | 2.5 | CDW | 94 | $V^{4.66+}$ ($3d^{0.33}$); double $T_c$ domes at $P$ = 0.6 and 2 GPa; $T_c$ ~ 0.9 K for the K and Rb analogues | [279, 280, 282] |
| $KMn_6Bi_5$ | 9 | AFM | 75 | $Mn^{2.33+}$ ($3d^{4.66}$); $P$ = 14 GPa | [476] |
| | | | | | |
| Mixed anions | | | | | |
| $Li_xZrNCl$ | 19.0 | Charge fluctuation? | | $T_c$ = 11.5 K ($x$ = 0.3) and 15.2 K（$x$ = 0.06） with Li intercalation; 19.0 K ($p$ = 0.011) by EDL doping | [226, 353, 355] |
| $Li_xHfNCl$ | 25.5 | Charge fluctuation? | | | [354] |
| $La_2IOs_2$ | 12 | | | 5d electrons of La and anionic Os; $La_2IRu_2$ with $T_c$ = 4.8 K | [376] |
| $Na_2CoSe_2O$ | 5.4 | | | | [381] |
| | | | | | |
| Ultrahigh pressure | | | | | |
| $H_3S$ | ~200 | Hydrgen phonons | | $P$ = 150 GPa | [388] |
| $LaH_{10}$ | ~240 | Hydrgen phonons | | $P$ = 150 GPa | [389] |

[a]logarhithmically averaged phonon frequency; [b]Debye temperature; [c]Pseudogap energy; [d]Antiferromagnetic interaction.